\RequirePackage{fix-cm}
\documentclass[natbib]{svjour3}
\smartqed  % flush right qed marks, e.g. at end of proof
\pdfoutput=1
\usepackage{graphicx}
\usepackage{amsmath}
\usepackage{natbib}
\def\BB{\hbox{\bf B}}
\def\fd{\hbox{\bf f}_{\rm d}}
\def\gamam{\gamma_{\rm m}}

\def\rf{r_{\rm f}}
\def\rhoc{\rho_{\rm c}}
\def\rhoi{\rho_{\rm i}}
\def\rhon{\rho_{\rm n}}
\def\tad{\tau_{\rm AD}}
\def\tdad{\tau_{\rm d,AD}}
\def\tdyn{\tau_{\rm dyn}}
\def\tturb{\tau_{\rm turb}}
\def\ui{\hbox{\bf u}_{\rm i}}
\def\un{\hbox{\bf u}_{\rm n}}
\def\va{v_{\rm A}}
\def\vd{\hbox{\bf v}_{\rm d}}

\def\ltsima{$\; \buildrel < \over \sim \;$}    % Use in text mode
\def\lesssim{\lower.5ex\hbox{\ltsima}}           % Use in math mode
\def\gtsima{$\; \buildrel > \over \sim \;$}    % Use in text mode
\def\gtrsim{\lower.5ex\hbox{\gtsima}}           % Use in math mode

\newcommand{\mss}{\scriptscriptstyle\mathsc}
\DeclareMathAlphabet{\mathsc}{OT1}{cmbr}{n}{sc}

\newcommand{\boldA}{{\bf A}}
\newcommand{\boldB}{{\bf B}}
\newcommand{\Em}{E_{\rm m}}
\newcommand{\kpcm}{{\rm ~kg~m}^{-3}}
\newcommand{\muo}{\mu_0}
\newcommand{\mul}{\mu_\ell}
\newcommand{\Msun}{M_{\odot}}
\newcommand{\nhat}{\hat n}
\newcommand{\pcc}{{\rm ~cm}^{-3}}
\newcommand{\pcm}{{\rm ~m}^{-3}}
\newcommand{\tff}{\tau_{\rm ff}}

\newcommand{\beq}{\begin{equation}}
\newcommand{\eeq}{\end{equation}}

 \newcommand{\vect}[1]{\vec{#1}}

\newcommand{\td}{\tau_{\mathrm{d}}}

\newcommand{\mutilde}{\tilde{\mu}}
\DeclareMathAlphabet{\mathsc}{OT1}{cmbr}{n}{sc}
\usepackage{graphicx}

 \journalname{Space Science Reviews}

\newcommand{\aj}{AJ}
\newcommand{\aap}{{Astron. Astrophys.}}
\newcommand{\apj}{{Astrophys. J.}}
\newcommand{\apjl}{{Astrophys. J. Lett.}}
\newcommand{\mnras}{{Month. Not. R. Astron. Soc.}}
\newcommand{\grl}{{Geophys. Res. Lett.}}
\newcommand{\solphys}{{Solar Phys.}}
\newcommand{\jgr}{{Journal of Geophys. Res.}}
\newcommand{\nat}{{Nature}}
\newcommand{\ssr}{{Space Sci. Rev.}}
\newcommand{\zap}{Z. f. Astrophysik}
\newcommand{\apss}{Astrophys. Space Sci.}
\newcommand{\sovast}{Sov. Astron.}
\newcommand{\araa}{Ann. Rev. Astron. Astrophys.}
\newcommand{\apjs}{ApJS}
\newcommand{\planss}{PLSS}
\newcommand{\icarus}{Icarus}
\newcommand{\memsai}{Mem. Soc. As. It.}
\begin{document}

\title{Partially Ionized Plasmas in Astrophysics}


\author{ Jos\'e Luis Ballester  \and Igor Alexeev  \and  Manuel Collados \and Turlough Downes \and Robert F. Pfaff \and Holly Gilbert \and Maxim Khodachenko \and Elena Khomenko \and Ildar F. Shaikhislamov \and Roberto Soler \and Enrique V\'azquez-Semadeni \and Teimuraz Zaqarashvili}

 \institute{         J. L. Ballester \at
              Departament de F\'\i sica $\&$ Institut d'Aplicacions Computacionals de Codi Comunitari (IAC$^3$) \\
              Universitat de les Illes Balears \\
              E-07122 Palma de Mallorca (Spain) \\
              Tel.: +34 971 173228 \\
              Fax: +34 971 173426 \\
              \email{joseluis.ballester@uib.es}
               \and
           I. I. Alexeev \at
              Lomonosov Moscow State University Skobeltsyn Institute of Nuclear Physics (MSU SINP) leninskie Gory, 119992, Moscow, Russia  \\
              Tel.: +7-495-9391036 \\
              Fax: +7-495-9393553  \\
              \email{iialexeev@mail.ru}
               \and
           M. Collados \at
              Instituto de Astrof\'\i sica de Canarias\\
              C/ V\'\i a L\'actea s/n \\
              La Laguna, 38205  Tenerife, (Spain) \\
              Tel.: +34 922  605317 \\
              Fax: +34 922 605210 \\
              \email{mcv@iac.es}        \and
           T. Downes \at
              School of Mathematical Sciences, Dublin City University, \\
              Glasnevin, Dublin 9. Ireland  \\
              Tel.: +353 1 700 5270 \\
              Fax: +353 1 700 5786 \\
              \email{turlough.downes@dcu.ie}    \and
            R. F. Pfaff \at 
              NASA/Goddard Space Flight Center \\
              Mail Code 674.0
             Greenbelt, MD 20771  USA \\
             Tel.: +1 301 286 6328\\
             Fax. +1 301 286 1648\\
             \email: Robert.F.Pfaff@nasa.gov  \and
             \newpage
            H. Gilbert \at
              Solar Physics Laboratory
              Heliophysics Science Division \\
              Goddard Space Flight Center \\
              Mail Code 671
              Greenbelt, MD 20771 (USA) \\
              Tel.: +1 301 286 3042 \\
              Fax: +1 301 286 5348 \\
              \email{holly.r.gilbert@nasa.gov}  \and
           M. Khodachenko \at
              Space Research Institute, Schmiedlstrasse 6, A-8042 \\
              Austrian Academy of Sciences, Graz (Austria) \\
              Tel: +43 +43-316-4120661 \\
              Fax: +43 +43-316-4120690 \\
              \email{maxim.khodachenko@oeaw.ac.at} \and
            E. Khomenko \at
              Instituto de Astrof\'\i sica de Canarias\\
              C/ V\'\i a L\'actea s/n \\
              La Laguna, 38205  Tenerife, (Spain) \\
              Tel.: +34 922  605319 \\
              Fax: +34 922 605210 \\
              \email{khomenko@iac.es}  \and
            I. F. Shaikhislamov \at
              Institute of Laser Physics SB RAS\\
              Novosibirsk (Russia)\\
              \email{ildars@ngs.ru} \and
           R. Soler \at
              Departament de F\'\i sica $\&$ Institut d'Aplicacions Computacionals de Codi Comunitari (IAC$^3$) \\
              Universitat de les Illes Balears\\
              E-07122 Palma de Mallorca (Spain) \\
              Tel.: +34 971 172391 \\
              Fax: +34 971 173426 \\
              \email{roberto.soler@uib.es}  \and
          E. V\'azquez-Semadeni \at
              Instituto de Radioastronomia y Astrof\'\i sica (IRyA), UNAM.
              Campus Morelia. Apdp Postal 3-72 (Xangari) Morelia Mich. 58089. M\'exico \\
              Tel.: +55  56232753 \\
              Fax: + 55 56232726 \\
              \email{e.vazquez@crya.unam.mx}  \and
          T. V. Zaqarashvili \at
              IGAM, Institute of Physics, University of Graz,\\
              Universit\"atsplatz 5, 8010 Graz (Austria) \\
              Tel.: :+43 316 4120 672 \\
              Fax: + 43 316 4120 690 \\
               Abastumani Astrophysical Observatory at Ilia State University, 3/5 Cholokashvili avenue, 0162 Tbilisi (Georgia) \\
              Space Research Institute, Austrian Academy of Sciences, Schmiedlstrasse 6, A-8042 Graz (Austria)\\
              \email{teimuraz.zaqarashvili@oeaw.ac.at}
              }

\date{Received: date / Accepted: date}

\maketitle

\begin{abstract}
Partially ionized plasmas are found across the Universe in many different astrophysical environments. They constitute an essential ingredient of the solar atmosphere, molecular clouds, planetary ionospheres and protoplanetary disks, among other environments, and display a richness of physical effects which are not present in fully ionized plasmas. This review provides an overview of the physics of partially ionized plasmas, including recent advances in different astrophysical areas in which partial ionization plays a fundamental role.  We outline outstanding observational and theoretical questions and discuss possible directions for future progress.

\end{abstract}

\newpage
%%%%%%%%%%%%%%%%%%%%%%%%%%%
\section{Introduction}
\label{intro_0}
Plasma pervades the Universe at all scales, and the term plasma universe was coined by Hannes Alfv\'en to point out the important role played by plasmas across the universe \citep{alfven86}. In general, the study of plasmas beyond Earth's atmosphere is called Plasma Astrophysics, and includes many different astrophysical environments such as the Sun, the heliosphere, magnetospheres of the Earth and the planets, the interstellar medium, molecular clouds, accretion disks, exoplanet atmospheres, stars and astrospheres, cometary tails, exoplanetary ionospheres, etc. In these environments, the ionization level varies from almost no ionization in cold regions to fully ionized in hot regions which, consequently, leads to a wide range of parameters being relevant to astrophysical plasmas. Furthermore, in some cases the plasma is influenced by, or coupled to, embedded dust, giving rise to dusty plasmas. While the use of non-ideal Magnetohydrodynamics (MHD for short) is relatively infrequent in solar physics, in recent years the study of partially ionized plasmas has become a hot topic because solar structures such as spicules, prominences, as well as layers of the solar atmosphere (photosphere and chromosphere), are made of partially ionized plasmas (PIP for short). On the other hand, considerable developments have taken place in the study of partially ionized plasmas applied to the physics of the interstellar medium, molecular clouds, the formation of protostellar discs, planetary magnetospheres/ionospheres, exoplanets atmospheres, etc. For instance, molecular clouds are mainly made up of neutral material which does not interact with magnetic fields. However, neutrals are not the only constituent of molecular clouds since there are also several types of charged species which do interact with magnetic fields. Furthermore, the charged fraction also interacts with the neutral material through collisions. These multiple interactions produce many different physical effects which may have a strong influence on star formation and molecular cloud turbulence. A further example can be found in the formation of dense cores in molecular clouds induced by MHD waves. Because of the low ionization fraction, neutrals and charged particles are weakly coupled and ambipolar diffusion plays an important role in the formation process. Even in the primeval universe, during the recombination era, when the plasma, from which all the matter of the universe was formed, evolved from fully ionized to neutral, it went through a phase of partial ionization. Partially ionized plasmas introduce physical effects which are not considered in fully ionized plasmas, for instance, Cowling's resistivity, isotropic thermal conduction by neutrals, heating due to ion/neutral friction, heat transfer due to collisions, charge exchange, ionization energy, etc., which are crucial to fully understand the behaviour of astrophysical plasmas in different environments. Therefore, in this comprehensive review we focus on the description of PIP in different astrophysical areas in which this type of plasmas play a key role. The scheme of the review is as follows: In the second, third and fourth sections, general equations for a multifluid magnetized plasma are introduced as well as MHD waves and numerical techniques suitable for numerical simulations including PIP; next, the remaining sections are devoted to describe the physics of PIP in astrophysical environments such as the solar atmosphere, planetary ionospheres, molecular clouds and exoplanets atmospheres and magnetospheres.

%%%%%%%%%%%%%%%%%%%%%%%%%%%%%%%%%%%%%%%%%%%%%%
\section{Formulation of the multifluid plasma description}
\label{Bsec}
%%%%%%%%%%%%%%%%%%%%%%%%%%%%%%%%%%%%%%%%%%%%%%%%%%%%%

%%%%%%%%%%%%%%%%%%%%%%%%%%%%%%%%%%%%%%%%%%%%%%%%%%%%%%
\subsection{Conservation equations for individual micro-states} \label{Bsec:eqs}
%%%%%%%%%%%%%%%%%%%%%%%%%%%%%%%%%%%%%%%%%%%%%%%%%%%%%%

It is assumed that a PIP is composed of multiple kind of particles such as electrons, ions that can have different ionization states $\mathsc{I} \ge 1$ and excited to states $\mathsc{E} \ge 0$, neutral particles with $\mathsc{I}=0$ and $\mathsc{E} \ge 0$, as well as dust grains, positively or negatively charged with $\mathsc{E} = 0$. The concept of a fluid can be applied separately to each of these components in all the environments of interest. Therefore the behavior of such a plasma can be described by a set of equations of mass, momentum and energy conservation for each of the components. Separate sets of equations can be written for particles in a given micro-state $\{ a \mathsc{IE} \}$ corresponding to a given chemical element or dust grain $a$ (the notation of a micro-state $\{ a \mathsc{IE} \}$ reduces to just $e$ for electrons).  The conservation equations are derived as usual from the moments of Boltzmann equation and have the following form:
\begin{equation}
\frac{\partial \rho_{a \mss{IE}}}{\partial t} + \vec{\nabla} (\rho_{a \mss{IE}}\vec{u}_{a \mss{I}}) = S_{a \mss{IE}}
\end{equation}
\begin{eqnarray}
\frac{\partial (\rho_{a \mss{IE}}\vec{u_{a \mss{I}}})}{\partial t} + \vec{\nabla}(\rho_{a \mss{IE}}\vec{u_{a \mss{I}}} \otimes \vec{u_{a \mss{I}}} + {\bf\hat{p}}_{a \mss{IE}}) &= & \rho_{a \mss{IE}}r_{a \mss{I}}(\vec{E} + \vec{u}_{a \mss{I}}\times\vec{B}) + \rho_{a \mss{IE}}\vec{g} +\vec{R}_{a \mss{IE}}
\end{eqnarray}
\begin{eqnarray}
&&\frac{\partial}{\partial t} \left(  e_{a \mss{IE}} +\frac{1}{2}\rho_{a\mss{IE}}u_{a\mss{I}}^2 \right) +  \vec{\nabla}\left( \vec{u}_{a \mss{I}} (e_{a \mss{IE}} + \frac{1}{2}\rho_{a\mss{IE}}u_{a\mss{I}}^2) + {\bf\hat{p}}_{a \mss{IE}}\vec{u}_{a \mss{I}} + \vec{q}_{a \mss{IE}} \right)  =  \nonumber \\
&&= M_{a \mss{IE}} + \frac{E_{a\mss{IE}}}{m_a}S_{a\mss{IE}} + \rho_{a\mss{IE}}\vec{u}_{a\mss{I}}\vec{g}+ \rho_{a\mss{IE}}r_{a\mss{I}}\vec{u}_{a\mss{I}}\vec{E}
\end{eqnarray}

\noindent In these equations,  ${\bf\hat{p}}_{a \mss{IE}} =\rho_{a \mss{IE}}\langle \vec{\tilde{c}}_{a \mss{I}} \otimes \vec{\tilde{c}}_{a \mss{I}}\rangle$ is the pressure tensor defined through random velocities, $\vec{\tilde{c}}_{a \mss{I}}$, taken with respect to the mean velocity of each individual component (note that this makes the system of reference for velocities different for all components). It is also assumed that velocities corresponding to different excitation states are the same for a given ionization state, but energies and number densities are different. The heat flow vector is given by $\vec{q}_{a \mss{IE}}= \frac{1}{2}\rho_{a \mss{IE}}\langle \tilde{c}_{a \mss{I}}^2\vec{\tilde{c}}_{a \mss{I}} \rangle$. The variable $e_{a \mss{IE}}$ is the internal energy that makes up of thermal energy by random motion and potential energy of ionization/excitation states
\begin{equation}
e_{a \mss{IE}}=\rho_{a \mss{IE}} \langle \tilde{c}_{a \mss{I}}^2 \rangle/2+ n_{a \mss{IE}} E_{a \mss{IE}}=\frac{3}{2}p_{a \mss{IE}} + n_{a \mss{IE}} E_{a \mss{IE}}.
\end{equation}
\noindent where  $p_{\alpha \mss{IE}}$ is the scalar pressure, $p_{\alpha \mss{IE}}=\frac{1}{3} \rho_{\alpha \mss{IE}}\langle c_{\alpha \mss{I}}^2\rangle$, and
\begin{equation} \label{Beq:q-m}
r_{a \mss{I}}=q_{a \mss{I}}/m_{a \mss{I}}
\end{equation}
\noindent is charge over mass ratio. The rest of the notation is standard.

The above equations are written in conservation form. The momentum conservation equation can be also rewritten as
\begin{equation}
\rho_{a \mss{IE}}\frac{D\vec{u}_{a \mss{I}}}{Dt}= \rho_{a \mss{IE}}r_{a \mss{I}}(\vec{E} + \vec{u}_{a \mss{I}}\times\vec{B}) + \rho_{a \mss{IE}}\vec{g} - \vec{\nabla}  {\bf\hat{p}}_{a \mss{IE}}+\vec{R}_{a \mss{IE}} - \vec{u}_{a \mss{I}}S_{a \mss{IE}},
\end{equation}
\noindent leading to appearance of the $- \vec{u}_{a \mss{I}}S_{a \mss{IE}}$ term on the right hand side. Similarly, the energy conservation equation can be written for the internal energy alone leading to
\begin{equation}
\frac{\partial e_{a \mss{IE}}}{\partial t} +  \vec{\nabla}(\vec{u}_{a \mss{I}} e_{a \mss{IE}}+ \vec{q}_{a \mss{IE}})+ {\bf\hat{p}}_{a \mss{IE}}\vec{\nabla}\vec{u}_{a \mss{I}}  = Q_{a \mss{IE}}
\end{equation}

Generally speaking, conservation equations as above can be written as well for photons as another type of fluid particles. However, since photons move at the speed of light and are mass-less, the conservation equations for them acquire the particular form of the radiative transfer equations \citep[see, e.g.,][]{Mihalas}.

The source terms on the right-hand side of these equations result from collisions of particles in micro-state $\{ a \mathsc{IE} \}$ with particles in other micro-states (generally including photons, or the radiation field), and are the mass collision term, $S_{a \mss{IE}}$, the momentum collision term, $\vec{R}_{a \mss{IE}}$, and the internal energy collision term, $Q_{a \mss{IE}}$. The source terms either lead to appearance/disappearance of new particles (in the case of the mass conservation) and bring/remove momentum and energy to/from micro-state $\{ a \mathsc{IE} \}$.  Expressions for these terms are  obtained through collisional integrals of the distribution function $f_{a {\mss {IE}}}$ of particles in a given micro-state and depend on the particular physical conditions of the medium:

\begin{equation}
S_{a {\mss {IE}}}=m_{a {\mss I}}\int_{V}{\left(\frac{\partial f_{a {\mss {IE}}}}{\partial t}\right)_{\rm coll} d^3 v} = \left( \frac{\partial \rho_{a {\mss {IE}}}}{\partial t} \right)_{\rm coll},
\end{equation}
\begin{equation}
\vec{R}_{a {\mss {IE}}}=m_{a {\mss I}}\int_V{\vec{v}\left(\frac{\partial f_{a {\mss {IE}}}}{\partial t}\right)_{\rm coll}d^3 v} = \left( \frac{\partial}{\partial t}
[\rho_{a {\mss {IE}}}\vec{u}_{a {\mss I}}] \right)_{\rm coll},
\end{equation}
\begin{equation}
\vec{R}_{a {\mss {IE}}} - \vec{u}_{a \mss{I}}S_{a \mss{IE}} =
\left( \frac{\partial}{\partial t} [\rho_{a {\mss {IE}}}\vec{u}_{a {\mss I}}] \right)_{\rm coll} -
 \vec{u}_{a \mss{I}} \left( \frac{\partial \rho_{a {\mss {IE}}}}{\partial t} \right)_{\rm coll}  =
\rho_{a {\mss {IE}}} \left( \frac{\partial \vec{u}_{a {\mss I}}}{\partial t} \right)_{\rm coll},
\end{equation}
\begin{equation}
M_{a {\mss {IE}}}=\frac{1}{2} m_{a {\mss I}} \int_V{v^2\left(\frac{\partial f_{a {\mss {IE}}}}{\partial t}\right)_{\rm coll} d^3 v} =
\left( \frac{\partial }{\partial t}
[\frac{1}{2}\rho_{a {\mss {IE}}} u_{a {\mss I}}^2] \right)_{\rm coll}  +
\left( \frac{\partial }{\partial t} [\frac{3}{2}p_{a {\mss {IE}}}] \right)_{\rm coll},
\end{equation}
and
\begin{eqnarray}
Q_{a {\mss {IE}}} & = &
M_{a {\mss {IE}}} - \vec{u}_{a {\mss I}}\vec{R}_{a {\mss {IE}}} +  \left(\frac{1}{2} u_{a {\mss I}}^2 + \frac{E_{a {\mss {IE}}}}{m_{a {\mss I}}}\right) S_{a {\mss {IE}}} = \left( \frac{\partial }{\partial t}
[\frac{1}{2}\rho_{a {\mss {IE}}} u_{a {\mss I}}^2] \right)_{\rm coll} + \nonumber \\
&+&
\left( \frac{\partial }{\partial t} [\frac{3}{2}p_{a {\mss {IE}}}] \right)_{\rm coll} - \vec{u}_{a {\mss I}} \left( \frac{\partial}{\partial t}
[\rho_{a {\mss {IE}}}\vec{u}_{a {\mss I}}] \right)_{\rm coll} +
\left(\frac{1}{2} u_{a {\mss I}}^2 + \frac{E_{a {\mss {IE}}}}{m_{a {\mss I}}}\right)
\left( \frac{\partial \rho_{a {\mss {IE}}}}{\partial t} \right)_{\rm coll} =
\nonumber \\
&=&
\frac{1}{2} u_{a {\mss I}}^2 \left( \frac{\partial \rho_{a {\mss {IE}}}}{\partial t} \right)_{\rm coll} +
\rho_{a {\mss {IE}}}\vec{u}_{a {\mss I}}
\left( \frac{\partial \vec{u}_{a {\mss I}}}{\partial t} \right)_{\rm coll}+
\left( \frac{\partial }{\partial t} [\frac{3}{2}p_{a {\mss {IE}}}] \right)_{\rm coll} -
\nonumber \\
&-&
u_{a {\mss I}}^2 \left( \frac{\partial \rho_{a {\mss {IE}}}}{\partial t} \right)_{\rm coll} -
\rho_{a {\mss {IE}}}\vec{u}_{a {\mss I}}
\left( \frac{\partial \vec{u}_{a {\mss I}}}{\partial t} \right)_{\rm coll} +
\frac{1}{2} u_{a {\mss I}}^2
\left( \frac{\partial \rho_{a {\mss {IE}}}}{\partial t} \right)_{\rm coll} +
\nonumber \\
&+& \frac{E_{a {\mss {IE}}}}{m_{a {\mss I}}}
\left( \frac{\partial \rho_{a {\mss {IE}}}}{\partial t} \right)_{\rm coll} =
\left( \frac{\partial }{\partial t} [\frac{3}{2}p_{a {\mss {IE}}}] \right)_{\rm coll} + E_{a {\mss {IE}}}
\left( \frac{\partial n_{a {\mss {IE}}}}{\partial t} \right)_{\rm coll} =
\nonumber \\
&=&
\left( \frac{\partial e_{a {\mss {IE}}}}{\partial t} \right)_{\rm coll}.
\end{eqnarray}
It is interesting to note the difference between the terms $\vec{R}_{a {\mss {IE}}} - \vec{u}_{a \mss{I}}S_{a \mss{IE}}$ and $\vec{R}_{a {\mss {IE}}}$,
as well as between $M_{a {\mss {IE}}}$ and $Q_{a {\mss {IE}}}$. $\vec{R}_{a {\mss {IE}}} - \vec{u}_{a \mss{I}}S_{a \mss{IE}}$ represents the momentum change exclusively due to velocity variation by collisions, while $\vec{R}_{a {\mss {IE}}}$ gives the total momentum exchange. Concerning the energy terms, $M_{a {\mss {IE}}}$ provides losses/gains of thermal ($\frac{3}{2}p_{a {\mss {IE}}}$) and kinetic ($\frac{1}{2}\rho_{a {\mss {IE}}} u_{a {\mss I}}^2$) energies due to collisions and $Q_{a {\mss {IE}}}$ is the rate of internal (i.e., thermal plus excitation) energy variation.

The above equations are written for particles at different excitation states. Such level of detail is necessary when considering interactions with the radiation field \citep[see][]{Khomenko+etal2014}, but is possibly too detailed for most of the practical astrophysical applications. Therefore, the general way to proceed is to add up equations for all excitation states of a particle $a$ in a given ionization state $\mathsc{I}$. This provides the following system of equations:

\begin{equation}  \label{Beq:ion-mass}
\frac{\partial \rho_{a \mss{I}}}{\partial t} + \vec{\nabla} (\rho_{a \mss{I}}\vec{u}_{a \mss{I}}) =S_{a \mss{I}},
\end{equation}
\begin{eqnarray} \label{Beq:ion-mom}
\frac{\partial (\rho_{a \mss{I}}\vec{u_{a \mss{I}}})}{\partial t} + \vec{\nabla}(\rho_{a \mss{I}}\vec{u_{a \mss{I}}} \otimes \vec{u_{a \mss{I}}} + {\bf\hat{p}}_{a \mss{I}}) &= & \rho_{a \mss{I}}r_{a \mss{I}}(\vec{E} + \vec{u}_{a \mss{I}}\times\vec{B}) + \rho_{a \mss{I}}\vec{g} +\vec{R}_{a \mss{I}},
\end{eqnarray}
\begin{eqnarray}  \label{Beq:ion-ene}
&&\frac{\partial}{\partial t} \left(  e_{a \mss{I}} +\frac{1}{2}\rho_{a\mss{I}}u_{a\mss{I}}^2 \right) +  \vec{\nabla}\left( \vec{u}_{a \mss{I}} (e_{a \mss{I}} + \frac{1}{2}\rho_{a\mss{I}}u_{a\mss{I}}^2) + {\bf\hat{p}}_{a \mss{I}}\vec{u}_{a \mss{I}} + \vec{q}_{a \mss{I}} \right)  =  \nonumber \\
&&= M_{a \mss{I}} + \frac{E_{a\mss{I}}}{m_a}S_{a\mss{I}} + \rho_{a\mss{I}}\vec{u}_{a\mss{I}}\vec{g}+ \rho_{a\mss{I}}r_{a\mss{I}}\vec{u}_{a\mss{I}}\vec{E}
\end{eqnarray}
with ${\bf\hat{p}}_{a \mss{I}}=\rho_{a \mss{I}}\langle \vec{\tilde{c}}_{a \mss{I}} \otimes \vec{\tilde{c}}_{a \mss{I}} \rangle$, and $\vec{q}_{a \mss{I}} = \frac{1}{2}\rho_{a \mss{I}}\langle \tilde{c}_{a \mss{I}}^2\vec{\tilde{c}}_{a \mss{I}} \rangle$.

This notation is usually simplified when dealing with particular cases of plasmas, e.g., those composed by hydrogen or hydrogen-helium or those containing grains.

%%%%%%%%%%%%%%%%%%%%%%%%%%%%%%%%%%%%%%%%%%%%%%%%%%%%%%
\subsection{Components of pressure tensor and viscosity}

%%%%%%%%%%%%%%%%%%%%%%%%%%%%%%%%%%%%%%%%%%%%%%%%%%%%%%

In the above equations the pressure tensor, ${\bf\hat{p}}_{a \mss{I}}$, has been defined. Its diagonal components provide the scalar pressure, which in a general situation can be anisotropic and depend on the direction parallel and perpendicular to the magnetic field. The non-diagonal components of ${\bf\hat{p}}_{a \mss{I}}$ provide viscosity. The expressions for the components of the complete tensor for electrons and ions of a fully ionized plasma can be found in \citet{braginski65}, see his equations (2.19--2.28), where he considered approximate expressions for the limiting cases of weak and strong magnetic field. For a PIP, \citet{khodachenko04, khodachenko06} propose to modify the expressions given in \citet{braginski65} to include ion-neutral and ion-electron collisions.

The heat flow vector $\vec{q}_{a \mss{I}}$, is given by:
\begin{equation}
\vec{q}_{a \mss{I}}=-\hat{\kappa}_{a \mss{I}}\vec{\nabla}T_{a \mss{I}},
\end{equation}
where $\hat{\kappa}_{a \mss{I}}$ is thermal conductivity tensor, and $T_{a \mss{I}}$ is the temperature of the plasma species $a$. Similarly, \citet{braginski65} provides expressions for the components of the electron and ion conductivity tensors of the fully ionized plasma (see his equations 2.10--2.16).

%%%%%%%%%%%%%%%%%%%%%%%%%%%%%%%%%%%%%%%%%%%%%%%%%%%%%%
\subsection{Collisional terms}
\label{Bsec:coll}
%%%%%%%%%%%%%%%%%%%%%%%%%%%%%%%%%%%%%%%%%%%%%%%%%%%%%%

Two different sub-sets of collisions are elastic and inelastic. If particle identity at the micro-state $\{ a \mathsc{IE} \}$ is maintained during the collision, such collision is called ``elastic''. For elastic collisions, the term $S_{a {\mss I}}$ is zero  and the term $\vec{R}_{a {\mss I}}$ simplifies to a large extent.

Collisions that lead to creation/destruction of particles are called ``inelastic''. Inelastic processes most relevant for the solar atmosphere are ionization, recombination, excitation and de-excitation. Charge-transfer processes, in which two colliding species modify their ionization state by exchanging an electron as a result of the interaction, also belong to this type of interaction. Chemical reactions, if they occur, are also inelastic collisions. In a general case, the collisional $S$ term summed over all excitation states can be written in the following form
\begin{equation} \label{Beq:S-term}
S_{a {\mss I}}=\sum_{{\mss E}}\sum_{{\mss {I' \neq I,E'}}}(\rho_{a {\mss {I'E'}}}P_{a {\mss {I'E' IE}}} - \rho_{a {\mss {IE}}}P_{a {\mss {IE I'E'}}}),
\end{equation}
where the $P$-terms are the probabilities of a transition between energy levels of the atom $a$ that include radiative and collisional contributions. Their expressions depend on the particular atom and transition and can be found in standard tutorials on radiative transfer, see \citet{Carlsson1986, Rutten2003}.

The sum of all $S_{a {\mss I}}$ terms over all possible micro-states and all possible species is zero if no nuclear reaction is taking place. Each particular $S_{a {\mss I}}$ term can also become zero if the plasma is in local thermodynamical equilibrium (LTE), i.e., when a detailed balance holds between the forward and backward transitions. The $S_{a {\mss I}}$ terms are usually neglected in applications to ionospheric and interstellar medium plasmas. In the solar atmosphere, LTE conditions are usually assumed in the photosphere. However, in the chromosphere the individual $S_{a {\mss I}}$ terms can not be neglected and contribute as source terms in their corresponding mass conservation equation \citep{Carlsson+Stein2002, Leenaarts2007}.

The term $\vec{R}_{a {\mss {IE}}}$ provides the momentum exchange due to collisions of particles in the micro-state $\{a \mathsc{IE} \}$ with other particles and is the result of the variation induced by elastic and inelastic collisions:
\begin{equation}
\vec{R}_{a {\mss {IE}}} = \vec{R}_{a {\mss {IE}}}^{\rm {el}} +
                          \vec{R}_{a {\mss {IE}}}^{\rm {inel}}.
\end{equation}
The expressions for elastic collisions between two particles of two different micro-states $\{a \mathsc{IE}\}$ and $\{b\mathsc{I'E'}\}$ (excluding photons) can be defined as:
\begin{equation}
\vec{R}_{a {\mss {IE}} ; b{\mss {I'E'}}}^{\rm {el}}  = - \rho_{a {\mss {IE}}} \rho_{b {\mss {I'E'}}} K_{a {\mss {I}} ; b{\mss {I'}}} (\vec{u}_{a {\mss {I}}} - \vec{u}_{b{\mss {I'}}}),
\end{equation}
so that the total momentum transfer due to elastic collisions between particles of kind $\{a \mathsc{IE}\}$ with all other particles is given by
\begin{equation}
\vec{R}_{a {\mss {IE}}}^{\rm {el}} = -\rho_{a {\mss {IE}}} \sum_{b{\mss {I'E'}}} \rho_{b {\mss {I'E'}}} K_{a {\mss {I}} ; b{\mss {I'}}} (\vec{u}_{a {\mss {I}}} - \vec{u}_{b{\mss {I'}}}),
\end{equation}
where
\begin{equation} \label{Beq:coll-rate}
K_{a {\mss {I}} ; b{\mss {I'}}} = \left< \sigma v \right>_{a {\mss {I}} ; b{\mss {I'}}}/(m_{a {\mss {I}}}+ m_{b {\mss {I'}}}) \,\,\,,
\end{equation}
and $\sigma$ is the cross-section of the interaction and $\left<\sigma v\right>_{a {\mss {I}} ; b{\mss {I'}}}$ the collisional rate between particles in microstate $\{a \mathsc{I}\}$ and particles in microstate $\{b \mathsc{I'}\}$, which can be reasonably assumed to be independent of the excitation state.

The total momentum transfer term due to elastic collisions, summing over all excitation states,  is trivially obtained (with $\nu_{a {\mss I} ; b{\mss I'}} = \sum_{\mss{E'}} \nu_{a {\mss {IE}}; b {\mss {I'E'}}} = \rho_{b {\mss {I'}}} K_{a {\mss {I}} ; b{\mss {I'}}}$):
\begin{eqnarray} \label{Beq:r-alpha-i}
\vec{R}_{a {\mss I}}^{\rm {el}} = -\sum_{{\mss E}}\rho_{a {\mss {IE}}}\sum_{b{\mss {I'E'}}} \nu_{a {\mss {IE}} ; b{\mss {I'E'}}}(\vec{u}_{a {\mss {I}}} - \vec{u}_{b{\mss {I'}}}) = -\rho_{a {\mss I}}\sum_{b {\mss I'}}\nu_{a {\mss I} ; b{\mss I'}}(\vec{u}_{a {\mss I}} - \vec{u}_{b {\mss I'}}).
\end{eqnarray}

The momentum transfer due to inelastic radiative collisions follows from Eq. \ref{Beq:S-term}, taking into account the velocity of the particle that has appeared or disappeared due to the absorption or emission of a photon:

\begin{equation}
\vec{R}_{a {\mss {I}}}^{\rm{inel,rad}} =
   \sum_{{\mss E}}\sum_{{\mss {I' \neq I,E'}}}(\rho_{a {\mss {I'E'}}}
      P_{a {\mss {I'E' IE}}} \vec{u}_{a {\mss {I'}}}
- \rho_{a {\mss {IE}}}P_{a {\mss {IE I'E'}}} \vec{u}_{a {\mss {I}}}),
\end{equation}

Inelastic collisional processes follow similar expressions for the momentum transfer as Eq. (\ref{Beq:r-alpha-i}) for chemical reactions or general charge transfer interactions \citep{Draine1986}. A special case of the latter collisions is that involving charge transfer between a neutral and a singly-ionized atom of the same chemical element (or, in general, charge transfer between two ionized states differing in one electron). In this case, there is no net mass variation (the corresponding S-term is zero for the species involved in the interaction) despite the change of nature of both particles (the ion becomes a neutral atom and vice versa) and the collision can be regarded as an elastic collision that also follows Eq. (\ref{Beq:r-alpha-i}) with $b=a$ and $I'= I \pm 1$. The cross-section of the charge-transfer interaction should be used in Eq. (\ref{Beq:coll-rate}), which in general is larger than the elastic collisional cross-section.

A collisional frequency of interaction between two types of particles is usually defined as
\begin{equation}
\nu_{a {\mss {IE}}; b {\mss {I'E'}}} = \rho_{b {\mss {I'E'}}} K_{a {\mss {I}} ; b{\mss {I'}}}.
\end{equation}
The notation containing explicit terms with the densities of the two interacting species leads to the symmetry condition $K_{a {\mss {I}} ; b{\mss {I'}}} = K_{b {\mss {I'}} ; a{\mss {I}}}$, while for collisional frequencies one has that, in general, $\nu_{a {\mss {IE}} ; b{\mss {I'E'}}} \neq \nu_{b {\mss {I'E'}} ; a{\mss {IE}}}$.

Collisional frequencies for momentum transfer between different species have been known for a long time.
The expression differs depending on the nature of the colliding particles.
The collisional frequency, $\nu_{a {\mss I} ; b{\mss 0}}$, between a neutral or a charged particle, in an ionization stage $\mathsc I$, with neutral particles (i.e., ${\mathsc I'}=0$) has the following expression for Maxwellian velocity distributions with, in general, different temperatures for both species, $T_{a {\mss I}}$ and $T_{b {\mss 0}}$, \citep{braginski65, Draine1986}:
\begin{equation} \label{Beq:nu_in}
\nu_{a {\mss I} ; b{\mss 0}} = n_{b{\mss 0}} \frac{m_{b {\mss 0}}}{m_{a {\mss I}}+m_{b {\mss 0}}}
\frac{4}{3}
\sqrt{
\frac{8 k_B T_{a {\mss I}}}{\pi m_{a {\mss I}}} +
\frac{8 k_B T_{b {\mss 0}}}{\pi m_{b {\mss 0}}}
}
\sigma_{a {\mss I} ; b{\mss 0}}\,\,\,,
\end{equation}

\noindent where $\sigma_{a {\mss I} ; b{\mss 0}}$ is the cross-section of the  interaction. The same equation can be applied for electrons as the charged particles. Temperature-dependent values of $\sigma_{a {\mss I} ; b{\mss 0}}$, including charge-transfer effects, have been calculated by \citet{Vranjes2013}. For temperatures below 1 eV ($\sim 10^4 K$), values range in the intervals $\sigma_{H^{+};H}=[0.5-2]\times10^{-18}$ m$^2$, $\sigma_{e^{-};H}={\rm few}\times 10^{-19}$ m$^2$ for the $H^{+}-H$ and the $e^{-}-H$ collisions, respectively \citep{khodachenko04, Leake+Arber2006, soler09a, khomenko12}. For large particles, such as interstellar medium dust grains, with sizes of the order of \hbox{$10^2 - 10^3$ \AA\ } \citep{Wardle+Ng1999}, a hard sphere model is taken with a cross-section equal to $\pi a^2$, where $a$ is the size is of the grains.

\noindent The expression corresponding to collisions between charged particles of species $\{ a \mathsc{I} \}$ and $\{ b \mathsc{I'} \}$ reads as \citep{braginski65, Lifschitz, Rozhansky}:
\begin{equation} \label{Beq:nu_ii}
\nu_{a {\mss I} ; b{\mss I'}}  = \frac{n_{b {\mss I'}} I^2 I'^2 e^4 \ln \Lambda }
{3\epsilon_0^2 m_{a {\mss {I} ; b{\mss I'}}}^2}
\left(
\frac{2 \pi k_B T_{a {\mss I}}}{m_{a {\mss I}}}
+ \frac{2 \pi k_B T_{b {\mss I'}}}{m_{b {\mss I'}}}
\right)^{-3/2},
\end{equation}
\noindent
where $m_{a {\mss {I} ; b{\mss I'}}}$ is the reduced mass and $\ln \Lambda$ is the Coulomb logarithm, where $\Lambda$ is given by the expression \citep{Bittencourt}
\begin{equation} \label{Beq:lambda}
\Lambda = \frac{12 \pi (\epsilon_0 k_B T_e)^{3/2}}{n_e^{1/2} e^3},
\end{equation}
which leads to the usual expression
\begin{equation} \label{Beq:logcoulomb}
\ln \Lambda = 23.4 - 1.15 \log_{10} \left[ n_e ({\rm cm}^{-3}) \right] +
   3.45 \log_{10} \left[ T_e ({\rm eV)} \right].
\end{equation}

\noindent
Small deviations from this expression depending on the colliding species can be found in \citet{Huba}.

The term $M_{a {\mss IE}}$ includes (bulk and thermal) kinetic energy losses/gains of the particles of micro-state $\{a  \mathsc{IE} \}$ due to collisions with other particles, and in general does not vanish. \citet{Draine1986} gives expressions for the term $Q_{a {\mss IE}}$ (total thermal energy exchange by collisions) for different cases. In general, $Q_{a {\mss IE}}$ is a function of the difference in temperature between the colliding species and of the squared difference of their velocities.

%%%%%%%%%%%%%%%%%%%%%%%%%%%%%%%%%%%%%%%%%%%%%%%%%%%%%%%%
\subsection{Ohm's law}
\label{Bsec:ohm}
%%%%%%%%%%%%%%%%%%%%%%%%%%%%%%%%%%%%%%%%%%%%%%%%%%%%%%%%

In the literature, the Ohm's law is frequently derived following \citet{braginski65} by using  the electron momentum equation
\begin{equation} \label{Beq:ele-mom}
\frac{\partial (\rho_e\vec{u}_e)}{\partial t} + \vec{\nabla}(\rho_e\vec{u}_e \otimes \vec{u}_e + {\bf\hat{p}}_e) =  \rho_e r_e(\vec{E} + \vec{u}_e\times\vec{B}) + \rho_e\vec{g} +\vec{R}_e,
\end{equation}
In this approach, from the very beginning one neglects the electron inertia terms, ${\partial (\rho_e\vec{u_e})}/{\partial t}$, $\vec{\nabla}(\rho_e\vec{u_e} \otimes \vec{u_e})$, and the gravity acting on electrons, $\rho_e\vec{g}$. Such an approach is recently used in, e.g.  \citet{Zaqarashvili11b}.

Here we provide a derivation of the Ohm's law in a more general situation for a plasma composed of an arbitrary number of positively and negatively charged and neutral species. Therefore, the strategy from \cite{Bittencourt} and \citet{Krall+Trivelpiece1973} is followed. In this approach the momentum equations of each species (Eq. \ref{Beq:ion-mom}) are multiplied by the charge over mass ratio (Eq. \ref{Beq:q-m}) and are summed up. So far, no force acting on particles is neglected, and electro-magnetic force, gravitational force, inertia terms and momentum exchange by collisions are all included. Since neutral species have zero charge, their contribution in such summation is null, and the final summation goes over $N+1$ charged components (where $N$ is the number of ions and charged grains plus one electron component).
\begin{eqnarray} \label{Beq:ohm-alfa-start}
&&\sum_{a, \mss{I} \ne 0}^{N+1} \left( r_{a \mss{I}} \frac{\partial (\rho_{a \mss{I}}\vec{u_{a \mss{I}}})}{\partial t} +r_{a \mss{I}} \vec{\nabla}(\rho_{a \mss{I}}\vec{u_{a \mss{I}}} \otimes \vec{u_{a \mss{I}}} ) \right) = \nonumber \\
&&=\sum_{a, \mss{I} \ne 0}^{N+1}\left(\rho_{a \mss{I}} r_{a \mss{I}}^2(\vec{E} + \vec{u}_{a \mss{I}} \times \vec{B}) + \rho_{a \mss{I}} r_{a \mss{I}} \vec{g}- r_{a \mss{I}} \vec{\nabla} {\bf\hat{p}}_{a \mss{I}} +r_{a \mss{I}} \vec{R}_{a \mss{I}} \right)
\end{eqnarray}

\noindent The left hand side of this equation can be manipulated using the continuity equations. For that, the total current density is defined
\begin{equation} \label{Beq:jota}
\vec{J} = \sum_{a, \mss{I} \ne 0}^{N+1} \rho_{a \mss{I}} r_{a \mss{I}}\vec{u}_{a \mss{I} }
\end{equation}
\noindent Another useful parameter is drift velocities of species, defined as a difference between the individual velocity of a species with respect to the center of mass velocity of all charged species
\begin{equation}
\vec{w}_{a \mss{I}}=\vec{u}_{a \mss{I}} - \vec{u}_c
\end{equation}
This way the general form of the Ohm's law is obtained:

\begin{eqnarray} \label{Beq:ohm-general}
&&\sum_{a, \mss{I} \ne 0}^{N+1}\rho_{a \mss{I}} r_{a \mss{I}}^2(\vec{E} + \vec{u}_{a \mss{I}} \times \vec{B})=  \frac{\partial\vec{J}}{\partial t} + \vec{\nabla}(\vec{J}\otimes \vec{u}_c + \vec{u}_c \otimes \vec{J}) + \nonumber \\
&& +\vec{\nabla} \sum_{a, \mss{I} \ne 0}^{N+1}(\rho_{a \mss{I}}r_{a \mss{I}}\vec{w}_{a \mss{I}}\otimes\vec{w}_{a \mss{I}}) +
 \sum_{a, \mss{I} \ne 0}^{N+1}r_{a \mss{I}}\vec{\nabla}{\bf\hat{p}}_{a \mss{I}}  -\sum_{a, \mss{I} \ne 0}^{N+1}r_{a \mss{I}} \vec{R}_{a \mss{I}}
\end{eqnarray}
The gravity term cancels out because of the charge neutrality. Up to now it is the most general form of the Ohm's law and no assumptions have been made except for the charge neutrality. Starting from this point particular version of the Ohm's law are to be considered.

%%%%%%%%%%%%%%%%%%%%%%%%%%%%%%%%%%%%%%%%%%%%%%%%%%%%%%%%
\subsubsection{Solar atmosphere}
\label{Bsec:solat}
%%%%%%%%%%%%%%%%%%%%%%%%%%%%%%%%%%%%%%%%%%%%%%%%%%%%%%%%

In the case of the solar atmosphere, Eq. (\ref{Beq:ohm-general}) can be simplified in several ways to make it of practical use. It is assumed that only singly ionized ions are present, since the abundance of the multiply ionized ones is small in the regions where the partial ionization of plasma is important. Separating the contribution of electrons with $r_e=-e/m_e$, the following equation is obtained:

\begin{eqnarray} \label{Beq:ohm-solar1}
&&\rho_er_e^2\left( \sum_{a, \mss{I}=1}^{N}\frac{\rho_{a \mss{I}}r_{a \mss{I}}^2}{\rho_e r_e^2}+ 1 \right)[\vec{E} + \vec{u}_c\times\vec{B}] + \rho_er_e^2\left( \sum_{a, \mss{I}=1}^N \frac{n_{a \mss{I}}\vec{w}_{a \mss{I}}}{n_e}(1 - \frac{r_{a \mss{I}}}{r_e}) \right)\times\vec{B}  + \nonumber \\
&&+ r_e[\vec{J}\times\vec{B}] = \frac{\partial\vec{J}}{\partial t} + \vec{\nabla}(\vec{J}\otimes \vec{u}_c + \vec{u}_c \otimes \vec{J}) + \vec{\nabla} \sum_{a, \mss{I} \ne 0}^{N+1}(\rho_{a \mss{I}}r_{a \mss{I}}\vec{w}_{a \mss{I}}\otimes\vec{w}_{a \mss{I}}) +
 \nonumber \\
 && + r_e\vec{\nabla}\left( \sum_{a, \mss{I}=1}^N{\bf\hat{p}}_{a \mss{I}}\frac{r_{a \mss{I}}}{r_e}  + {\bf\hat{p}}_e\right) - \sum_{a, \mss{I} \ne 0}^{N}r_{a \mss{I}} \vec{R}_{a \mss{I}}
\end{eqnarray}

The collisional term (see sect.~\ref{Bsec:coll})needs further simplification as the particular velocities of species have to be removed and expressed in terms of the average reference velocity of neutral and charge species. This brings us to the following expression:
\begin{eqnarray}
&&\sum_{a, \mss{I} \ne 0}r_{a \mss{I}} \vec{R}_{a \mss{I}} \approx - \vec{J}\left(\sum_{a, \mss{I}=1} \nu_{e; a\mss{I}} + \sum_{b} \nu_{e; b0} \right) +\nonumber \\
&& + en_e (\vec{u}_c  -\vec{u}_n)\left( \sum_{b}\nu_{e; b0} - \sum_{a, \mss{I}=1}\sum_{b}\nu_{a\mss{I}; b0} \right)
\end{eqnarray}

\noindent where $\vec{u}_n$ is used to denote the average center of mass velocity of the neutral species and $\nu_{e; a\mss{I}}$, $\nu_{e; b0}$, $\nu_{a\mss{I}; b0}$ for collisional frequencies between electrons, neutral and ions as defined in the section above. In the expression above the terms containing $\sum_a n_{a \mss{I}}/n_e\vec{w}_{a \mss{I}}$ were neglected, and it was assumed that the variation of the coefficient containing atomic mass of the colliding particles, $\sqrt{(A_a + A_b)/A_a A_b}$ is weak and can also be neglected. This can be done in the particular case of the solar atmosphere since the abundance of the heavy atoms and their probability of collisions is not large (see \citet{Khomenko+etal2014}).

Finally, Eq. (\ref{Beq:ohm-solar1}) is additionally simplified by assuming small electron to ion mass ratio $r_{a \mss{I}}/r_e=m_e/m_{a \mss{I}} \approx 0$ and the following Ohm's law is obtained:

\begin{eqnarray} \label{Beq:ohm-solar}
&&[\vec{E} + \vec{u}_c\times{\vec{B}}] = \eta_H\frac{[\vec{J}\times \vec{B}]}{|B|} - \eta_H\frac{\vec{\nabla}{\bf\hat{p}}_e}{|B|}
+ \eta\vec{J}  - \chi(\vec{u}_c - \vec{u}_n) + \nonumber \\
&&+ \frac{\rho_e}{(en_e)^2}\left( \frac{\partial\vec{J}}{\partial t} + \vec{\nabla}(\vec{J}\otimes \vec{u}_c + \vec{u}_c \otimes \vec{J}) \right)
\end{eqnarray}
\noindent were magnetic resistivities are defined in units of $ml^3/tq^2$ as
\begin{eqnarray} \label{Beq:eta-H}
\eta=\frac{\rho_e}{(en_e)^2}\left(\sum_{a, \mss{I}=1} \nu_{e; a\mss{I}} + \sum_{b} \nu_{e; b0} \right);    \,\,\,\,  \eta_H=\frac{|B|}{en_e}
 \end{eqnarray}
\noindent and the coefficient $\chi$ has the same units as magnetic field, $m/tq$,
\begin{equation}
\chi=\frac{\rho_e}{en_e}\left( \sum_{b}\nu_{e; b0} - \sum_{a, \mss{I}=1}\sum_{b}\nu_{a\mss{I}; b0} \right).
\end{equation}
The charge velocity $\vec{u}_c$ in the Ohm's law (see sect.~\ref{Bsec:ohm}) can be substituted for the ion velocity $\vec{u}_i$, by neglecting the contribution of electrons to the center of mass velocity of charged species.

Eq. (\ref{Beq:ohm-solar}) can be further modified for the cases of hydrogen and hydrogen-helium plasma. In the case of pure hydrogen plasma, the summation of the collision coefficients is not necessary since only one type of particle is present. The Ohm's law in \citet{Pandey+Wardle2008, Zaqarashvili2011a} and \citet{leake14} is obtained as above, with simplified expressions for some of the coefficients
\begin{equation}
\eta=\frac{\rho_e(\nu_{ei} + \nu_{en})}{(en_e)^2}; \,\,\, \chi=\frac{\rho_e\nu_{en}}{en_e}
\end{equation}

In the case of hydrogen-helium plasma one gets the Ohm's law similar to \citet{Zaqarashvili11b} (except that terms $D\vec{J}/Dt$ and the one proportional to $(\vec{\vec{u}_i - \vec{u}_n} )$ are not present because they were specifically neglected by the authors in that particular application) with $\eta$ given by:
\begin{eqnarray}
\eta= \frac{\rho_e(\nu_{eH^+} + \nu_{e He^+})}{(en_e)^2}
\end{eqnarray}

The Ohm's law (Eq. \ref{Beq:ohm-solar}) explicitly contains velocities of the charged and neutral components, $\vec{u}_c$ (or $\vec{u}_i$) and $\vec{u}_n$. Therefore it can be only applied in combination with the motion equations providing the velocities of these components. In the case of strongly collisionally coupled plasma, as in the case of the solar photosphere and possibly low chromosphere, a more approximate and less general form of the Ohm's law is often used, where the individual velocities are eliminated in the favor of the center of mass velocity of the whole plasma, by applying an approximate expression using the motion equations. The relation between both reference systems is
\begin{equation} \label{Beq:ref-u}
[\vec{E} + \vec{u}\times\vec{B}] = [\vec{E} + \vec{u}_c\times\vec{B} - \xi_n\vec{w}\times\vec{B}]
\end{equation}
where $\xi_n=\rho_n/\rho$ is neutral fraction.

By neglecting the term $\rho\xi_n(1-\xi_n)\partial \vec{w}/\partial t$ for processes slow compared to the typical collisional times, the following expression for the relative charge-neutral velocity is obtained,
\begin{equation}
\label{Beq:w_bis}
\vec{w} =\vec{u}_c - \vec{u}_n \approx \frac{\xi_n}{\alpha_n} \left[\vec{J} \times\vec{B} \right] - \frac{\vec{G}}{\alpha_n} + \sum_{b}\frac{\rho_e\nu_{e; b0}}{en_e }\frac{\vec{J}}{\alpha_n}
\end{equation}
\noindent where
\begin{equation}
\vec{G} = \xi_n \vec{\nabla}{\bf\hat{p}}_{c} - \xi_i \vec{\nabla}  {\bf\hat{p}}_n
\end{equation}
\noindent and
\begin{eqnarray}
\alpha_n &=& \sum_b\rho_e\nu_{e; b0} + \sum_{a, \mss{I}=1}\sum_b\rho_{a \mss {I}}\nu_{a\mss{I}; b0}
\end{eqnarray}
\noindent being ${\bf\hat{p}}_{c}$ and ${\bf\hat{p}}_{n}$ charged and neutral pressure tensors. Then, neglecting the terms proportional to the ratio of the electron to ion mass and the term $D\vec{J}/Dt$, the following Ohm's law is obtained:
\begin{eqnarray} \label{Beq:ohm-single}
[\vec{E} + \vec{u}\times{\vec{B}}]&=&\eta_H\frac{[\vec{J} \times \vec{B}]}{|B|} - \eta_H\frac{\vec{\nabla}{\bf\hat{p}}_e}{|B|}  + \eta\vec{J}  - \eta_A\frac{[(\vec{J} \times \vec{B}) \times \vec{B}]}{|B|^2} \nonumber \\
 &+&  \eta_p\frac{[\vec{G} \times \vec{B}]}{|B|^2}
\end{eqnarray}
\noindent where additional resistivity coefficients are defined as
\begin{equation} \label{Beq:eta-A}
\eta_A=\frac{\xi_n^2 |B|^2}{\alpha_n}; \,\,\, \eta_p=\frac{\xi_n |B|^2}{\alpha_n}
\end{equation}

\noindent  In the Ohm's laws given by Eq. (\ref{Beq:ohm-solar})  and (\ref{Beq:ohm-single}), the first three terms in common are Hall, battery and Ohmic terms. The fourth term in Eq. (\ref{Beq:ohm-single}) proportional to $[(\vec{J} \times \vec{B}) \times \vec{B}]$ is the ambipolar diffusion term that appears due to substitution of the charge velocity, $\vec{u}_c$, in the expression for electric field by the centre of mass velocity of the whole plasma, $\vec{u}$ (see Eq. \ref{Beq:ref-u}).

Various terms of the Ohm's law containing $\vec{J}$ can be elegantly combined in a form of the conductivity tensor.

%%%%%%%%%%%%%%%%%%%%%%%%%%%%%%%%%%%%%%%%%%%%%%%%%%%%%%%%
\subsubsection{Ionosphere}
%%%%%%%%%%%%%%%%%%%%%%%%%%%%%%%%%%%%%%%%%%%%%%%%%%%%%%%%

The magnetosphere, ionosphere and thermosphere of the Earth are also characterized by similar transitions as solar atmosphere, ranging from highly ionized collisionless plasma in the magnetosphere (above approximately 400 km), to progressively more collisional (below 400 km) and neutral (below 150 km) plasma in the ionosphere \citep{Song-et-al-2001, leake14}. The ionosphere is therefore usually described as a three-fluid system. The chemical composition of the ionosphere, (usually divided into by F, E, and D layers), changes with height. In the lowest, D, layer the dominant neutral is molecular nitrogen ($N_2$), while its dominant ion is nitric oxide ($NO^+$); in the E layer in addition to nitrogen there appears molecular oxygen ions $O_2^+$; in the highest F layer, atomic oxygen is dominant both in neutral and ionized states \citep{leake14}. Therefore, only singly ionized contributors can be retained in Eq. (\ref{Beq:ohm-general}), and approximations of $m_e/m_{a \mss{I}}$ and $m_{a \mss{I}} \approx m_{b 0}$ can be made,  leading to an equation similar to Eq. (\ref{Beq:ohm-solar}).

However, as discussed in many works, system of reference of the ions (i.e. using $\vec{u}_i$  in Eq. \ref{Beq:ohm-solar}) generally is not the best choice for a frame of reference in a weakly ionized ionospheric mixture with weak collisions \citep{Song-et-al-2001, Vasyliunas-2012, leake14}. Therefore, for ionospheric applications, the plasma velocity in the Ohm's law (see sect.\ref{Bsec:ohm}) is replaced by the velocity of neutral particles.  The relation between both reference systems is
\begin{equation}
[\vec{E} + \vec{u}_n\times\vec{B}]=[\vec{E} + \vec{u}_i\times\vec{B} - \vec{w}\times\vec{B}]
\end{equation}

\noindent The following Ohm's law is obtained

\begin{eqnarray} \label{Beq:ohm-neutral}
&&[\vec{E} + \vec{u}_n\times{\vec{B}}] = \eta_H\frac{[\vec{J} \times \vec{B}]}{|B|} - \eta_H\frac{\vec{\nabla}{\bf\hat{p}}_e}{|B|}  + \eta\vec{J}  - \eta_A\frac{[(\vec{J} \times \vec{B}) \times \vec{B}]}{|B|^2} \nonumber \\
&&+  \eta_p\frac{[\vec{G} \times \vec{B}]}{|B|^2}
\end{eqnarray}
\noindent with
\begin{equation} \label{Beq:eta-A-ionosph}
\eta_A=\frac{\xi_n |B|^2}{\alpha_n}; \,\,\, \eta_p=\frac{|B|^2}{\alpha_n}
\end{equation}
and the rest of the resistive coefficients given by Eq. (\ref{Beq:eta-H}). This equation is similar to the Ohm's law in a neutral reference frame given in \citet{leake14} (see their Eq. 33), when reduced to our notation and removing time derivatives of currents and drift velocity, with the resistivities given by:

\begin{eqnarray} \label{Beq:eta-A-leake}
\eta=\frac{\rho_e}{(en_e)^2} \left(\nu_{ei} + \nu_{en} \right); \,\,\,  \eta_A=\frac{2\xi_n|B|^2}{\rho_i\nu_{in}}; \,\,\, \eta_p=\frac{2|B|^2}{\rho_i\nu_{in}}
\end{eqnarray}
taking into account that $\alpha_n \approx \rho_i\nu_{in}$ (i.e. ignoring collisions with electrons), so the only difference are the two coefficients before the ambipolar and $[\vec{G} \times \vec{B}]$ terms. Similar equation is also provided in \citet{Vasyliunas-2012}.

Note, that the choice of the system of reference affects the expression for ambipolar diffusion and the $[\vec{G} \times \vec{B}]$ terms that can be seen by comparing the corresponding coefficients between Eqs. (\ref{Beq:eta-A}) and (\ref{Beq:eta-A-leake}). The choice of the system of reference also affects the Joule heating, as discussed in  \citet{Vasyliunas+Song2005, leake14}.

%%%%%%%%%%%%%%%%%%%%%%%%%%%%%%%%%%%%%%%%%%%%%%%%%%%%%%%%
\subsubsection{Interstellar medium (ISM)}
%%%%%%%%%%%%%%%%%%%%%%%%%%%%%%%%%%%%%%%%%%%%%%%%%%%%%%%%

In the case of the interstellar medium, the ionization degree may be very low and the center of mass velocity of the plasma is assumed to be that of neutrals. It is also assumed that the majority of collisions experienced by each charged particle will be with the neutral fluid, and so all other collisions can be neglected. In addition, inertia and pressure forces acting on charged particles is also neglected. In this case, the momentum equation for the charged species (Eq. \ref{Beq:ion-mom}) reduces to:
\begin{eqnarray}
\rho_{a \mss{I}}r_{a \mss{I}}(\vec{E} + \vec{u}_{a \mss{I}}\times\vec{B}) + \vec{R}_{a \mss{I}} -\vec{u}_{a \mss{I}}S_{a \mss{I}}=0
\end{eqnarray}

\noindent In applications of weakly ionized plasmas of molecular clouds, the $S_{a \mss{I}}$ terms are usually neglected assuming there is no mass charges between the species, which results in neglecting the charge exchange process.

According to Eq. (\ref{Beq:r-alpha-i}) the total collisional term is equal to
\begin{equation}
\vec{R}_{a {\mss I}} = -\rho_{a {\mss I}}\rho_n K_{a {\mss {I}} ; n}(\vec{u}_{a {\mss I}} - \vec{u}_n)
\end{equation}
where only the collisions with the neutral species with its average speed $\vec{u}_n$ are taken into account.

Similarly to the previous section, the electric field in ISM applications is expressed in the frame of reference of the neutral component, by defining
\begin{equation}
\vec{w}_{a \mss{I}}^{'}=\vec{u}_{a \mss{I}} - \vec{u}_n
\end{equation}

\noindent leading to the following momentum equation:
\begin{eqnarray}
\rho_{a \mss{I}}r_{a \mss{I}}(\vec{E} + \vec{u}_n\times\vec{B}  + \vec{w}_{a \mss{I}}^{'}\times\vec{B}) - \rho_{a {\mss I}}\rho_n K_{a {\mss {I}} ; n}\vec{w}_{a {\mss I}}^{'} =0
\end{eqnarray}

\noindent Defining the coefficients
\begin{equation}
\beta_{a \mss{I}}=\frac{r_{a \mss{I}}|B|}{\rho_n K_{a {\mss {I}} ; n}}
\end{equation}
\noindent it becomes
\begin{equation}
(\vec{E} + \vec{u}_n\times\vec{B}  + \vec{w}_{a \mss{I}}^{'}\times\vec{B}) - \frac{|B|}{\beta_{a \mss{I}}}\vec{w}_{a {\mss I}}^{'} =0
\end{equation}
and the coefficients $\beta_{a \mss{I}}$ are known as the Hall parameters.
Manipulating this equation, summing it up over all the charged species, using the definition of current $\vec{J}=\sum_a\rho_{a \mss{I}}r_{a \mss{I}}\vec{w}_{a \mss{I}}^{'}$, and taking into account charge neutrality, leads to the following Ohm's law \citep{Falle03, Ciolek02}:

\begin{equation} \label{Beq:ohm-ISM}
[\vec{E} + \vec{u}_n \times \vec{B}] = r_0 \frac{(\vec{J} \cdot\vec{B})\vec{B}}{|B|^2} - r_1 \frac{\vec{J} \times \vec{B}}{|B|} + r_2 \frac{\vec{B} \times (\vec{J} \times \vec{B})}{|B|^2}
\end{equation}

\noindent where the following resistive coefficients have been defined

\begin{equation} \label{Beq:eta-ISM}
r_0 \equiv \frac{1}{\sigma_\|} \mbox{ , }   r_1 \equiv \frac{\sigma_H}{\sigma_\bot^2 + \sigma_H^2}  \mbox{ , }   r_2 \equiv \frac{\sigma_\bot}{\sigma_\bot^2 + \sigma_H^2}
\end{equation}

\begin{eqnarray}
\sigma_H & \equiv &  \frac{1}{|B|}  \sum_{a, \mss{I} \ne 0}{\frac{r_{a \mss{I}}\rho_{a \mss{I}}\beta_{a \mss{I}}^2}{(1 + \beta_{a \mss{I}}^2)}} = - \frac{1}{|B|}  \sum_{a, \mss{I} \ne 0}{\frac{r_{a \mss{I}}\rho_{a \mss{I}} }{(1 + \beta_{a \mss{I}}^2)}} \label{Beq:sigmahall} \\
\sigma_\bot & \equiv & \frac{1}{|B|}  \sum_{a, \mss{I} \ne 0}{\frac{r_{a \mss{I}}\rho_{a \mss{I}}\beta_{a \mss{I}}}{(1 + \beta_{a \mss{I}}^2)}} \\
\sigma_\| & \equiv & \frac{1}{|B|}  \sum_{a, \mss{I} \ne 0}{r_{a \mss{I}}\rho_{a \mss{I}}\beta_{a \mss{I}}}
\end{eqnarray}

The equality of both definitions of $\sigma_H$ can be verified using the condition of charge neutrality, $\sum r_{a \mss{I}}\rho_{a \mss{I}}=0$. These coefficients can be rewritten in the form similar to those defined above (see Eq. \ref{Beq:eta-H} and first equation in \ref{Beq:eta-A}) in the case only one type of positively charged particles is present.

%%%%%%%%%%%%%%%%%%%%%%%%%%%%%%%%%%%%%%%%%%%%%%%%%%%%%%%%
\subsection{Induction equation}
\label{Bsec:ind}
%%%%%%%%%%%%%%%%%%%%%%%%%%%%%%%%%%%%%%%%%%%%%%%%%%%%%%%%

To get the induction equation, Ohm's law, together with the Faraday's law and Ampere's law are used, neglecting Maxwell's displacement current:
\begin{equation} \label{Beq:fa}
\frac{\partial\vec{B}}{\partial t} = -\vec{\nabla} \times \vec{E}; \,\,\, \vec{J} = \frac{1}{\mu}\vec{\nabla} \times \vec{B}
\end{equation}
A particular form of the induction equation depends on the frame of reference for the electric field. In the case the center of mass fluid velocity, $\vec{u}$,  is taken as a reference, by using Eq. (\ref{Beq:ohm-single}) one gets:

\begin{equation} \label{Beq:induction-single}
\begin{split}
\frac{\partial\vec{B}}{\partial t}  =  \vec{\nabla}\times \left[(\vec{u}\times\vec{B}) - \eta_H\frac{[\vec{J} \times \vec{B}]}{|B|} + \eta_H\frac{\vec{\nabla}{\bf\hat{p}}_e}{|B|}  - \eta\vec{J}  + \eta_A\frac{[(\vec{J} \times \vec{B}) \times \vec{B}]}{|B|^2} \right. \\
\left.-  \eta_p\frac{[\vec{G} \times \vec{B}]}{|B|^2} \right]
\end{split}
\end{equation}
\noindent where the coefficients are defined by equations Eq. (\ref{Beq:eta-H}) and (\ref{Beq:eta-A}).
For ionospheric applications, applying Eq. (\ref{Beq:fa}) to Eq. (\ref{Beq:ohm-neutral}), replacing $\vec{u}$ by $\vec{u}_n$, the average neutral velocity, and using the resistive coefficients given by Eq. (\ref{Beq:eta-H}) and Eq. (\ref{Beq:eta-A-ionosph}), a similar induction equation can be obtained.
   
In the case of extremely weakly ionized plasmas such as the ISM, Ohm's law in the form of Eq. (\ref{Beq:ohm-ISM}) is used leading to,
\begin{equation} \label{Beq:ohm-ISM1}
\frac{\partial\vec{B}}{\partial t}  = \vec{\nabla}\times \left[ (\vec{u}_n \times \vec{B}) - r_0 \frac{(\vec{J} \cdot\vec{B})\vec{B}}{|B|^2} + r_1 \frac{\vec{J} \times \vec{B}}{|B|} - r_2 \frac{\vec{B} \times (\vec{J} \times \vec{B})}{|B|^2} \right]
\end{equation}
\noindent where the coefficients are defined by Eq. (\ref{Beq:eta-ISM}) above.

Finally, using the average charged velocity, $\vec{u}_c$, as a reference, the use of Eq. \ref{Beq:ohm-solar} (neglecting $D\vec{J}/Dt$ term) leads to

\begin{eqnarray} \label{Beq:induction-solar}
\frac{\partial\vec{B}}{\partial t}  = \vec{\nabla}\times \left[ (\vec{u}_c\times{\vec{B}}) - \eta_H\frac{[\vec{J}\times \vec{B}]}{|B|} + \eta_H\frac{\vec{\nabla}{\bf\hat{p}}_e}{|B|} - \eta\vec{J}  - \chi(\vec{u}_c - \vec{u}_n) \right]
\end{eqnarray}

%%%%%%%%%%%%%%%%%%%%%%%%%%%%%%%%%%%%%%%%%%%%%%%%%%%%%%%%
\subsection{Two-fluid and single-fluid description}
\label{Bsec:2-fluid}
%%%%%%%%%%%%%%%%%%%%%%%%%%%%%%%%%%%%%%%%%%%%%%%%%%%%%%%%

Sect. \ref{Bsec:eqs} defined equations for particular charged and neutral components composing the plasma, see Eqs. (\ref{Beq:ion-mass}), (\ref{Beq:ion-mom}), and (\ref{Beq:ion-ene}). In many applications it is convenient to sum these equations separately for charges and neutrals, leading to a two-fluid system of equations,

\begin{equation}
\frac{\partial \rho_n}{\partial t} + \vec{\nabla} (\rho_n\vec{u}_n) = S_n
\end{equation}
\begin{equation}
\frac{\partial \rho_c}{\partial t} + \vec{\nabla} (\rho_c\vec{u}_c) = -S_n
\end{equation}
\begin{equation}
\frac{\partial (\rho_n\vec{u_n})}{\partial t} + \vec{\nabla}(\rho_n\vec{u_n} \otimes \vec{u_n}+{\bf\hat{p}}_n)= \rho_n\vec{g} +\vec{R}_n
\end{equation}
\begin{equation} \frac{\partial (\rho_c\vec{u_c})}{\partial t} + \vec{\nabla}(\rho_c\vec{u}_c\otimes\vec{u}_c+{\bf\hat{p}}_{c})=[\vec{J}\times\vec{B}] + \rho_c\vec{g}  -\vec{R}_n
\end{equation}
\begin{eqnarray}
\frac{\partial}{\partial t} \left( e_n+\frac{1}{2}\rho_n u_n^2 \right) +  \vec{\nabla}\left( \vec{u}_n (e_n + \frac{1}{2}\rho_n u_n^2) + {\bf\hat{p}}_n\vec{u}_n + \vec{q}_{n}^{\prime} + \vec{F}_R^n \right)  = \\ \nonumber
= \rho_n\vec{u}_n\vec{g}  + M_n
\end{eqnarray}
\begin{eqnarray}
\frac{\partial}{\partial t} \left(  e_c+\frac{1}{2}\rho_c u_c^2 \right) +  \vec{\nabla}\left( \vec{u}_c (e_c + \frac{1}{2}\rho_c u_c^2) + {\bf\hat{p}}_c\vec{u}_c + \vec{q}_{c}^{\prime} + \vec{F}_R^c \right)  = \\ \nonumber
= \rho_c\vec{u}_c\vec{g} + \vec{J}\vec{E} -M_n
\end{eqnarray}
\noindent where $\vec{F}_R^n$ and $\vec{F}_R^c$ are radiative energy fluxes for neutrals and charges, and $\vec{q}_n^{\prime}$  and $\vec{q}_{c}^{\prime}$ are heat flow vectors, corrected for ionization-recombination effects, see \citet{Khomenko+etal2014}. The $R$ terms for elastic collisions are given by:
\begin{eqnarray}
\vec{R}_n \approx -\rho_e(\vec{u}_n - \vec{u}_e) \sum_b^N \nu_{e; b0} -\rho_i(\vec{u}_n - \vec{u}_i) \sum_{a, \mss{I}=1}\sum_{b}^N\nu_{a\mss{I}; b0}
\end{eqnarray}
\noindent and the rest of the collisional terms depend on the particular application.

This approach is valid when the the difference in behavior between neutrals and charges is larger than between the neutrals/charges of different kind themselves. The latter is a reasonable assumption given that only charges feel the presence of the magnetic field. However, different kinds of neutral (or charged) components themselves can also behave differently from one another because of the different inertia, which happens for the cases of weak collisional coupling \citep {Zaqarashvili2011a}.  In that case, multi-fluid equations as in previous sections can be used.

The above equations are useful for solar and ionospheric applications. In the case of the ISM, the assumption of extremely weak ionization fraction allows us to simplify the system above.  The ion inertia, energy and pressure are neglected. The radiation field-related effects and thermal conduction are also usually neglected, leading to the drop of the $S$ terms, $\vec{F}_R$ terms and $\vec{q}$ terms. Gravity is usually not taken into account. Nevertheless, separate equations for different ions and grains are maintained, so strictly speaking the system of equations is multi-fluid, and not just two-fluid \citep{Falle03, Ciolek02, OSD06},

\begin{equation}
\frac{\partial \rho_n}{\partial t} + \vec{\nabla}(\rho_n\vec{u}_n)  = 0 ,
\end{equation}
\begin{equation}
\frac{\partial \rho_{a \mss{I}}}{\partial t} + \vec{\nabla}(\rho_{a \mss{I}}\vec{u}_{a \mss{I}})  = 0 ,
\end{equation}
\begin{equation}
\frac{\partial \rho_n \vec{u}_n}{\partial t}  + \vec{\nabla}( \rho_n\vec{u_n} \otimes \vec{u_n}+{\bf\hat{p}}_n )  =  \vec{J}\times\vec{B} ,
\end{equation}
\begin{eqnarray}
r_{a \mss{I}}(\vec{E} + \vec{u}_{a\mss{I}}\times\vec{B} ) + \rho_n K_{a {\mss {I}} ; n}(\vec{u}_{a\mss{I}} - \vec{u}_n)=0
\end{eqnarray}
\begin{equation}
\frac{\partial}{\partial t} \left( e_n+\frac{1}{2}\rho_nu_n^2 \right)+ \vec{\nabla} \left (\vec{u}_n(e_n + \frac{1}{2}\rho_n u_n^2) + {\bf{\hat{p}_n}\vec{u}}\right) =\vec{J}\vec{E}  + M_n
\end{equation}
\begin{equation}
M_{a\mss{I}} + \rho_{a \mss{I}} r_{a \mss{I}} \vec{u}_{a \mss{I}} \vec{E} =0 ,
\end{equation}

Finally, in the case that collisional coupling is strong enough, a single fluid approach can be used, with an appropiate induction equation and Ohm's law:

\begin{eqnarray}
\frac{\partial \rho}{\partial t} + \vec{\nabla}\left(\rho\vec{u}\right) =  0
\end{eqnarray}
\begin{equation}
\frac{\partial (\rho\vec{u})}{\partial t} + \vec{\nabla}(\rho\vec{u} \otimes \vec{u} +{\bf\hat{p}})  = \vec{J}\times\vec{B} + \rho\vec{g}
\end{equation}
\begin{equation}
\label{Beq:energy-single-p}
\frac{\partial }{\partial t}\left(e + \frac{1}{2}\rho u^2 \right) + \vec{\nabla}  \left( \vec{u}\, ( e + \frac{1}{2}\rho u^2) +{\bf\hat{p}}\vec{u} +  \vec{q}^{\prime} + \vec{F}_R \right)   = \vec{J} \vec{E}   + \rho\vec{u} \vec{g}
\end{equation}

\newpage

%%%%%%%%%%%%%%%%%%%%%%%%%%%%%%%
\section{Numerical approaches for partially ionized systems}
\label{sect:D}
%%%%%%%%%%%%%%%%%%%%%%%%%%%%%%%%%

There are significant challenges when moving from numerically solving the
equations of ideal MHD to those of non-ideal, or multi-fluid, MHD (see sect.~\ref{Bsec}).  The most
well-known is that of the introduction of diffusive (parabolic) terms in the
induction equation.  However, depending on the philosophy of the approach
being used one may also have to deal with very large signal speeds in the
ions (for example in the case of weakly ionised, two-fluid approaches), and the Hall
effect.  The latter is a dispersive term which leads to a class of waves, known
as Whistler waves, which have higher speeds for shorter wavelengths and
for which $\displaystyle \lim_{k \to \infty} \frac{\omega}{k} = \infty$.  All
of these issues lead to short stable time-steps, at least for explicit schemes.
Further, in many systems of astrophysical interest, studying the impact
of these non-ideal effects leads to a requirement for high spatial resolution.
The computational cost of high resolution simulations may mean the
necessity of using massively parallel compute environments, and implicit
numerical schemes are notoriously difficult to implement efficiently in such
environments.

Typically, then, the goal will be to have an explicit scheme to deal with
non-ideal, or multi-fluid, effects.  The rest of this section will focus
primarily on such schemes, although we will also point out possibilities for
using implicit or semi-implicit schemes.

We begin our discussion of numerical methods for non-ideal and/or multi-fluid
MHD systems by exploring the possibility of directly solving the full
multi-fluid MHD equations, pointing out that in many systems of interest this
would be an extremely computationally challenging task to perform accurately.
We then point out the convenience of operator splitting approaches for
extending any of the well-tested schemes for ideal MHD to non-ideal and
multi-fluid MHD.  If a sufficiently simple generalised Ohm's law can be 
written for the system
of interest (see Sect.\ \ref{Bsec:ohm}) we do not need to solve the Poisson
equation for the electric field and so the non-ideal terms arise as either
diffusive (parabolic) or dispersive (hyperbolic) terms in our equations.  We
thus discuss numerical approaches to diffusive terms.  These are
applicable to systems in which Ohmic and/or ambipolar diffusion are of
interest.  We go on to discuss methods of dealing with the Hall term.  Finally
we discuss the heavy ion approximation which is a rather different approach
and can be used in situations where ambipolar diffusion is the only non-ideal
process of interest.

\subsection{The full system of multi-fluid MHD equations}
\label{Dsec:mf-fullsystem}

One can use existing standard discretisations to solve the full system
of multi-fluid MHD equations, after making suitable approximations to enable
evaluation of the induction equation (see sect.~\ref{Bsec:ind}) combined with a generalised Ohm's law (see sect.~\ref{Bsec:ohm}).
One might reasonably ask why this approach should not be used generally.
Two of the most significant disadvantages of such an approach are the
possibility of extremely high magnetic field mediated wave speeds in the
charged species in the case where the ionisation fraction is low, and the
stiffness of the system when the interaction terms in the momentum equations
are large due to high collision frequencies or cross-sections.  Both of these
considerations lead to short time-steps being necessary for both accuracy and
stability reasons, meaning that both implicit and explicit schemes will
lead to challenging computational requirements.  As a general approach
this is not promising.

In the case of protoplanetary disks, where the ionisation fraction can be
of order $10^{-8}$ or lower, the Alfv\'en speed in the charged species
can be of order $10^6$\, m/s.  Coupling between the charged species and the
neutrals through collisions reduces the effective Alfv\'en speed by 4
orders of magnitude.  Thus, for these systems, one has a very high signal
speed and one relies on the stiff source terms in the system in order to
reduce the propagation speed of waves to the correct speed.  Numerically,
this is fraught with danger arising from finite digit arithmetic in addition
to any truncation errors associated with the scheme.  The situation is not
quite so difficult in the solar photosphere where the effective
Alfv\'en speed and the Alfv\'en speed in the charged species differ by only
2 orders of magnitude, but it is nonetheless advisable to pursue other
approaches which have a higher possibility of yielding computationally
tractable equations.  We also refer the reader to Sect.\ \ref{Dsec:heavy-ion}
for a description of a different approach in which the ion density is
arbitrarily raised in order to reduce the Alfv\'en speed in the ions, making
numerical solution of the two-fluid equations tractable in this case.

\subsection{Operator Splitting}

A most convenient way to extend an ideal MHD code to deal with non-ideal
effects is through operator splitting.  In this case we have a state vector
at time level $n$ and grid zone $(i,j,k)$, $U^{n}_{ijk}$ say, which we wish to
update to time level $n+1$.  If we split our update into the effect of $N$
different operators then we can write our update as
\begin{equation}
\label{Deqn:strang-splitting}
U_{ijk}^{n+1} = L_N L_{N-1} \cdots L_1 U_{ijk}^n
\end{equation}
If we update $U$ repeatedly in this fashion we have a truncation error of
order $\Delta t$ - i.e.\ a scheme which is first order in time.  One can
improve on this by permuting the operators appropriately
\citep{Strang68, Ryu95}.  In this way one can achieve a scheme which is second
order in time.  In the situation under discussion here we can apply a standard
ideal MHD scheme to our system as one operator, and then add in the non-ideal
effects as one or more other operators.

\subsection{Dealing with diffusive terms}

It is well-known that the stable time-step for explicit schemes for diffusion
(parabolic) equations is proportional to $\Delta x^2$.  This makes
the use of these schemes to investigate many systems of astrophysical
interest impractical.  As mentioned, it is natural to consider implicit
schemes for such calculations, but these are computationally expensive
and unsuited to massively parallel compute environments.  Instead we pursue
explicit schemes which are accelerated in some way so as to make their
use practical.

\paragraph{Super-Time-Stepping}

\cite{AA96} drew attention to a widely overlooked method of accelerating
explicit schemes for diffusion equations.  First applied in an
astrophysical context by \cite{OSD06, ODS07}, the method results in a speed-up
of the underlying scheme by using unstable (large) time-steps and stabilising
the result by taking a sequence of short time-steps afterwards.  The
lengths of the time-steps are chosen using Chebyshev polynomials, making this
scheme an example of a Chebyshev-Runge-Kutta scheme.  If we write our
numerical scheme as
\begin{equation}
\vec{U}^{n+1} = \vec{U}^n -\Delta t \vec{A}\vec{U}^n
\end{equation}
then we have the usual time-step restriction of
\begin{equation}
\rho(\vec{I}- \Delta t \vec{A}) < 1
\end{equation}
where $\rho$ is the spectral radius of its argument.  Now let us consider
a slightly different approach.  Writing $\Delta t = \sum_{i=1}^N \tau_i$ and
only requiring stability over the time-step $\Delta t$ and {\em not} over
the individual time-steps $\tau_i$, we can write
\begin{equation}
\vec{U}^{n+1} = \left[\prod_{i=1}^N (I - \tau_i A)\right]\vec{U}^n.
\end{equation}
Our stability condition becomes
\begin{equation}
\rho\left(\prod_{i=1}^N (I - \tau_i A)\right) < 1,
\end{equation}
and we can satisfy this inequality if
\begin{equation}
\left|\prod_{i=1}^N (I - \tau_i \lambda)\right| < 1,
\end{equation}
for every eigenvalue, $\lambda$, of $\mathbf{A}$.  We wish to have
this condition satisfied while at the same time maximising the value of
$\Delta t$.  Making the choice
\begin{equation}
\tau_i = \Delta t_{\rm exp} \left\{(\nu -1) \cos\left[\frac{2i-1}{N}
\frac{\pi}{2}\right] + \nu +1 \right\}^{-1} \quad i = 1\ldots N
\end{equation}
where $\nu$ and $N$ are then free parameters which can be tuned to optimise
the performance as desired and $\Delta t_{\rm exp}$ is the maximum stable
time-step of the underlying scheme (for our purposes this would be the ideal 
MHD time-step).  Letting $\nu$ become small we find
$\lim_{\nu \to 0} \Delta t = N^2 \Delta t_{\rm exp}$ and hence, recalling
that we must take $N$ time-steps to integrate from $t$ to $t + \Delta t$,
we get an overall speed-up of a factor $N$, the number of steps in our
super-time-step.  For good accleration of the underlying scheme it is
important to choose $\nu$ small, but if it is too small then the scheme
destabilises and produces meaningless results.  Some sample values of $\nu$
are given in Sect.\ \ref{Dsec:tests} for specific systems.  It should be
realised that the values of $\vec{U}^n$ at the intermediate time-steps do
not have any approximation properties (i.e.\ they do not approximate the
solution of the original differential equation in an obvious way).  It is only
the result after the composite ``super-time-step'',  $N^2 \Delta t_{\rm exp}$,
which has the usual properties such as convergence.  
For appropriate choices of $N$ and $\nu$ (and double precision arithmetic) the
inherently unstable nature of the intermediate values of $\vec{U}$ has not 
proven to be an issue in a wide range of simulations \citep{OSD06, ODS07,Jones11,
Downes11, Downes12, Gressel13}.

This scheme can be applied easily to the induction equation with either
ambipolar or Ohmic resistivity, but has the drawback of being first order
accurate in time.  This can be dealt with using Richardson extrapolation
\citep[e.g.][]{Press92}.  This approach has since been used in many works
\citep[e.g.][]{Mignone07, Downes11, Jones11, Lee11, Jones12, Downes12,
Tsukamoto13, Gressel13}.

An attempt to extend STS technique to the higher order was presented in 
\cite{meyer12} and used in, for example, \cite{Gressel15}. In the latter work 
a super-time-stepping scheme second order accurate in time was achieved.  In 
this case, the technique relies on Legendre polynomials rather than Chebyshev 
polynomials and so is a Runge-Kutta-Legendre method.  In common with the 
super-time-stepping algorithm above, this method is easy to implement on top 
of any existing scheme.  A drawback is that it requires the simultaneous 
storage of 4 copies of the state vector throughout the grid.  For large-scale 
simulations this requirement may be prohibitive.  It also requires two 
evaluations of the diffusion operator for each sub-step (the first order 
method requires one, but this should be balanced with the need for Richardson 
extrapolation for the first order scheme), making it less efficient than the 
first order scheme as noted by \cite{Tsukamoto13}.  An advantage is that it 
is possible to automate the choice of the number of sub-steps to be taken, 
rather than setting it as a parameter and hoping it is appropriate for the 
system under scrutiny.

\paragraph{The Heavy Ion approximation}
\label{Dsec:heavy-ion}

In systems such as molecular clouds (see sect.\ref{mc}) where ambipolar diffusion is believed to
be the dominant non-ideal MHD mechanism one can represent the system as
a two-fluid one: a neutral fluid and a charged fluid (see the discussion
in sect.\ \ref{Bsec:2-fluid}).  In the case of molecular cloud
simulations it is typical to write the induction equation as
\begin{equation}
\label{Deq:2fluid-ind}
\frac{\partial \vec{B}}{\partial t} = \vec{\nabla}\times(\vec{u}_c
		\times \vec{B}).
\end{equation}
\noindent with the resulting implication that the magnetic field and
charged fluid are perfectly coupled.  The lack of distinction between
various charged fluids in this approach prevents detailed modeling of the
Hall effect.  In weakly ionised systems it can also result in very high
Alfv\'en speeds since the density of the charged species is low and the
Alfv\'en speed is given by $v_{\rm A} = \sqrt{\frac{B^2}{\mu \rho_c}}$
in appropriate MKS units.  One then relies on the coupling terms,
$\vec{R}_n$, to reduce the effective Alfv\'en speed through the bulk
fluid.  Numerically modeling such a system is extremely challenging: the high
Alfv\'en speed leads to a very short time-step, while the strong coupling
terms lead to a stiff set of equations.

In order to work around this one can adopt the following approach
\citep[introduced by][]{LMK06}.  Since ambipolar diffusion occurs as a result
of imperfect collisional coupling between the neutral and charged species, one
might imagine that if the momentum transfer terms $\vec{R}_n$ are calculated
correctly then we should capture the evolution of the system properly.  To
reduce the Alfv\'en speed in the ions we would like them to have a higher
mass density so we arbitrarily increase the mass of the ions, and reduce
the cross section for collisions between neutrals and ions by the same
factor.  Then the size of $\vec{R}_n$ is unaffected, while the Alfv\'en
speed is reduced.  Thus we adopt the ``Heavy ion approximation''.  This
approach has been used in, for example, \cite{MLK08}, \cite{MLK08} and
\cite{LMM12}.

There are some limitations of this approach.  In a weakly ionised medium
one can generally assume that the ion inertia is small.  It must be
ensured that when the mass of the ions is increased, it is done in such
a way that the inertia of these ``heavy'' ions is still negligible.
This yields the requirement that $M^2_{{\rm A}_c} << R_{\rm AD}(l_{u_c})$,
where $M_{{\rm A}_c}$ is the Mach number based on the Alfv\'en speed in the charged fluid and
$R_{\rm AD}(l_{u_c})$ is the ambipolar Reynolds number at a length-scale
$l_{u_c}$.  As noted by \cite{LMK06}, for flows with large gradients as
in the case of turbulence this can be difficult to ensure \citep[see, for
example,][where it is only marginally satisfied in some of the
simulations]{LMKF08}.  The values of $M^2_{{\rm A}_i} / R_{\rm AD}(l_{u_c})$
in this latter work are root-mean-square values and therefore it is difficult
to tell whether this condition is satisfied throughout the grid at all times,
as would be necessary for the approximation to be valid.

\paragraph{Other methods}

In situations where implicit schemes are acceptable, such as in cases where
the use of massively parallel compute systems is not anticipated, one can
adopt the approach of updating the ideal MHD system of equations using an
explicit scheme and then performing an implicit update for the non-ideal
terms \citep[see, e.g.,][]{Falle03}.  Thus, denoting the ideal MHD operator as
$L_{\rm mhd}$ and the operator for the non-ideal, or multi-fluid, effects as
being $L_{\rm mf}$, and using Strang splitting for two operators, the
algorithm is
\begin{eqnarray}
\vec{U}^{n+1/2} = L_{\rm mhd}(\vec{U}^n), \\
\vec{U}^{n+3/2} = L_{\rm mf}(\vec{U}^{n+1/2},\vec{U}^{n+3/2}),
	\label{Deqn:imp-mf}\\
\vec{U}^{n+2} = L_{\rm mhd}(\vec{U}^{n+3/2}).
\end{eqnarray}
where we note the implicit update in equation \ref{Deqn:imp-mf}.  This avoids
all the time-step restrictions alluded to above although one is still
required to restrict the time-step for accuracy, rather than stability,
reasons.  It is worth noting that \citet{OSD06} showed that for
dynamically evolving systems, where accuracy constraints are non-negligible,
the first order super-time-stepping method using Richardson extrapolation to
achieve second order temporal accuracy, is faster than the mixed implicit,
explicit scheme above.

\subsection{The Hall term}

The Hall term in the induction equation is one worthy of considerable
attention when attempting to produce numerical solutions for systems in
which it dominates over the other terms. As discussed in 
sections~\ref{Bsec:solat} and \ref{Bsec:ind},
this is a dispersive term which does not remove energy from the system and
leads to waves with near infinite propagation speed for near zero wavelengths.
\citet{Falle03} showed that a standard, centred difference approach to this
term leads to a stable time-step, $\Delta t_{\rm exp}$, of zero if the Hall
term dominates other terms in the induction equation.  Such dominance of the
Hall term is thought to happen in, for example, certain regions of
protoplanetary disks, near the surfaces of
neutron stars \citep{Hollerbach02,Hollerbach04} and potentially in the solar
photosphere (Sect.\ \ref{solat}).  Works prior to this in
the field of star formation, such as \cite {Hollerbach02}, \cite{Sano02a}, \cite{Sano02b},
\cite{Ciolek02}, all used schemes which were subject to the severe stable time-step
restriction noted by \citet{Falle03}.  It is likely that many of these schemes
successfully simulated the required systems only as a result of the presence
of terms, such as Ohmic diffusion, in the governing equations which
stabilised the numerical schemes employed.

Note that the stability limit above is much {\em worse} than what would be
expected if one were to calculate the stable time-step on the basis of the
propagation speed of the fastest wave in the system, as one is tempted to do
based on the famous Courant-Friedrich-Lewy condition.   Following this
philosophy, since the Hall term gives rise to Whistler waves which have speed
approximately inversely proportional to their wavelength and since the minimum
wavelength representable on a grid with resolution $\Delta x$ is
$2 \Delta x$ one ends up with a stable time-step, $\Delta t_{\rm exp}
\propto (\Delta x)^2$ and not identically zero.  Thus the Hall term,
when differenced naively results in an unexpectedly pathological difference
equation.  Nonetheless, two approaches have been devised which allow for
explicit differencing of the Hall term while still retaining the usual
(parabolic) time-step restriction.

\paragraph{Hyper-diffusivity}

In the field of space science particularly, the Hall term has been dealt with
using standard explicit discretisations which are then stabilised using
$4^{\rm th}$ or $6^{\rm th}$ order hyper-diffusivity, an analogue of the
$2^{\rm nd}$ order artificial viscosity employed in shock-capturing advection
schemes for inviscid systems \citep{Yin01, Ma01}.  In this case, the term
$(\eta_{\rm hyp} \nabla^2 \vec{J})$ is added onto $\vec{E} + \vec{u}_c \times
\vec{B}$.  While this can be argued to have physical origins, as indeed
can the artificial viscosity used in shock-capturing schemes, it is
nonetheless used to stabilise the underlying MHD scheme and not to reflect
underlying physics.

In \cite{Toth08} a similar approach is put forward, and applied to
block-adaptive mesh simulations, in which this hyper-diffusivity is
effectively incorporated into the usual limiters used in TVD schemes so that it
does not appear explicitly in the equations being solved. In solar physics, 
hyper-diffusivity is frequently used in magneto-convection simulations, as 
in \citet{Stein+Nordlund1998}, \citet{Caunt+Korpi2001} and 
\citet{Vogler+etal2005}. There have been attempts to use it to deal with 
simulations including Hall term in the photosphere by \cite{Cameron+etal2012} 
and \citet{martinez12}.

\paragraph{The Hall Diffusion Scheme}

Another approach which can be adopted is known as the Hall Diffusion Scheme,
first introduced by \cite{OSD06} and extended to the 3D case in
 \cite{ODS07}.  While this was applied specifically to the Hall effect,
	the general philosophy is applicable to any equation of the form
\begin{equation}
\frac{\partial \vec{U}}{\partial t} = \frac{\partial}{\partial x}
\left\{\mathbf{R}\frac{\partial \vec{U}}{\partial x}\right\},
\end{equation}
where $\mathbf{R}$ has zeroes on the diagonal.  In the case of the Hall
effect, $\mathbf{R}$ is skew-symmetric and this is a special case of
such operators.

Noting that $\mathbf{R}$ has zeroes on its diagonal, we see that
the instantaneous rate of change of any one component of $\vec{B}$
depends only on the spatial gradients of the other two components.  In
order to illustrate the idea, let us take
\begin{equation}
\mathbf{R} = (-\mathbf{I})^{1/2} = \left(\begin{array}{cc} 0 & 1 \\
		                                 -1 & 0 \end{array}\right),
\end{equation}
for simplicity and assume our system is a 1D flow (with variation in the
$x$ direction only).  Hence we can write
\begin{eqnarray}
\frac{\partial B_y}{\partial t} = \frac{\partial^2 B_z}{\partial
	x^2} \\
\frac{\partial B_z}{\partial t} = -\frac{\partial^2 B_y}{\partial
	x^2}
\end{eqnarray}
and ºwe can difference this as
\begin{eqnarray}
\frac{(B_y)_i^{n+1} - (B_y)_i^n}{\Delta t} = \frac{(B_z)_{i+1}^n - 2
	(B_z)_i^n + (B_z)_{i-1}^n}{(\Delta x)^2} \\
\frac{(B_z)_i^{n+1} - (B_z)_i^n}{\Delta t} = -\frac{(B_y)_{i+1}^{n+1} - 2
	(B_y)_i^{n+1} + (B_y)_{i-1}^{n+1}}{(\Delta x)^2},
\end{eqnarray}
yielding
\begin{eqnarray}
(B_y)_i^{n+1}=  (B_y)_i^n + \frac{\Delta t}{(\Delta x)^2}\left[(B_z)_{i+1}^n
		- 2 (B_z)_i^n + (B_z)_{i-1}^n\right], \\
(B_z)_i^{n+1} = (B_z)_i^n -\frac{\Delta t}{(\Delta x)^2}
\left[(B_y)_{i+1}^{n+1} - 2 (B_y)_i^{n+1} + (B_y)_{i-1}^{n+1}\right].
\end{eqnarray}

This appears to be a scheme which is implicit in some sense for $B_z$,
but explicit for $B_y$.  If the values of $B_y$ are updated throughout
the computational grid prior to updating $B_z$ then the difference
equations are explicit in the sense that no matrix inversions, or
approximations of matrix inversions, are required.  Thus this scheme, in
common with the hyper-diffusivity approach, fulfills the requirement that it
be easily and efficiently parallelisable to enable large-scale simulations on
massively parallel systems.  A potential advantage of the Hall Diffusion
Scheme, though, is that no arbitrary parameters are required unlike the
hyper-diffusivity schemes where $\eta_{\rm hyp}$ must be chosen.

\cite{Toth08} suggested that the Hall Diffusion Scheme is
simply a two-step version of the usual one-step hyper-diffusivity approach of,
for example, \citet{Ma01}. This, however, is a misinterpretation since in the
hyper-diffusive approach one introduces the free parameter $\eta_{\rm hyp}$,
which must be chosen in a somewhat subjective manner, in order to
over-power the instability arising from the nature of the truncation
error in the underlying scheme.  The Hall Diffusion Scheme is successful
because its truncation error does not lead to instability in the first place.

\begin{figure}
\vspace*{1mm}
\begin{center}
\includegraphics[width=9.5cm]{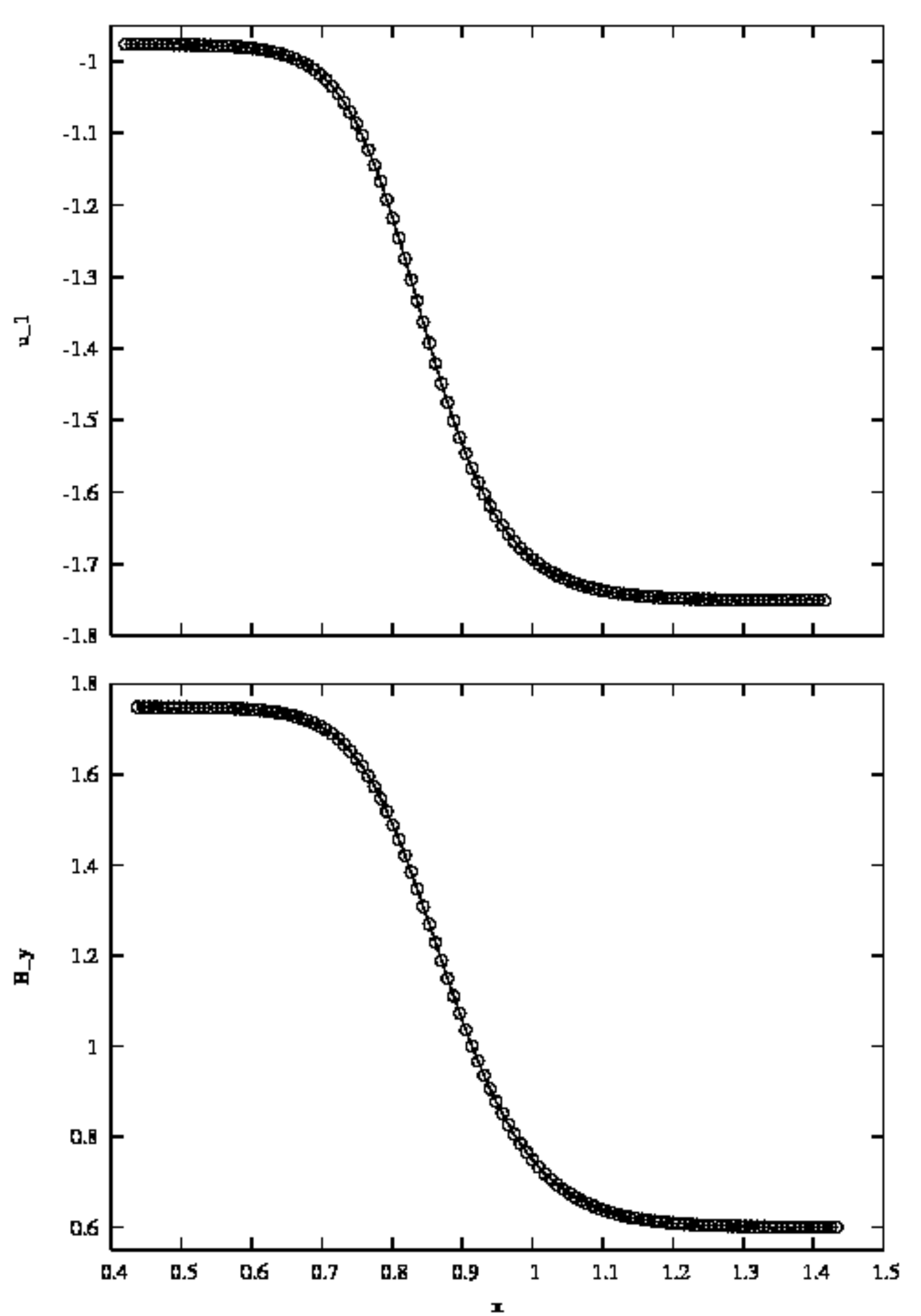}
\end{center}
\caption{\label{fig:casea_test} Numerical shock tube test, case A. Plots of the $x$ component of the velocity and the $y$ component of the magnetic field as functions of $x$.  The solid
line is a semi-analytic solution for the problem, while the circles are 
solution values from the HYDRA code.}
\end{figure}

\begin{figure}
\vspace*{1mm}
\begin{center}
\includegraphics[width=9.5cm]{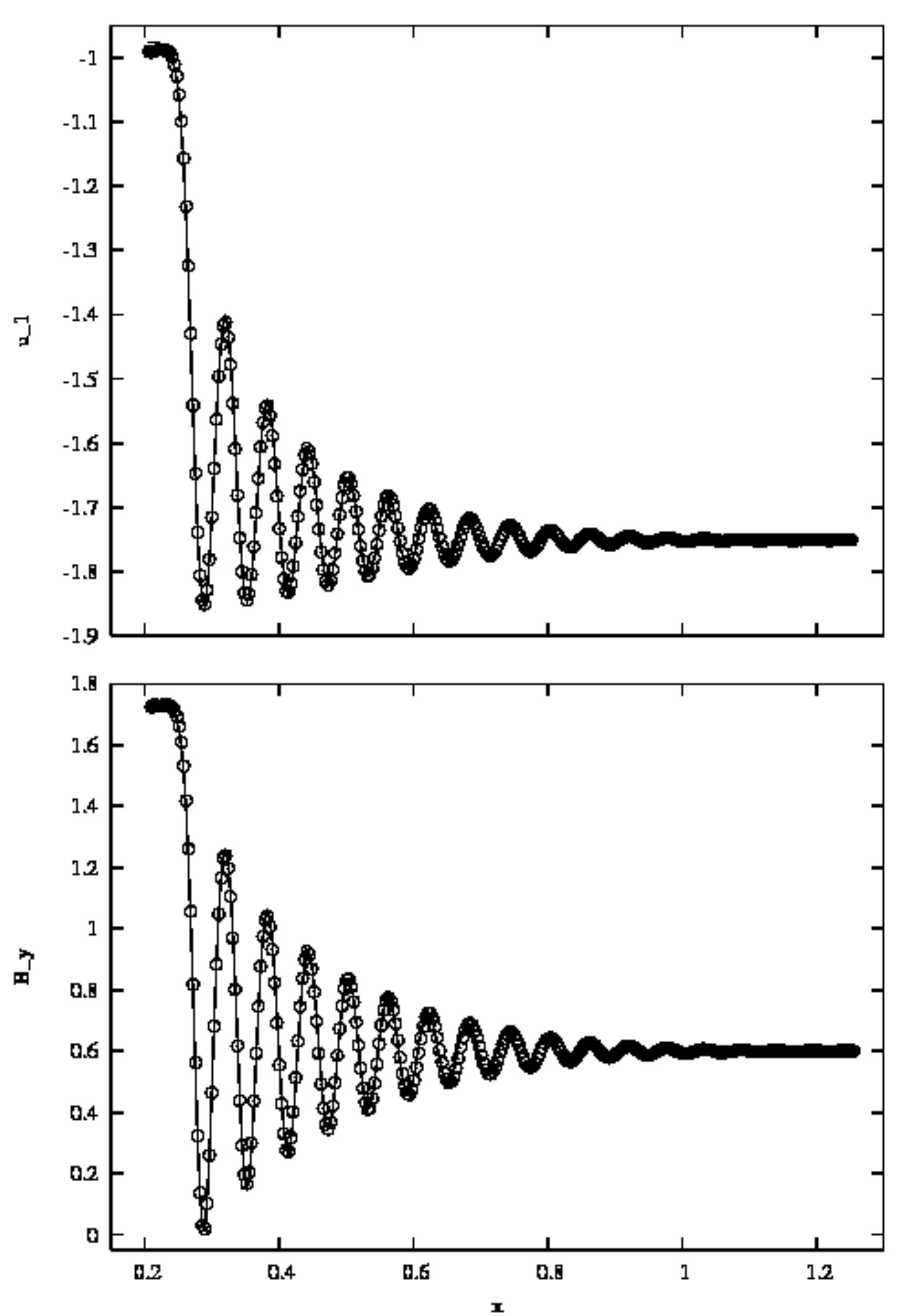}
\end{center}
\caption{\label{fig:caseb_test}  Numerical shock tube test, case B.  The format of the figure is the same as Fig \ref{fig:casea_test}. The influence of the Hall effect, through the presence of a Whistler wave, is
clearly visible in the solution.}
\end{figure}

\begin{figure}
\vspace*{1mm}
\begin{center}
\includegraphics[width=9.5cm]{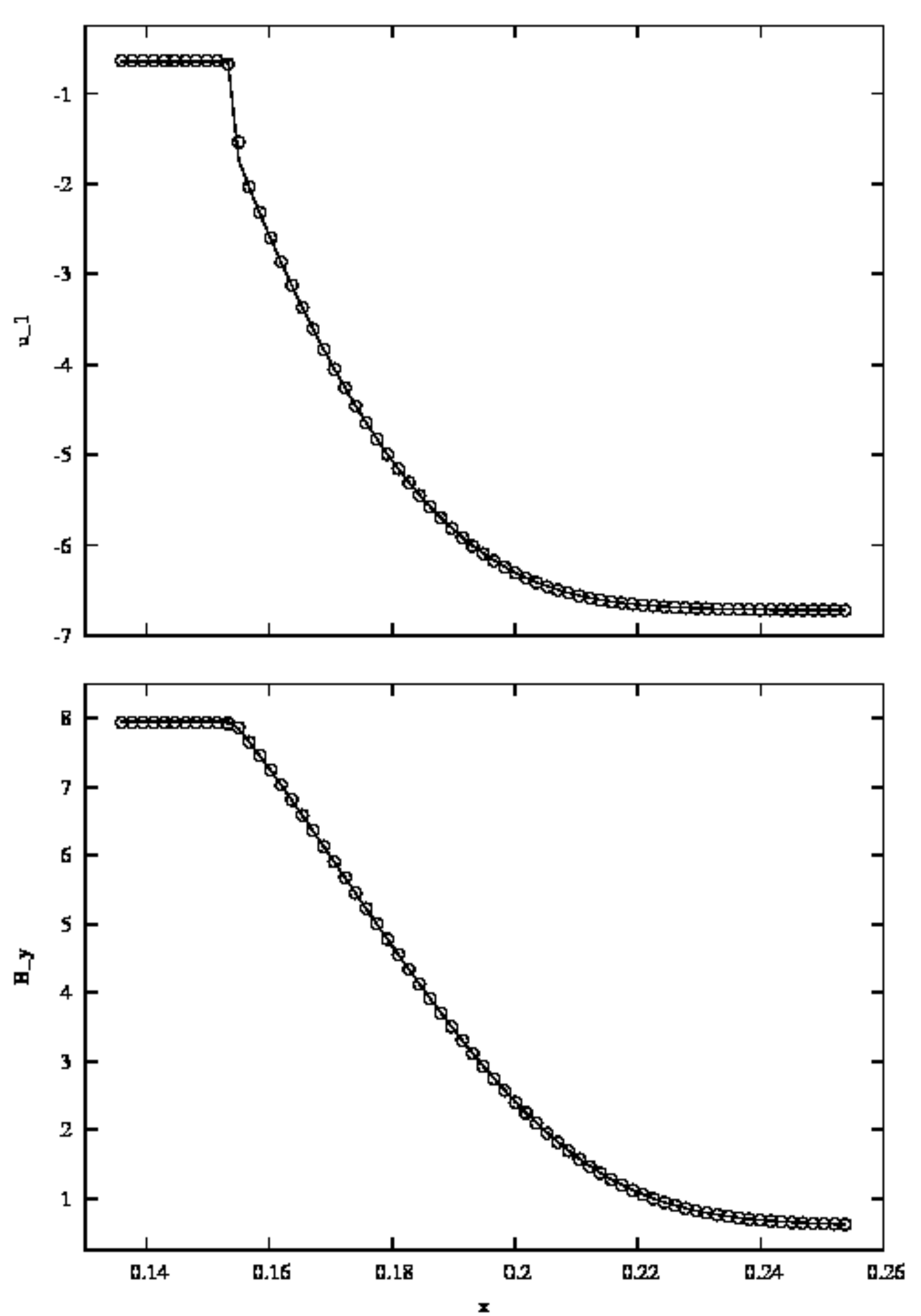}
\end{center}
\caption{\label{fig:casec_test}  Numerical shock tube test, case C.  The format of the figure is the same as Fig \ref{fig:casea_test}. A sub-shock (discontinuity) is visible in the velocity, but not in the magnetic
field.}
\end{figure}

\begin{figure}
\vspace*{1mm}
\begin{center}
\includegraphics[width=9.5cm]{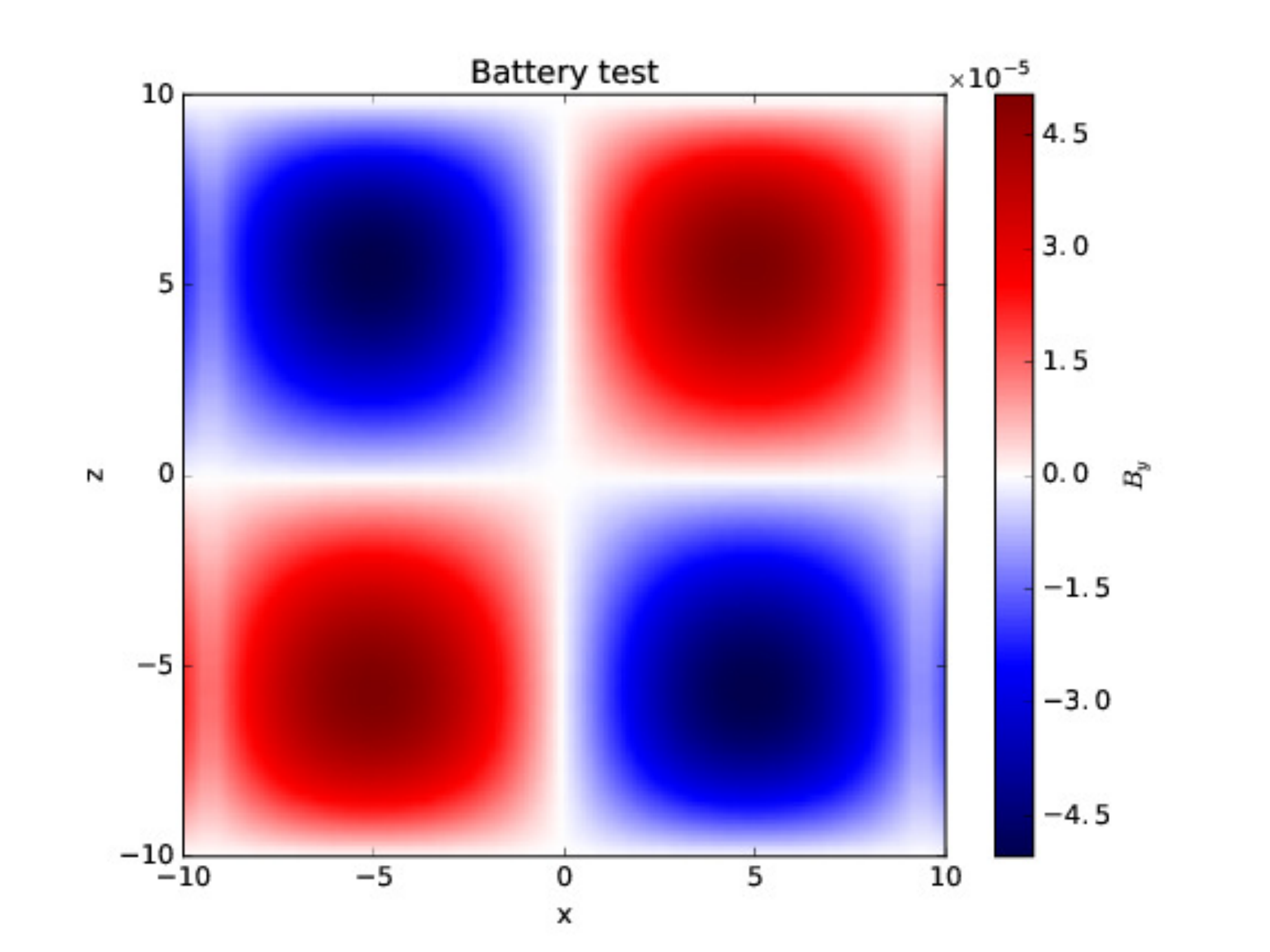}
\end{center}
\caption{\label{fig:biermann_test} Numerical test for Biermann battery effect. 
Plot of the $y$ component of the magnetic field after one time-step of the 
MANCHA code. The influence of the battery term is clearly reflected by the 
presence of a magnetic field.}
\end{figure}

\begin{figure}
\vspace*{1mm}
\begin{center}
\includegraphics[width=9.5cm]{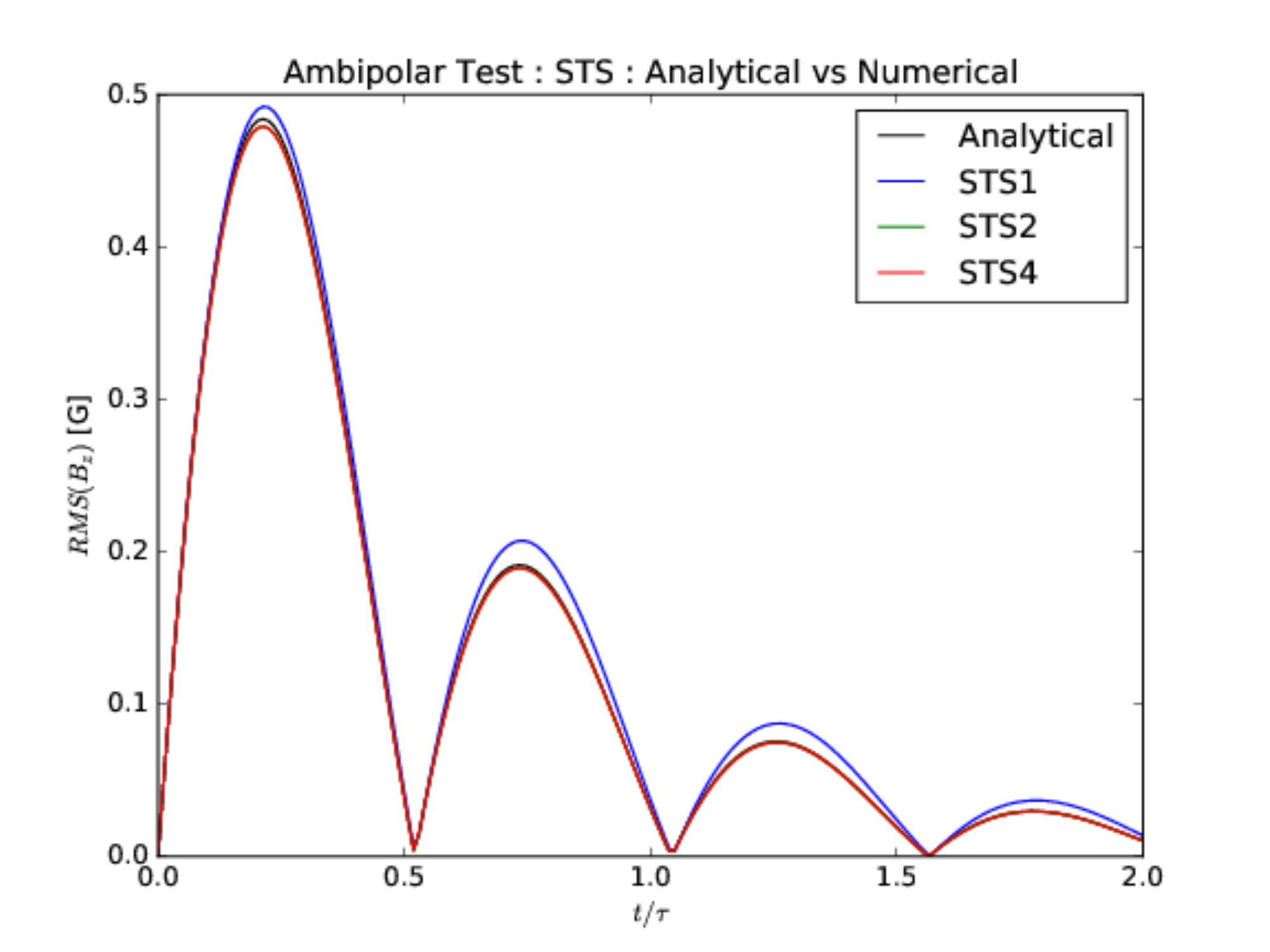}
\end{center}
\caption{\label{fig:ambi_test} Numerical test for Alfv\'en wave decay in the 
presence of ambipolar diffusion. Plots of the RMS value of the $z$ component 
of the magnetic field as a function of time.  Plots are shown for both the 
analytic solution, as well as super-time-stepping combined with Runge-Kutta 
schemes of order 1, 2 and 4.}
\end{figure}

\begin{figure}
\vspace*{1mm}
\begin{center}
\includegraphics[width=9.5cm]{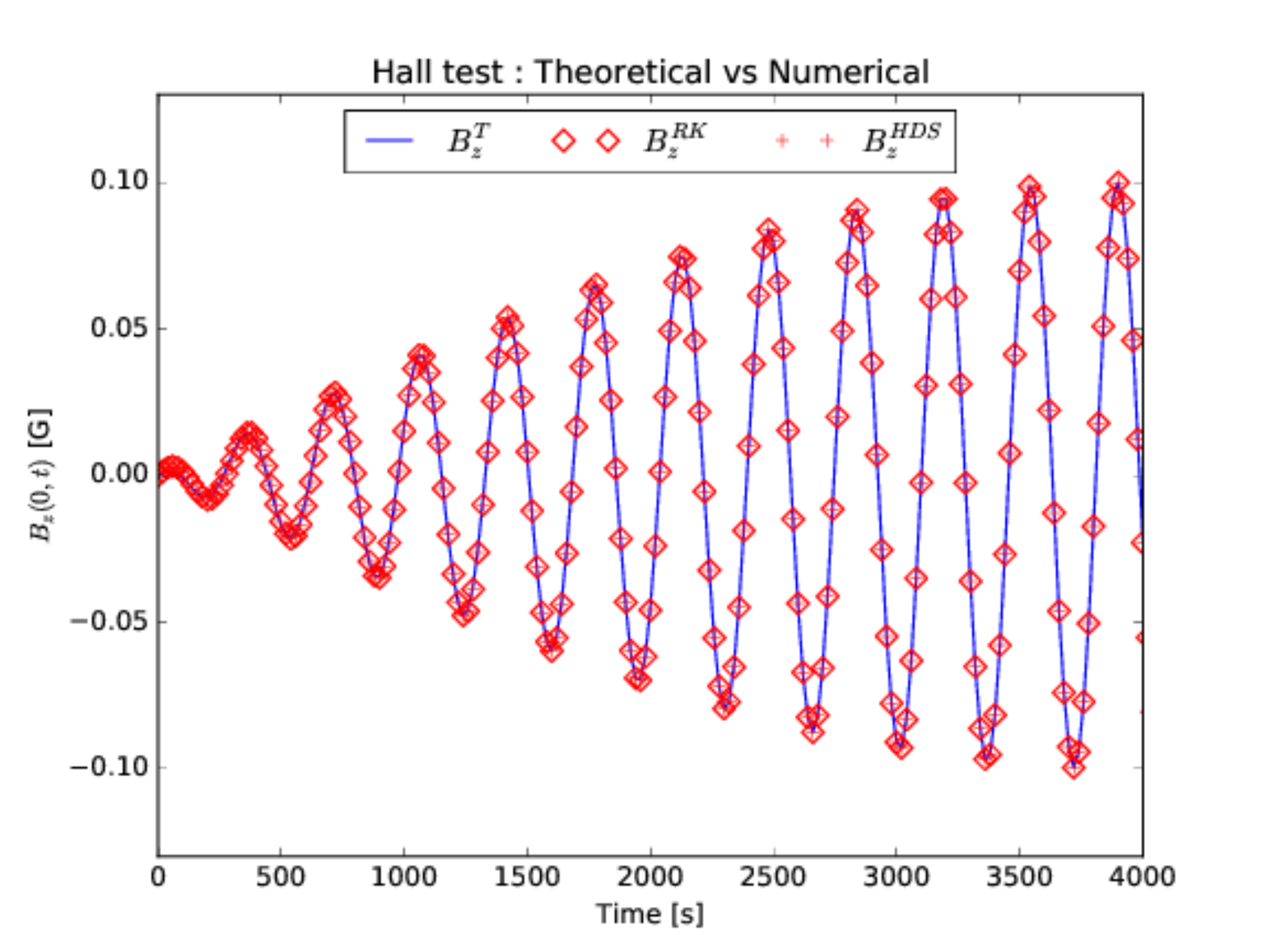}
\end{center}
\caption{\label{fig:hall_test} Numerical test for plane-polarized Alfv\'en wave in 
the presence of the Hall effect. The plot shows the $z$ component of the 
magnetic field as a function of time.  It can be seen that HDS and RK2 agree 
remarkably well both with each other and the analytic solution.}
\end{figure}

\subsection{A Numerical Test Suite}
\label{Dsec:tests}

In this section we suggest a suite of numerical tests which might be
useful in determining whether a code is producing sufficiently accurate
results for a multi-fluid, or non-ideal, MHD system.  We first suggest
a suite of shock tube tests, the conditions for which are given in Table 
\ref{Dsec_table:shocktube}.

\paragraph{Shock tube tests}

\begin{table}
\caption{\label{Dsec_table:shocktube} Conditions for the shock tube tests
encompassing a flow containing a C-shocks a shock disrupted by the Hall effect
and a flow containing a sub-shock.  In each case the flow is assumed to have
3 fluids (one neutral, one positively charged and one negatively charged).
Here we are assuming weak ionisation and so we ignore collisions between
charged fluids, and only take account of collisions between charged and neutral fluids. We also explicitly ignore the ionisation state for the collision
coefficients, $K_{i;j}$, and so we do not include them in the notation here for clarity.  We assume that each charged species has a single ionisation state leading to a single charge-to-mass ratio, $r_a$, and so do not label this in the
notation below.  The units are chosen for numerical simplicity to be Gaussian
scaled so that the speed of light and factors of $4\pi$ do not appear.}
\begin{tabular}{llllll}
\hline\noalign{\smallskip}
Case A & & & & & \\
Right State & $\rho_1=1$ & $\mathbf{u}_1 = (-1.751,0,0)$ & $\mathbf{B} =
(1,0.6,0)$ & $\rho_2=5\times10^{-8}$ & $\rho_3 = 1\times10^{-3}$ \\
Left State & $\rho_1=1.7942$ & $\mathbf{u}_1 = (-0.9759,-0.6561,0)$ & $\mathbf{B} =
(1,1.74885,0)$ & $\rho_2=8.9712\times10^{-8}$ &
$\rho_3 = 1.7942\times10^{-3}$ \\
 & $r_2=-2\times10^{12}$ & $r_3 = 1\times 10^8$ & $K_{2;1} = 4
\times 10^5$ & $K_{3;1} = 2 \times 10^4$ & $a=0.1$ \\
 & $\nu = 0.05$ & $N_{\rm STS} = 5$ & $N_{\rm HDS} = 0$ & & \\
Case B & & & & & \\
Right State & As case A & & & & \\
Left State & As case A & & & & \\
 & $r_2=-2\times10^{9}$ & $r_3 = 1\times 10^5$ & $K_{2;1} = 4
\times 10^2$ & $K_{3;1} = 2.5 \times 10^6$ & $a=0.1$ \\
 & $\nu = 0$ & $N_{\rm STS} = 1$ & $N_{\rm HDS} = 8$ & & \\
Case C & & & & & \\
Right State & $\rho_1=1$ & $\mathbf{u}_1 = (-6.7202,0,0)$ & $\mathbf{B} =
(1,0.6,0)$ & $\rho_2=5\times10^{-8}$ & $\rho_3 = 1\times10^{-3}$ \\
Left State & $\rho_1=10.421$ & $\mathbf{u}_1 = (-0.6449,-1.0934,0)$ & $\mathbf{B} =
(1,7.9481,0)$ & $\rho_2=5.2104\times10^{-7}$ &
$\rho_3 = 1.0421\times10^{-2}$ \\
 & $r_2=-2\times10^{12}$ & $r_3 = 1\times 10^8$ & $K_{2;1} = 4
\times 10^5$ & $K_{3;1} = 2 \times 10^4$ & $a=1$ \\
 & $\nu = 0.05$ & $N_{\rm STS} = 15$ & $N_{\rm HDS} = 0$ & & \\
\noalign{\smallskip}\hline
\end{tabular}
\end{table}

The first test presented suggested here is Case A from \cite{Falle03}, also
published in \cite{OSD06} and \cite{ODS07}.  It is an MHD shock tube test
in a system in which ambipolar diffusion is significant.  Under the particular
conditions for this test we expect the shock to be fully smoothed out to a 
C-shock and there to be no discontinuities present.  Since the solution is 
a single C-shock, the initial conditions are set to the left- and right-
states, separated by a region in which a $\tanh$ function is used to 
interpolate smoothly between the two states.  Recall that in multifluid MHD it 
is not possible to have discontinuities in the magnetic field and hence this 
interpolation is a rational thing to do.  The system is then allowed to 
evolve until it reaches a steady state.  Results are plotted for the HYDRA
code, as presented in \cite{ODS07}, compared with a semi-analytic solution
of the same set of equations in figure \ref{fig:casea_test}.

The second test, Case B, is a Hall dominated system with the same left
and right neutral states.  In this case we expect the usual MHD shock to be
broken down into a set of Whistler waves.  The steady state solution will then
only contain the whistler wave which has the same velocity as the MHD shock 
(see Fig.\ \ref{fig:caseb_test}).  The test presented here was also presented 
in \cite{ODS07} and is more highly Hall dominated than the Hall dominated test 
presented in \cite{Falle03}.

The third case, Case C, involves a flow which has a shock precursor and a 
sub-shock, in contrast to Case A in which the ambipolar diffusion is strong 
enough to completely smear out all the fluid variables.  Since this test
contains a discontinuity (see Fig.\ \ref{fig:casec_test}) it acts as a test of 
the shock-capturing abilities of the numerical scheme being employed.

\paragraph{Test for Battery effect. }
The Biermann battery effect is different from the other effects in the 
generalized Ohm's law. The presence of this term (usually small in most of the 
systems) does not present significant numerical problems, since it is 
only acting as a small source term. Nevertheless, it represents an interesting 
physical effect and allows the creation of magnetic field from misaligned 
gradients in density and pressure. It was used recently in the context of 
solar physics to seed the solar local dynamo by \citet{Khomenko+etal2017}.
To test how well the influence of the Biermann battery is captured we
suggest a test from \cite{toth12}.  In this test a fluid is at rest without
any magnetic field present.  A perturbation in the electronic number density
and pressure is imposed following
\begin{eqnarray}
n_{\rm e} = n_0 + n_1 \cos(k_x x), \\
p_{\rm e} = p_0 + p_1 \cos(k_z z).
\end{eqnarray}

The misalignment between the pressure and density gradients will give rise to 
a current which will produce a magnetic field in the system.  In the results
presented in Fig \ref{fig:biermann_test} we choose $n_0 = p_0 = 1$, 
$n_1 = p_1 = 0.1$ and $k_x = k_z = \pi/L$ with $L = 10$.

\paragraph{Test for Alfv\'en waves decay}. It is well known that Alfv\'en waves decay in the presence of ambipolar 
diffusion.  Fortunately this decay is relatively easy to determine
analytically and so this serves as a good test of the treatment of 
ambipolar diffusion.  Following \cite{balsara96}, the dispersion relation
for Alfv\'en waves in this system is
\begin{equation}
\omega^2 + i k^2 \eta_{\rm A} \omega - k^2 c^2_{\rm A} = 0,
\end{equation}
where $c_{\rm A}$ is the usual Alfv\'en speed and $k^2 = k_x^2 + k_y^2$.
This can easily be solved in the case where $k_x = k_y$ to find that the
decay rate, $\omega_{\rm I} = - k_x^2 \eta_{\rm A}^2/2$.  Figure 
\ref{fig:ambi_test} contains the RMS of the $z$ component of the magnetic 
field as a function of time.

\paragraph{Test for Alfv\'en waves in the presence of Hall effect. } Finally, in addition to the shock tube proposed, we suggest a further test of the Hall effect as follows.  The magnetic field associated with a 
plane-polarised Alfv\'en wave propagating along the $x$-axis in a fluid with 
Hall coefficient $\eta_{\rm H}$ obeys
\begin{eqnarray}
B_x(x,t) = B_0,\\
B_y(x,t) = b \cos(\sigma t)\cos(kx)\cos(\omega t),\\
B_z(x,t) = b \sin(\sigma t)\cos(kx)\cos(\omega t),
\end{eqnarray}
\noindent where $B_0$ is the magnitude of a guide field in the $x$ direction,
$b$ is the (small) amplitude of the Alfv\'en wave, and $\sigma = \eta_{\rm H}
k^2/2$.  We set $\eta_{\rm H} = 2.03\times10^7$\,m$^2$\,s$^{-1}$, $B_0
= 100$\,G, $T = 4000$\,K, $b=0.1$\,G, $p_{\rm gas} = 10$\,N\,m$^{-2}$,
$\rho = 10^{-6}$\,kg\,m$^{-3}$ and finally $L = 10^5$\,m.  Figure 
\ref{fig:hall_test} contains a plot of $B_z$ for two different schemes, a
Runge-Kutta 2 scheme and the Hall Diffusion Scheme.

%%%%%%%%%%%%%%%%%%%%%%%%%%%%%%%%%%%%%%%%
\newpage
\section{Magnetohydrodynamic waves in partially ionized plasmas}
\label{sec:waves}

Since the pioneering  works by, e.g., \citet{piddington1956}, \citet{osterbrock1961}, and \citet{KP69} it is known that the presence of neutral particles in the plasma affects the dynamics of magnetohydrodynamic waves. In this section, we review some basic properties of Alfv\'en and magneto-acoustic waves in partially ionized plasmas. To do so, we consider the simplest model possible: a homogeneous partially ionized plasma permeated by a straight and constant magnetic field.

We consider a partially ionized plasma, which consists of electrons, protons and neutral hydrogen atoms. The unperturbed state is described by a uniform magnetic field $\vec{B}$, electron ($P_e$), proton (($P_i$)) and hydrogen ($P_n$) thermal pressures, and electron ($\rho_e$), proton ($\rho_i$) and hydrogen ($\rho_n$) mass densities. We suppose that each sort of species has a Maxwell velocity distribution and, therefore, they can be described as separate fluids. For time scales longer than proton-electron collision time, the electron and ion gases can be considered as a single fluid. This significantly simplifies the equations taking into account the smallness of electron mass with regards to the masses of ion and neutral atoms. Then the three-fluid description can be replaced by two-fluid description (Eqs. 66-71), where one component of the plasma is the on-electron fluid and the second component is a fluid made of neutral atoms. The linearized two-fluid MHD equations in a homogeneous plasma can be easily derived from Eqs. (66)-(71)  as

\begin{equation}\label{waves1}
\rho_i\frac{\partial \vec{u_i}}{\partial t}= -{\vec{\nabla}{p_i}}-\frac{\vec{\nabla}(\vec{B}\cdot\vec{b})}{\mu} + \frac{(\vec{B}\cdot \vec{\nabla})\vec{b}}{\mu}-\alpha_{in}(\vec{u_i}-\vec{u_n})
\end{equation}
\begin{equation}\label{waves2}
\rho_n\frac{\partial \vec{u_n}}{\partial t}= -{\vec{\nabla}{p_n}}+\alpha_{in}(\vec{u_i}-\vec{u_n})
\end{equation}
\begin{equation}\label{waves3}
\frac{\partial \vec{b}}{\partial t}= (\vec{B}\cdot\vec{\nabla})\vec{u_i}-\vec{B}\vec{\nabla}\cdot\vec{u_i}-\frac{1}{\mu en_e}\vec{\nabla}\times\left [\left (\vec{\nabla}\times\vec{b}\right )\times\vec{B} \right ]
\end{equation}
\begin{equation}\label{waves4}
\frac{\partial p_i}{\partial t} + \gamma P_i\vec{\nabla}\cdot\vec{u}_i = 0
\end{equation}
\begin{equation}\label{waves5}
\frac{\partial p_n}{\partial t} + \gamma P_n\vec{\nabla}\cdot\vec{u}_n = 0,
\end{equation}
where $\vec{b}$, $\vec{u_i}$ ($\vec{u_n}$) and $p_i$ ($p_n$) are the perturbations of magnetic field, proton (neutral hydrogen) velocity and proton (neutral hydrogen) pressure respectively. For simplicity, in these equations we have neglected electron inertia, viscosity, magnetic diffusion, and electron-neutral collisions. These equations describe MHD waves in a homogeneous medium, which are modified by collisions of ions with neutral hydrogen atoms (see sect.~\ref{Bsec:coll}). We shall consider Alfv\'en and magneto-acoustic waves separately. Alfv\'en waves are incompressible and produce vorticity perturbations. On the contrary, magneto-acoustic waves are compressible and have no fluid vorticity associated with them.

\subsection{Alfv\'en waves}

We consider the unperturbed magnetic field, $B_z$, directed
along the $z$ axis and the wave propagation along the magnetic field. The perpendicular components of magnetic field ($\vec b_{\perp}$) and velocity ($\vec u_{i\perp}, \vec u_{n\perp}$) perturbations describe the Alfv\'en waves. Then,  Eqs. (\ref{waves1})-(\ref{waves5}) give
\begin{equation}\label{alfen-viy}
{{\partial \vec u_{i\perp}}\over {\partial t}}={B_z\over {\mu
\rho_{i}}}{{\partial \vec b_{\perp}}\over {\partial z}} -
{{\alpha_{in}}\over \rho_{i}} (\vec u_{i\perp} - \vec u_{n\perp}),
\end{equation}
\begin{equation}\label{alfven-vny}
{{\partial \vec u_{n\perp}}\over {\partial t}}={{\alpha_{in}}\over \rho_{n}}
(\vec u_{i\perp} - \vec u_{n\perp}),
\end{equation}
\begin{equation}\label{alfven-by}
{{\partial b_{\perp}}\over {\partial t}}=B_z{{\partial \vec u_{i\perp}}\over {\partial z}}-\frac{1}{\mu en_e}\vec{\nabla}_{\perp}\times\left [\left (\vec{\nabla}\times\vec{b}_{\perp}\right )\times\vec{B} \right ].
\end{equation}
First we neglect the Hall term in the induction equation and perform a
Fourier analysis assuming disturbances to be proportional to $exp[i(k_z z -\omega t)]$. This gives the dispersion relation (see, e.g., Zaqarashvili et al. 2011a, Soler et al. 2013a)

\begin{equation}\label{alfev-disp2}
\omega^3 +i \frac{\alpha_{in}}{\rho \xi_i \xi_n}\omega^2 - v^2_A k^2_z\omega - i\frac{\alpha_{in}}{\rho \xi_n}v^2_A k^2_z=0,
\end{equation}
where
$$
\rho=\rho_i+\rho_n,\,\,\xi_i={\rho_i\over \rho}, \,\, \xi_n={\rho_n\over
\rho}, \,\,  v_A= {B_z\over {\sqrt{\mu \rho_i}}}.
$$
Note that only the ion density appears in the present definition of the Alfv\'en velocity, $v_A$.

We assume a real wavenumber, $k_z$, and solve the dispersion relation (Equation~(\ref{alfev-disp2})) to obtain the complex frequency, $\omega= \omega_{\rm R} + i \omega_{\rm I}$, with $\omega_{\rm R}$ and $\omega_{\rm I}$ the real and imaginary parts of $\omega$, respectively. Since $\omega$ is complex the amplitude of perturbations is multiplied by the factor $\exp(\omega_{\rm I} t)$, with $\omega_{\rm I} < 0$. Therefore the perturbations are damped in time. Physically, the damping is caused by the dissipation associated with ion-neutral collisions.

Equation~(\ref{alfev-disp2}) is a cubic equation in $\omega$ so it has three solutions. 
 The exact analytic solution to Equation~(\ref{alfev-disp2}) is too complicated to shed any light on the physics. However we can investigate the nature of the solutions using the concept of the polynomial discriminant, as done by Soler et al. (2013a). We perform the change of variable $\omega = i s$, so that Equation~(\ref{alfev-disp2}) becomes
\begin{equation}
 s^3 +  \left( 1+\chi \right) \nu_{\rm ni} s^2 + k_z^2 v_{\rm A}^2 s + \nu_{\rm ni} k_z^2 v_{\rm A}^2 = 0. \label{eq:relalf1}
\end{equation}
with 
\begin{displaymath}
\chi = \frac{\rho_n}{\rho_i}, \qquad \nu_{\rm ni} = \frac{\alpha_{in}}{\rho \xi_n}.
\end{displaymath}
Equation~(\ref{eq:relalf1}) is a cubic equation and all its coefficients are real. From Equation~(\ref{eq:relalf1}) we  compute the polynomial discriminant, $\Lambda$, namely 
\begin{equation}
 \Lambda = - k_z^2 v_{\rm A}^2 \left[ 4\left(1+\chi \right)^3 \nu_{\rm ni}^4 - \left(  \chi^2 + 20\chi -8\right)\nu_{\rm ni}^2 k_z^2 v_{\rm A}^2 +4 k_z^4 v_{\rm A}^4\right],
  \label{eq:relalf2u}
\end{equation}
The discriminant, $\Lambda$, is defined so that (i) Equation~(\ref{eq:relalf1}) has one real zero and two complex conjugate zeros when $\Lambda <0$, (ii) Equation~(\ref{eq:relalf1}) has a multiple zero and all the zeros are real when $\Lambda = 0$, and (iii) Equation~(\ref{eq:relalf1}) has three distinct real zeros when $\Lambda >0$. This classification is very relevant because the complex zeros of Equation~(\ref{eq:relalf1}) result in damped oscillatory  solutions of Equation~(\ref{alfev-disp2}) whereas the real zeros of Equation~(\ref{eq:relalf1})  correspond to  evanescent solutions of Equation~(\ref{alfev-disp2}).

It is instructive to consider again the situation in which there are no collisions between the two fluids, so we set $\nu_{\rm ni}=0$. The discriminant becomes $\Lambda= -4k_z^6 v_{\rm A}^6<0$, which means that Equation~(\ref{eq:relalf1}) has one real zero and two complex conjugate zeros. Indeed, when $\nu_{\rm ni}=0$ the zeros of Equation~(\ref{eq:relalf1}) are
\begin{equation}
 s = \pm i k_z v_A, \qquad s=0, 
\end{equation}
which correspond to the following values of $\omega$,
\begin{equation}
 \omega = \pm k_z v_A, \qquad \omega=0. \label{eq:uncoupled}
\end{equation}
The two non-zero solutions correspond to the classic, ideal Alfv\'en waves, as expected, while the third solution vanishes.

We go back to the general case $\nu_{\rm ni} \neq 0$. To determine the location where the nature of the solutions changes we set $\Lambda = 0$ and find the corresponding relation between the various parameters. For given $\nu_{\rm ni}$ and $\chi$ we find two different values of $k_z$, denoted by $k_z^+$ and $k_z^-$, which satisfy $\Lambda = 0$, namely
\begin{equation}
k_z^\pm = \frac{\nu_{\rm ni}}{v_{\rm A}}\left[\frac{\chi^2+20\chi-8}{8\left( 1+\chi \right)^3} \pm \frac{\chi^{1/2} \left(\chi-8 \right)^{3/2}}{8\left( 1+\chi \right)^3}\right]^{-1/2}. \label{eq:rmasmenos}
\end{equation}
Since we assume that $k_z$ is real, Equation~(\ref{eq:rmasmenos}) imposes a condition on the minimum value of $\chi$ which allows $\Lambda = 0$. This minimum value is $\chi = 8$ and the corresponding critical $k_z$ is $k_z^+ = k_z^- = 3\sqrt{3} \nu_{\rm ni}/v_A$. When $\chi > 8$, Equation~(\ref{eq:rmasmenos}) gives $k_z^+ < k_z^-$. For $k_z$ outside the interval $(k_z^+,k_z^-)$ we have $\Lambda <0$ so that there are two damped Alfv\'en waves, while the remaining solution is evanescent.  For $k_z\in(k_z^+,k_z^-)$ we have $\Lambda >0$ and so all three zeros of Equation~(\ref{eq:relalf1}) are real, i.e., they correspond to purely imaginary solutions of Equation~(\ref{alfev-disp2}). There are no propagating waves for $k_z\in(k_z^+,k_z^-)$. We call this interval the cut-off region.  \citet{KP69} were the first to report on the existence of a cut-off region of wavenumbers for Alfv\'en waves in a partially ionized plasma, when studying the propagation of cosmic rays.

The physical reason for the existence of a range of cut-off wavenumbers in partially ionized plasmas can be understood as follows. When $k_z > k_z^-$ magnetic tension drives ions to oscillate almost freely, since the friction force is not strong enough to transfer significant inertia to neutrals. In this case, disturbances in the magnetic field affect only the ionized fluid as happens for classic Alfv\'en waves in fully ionized plasmas. Conversely, when $k_z < k_z^+$ the ion-neutral friction is efficient enough for neutrals to be nearly frozen into the magnetic field. After a perturbation, neutrals are dragged by ions almost instantly and both species oscillate together as a single fluid. The intermediate situation occurs when $k_z\in(k_z^+,k_z^-)$. In this case, a disturbance in the magnetic field decays due to friction before the ion-neutral coupling has had time to transfer the restoring properties of magnetic tension to the neutral fluid. In other words, neutral-ion collisions are efficient enough to dissipate  perturbations in the magnetic field but, on the contrary, they are not efficient enough to transfer significant inertia to neutrals before the magnetic field perturbations have decayed. Hence, oscillations of the magnetic field are suppressed when $k_z\in(k_z^+,k_z^-)$.

For wavenumbers outside the cut-off region, two of the solutions of the dispersion relation are complex with real and imaginary parts, which represent usual Alfv\'en waves damped by ion-neutral collisions. The third solution is purely imaginary and is  connected to the so-called vortex mode (with $Re(\omega)=0$) of neutral hydrogen fluid that damps through ion-neutral collisions. The vortex modes are solutions of fluid equations and they correspond to the fluid vorticity. The vortex modes have zero frequency in the ideal fluid, but may gain a purely imaginary frequency if dissipative processes are evolved. The reason that the vortex mode is a purely imaginary solution is that, for incompressible perturbations, there is no restoring force in the neutral fluid. Consequently,  vortex modes are unable to propagate in the form of waves. 

\begin{figure}[t]
\vspace*{1mm}
\begin{center}
\includegraphics[width=9.5cm]{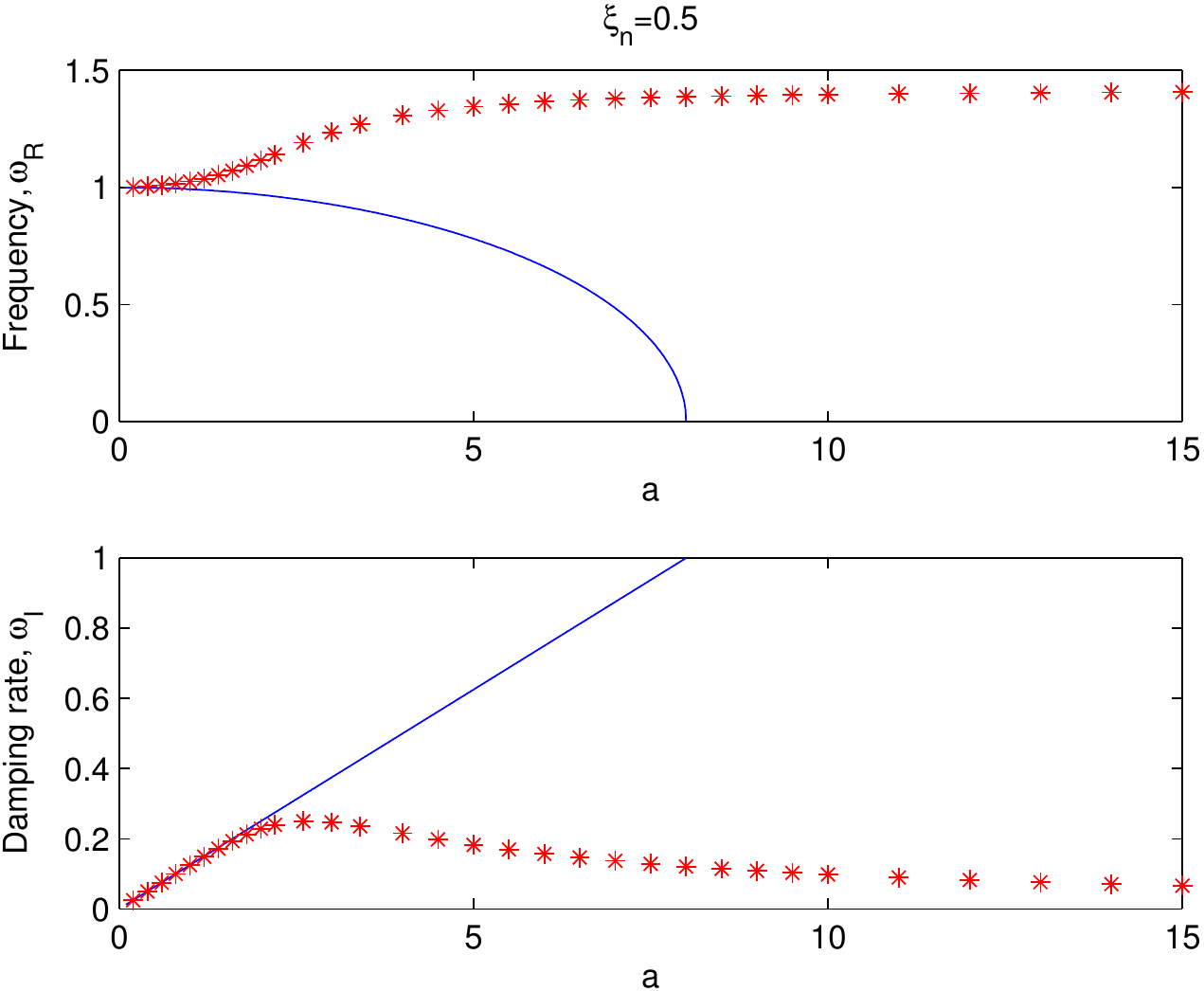}
\end{center}
\caption{Real ($\omega_R$) and imaginary ($\omega_I$) parts of
Alfv\'en wave frequency normalized by $v_A k_z \sqrt{\xi_i}$ vs $a={{k_z
v_A \rho \sqrt{\xi_i}}/ {\alpha_{in}}}$. The blue line corresponds to the solution of
single-fluid dispersion relation, i.e. Eq. (\ref{alfev-disp1}) and
red asterisks are the solutions of two-fluid dispersion relation,
Eq. (\ref{alfev-disp2}). The values are calculated for 50\% of
neutral hydrogen, $\xi_n$=0.5. Adapted from Zaqarashvili et al. (2011a).}
\label{fig:alf}
\end{figure}

Now we may go a step further and derive the single-fluid MHD equations (see sect.~\ref{Bsec:2-fluid}). Comparing the two-fluid and single-fluid results can shed light on the accuracy of the single-fluid model. We define the total velocity (i.e. velocity of center of mass)
\begin{equation}
\vec u_{\perp}= {{\rho_{i}\vec u_{i\perp}+\rho_{n}\vec u_{n\perp}}\over
{\rho}},\,\
\end{equation}
and relative velocity
\begin{equation}\label{w}
\vec w_{\perp}= \vec u_{i\perp} - \vec u_{n\perp}.
\end{equation}
Then, Eqs. (\ref{alfen-viy}-\ref{alfven-by}) are rewritten as:
\begin{equation}\label{V1}
\rho{{\partial \vec u_{\perp}}\over {\partial t}}={B_z\over {\mu}}{{\partial \vec b_{\perp}}\over {\partial z}},
\end{equation}
\begin{equation}\label{w1}
{{\partial \vec w_{\perp}}\over {\partial t}}={B_z\over {\mu \rho_i}}{{\partial \vec b_{\perp}}\over {\partial z}} - \alpha_{in}\left (\frac{1}{\rho_i}+\frac{1}{\rho_n} \right )\vec w_{\perp},
\end{equation}
\begin{equation}\label{B1}
{{\partial \vec b_{\perp}}\over {\partial t}}=B_z{{\partial \vec u_{\perp}}\over {\partial z}}+\xi_n B_z{{\partial \vec w_{\perp}}\over {\partial z}}-\frac{1}{\mu en_e}\vec{\nabla}_{\perp}\times\left [\left (\vec{\nabla}\times\vec{b}_{\perp}\right )\times\vec{B} \right ].
\end{equation}
The single-fluid linearized Hall MHD equations are obtained from Eqs. (\ref{V1}-\ref{B1}) as follows.
The inertial term (the left hand-side term in Eq. \ref{w1}) is neglected, which is a good approximation for time scales longer than ion-neutral collision time, but fails for the shorter time scales (Zaqarashvili et al. 2011a). Then $\vec w_{\perp}$ defined from Eq. (\ref{w1}) is substituted into Eq. (\ref{B1}) and one can obtain the linearized Hall MHD equations
\begin{equation}\label{V2}
\rho{{\partial \vec u_{\perp}}\over {\partial t}}={B_z\over {\mu}}{{\partial \vec b_{\perp}}\over {\partial z}},
\end{equation}
\begin{equation}\label{B2}
{{\partial \vec b_{\perp}}\over {\partial t}}=B_z{{\partial \vec u_{\perp}}\over {\partial z}}+\frac{\xi^2_n B^2_z}{\mu \alpha_{in}}{{\partial^2 \vec{b}_{\perp}}\over {\partial z^2}}-\frac{1}{\mu en_e}\vec{\nabla}_{\perp}\times\left [\left (\vec{\nabla}\times\vec{b}_{\perp}\right )\times\vec{B} \right ].
\end{equation}
The usual single-fluid MHD equations, which are widely used for description of Alfv\'en waves in partially ionized plasmas, are obtained from Eqs. (\ref{V2}-\ref{B2}) after neglecting the Hall term in Eq. (\ref{B2})
\begin{equation}\label{V3}
\rho{{\partial \vec u_{\perp}}\over {\partial t}}={B_z\over {\mu}}{{\partial \vec b_{\perp}}\over {\partial z}},
\end{equation}
\begin{equation}\label{B3}
{{\partial \vec b_{\perp}}\over {\partial t}}=B_z{{\partial \vec u_{\perp}}\over {\partial z}}+\frac{\xi^2_n B^2_z}{\mu \alpha_{in}}{{\partial^2 \vec{b}_{\perp}}\over {\partial z^2}}.
\end{equation}
As before, Fourier analysis with $exp[i(k_z z -\omega t)]$ gives the dispersion relation
\begin{equation}\label{alfev-disp1}
\omega^2 +i \frac{v^2_A k^2_z \rho \xi_i \xi^2_n }{\alpha_{in}}\omega - v^2_A k^2_z\xi_i=0.
\end{equation}
This is the usual dispersion relation of Alfv\'en waves, which is obtained in single fluid partially ionized plasmas \citep{haerendel1992, depontieu01}. For $v_A k_z \rho \sqrt{\xi_i} \xi^2_n/{\alpha_{in}}<2$, it gives the damping rate
\begin{equation}\label{sol-alfev}
2{\omega_i}={{\xi^2_n B^2_z}\over {\mu \alpha_{in}}}k^2_z
\end{equation}
in full agreement with previous works (e.g. Braginskii 1965, etc.).  In order to understand the wave damping due to ion-neutral
collision, we rewrite Eq. (\ref{sol-alfev}) as follows
\begin{equation}\label{sol-alfev1}
\frac{\omega_i}{k_z v_A}=\frac{k_z v_A}{\nu_{in}}\frac{\xi^2_n}{2}.
\end{equation}
This expression clearly indicates that the normalized
damping rate depends on the ratio of Alfv\'en and ion-neutral collision
frequencies and plasma ionization rate.
The earlier statement that the damping rate depends on the magnetic
field strength can be translated as follows: the increase in
the magnetic field leads to an increase in the Alfv\'en speed, therefore
the waves with the same wavenumber have a higher frequency,
which is closer to ion-neutral collision frequency, so the
normalized damping rate increases. In fact, it is the ratio of neutral
and total (proton+hydrogen) fluid densities ($\xi_n$) that determines the damping rate of
a particular wave harmonic. High-frequency waves are damped
quickly. However, this statement is valid for a low-frequency
wave spectrum below the ion-neutral collision frequency since the single-fluid model has been adopted.

On the
other hand, the condition $v_A k_z \rho \sqrt{\xi_i} \xi^2_n/{\alpha_{in}}<2$ in Eq. (\ref{sol-alfev})
retains only the imaginary part, which gives the cut-off wave number
\begin{equation}\label{cut-off}
k_{c}={{2 \alpha_{in}}\over {\rho \xi^2_n v_A \sqrt{\xi_i}}},
\end{equation}
which means that the waves with higher wave number than ${k_{c}}$ are evanescent. We note that, of the two cutoff wavenumbers that appear in the two-fluid description, only the lowest one, i.e., $k_z^+$, remains in the single-fluid approximation. Also, the threshold value $\chi = 8$ is absent in the single-fluid approximation. The second propagating window that is present in the two-fluid model when $k_z > k_z^-$ is absent from the single-fluid model. This result points out that the single-fluid model breaks down at  very small scales.

\begin{figure}[t]
\vspace*{1mm}
\begin{center}
\includegraphics[width=9.5cm]{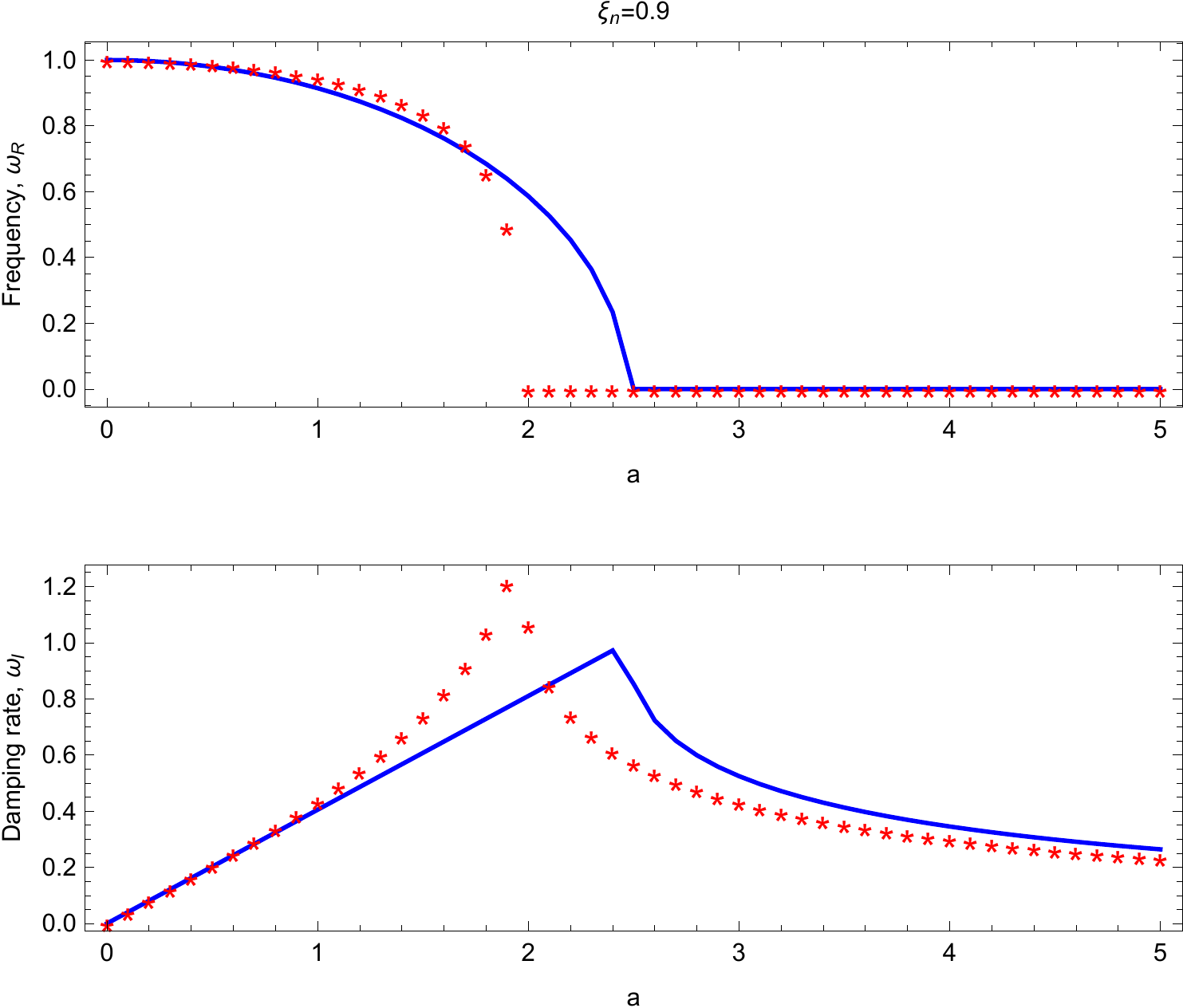}
\end{center}
\caption{Same as in Figure~\ref{fig:alf} but with $\xi_n$=0.9. Adapted from Zaqarashvili et al. (2011a).}
\label{fig:alf1}
\end{figure}

Figure~\ref{fig:alf} displays the solutions of the single-fluid (Eq.
\ref{alfev-disp1}, blue lines) and two-fluid (Eq. \ref{alfev-disp2},
red asterisks) dispersion relations for $\xi_n=0.5$. We see that the
frequencies and damping rates of Alfv\'en waves are same in
single-fluid and two-fluid approaches for the low-frequency branch of
the spectrum (small $a={{k_z
v_A \rho \sqrt{\xi_i}}/ {\alpha_{in}}}$). But, as expected, the behaviour is dramatically changed
when the wave frequency becomes comparable to or higher than the
ion-neutral collision frequency, $\nu_{in}$, i.e. for $a>1$. The
damping time linearly increases with $a$  and the wave frequency
becomes zero at some point in single-fluid case (blue lines). The
point where the wave frequency becomes zero corresponds to the single-fluid
cut-off wave number ${k_{c}}$. However, for the parameters used in Figure~\ref{fig:alf} there is no cut-off wave number \citep{zaqarashvili2012} in solutions of the two-fluid dispersion relation: Eq. (\ref{alfev-disp2}) always has a solution with a real part. But the situation is changed when lower values of the ionization degree are considered, i.e., for $\chi  > 8$, for which the two-fluid cutoff wavenumbers appear (see Soler et al. (2013, 2015) and the subsection 4.3.1). Figure~\ref{fig:alf1} shows equivalent results to those displayed in Figure 1 but with $\xi_n = 0.9$. In this case, the cutoff appears in both single-fluid and two-fluid cases, and the single-fluid results are now much more accurate than for $\xi_n = 0.5$. 

Damping rates derived in the two-fluid approach reach
a peak near the ion-neutral collision frequency and then decrease
for higher frequencies unlike the single-fluid approach,
where the damping increases linearly with increasing frequency
(Fig.~\ref{fig:alf}). This result can be checked analytically. To do so, we look for approximate analytic solutions to the two-fluid dispersion relation (Equation~(\ref{alfev-disp2})) assuming that $k_z$ is outside the cut-off interval $(k_z^+,k_z^-)$, so that there are two complex solutions and one purely imaginary solution.  First we look for an approximate expression for the two oscillatory solutions. To do so we write $\omega = \omega_{\rm R} + i \omega_{\rm I}$ and insert this expression in Equation~(\ref{alfev-disp2}). We assume $|\omega_{\rm I}|\ll |\omega_{\rm R}|$ and neglect terms with $\omega_{\rm I}^2$ and higher powers. After some algebraic manipulations we derive approximate expressions for $\omega_{\rm R}$ and $\omega_{\rm I}$, namely
\begin{eqnarray}
 \omega_{\rm R} &\approx &  \pm k_z v_{\rm A} \sqrt{\frac{k_z^2 v_{\rm A}^2+\left( 1 +\chi \right)\nu_{ni}^2}{ k_z^2 v_{\rm A}^2+\left( 1 +\chi \right)^2\nu_{ni}^2}}, \label{eq:wr} \\
\omega_{\rm I}&\approx &   -\frac{\chi \nu_{ni}}{2\left[  k_z^2 v_{\rm A}^2 +\left( 1+\chi \right)^2 \nu_{ni}^2\right]}  k_z^2 v_{\rm A}^2. \label{eq:wi}
\end{eqnarray}
On the other hand, the remaining purely imaginary, i.e., evanescent, solution is $\omega = i \epsilon$, with the approximation to $\epsilon$ given by
\begin{equation}
 \epsilon \approx - \nu_{ni} \frac{k_z^2 v_{\rm A}^2+\left( 1 +\chi \right)^2 \nu_{in}^2}{k_z^2 v_{\rm A}^2+\left( 1 +\chi \right)\nu_{in}^2}. \label{eq:gamma}
\end{equation}
When $\nu_{in}=0$, we find $\omega_{\rm R} =  \pm k_z v_{\rm A}$, $\omega_{\rm I} = 0$ and $\epsilon=0$, hence we recover the solutions in the ideal case. It is useful to investigate the behavior of the solutions in the various limits of $\nu_{in}$. First we consider the limit $\nu_{in} \ll k_z v_{\rm A}$, i.e., the case of low collision frequency, which means that the coupling between fluids is weak. Equations~(\ref{eq:wr}) and (\ref{eq:gamma}) simplify to
\begin{eqnarray}
 \omega_{\rm R} &\approx &  \pm k_z v_{\rm A}, \label{eq:wrlow} \\
\omega_{\rm I}&\approx &   -\frac{\chi \nu_{in}}{2}, \label{eq:wilow} \\
\epsilon & \approx & -\nu_{in}.
\end{eqnarray}
In this limit $\omega_{\rm R}$ coincides with its value in the ideal case and $\omega_{\rm I}$ is independent of $k_z$. Hence, the the damping of Alfv\'en waves does not depend on the wavenumber. On the other hand, when $\nu_{in} \gg k_z v_{\rm A}$, i.e., the case of strong coupling between fluids,  we find
\begin{eqnarray}
 \omega_{\rm R} &\approx &  \pm \frac{k_z v_{\rm A}}{\sqrt{1+\chi}}, \label{eq:wrhigh} \\
\omega_{\rm I}&\approx &   -\frac{\chi}{2\left( 1+\chi \right)^2 } \frac{k_z^2 v_{\rm A}^2 }{\nu_{in}}, \label{eq:wihigh} \\
\epsilon & \approx & -(1+\chi)\nu_{ni}.
\end{eqnarray}
Now the expression of $\omega_{\rm R}$ involves the factor $\sqrt{1+\chi}$ in the denominator, so that the larger the amount of neutrals, the lower $\omega_{\rm R}$ compared to the value in the fully ionized case. Now $\omega_{\rm I}$ is proportional to $k_z^2$, meaning that the shorter the wavelength, the more efficient damping. In this last limit, the two-fluid results agree with those obtained in the single-fluid model.

Finally, Eqs. (\ref{V2}-\ref{B2}) give the dispersion relation of Alfv\'en waves in Hall MHD, which has an expression (Pandey \& Wardle 2008, Zaqarashvili et al. 2012)
\begin{equation}\label{dispi1}
\omega^2 + \left [ i\frac{\xi^2_n B^2_z k^2_z}{\mu \alpha_{in}} \pm \frac{v^2_A k^2_z}{\Omega_{cp}} \right ]\omega - v^2_A k^2_z\xi_i=0,
\end{equation}
where $\Omega_{cp} = eB/m_{\rm p}$ is the proton cyclotron frequency. Contrary to the case in which Hall's term is not included, the solutions of Equation~(\ref{dispi1}) always have a real part. Therefore the single-fluid Hall MHD approach does not include a cut-off wavenumber for Alfv\'en waves. Multi-fluid results obtained by Soler et al. (2015) also show that the strict two-fluid cut-offs disappear when Hall's term is included. However, as discussed in Soler et al. (2015), the absence of strict cut-offs makes no practical difference concerning the behavior of the waves. The presence of Hall's current and electron inertia cause the strict frequency cut-offs  to be replaced by zones where Alfv\'en waves are overdamped, i.e., the waves are unable to propagate. In the absence of Hall's current and electron inertia, electrons can be considered as tightly coupled to ions, in the sense that electrons just follow the behavior of ions. Both ions and electrons are frozen into the magnetic field. In this case, ion-neutral collisions can completely suppress the magnetic field perturbations and so cause the strict cut-offs. However, when either Hall's current or electron inertia are included, electrons can have a different dynamics than that of ions. Ions may not be able follow the magnetic field fluctuations due to the effect of ion-neutral collisions, but it is easier for electrons to remain coupled to magnetic field, i.e., magnetized. Therefore, ion-neutral collisions cannot completely suppress the fluid oscillations because of the distinct behavior of electrons when Hall's current and/or electron inertia are included (see also the discussion in Pandey \& Wardle 2008).

\subsection{Magneto-acoustic waves}

Now we move to the case of compressible magneto-acoustic waves. As before, we consider the unperturbed magnetic field, $B_z$, directed along the $z$-axis and perform a Fourier analysis as $\exp[-i\omega t+ik_x x +ik_y y + ik_z z]$. We define $\Delta_i$ and $\Delta_n$ as the compressibility perturbations of the electron-ion and the neutral fluids, respectively,
\begin{eqnarray}\label{magneto-acoustic1}
\Delta_i &=& \vec \nabla \cdot \vec u_i=i k_x u_{ix}+i k_y u_{iy}+i k_z u_{iz},\\
\Delta_n &=& \vec \nabla \cdot \vec u_n=i k_x u_{nx}+i k_y u_{ny}+i k_z u_{nz}.
\end{eqnarray}
We combine Eqs. (\ref{waves1})-(\ref{waves5}) and after some algebraic manipulations
we obtain the two following coupled equations involving
$\Delta_i$ and $\Delta_n$ only, namely (Soler et al. 2013b)
$$
(\omega^4-k^2(v^2_A+c^2_i)\omega^2+k^2k^2_zv^2_Ac^2_i)\Delta_i=-i\nu_{in}\omega^3(\Delta_i-\Delta_n)+
$$
\begin{equation}\label{magneto-acoustic2}
+\frac{i\nu_{in}}{\omega+i(\nu_{in}+\nu_{ni})}k^2k^2_zv^2_A(c^2_i\Delta_i-c^2_n\Delta_n),
\end{equation}
\begin{equation}\label{magneto-acoustic3}
(\omega^2-k^2c^2_n)\Delta_n=-i\nu_{ni}\omega(\Delta_n-\Delta_i),
\end{equation}
where $k^2=k^2_x+k^2_y+k^2_z$ and $c_i$ ($c_n$) is ion-electron (neutral hydrogen) fluid acoustic speed defined as follows
\begin{equation}\label{acoustic}
c^2_i=\frac{\gamma P_i}{\rho_i},\,\,c^2_n=\frac{\gamma P_n}{\rho_n}.
\end{equation}
Eqs. (\ref{magneto-acoustic2})-(\ref{magneto-acoustic3}) give the dispersion relation of magneto-acoustic waves
$$
[(\omega^4+i\nu_{in}\omega^3-k^2(v^2_A+c^2_i)\omega^2)(\omega+i(\nu_{in}+\nu_{ni}))+
$$
$$
k^2k^2_zv^2_Ac^2_i(\omega+i\nu_{ni})](\omega^2-k^2 c^2_n+i\nu_{ni}\omega)+
$$
\begin{equation}\label{magneto-acoustic-disp}
+\nu_{in}\nu_{ni}\omega [\omega^3 (\omega+i(\nu_{in}+\nu_{ni})) - k^2k^2_zv^2_Ac^2_n]=0.
\end{equation}
The dispersion relation Eq. (\ref{magneto-acoustic-disp}) is a seventh order equation with
$\omega$, therefore it has seven different solutions. If we neglect the collision between neutrals and protons, then the dispersion relation is transformed into the expression
\begin{equation}\label{magneto-acoustic-disp1}
\omega(\omega^4-k^2(v^2_A+c^2_i)\omega^2 + k^2k^2_zv^2_Ac^2_i)
(\omega^2-k^2 c^2_n)=0.
\end{equation}
Here we see that the seven different modes are four magneto-acoustic modes of ion-electron fluid, two magneto-acoustic modes of neutral hydrogen and one $\omega=0$ solution, which is associated with the so-called entropy mode (Soler et al. 2013b). As expected, in the absence of ion-neutral collisions we consistently recover the classic (forward and backward) magnetoacoustic waves in the ion-electron fluid and the classic (forward and backward) acoustic waves in the neutral fluid. Therefore, three distinct waves are present in the uncoupled, collisionless case, in addition to the non-propagating entropy mode. These modes do not interact and are undamped in the absence of collisions.

Conversely to the uncoupled case, the strongly coupled  limit represents the situation in which ion-electrons and neutrals  behave as one fluid. To study this case, we take the limits $\nu_{in} \to \infty$ and $\nu_{ni} \to \infty$ in Equations~(\ref{magneto-acoustic2}) and (\ref{magneto-acoustic3}). We realize that, if $\omega \neq 0$, it is necessary that $\Delta_i = \Delta_n$ for the equations to remain finite. This is equivalent to assume that the two fluids move as a whole. Then,  when $\nu_{in} \to \infty$ and $\nu_{ni} \to \infty$, we find a single equation for the compressibility perturbations, namely
\begin{equation}
 \omega^3 \left( \omega^4 - \omega^2 k^2 \frac{v_A^2 + c_i^2 + \chi c_n^2}{1+\chi} + k^4 \frac{v_A^2 \left( c_i^2 +\chi c_n^2\right)}{\left( 1+\chi \right)^2} \cos^2\theta \right) \Delta_{\rm i,n} = 0,  \label{eq:high}
\end{equation} 
where we use $\Delta_{\rm i,n}$ to represent either $\Delta_i$ or $\Delta_n$ and  $\theta$ is the angle that forms the wavevector, $\bf k$, with the equilibrium magnetic field, $\bf B$. For nonzero $\Delta_{\rm i,n}$, the solutions to Equation~(\ref{eq:high}) must satisfy
\begin{equation}
 \omega^3\left(\omega^4 - \omega^2 k^2 \frac{v_A^2 + c_i^2 + \chi c_n^2}{1+\chi} + k^4 \frac{v_A^2 \left( c_i^2 +\chi c_n^2\right)}{\left( 1+\chi \right)^2} \cos^2\theta  \right) = 0.  \label{eq:reldisperhigh}
\end{equation} 
This is the wave dispersion relation in the strongly coupled case. Its solutions are, on the one hand,
\begin{equation}
\omega^2 = k^2 \frac{v_A^2+c_i^2+\chi c_n^2}{2(1+\chi)} \pm k^2\frac{v_A^2+c_i^2+\chi c_n^2}{2(1+\chi)} \left[ 1 - \frac{4v_A^2(c_i^2+\chi c_n^2)\cos^2\theta}{\left(v_A^2 + c_i^2+\chi c_n^2\right)^2} \right]^{1/2}, \label{eq:magnetolim}
\end{equation}
where the $+$ sign is for the modified (forward and backward) fast wave and the $-$ sign is for the modified (forward and backward) slow wave. We use the adjective `modified' to stress that these waves are the counterparts of the classic fast and slow modes but modified by the presence of the neutral fluid. On the other hand, $ \omega^3 = 0$ is also a solution. 

The first important difference between the uncoupled and strongly coupled cases is in the number of independent solutions. In the uncoupled case there are three distinct waves, namely the classic slow and fast magnetoacoustic waves and the neutral acoustic wave, plus a non-propagating mode. However, in the strongly coupled  limit we only find two propagating waves, i.e., the modified version of the slow and fast magnetoacoustic modes, plus three non-propagating modes. The modified counterpart of the classic neutral acoustic wave is apparently absent, although the truth is that the neutral acoustic wave has become a non-propagating solution. The classic neutral acoustic mode and the classic ion-electron magneto-acoustic modes interact heavily when ion-neutral collisions are at work. The two resulting modes in the strongly coupled regime (modified fast mode and modified slow mode) have mixed properties in general and are affected by the physical conditions in the two fluids. The degree to which the properties of the classic waves are present in the resulting waves depends on the relative values of the Alfv\'en and sound velocities and on the ionization fraction of the plasma.
In addition, the wave frequencies in the strongly coupled limit are real as in the uncoupled case. This means that the waves are undamped in the limit of high collision frequencies as well. The damping of magnetoacoustic waves due to ion-neutral collisions takes place for intermediate collision frequencies.

\begin{figure*}
	\centering
	\includegraphics[width=.49\columnwidth]{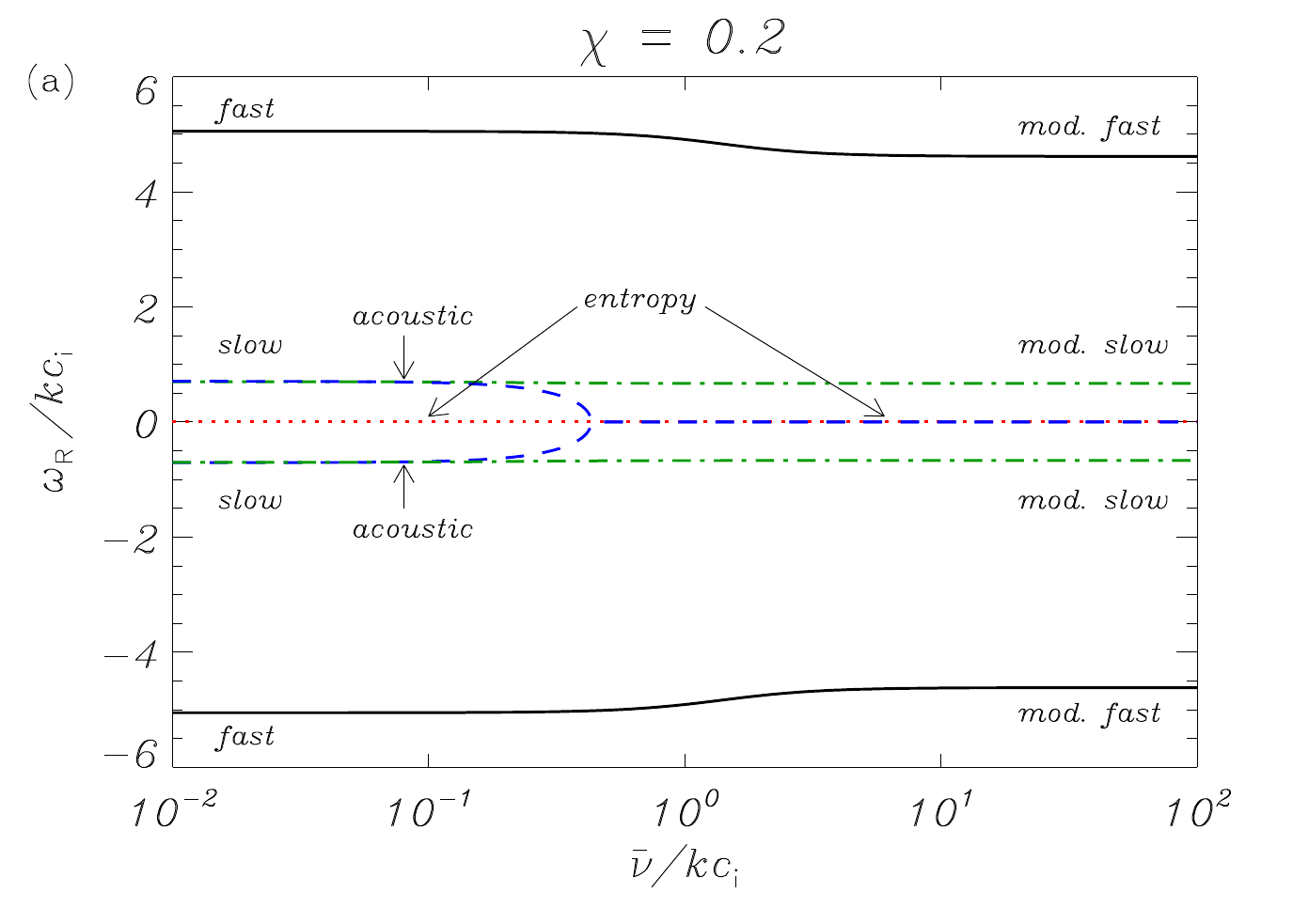}
	\includegraphics[width=.49\columnwidth]{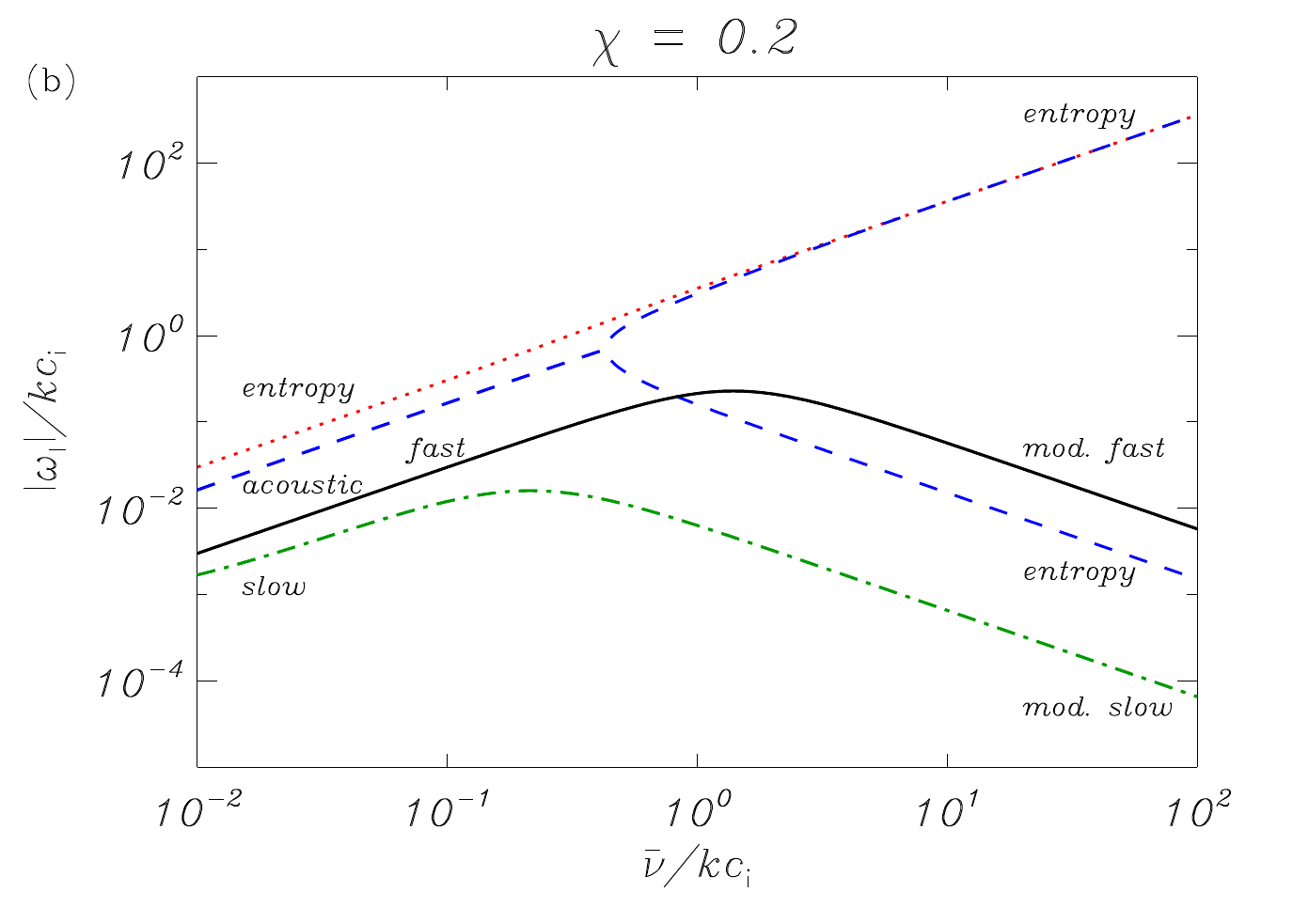}
	\includegraphics[width=.49\columnwidth]{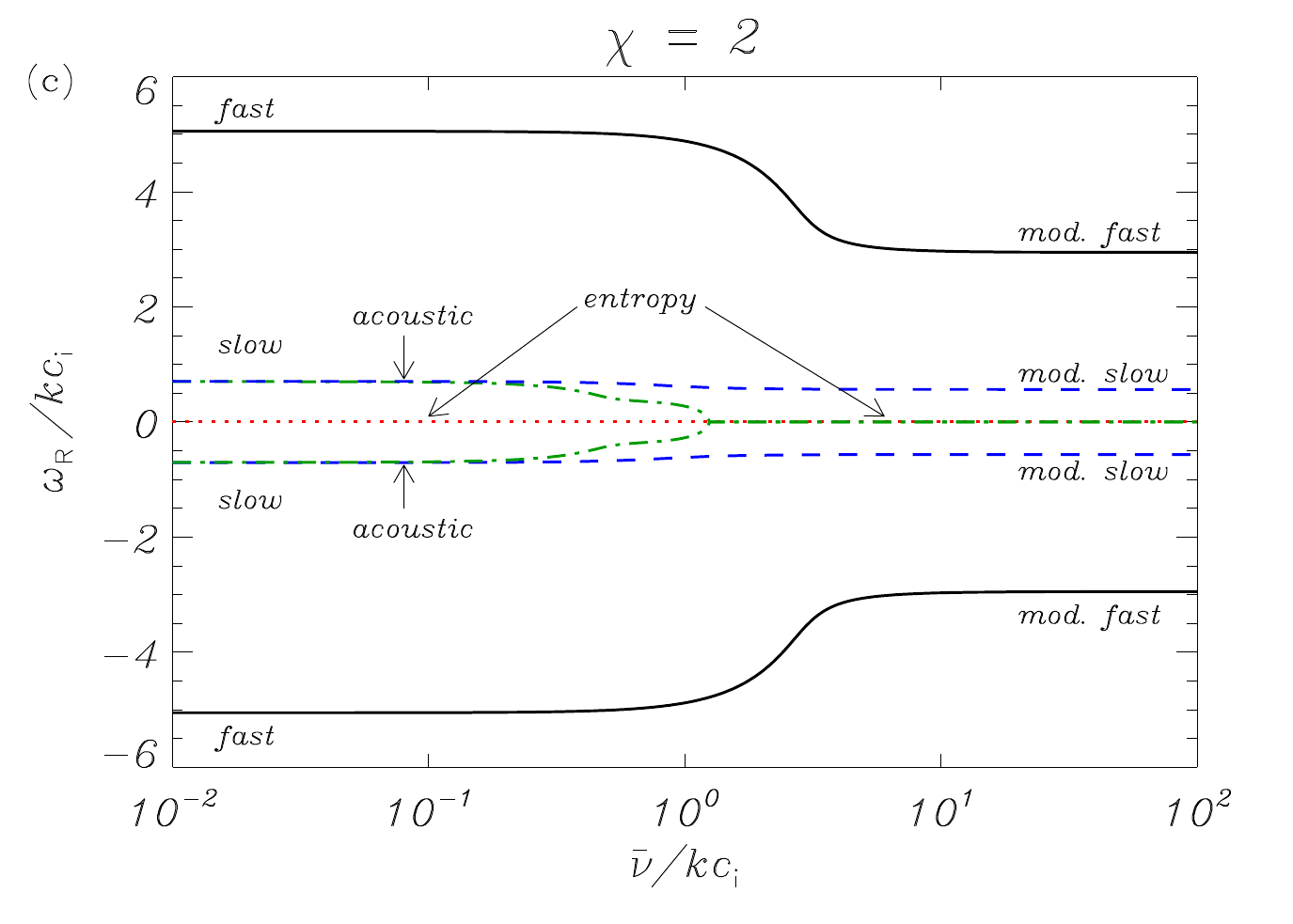}
	\includegraphics[width=.49\columnwidth]{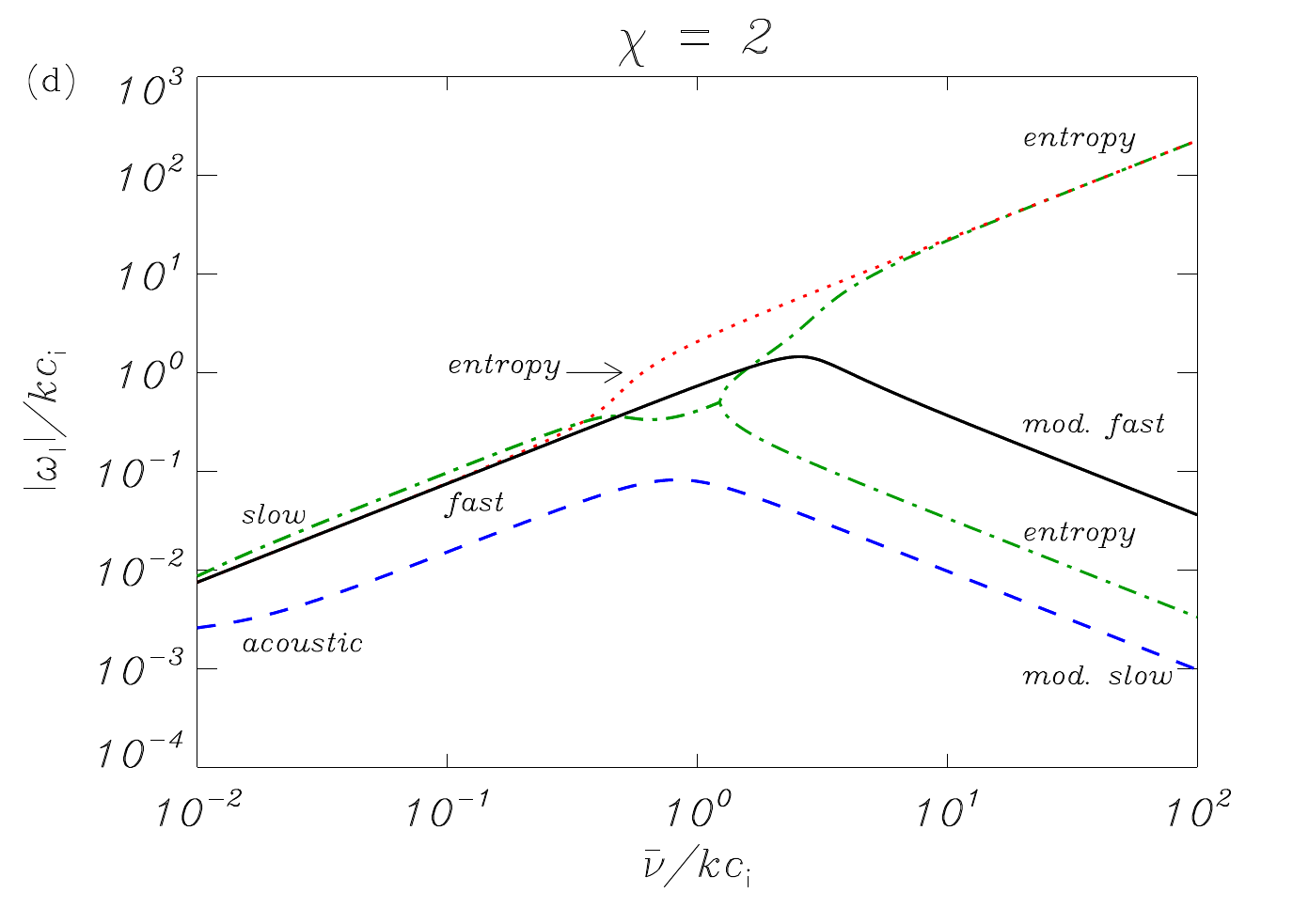}
	\includegraphics[width=.49\columnwidth]{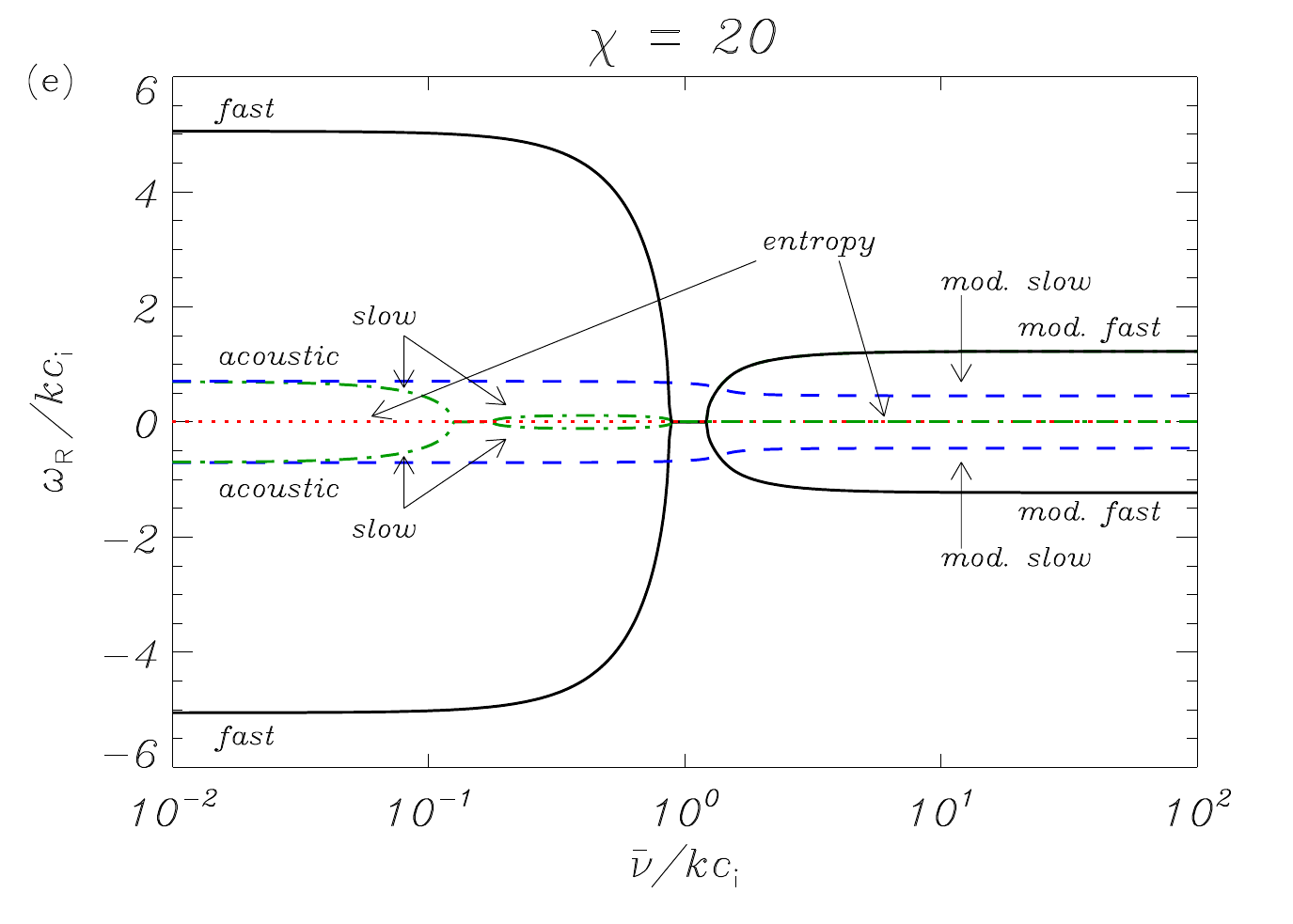}
	\includegraphics[width=.49\columnwidth]{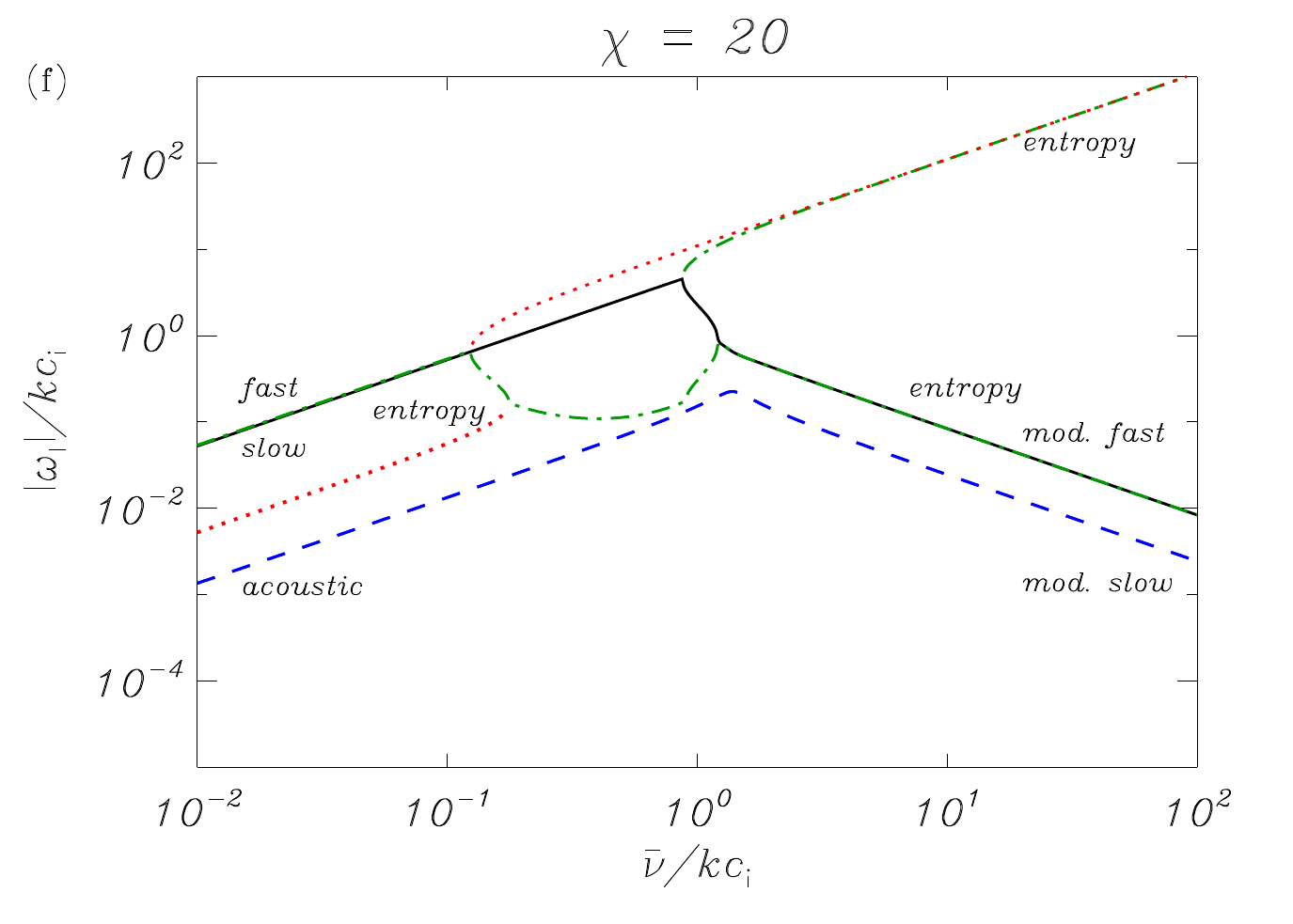}
	\caption{Real (left) and imaginary (right) parts of the frequency of the various waves versus the averaged collision frequency (in logarithmic scale) for oblique propagation to the magnetic field, i.e., $\theta=\pi/4$, in a strongly magnetized plasma with $c_{\rm i}^2 = 0.04 c_A^2$. Panels (a)--(b) are for $\chi=0.2$, panels (c)--(b) for $\chi = 2$, and panels (c)--(b) for $\chi = 20$. All frequencies are expressed in units of $k c_i$. Note that the absolute value of $\omega_{\rm I}$ is plotted. Adapted from Soler et al. (2013b).}
	\label{fig:ma}
\end{figure*}

Parameter studies of the solution to the general dispersion relation (Eq. (\ref{magneto-acoustic-disp})) were performed in Soler et al. (2013b). The ion-neutral coupling strongly affects the  behavior and properties of the waves. There is a large number of possible scenarios for wave propagation depending on the plasma physical properties. The various waves supported by a partially ionized plasma display complex interactions and couplings that are not present in the fully ionized case (see also Mouschovias et al. (2011)). Among these complex interactions, the presence of cut-offs and forbidden intervals has a strong impact  since the allowed wavelengths of the propagating modes become heavily constrained. An example of this complex coupling is seen in Figure~\ref{fig:ma}, which shows the dispersion diagram of obliquely propagating waves in  a strongly magnetized plasma as a function of the normalised ion-neutral collision frequency, $\bar{\nu}=2\alpha_{in}/(\rho_i + \rho_n)$, and different values of the ionization ratio, $\chi$. 

\subsection{General remarks}

Investigation of the  modes present in two-fluid and single-fluid models reveals the complexity that ion-neutral collisions impose on the behavior of the waves. Several general conclusions can be drawn. First, the efficiency of the damping, and so that of the wave energy dissipation, is determined by the relative values of the wave frequency and the collision frequency. The damping is most efficient when these two frequencies are of the same order of magnitude. Second, dissipation can be so strong for some parameter combinations that  wave propagation is suppressed. This leads to the presence of cut-offs and forbidden windows in the dispersion diagrams. Third, the number of possible waves that exist in the partially ionized plasma is determined by the strength of the coupling between species. Hence, the neutral fluid may be able to sustain its own acoustic waves when the coupling is weak, but the neutral waves and the ion-electron waves become unavoidably entangled when the coupling between fluids is strong.

Apart from being of academic interest, the theoretical properties of waves in partially ionized plasmas reveal interesting features that may be important for applications in real astrophysical environments. For instance, the deposition of wave energy associated to ion-neutral damping may be necessary to explain plasma heating in the solar chromosphere (see sect.~\ref{heatchro}).

%%%%%%%%%%%%%%%%%%%%%%%%%%%%%
\newpage
%%%%%%%%%%%%%%%%%%%%%%%%%%%%%
\section{Applications to Astrophysical Environments}
\label{AFenv}
%%%%%%%%%%%%%%%%%%%%%%%%%%%%%%
\subsection{Solar atmosphere}
\label{solat}
\subsubsection{Introduction}
\label{introsolphys}
%Add references missing

 Coupling between ions and neutrals in magnetized plasmas is of great importance in many aspects of heliophysics. The plasma is weakly ionized in the photosphere/lower chromosphere, but temperature increase with height leads to the ionization of neutral atoms, which become fully ionized in the solar corona. Therefore, the transition between partially and fully ionized plasma states occurs near the transition region with sharp temperature gradient. Usually, the ideal MHD approximation is used to study the dynamical processes taking place in magnetized solar plasmas from the photosphere to the corona. However, in some layers and structures of the solar atmosphere, the effects of partial ionization are of particular importance. For instance, in the lower solar atmosphere where the temperature is relatively cool, the degree of ionization is very small. This fact, together with the decrease in the collision frequency with height, produces conditions under which the assumptions supporting the ideal MHD approximation are not valid. Therefore, in these layers other additional effects such as the Hall effect and ambipolar diffusion must be taken into account, and could produce interesting effects related, for instance, to chromospheric heating \citep{khomenko12}. Furthermore, it has been also shown that the single-fluid MHD approximation fails in some circumstances, therefore multi-fluid theory must be adopted. Another region in the solar atmosphere where ion-neutral coupling needs to be considered is in the physics of solar prominences, which are dense cold plasma clouds supported by coronal magnetic field against gravity, where cross-field diffusion of neutrals may play a critical role in determining the cross-field structure and mass variation of prominences (terradas15). Recently, high-resolution observations made by the HINODE satellite have revealed a lot of dynamical processes taking place in limb prominences\citep{berger08}. Most of these dynamical features have been interpreted and modeled in terms of different instabilities (Rayleigh-Taylor, Kelvin-Helmholtz, etc.) which develop in the prominence plasma considered as fully ionized. However, because of low temperature, prominence plasmas are PIP, with physical properties similar to those of chromosphere, and the study of thermal, Rayleigh-Taylor and Kelvin-Helmholtz instabilities in PIP has been already started \citep{soler12KHI, diaz12, khomenko14} with the aim of determining the modifications in the instability thresholds and growth rates produced by partial ionization effects. For all these reasons, the solar atmosphere is an excellent laboratory to study the effects of partial ionization on plasma dynamics.

\subsubsection{Ionization, collisions and magnetization in the lower solar atmosphere}

\begin{figure}
\center
  \includegraphics[width=0.9\columnwidth]{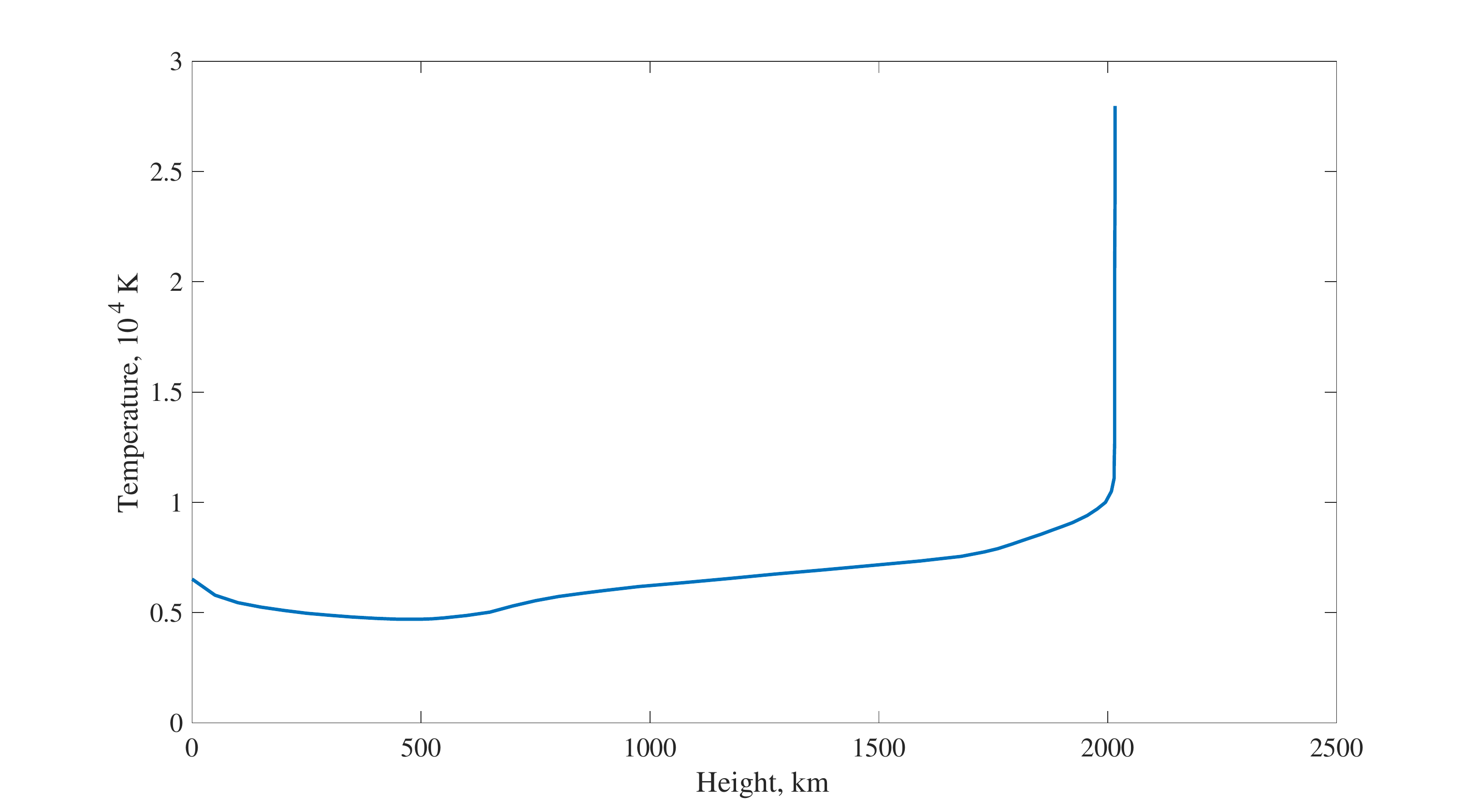}
  \includegraphics[width=0.9\columnwidth]{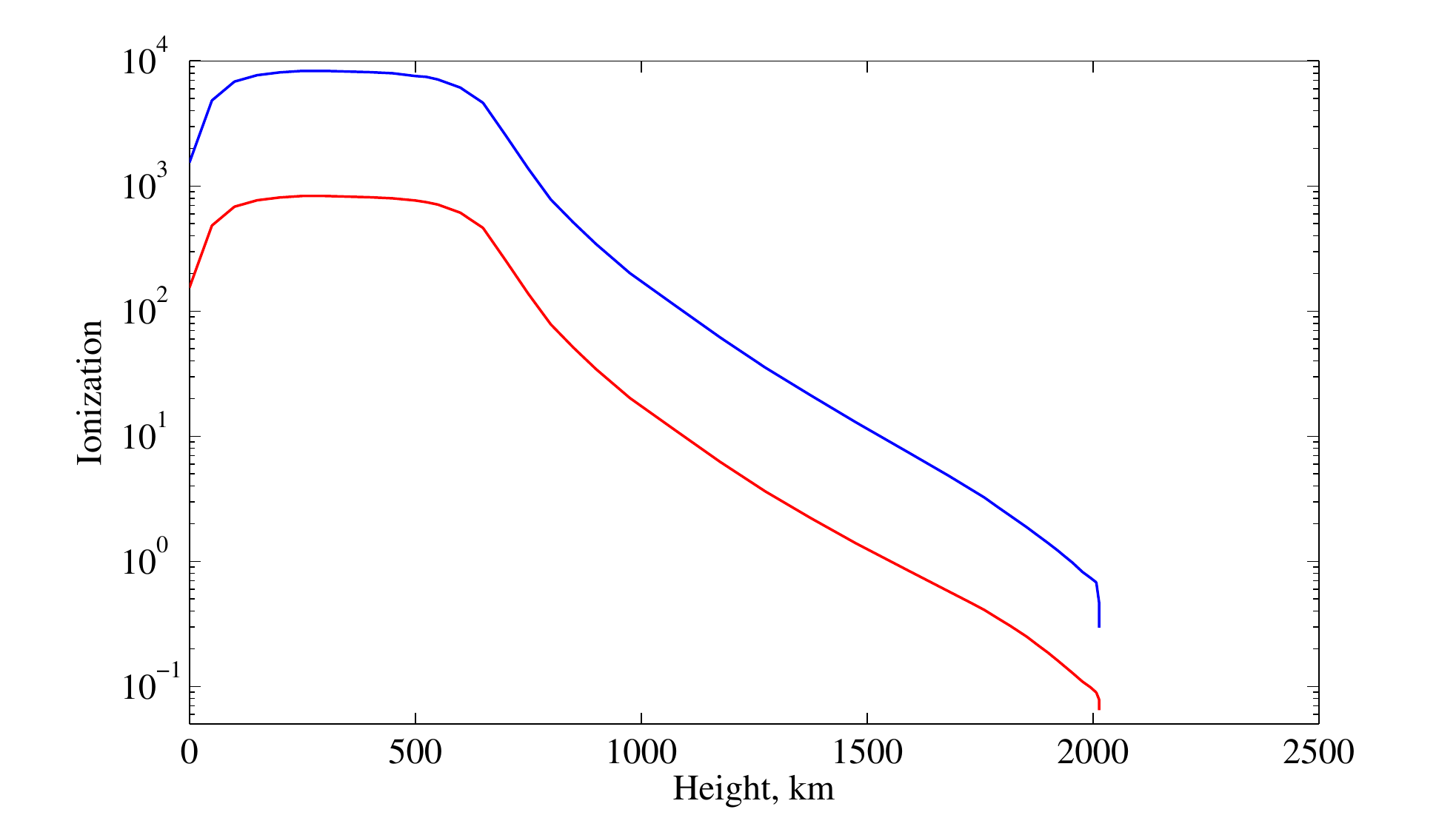}
\caption{Height dependence of atmospheric parameters according to FAL93-F model \citep{fontenla93}. Upper panel: the plasma temperature.
Lower panel: blue (red) solid line -– the ratio of neutral hydrogen (neutral helium) and electron number densities.}
\label{fig:falF}
\end{figure}

Figure~\ref{fig:falF} shows the dependence of plasma parameters on height according to the FAL93-F model \citep{fontenla93}.
This model includes the dependence of ionization degree on heights for both hydrogen and helium atoms. The upper panel shows the plasma temperature vs. height. The temperature minimum is located near 500 km above the base of the photosphere, while the transition region is just above 2000 km height. The neutral hydrogen number density is much higher than the electron number density at the lower heights, but becomes comparable near $\approx$ 1900 km, which corresponds to the temperature of 9400 K (lower panel). Hydrogen atoms quickly become ionized above this height. The neutral helium number density is also higher than the electron number density at the lower heights (lower panel). They become comparable near $\approx$ 1600 km, which corresponds to the temperature of $\approx$ 7300 K. The ratio becomes smaller and smaller just above this height. The ratio of neutral helium and neutral hydrogen number densities stays nearly constant ($\approx$ 0.1) up to 1500 km height, then it quickly increases up to 0.2 at height of 2000 km \citep{Zaqarashvili11b}.

The change of plasma parameters with height leads to variation of cyclotron and collision frequencies as well as transport coefficients. These values are important in understanding the fundamental processes in the solar atmospheric plasma.
The fluid theory implicitly assumes that collisions between species are frequent enough to keep the Maxwell-Boltzmann velocity distribution. This imposes time scales where the fluid theory is either applicable or breaks down. Magnetization of ions and electrons is also important for various instabilities and wave dynamics.

The collision frequencies between charged species are essentially determined by the densities and the temperature, while the collision frequencies involving neutrals also depend on the neutral collision cross section. The classical approach to compute the collision cross section is the so-called model of hard spheres, from here on HS. In the HS model the  particles are considered as solid spheres that interact by means of direct impacts only \citep[e.g.,][]{chapman1970}. The HS cross section is usually computed as $\sigma_{\beta\beta'}=\pi \left( r_\beta + r_{\beta'} \right)^2$, where $r_\beta$ and $r_{\beta'}$ are the radii of particles $\beta$ and $\beta'$, respectively. In the case of ion-neutral and electron-neutral collisions, the radii of both ions and electrons are much smaller than the radius of neutral atoms, so that their HS cross sections are approximately the same, namely $\sigma_{\rm in} \approx \sigma_{\rm en} \approx 10^{-20}$~m$^{2}$. Likewise, the HS cross section for neutral-neutral collisions is $\sigma_{\rm nn} \approx 4\times 10^{-20}$~m$^2$.

Recently,  \citet{Vranjes2013}, from here on VK, presented quantum-mechanical computations of collision cross sections that include several important ingredients missing from the classic HS model. For instance, VK  considered variations of the cross section with temperature, quantum indistinguishability corrections, and charge transfer. The VK cross sections coincide with the classical ones at high temperatures, but are different at low temperatures akin to those in the chromosphere. In their paper, VK plot the computed cross section as a function of the energy of the colliding species, which is related to the temperature. For chromospheric temperatures of interest here, we have to consider the results at low energies. The ion-neutral collision cross section from Figure~1 of VK is $\sigma_{\rm in} \approx 10^{-18}$~m$^{2}$, while the electron-neutral collision cross section from their Figure~4 is $\sigma_{\rm en} \approx 3\times 10^{-19}$~m$^{2}$. Concerning neutral-neutral collisions, VK provide in their Figure~3 different cross sections for momentum transfer and viscosity. The cross section for momentum transfer is $\sigma_{\rm nn} \approx 10^{-18}$~m$^{2}$, while the cross section for viscosity is $\sigma_{\rm nn} \approx 3 \times 10^{-19}$~m$^{2}$. In summary, the VK cross sections are between one and two orders of magnitude larger than the classic HS cross sections.

\begin{figure}
  \includegraphics[width=0.6\columnwidth]{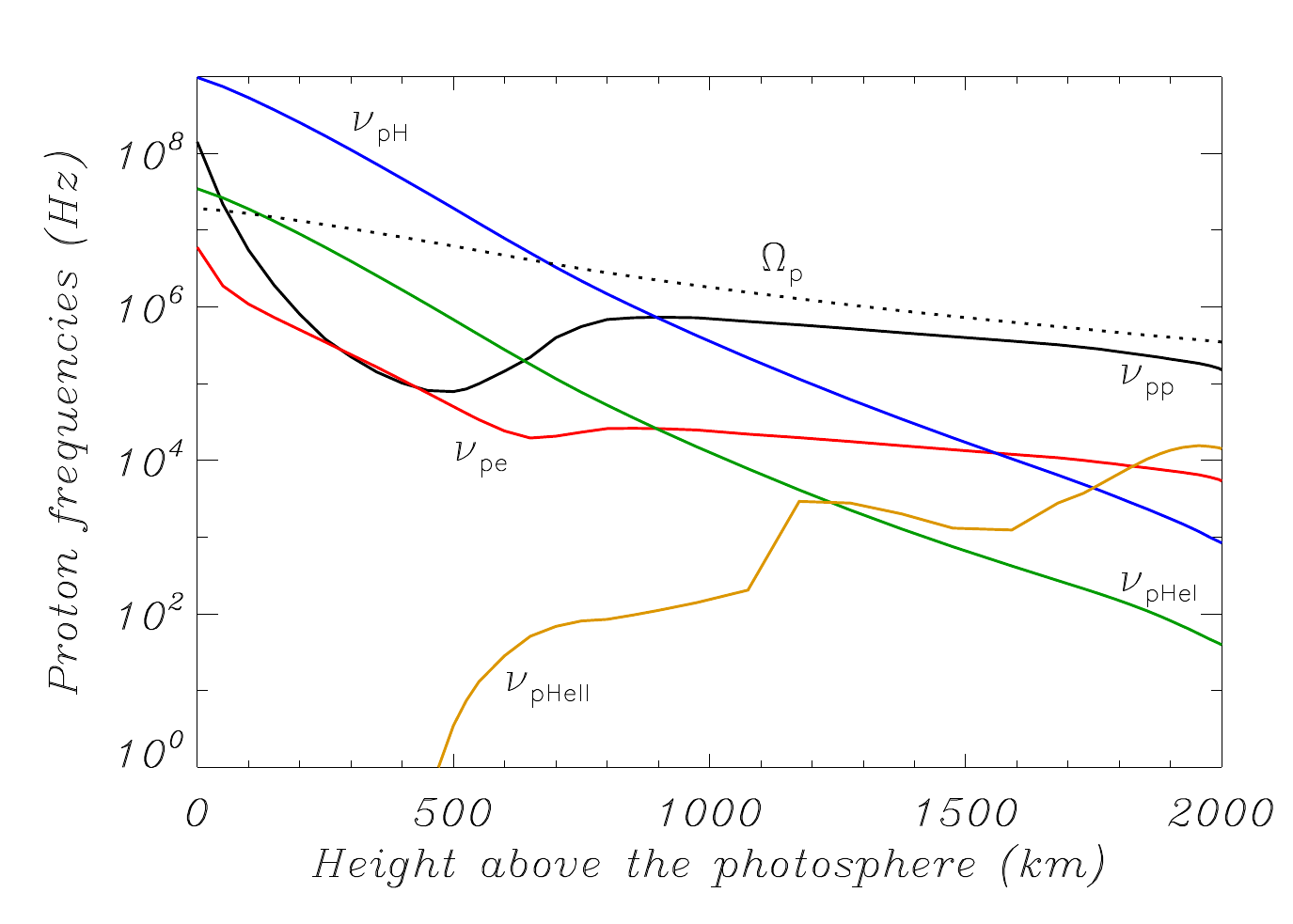}
  \includegraphics[width=0.6\columnwidth]{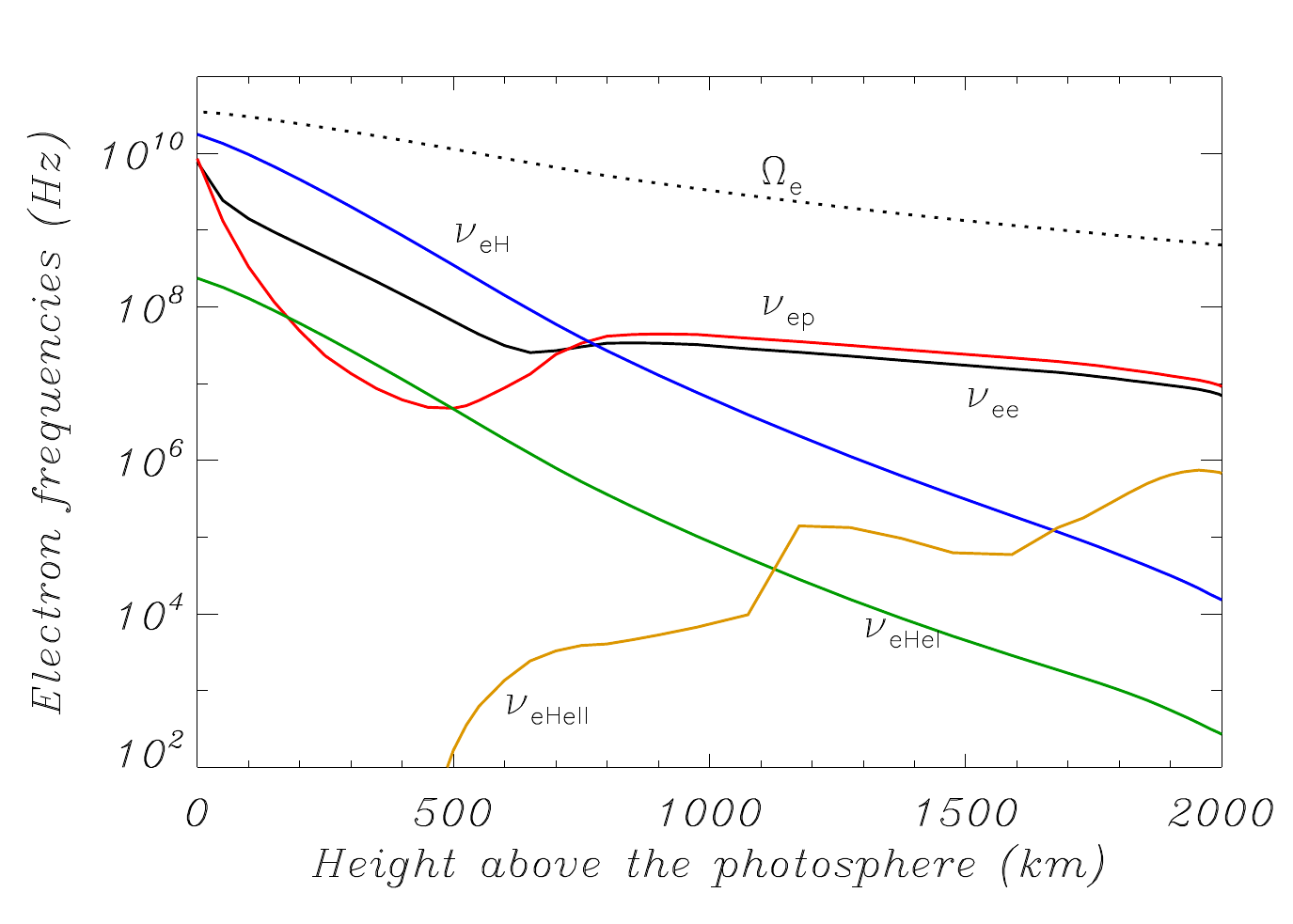}
    \includegraphics[width=0.6\columnwidth]{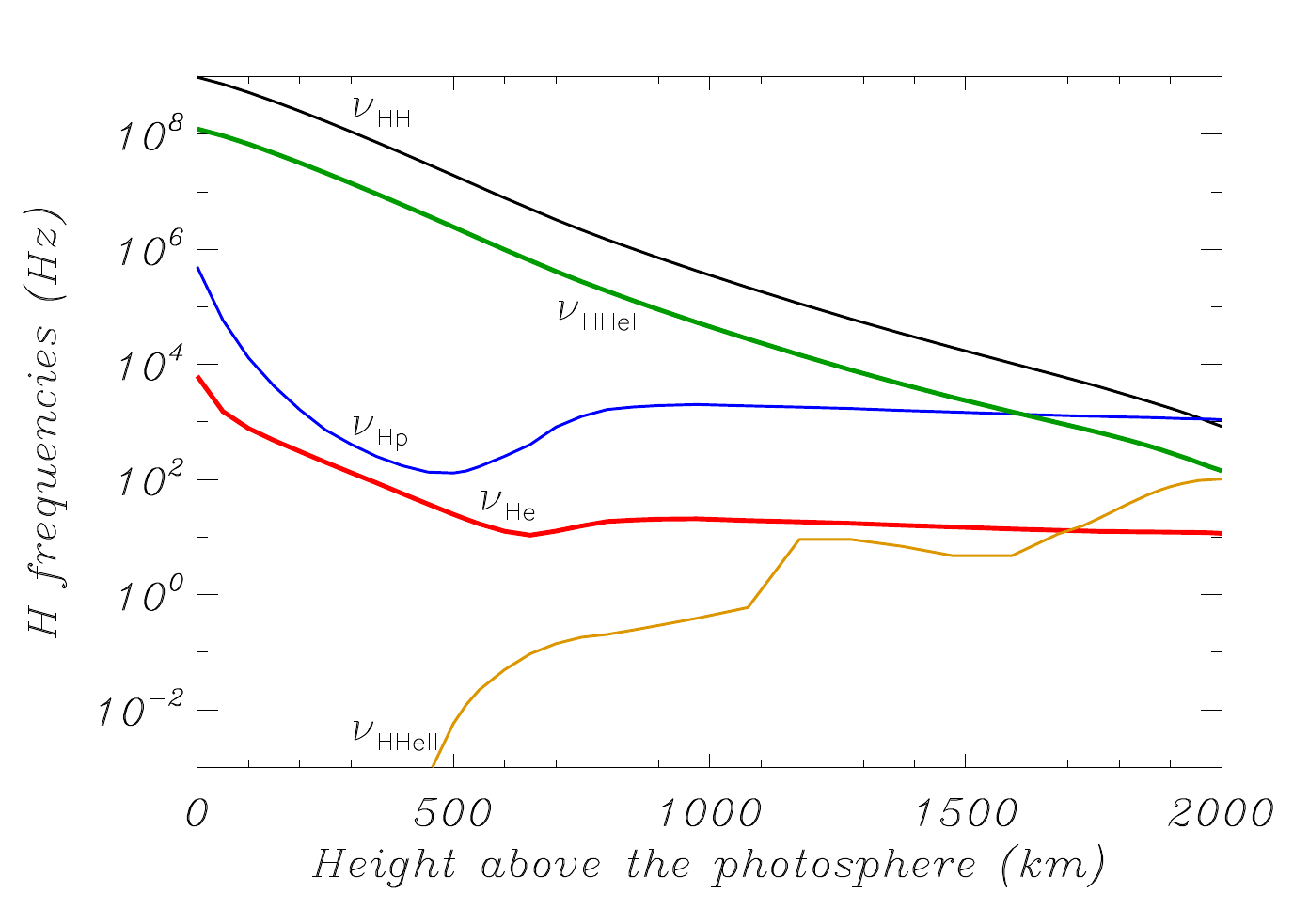}
  \includegraphics[width=0.6\columnwidth]{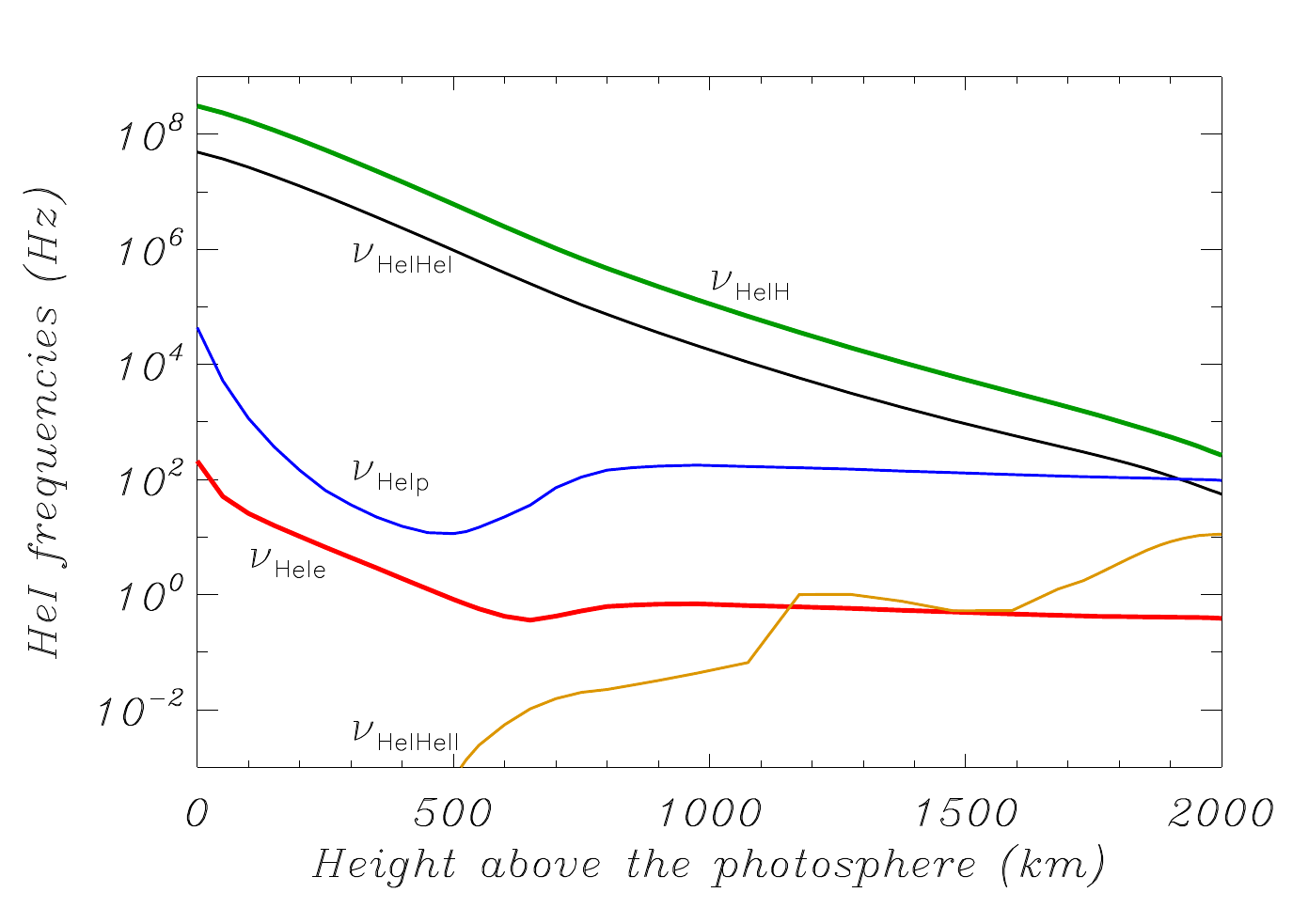}
   \includegraphics[width=0.6\columnwidth]{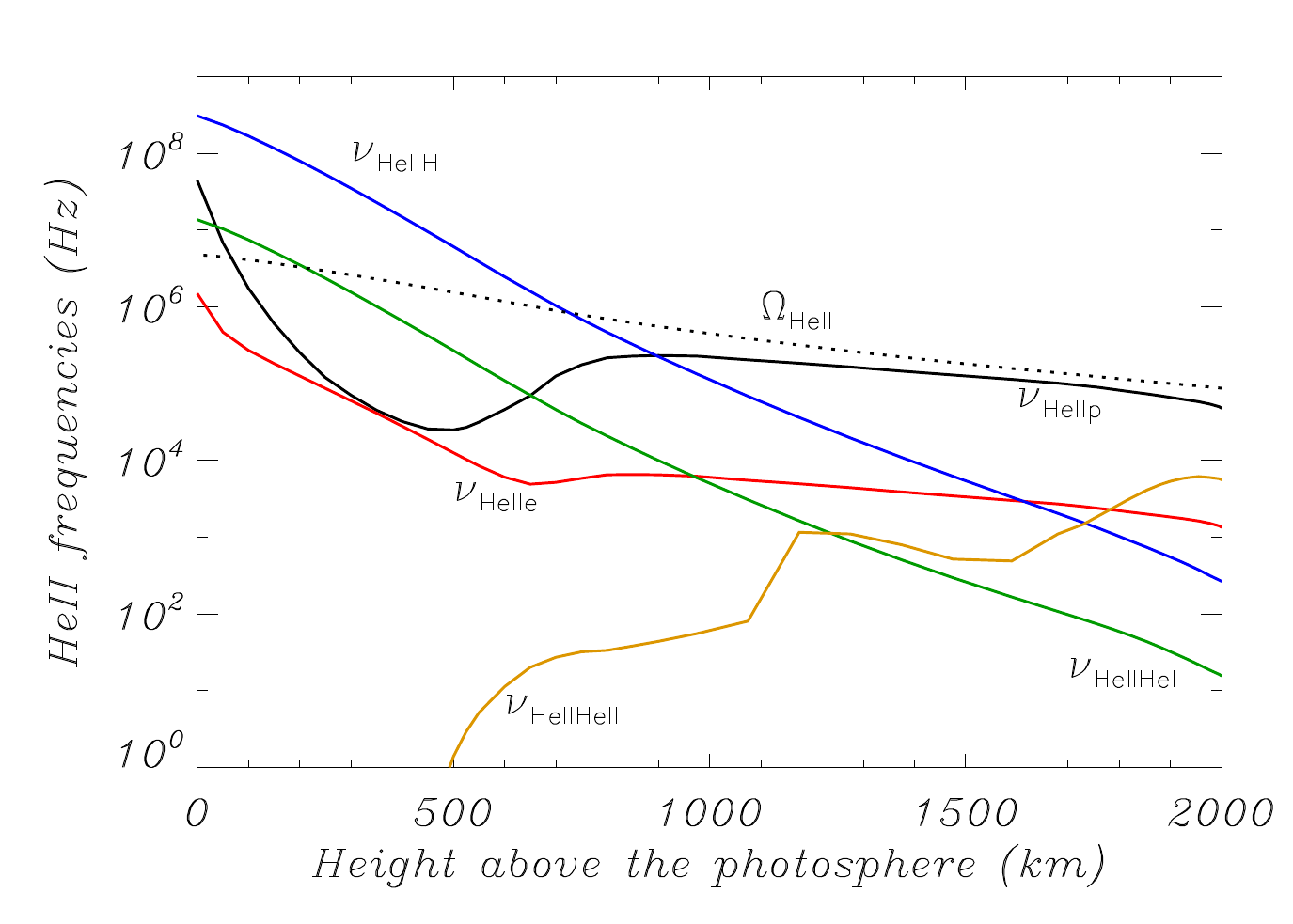}
\caption{Collision frequencies for (a) protons, (b) electrons, (c) neutral hydrogen, (d) neutral helium and (e) singly ionized helium vs height above the photosphere. Here we use the collision cross sections from \citet{Vranjes2013}. In panels (a), (b) and (e) $\Omega_{cp}$, $\Omega_{ce}$, and $\Omega_{cHeII}$
denote the proton, electron and singly ionized helium cyclotron frequencies, respectively, and are represented by dotted lines.}
\label{fig:falc2}
\end{figure}

The value of the cross section is not only important from the academic point of view but also from the practical one. The reason is that the value of the cross section directly affects the collision frequency and, therefore, it is important to determine both the applicability of the fluid theory and the optimal frequency for wave damping in the multi-fluid description. Uncertainties in the collision frequency also cause uncertainties in the various transport coefficients that govern basic collisional phenomena in the plasma \citep[e.g.,][]{martinez12}. The value of the collision cross section may play an important role in theoretical computations.

Figure~\ref{fig:falc2} shows the dependence on height of the relevant collision frequencies for protons, electrons, neutral hydrogen, neutral helium and singly ionized helium taking into account collision cross sections from \citet{Vranjes2013}. Concerning protons (Figure~\ref{fig:falc2}(a)), we find that $\nu_{\rm pH} \gg \nu_{\rm pp}$ in the low chromosphere. Collisions with neutrals are too frequent for protons to reach a Maxwell-Boltzmann distribution independently. In other words, protons collide too frequently with neutral hydrogen so that proton-proton collisions do not have enough time to make the proton distribution Maxwellian on their own without the influence of neutrals. Protons are too coupled with neutrals. This means that the condition for protons to be treated as an individual fluid is not satisfied. The multi-fluid theory breaks down  for $h \sim 900$~km. We also see that protons would not be magnetized for $h \sim 700$~km in the VK case since $\nu_{\rm pH} > \Omega_{cp}$ at those low heights, where $\Omega_{cp} = eB/m_{\rm p}$ is the proton cyclotron frequency \citep[see a discussion on this issue in][]{vranjes2008}.

In the case of electrons (Figure~\ref{fig:falc2}(b)), we also see that electrons should not be treated as a separate fluid at low heights due to the very frequent collisions with neutral hydrogen, i.e., $\nu_{\rm eH} \gg \nu_{\rm ee}$. At large heights, however, it is found that $\nu_{\rm ee} \gg \nu_{\rm eH}$ and $\nu_{\rm ee}\sim\nu_{\rm ep}$, so that electrons are strongly coupled to protons (they effectively behave as an ion-electron single fluid) but are weakly affected by neutrals.  In all cases, the electron cyclotron frequency, $\Omega_{ce} = e B/m_{\rm e}$, remains larger than the electron-electron collision frequency.

It is seen from Figures ~\ref{fig:falc2}(a) and ~\ref{fig:falc2}(e) that protons and singly ionized heliums are magnetized below the height of $\sim 700$~km, but become unmagnetized above this height. On the other hand, electrons are magnetized through whole solar chromosphere.

Finally, the collision frequencies for neutral hydrogen and neutral helium (Figure~\ref{fig:falc2}(c) and (d)) show that $\nu_{\rm nn}$ is always the largest frequency at all heights, so that treating neutrals as a separate fluid is a valid assumption. This last result remains the same for both HS and VK cross sections.

The discussion in previous paragraphs implies that for the waves near the proton-neutral collision frequency, protons should not be considered as an individual fluid separate from neutrals at low heights in the chromosphere. This was previously noted by \citet{Vranjes2013}. The same restriction applies to electrons. The multi-fluid theory is applicable for the height of $\sim 600$~km using the HS cross sections and for $h \sim 900$~km using the VK cross sections.  Hence, the use of the more accurate VK cross sections results in a more restrictive criterion for the applicability of the multi-fluid theory than that obtained with the classical HS cross sections. The correct study of high-frequency waves in the low chromosphere should be done using hybrid fluid-kinetic models, or even fully kinetic models, since standard fluid theory is not applicable for those high-frequencies. Of course, for wave frequencies lower than all the collision frequencies the plasma dynamics can be studied using the single-fluid approximation \citep[e.g.,][]{depontieu01,khodachenko04}, which assumes that ions, electrons, and neutrals are strongly coupled.

It is also important to realize that the solar atmosphere is not only strongly stratified but is also strongly horizontally inhomogeneous. Therefore, even at the same heights, the typical values of the ionisation fraction, magnetisation and transport coefficients strongly vary. \citet{Khomenko+etal2014} evaluated the magnitude of the different terms in the Ohm's law (as Ohmic, Hall, Ambipolar and battery terms, see Sect. 2) for several representative models of the solar atmosphere, covering quiet and active regions. This evaluation has shown that in the quiet solar regions (those covering at least 90\% of solar surface at any time),  Hall and Ambipolar terms dominate the electric current, being largest at the borders of inter granular lanes with strongest gradients of all parameters. In the sunspot model, the Ambipolar term dominates by 3 orders of magnitude over the next in magnitude Hall term. These order of magnitude estimates can be taken as a guidance.

\subsubsection{Magnetic reconnection in partially ionized photospheric plasma}

Magnetic reconnection is a fundamental process in magnetized plasmas, when the magnetic energy of oppositely directed magnetic field is transformed into radiation and heat (Priest and Forbes 2000).
Martin et al. (1985) observed that the photospheric magnetic fragments with opposite polarity, which
were characterized by fluxes exceeding $10^{17}$ Mx and sizes of about 1 arc sec or larger, approached each other and disappeared. Litvinenko (1999) showed that the photospheric magnetic reconnection is a mechanism for the cancellation. The reconnection process should be most efficient around the temperature minimum region about 600 km above the lower photospheric boundary. The observation of Martin et al. (1985) is believed to be the first observational evidence of magnetic reconnection at the photosphere.

More recently, Katsukawa et al. (2007) reported the discovery of small jetlike features in sunspot penumbra, using Ca II H observations from the Hinode satellite. The observed penumbral microjets are highly transient events with lifetimes of less than 2 minutes, lengths of 1000-4000 km, and widths of $\approx$ 400 km. Sakai and Smith (2008) performed 2.5D numerical simulation of two horizontal penumbral filaments using a two-fluid model of partially ionized plasmas, where one fluid is neutral hydrogen atoms and second one is charged particles (protons+electrons). They showed that inclined bidirectional jet-like flows, driven by the magnetic reconnection, propagate along the
vertical magnetic flux tube, which exists between the filaments. Strong proton heating, up to 25 times their original temperature, was observed in these generated jets. Conversely, the neutral-hydrogen particles are only very weakly heated. Sakai and Smith (2008) proposed that these plasma jets may explain the phenomenon of penumbral microjets observed by Katsukawa et al. (2007).

Sakai and Smith (2009) found that the magnetic reconnection rate of the coalescing penumbral filaments is strongly enhanced by an initial velocity of filaments. An initial collision velocity corresponding to 10 \% of the sound speed increases the magnetic reconnection rate by a factor of 50 when compared to spontaneous coalescence. They concluded that the magnetic reconnection rate of coalescing penumbra filaments can be strongly enhanced by weak photospheric neutral-hydrogen flows.

\subsubsection{Energy flux of transverse waves in the photosphere}

Mechanical energy of photospheric motions has been considered to be a source for chromospheric and coronal heating. Granular cells may excite transverse waves (MHD kink or torsional Alfv\'en waves) in intense magnetic flux tubes, which carry energy up and heat the surrounding plasma. These waves can be observed in the chromosphere by a transverse displacement or Doppler shift in magnetic structures \citep[see recent review by][]{zaqarashvili2009}. \citet{vranjes2008} suggested that the energy flux of Alfv\'en waves in the weakly ionized photospheric plasma is orders of magnitude smaller than in the ideal case, which makes ineffective the generation of the waves near the solar surface. Consequently, they raised a question of applicability of photospheric Alfv\'en waves as a possible energy source for chromospheric and coronal heating. On the other hand, \citet{tsap2011} concluded that the energy flux of Alfv\'en waves in the weakly ionized photospheric plasma is the same as in the ideal case. This discrepancy was clarified by \citet{soler13}, who showed that the energy flux depends on initial velocities of both, ion and neutral fluids. \citet{vranjes2008} considered zero initial velocity of neutral fluid putting initial energy only into the ion fluid, therefore they obtained much smaller energy flux due to the small density of ions with regards to neutrals at the photosphere. But \citet{tsap2011} assumed that both ions and neutrals are initially perturbed with the same velocity due to the strong coupling, although this condition is not explicitly stated in the paper. Therefore, they obtained the energy flux of Alfv\'en waves similar to the ideal plasma.

Transverse (kink or Alfv\'en) waves are excited in magnetic flux tubes by buffeting of granular cells which means that initial energy is mostly stored in the neutral fluid. Therefore, the initial conditions assumed by \citet{vranjes2008} are not adequate to the photosphere. Consequently, the energy flux of generated Alfv\'en waves can be considered as a source for chromospheric and coronal heating.

\subsubsection{Damping of Alfv\'en waves and associated heating in the solar chromosphere}
\label{heatchro}
There is extensive observational evidence of Alfv\'enic waves propagating in the solar chromosphere \citep[e.g.,][]{kukhianidze2006,zaqarashvili2007a,depontieu07,depontieu12,okamoto11,kuridze12,morton13}. The plasma heating of the solar atmosphere is one of the long-standing problems in solar physics, and it is believed that energy transport by Alfv\'en waves and its dissipation may play a relevant role \citep[see,e.g.,][]{erdelyi07,cargill11,mcintosh11,hahn14,arregui15,jess15}. The relatively cool temperature in the chromosphere causes the plasma to be partially ionized, with a predominance of neutrals at low heights in the chromosphere. It has been shown that partial ionization effects have a strong impact on chromospheric dynamics \citep[see, e.g.,][]{martinez12,leake14}. Ion-neutral collisions may play a crucial role in the release of magnetic energy in the form of heat \citep{khomenko12, shelyag16}. In this context, ion-neutral collisions have been invoked as a viable energy dissipation mechanism for Alfv\'en waves by, e.g., \citet{depontieu01,khodachenko04,leake05,russell13}, among others. Strongly damped waves are good candidates to produce significant heating of the chromospheric plasma via conversion of wave energy into thermal energy. Estimations of the heating rate due to Alfv\'en waves damped by ion-neutral collisions computed by \citet{song11} and \citet{goodman11} indicate that this mechanism may generate sufficient heat to compensate the radiative losses at low altitudes in the solar atmosphere. Ion-neutral interaction can lead to dissipation of perpendicular currents produced by waves, converting the magnetic energy of waves into thermal energy and producing an important heating of the magnetized chromosphere above magnetic elements \citep{khomenko12, shelyag16}.

\begin{figure}
  \includegraphics[width=1.1\columnwidth]{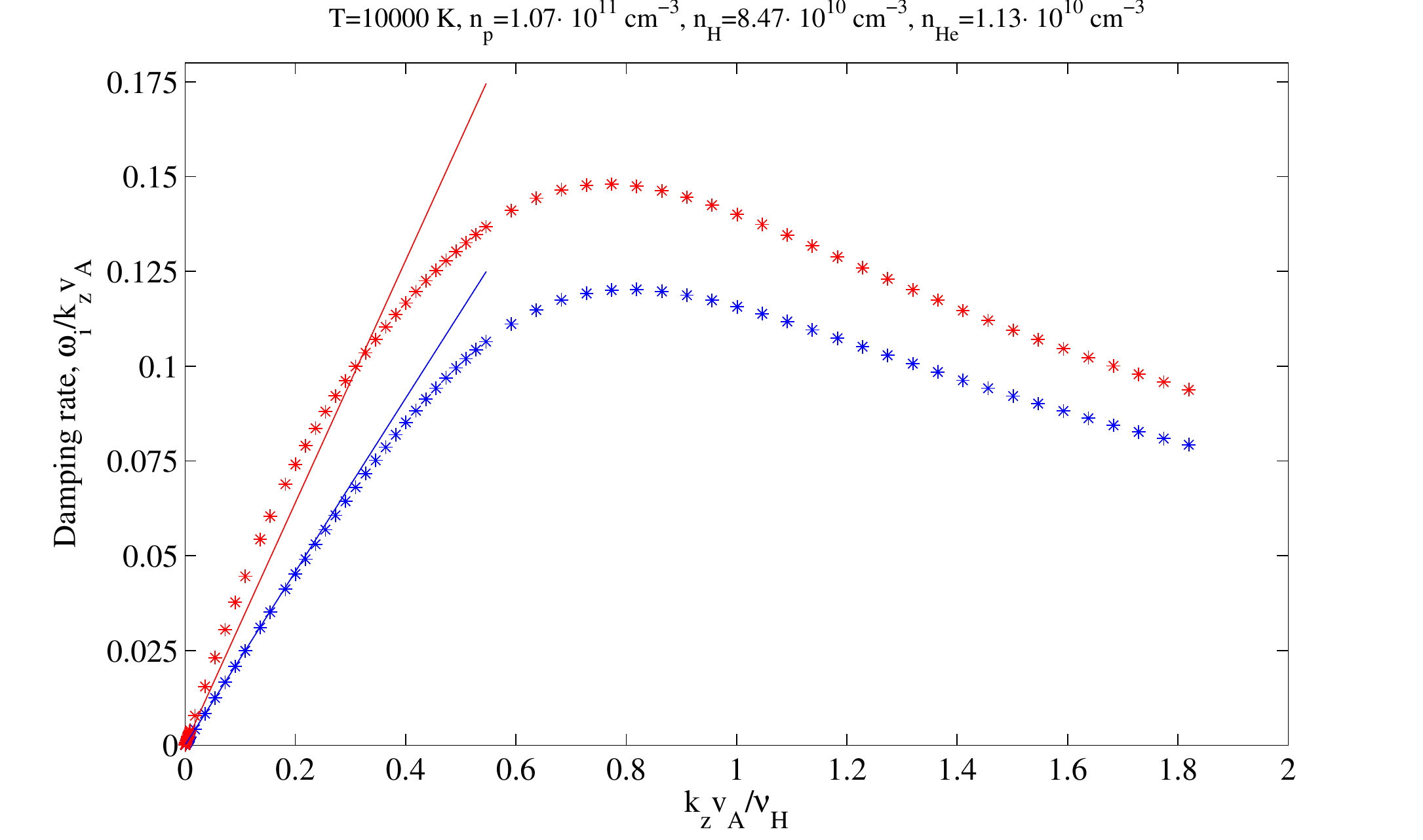}
\caption{Damping rate (imaginary part of frequency) of Alfv\'en waves (normalized by Alfv\'en frequency $k_zv_A$), vs Alfv\'en frequency normalized by hydrogen collision frequency $\nu_{\mathrm {H}}$ for the temperature of 10000 K. Blue asterisks correspond to the damping rates due to ion collision with neutral hydrogen atoms only, while red asterisks correspond to the damping rates due to ion collision with both, neutral hydrogen and neutral helium atoms. Red (blue) solid line corresponds to the damping rate derived in the single-fluid approach with (without) neutral helium. The collision frequency is defined as $\alpha_{\mathrm {H}}/\rho_{\mathrm {H}}+\alpha_{\mathrm {H}}/\rho_{\mathrm {H^+}}$ for the case of neutral hydrogen and $\alpha_{\mathrm {H^+H}}/\rho_{\mathrm {H}}+\alpha_{\mathrm {H^+H}}/\rho_{\mathrm {H^+}}+\alpha_{\mathrm {He^+H}}/\rho_{\mathrm {H}}+\alpha_{\mathrm {He^+H}}/\rho_{\mathrm {He^+}}$. The values of temperature and number densities are taken from the FAL93-3 model \citep{fontenla93}. }
\label{fig:helium-2}
\end{figure}

Zaqarashvili et al. (2011a) used a two-fluid MHD model of partially ionized plasma, where ion-electron plasma and neutral hydrogen atoms are considered as separate fluids, to obtain the dispersion relations of linear MHD waves in the simplest case of a homogeneous medium for different parameters of background plasma. The dispersion relation of Alfv\'en waves in the two-fluid approach is
\begin{equation}
\omega^3+i{{\alpha_{\mathrm {H}}}\over {\rho_{\mathrm {H}}}}\left (1+{{\rho_{\mathrm {H}}}\over {\rho_{\mathrm {H^+}}}} \right )\omega^2-k^2_{\mathrm {z}}v^2_{\mathrm {A}}\omega-i{{\alpha_{\mathrm {H}}}\over {\rho_{\mathrm {H}}}}k^2_zv^2_A=0,
\label{eq:two-fluid}
\end{equation}
where $\omega$ and $k_{\mathrm {z}}$ are the frequency and wave number of Alfv\'en waves, $\rho_{\mathrm {H}}$ and $\rho_{\mathrm {H^+}}$ are the density of neutral hydrogen atoms and protons, respectively, $v_{\mathrm {A}}=B_0/\sqrt{\mu \rho_{\mathrm {H^+}}}$ is the Alfv\'en speed of electron-proton fluid and $\alpha_{\mathrm {H}}$ is the coefficient of friction between protons and neutral hydrogen atoms. This equation has three different solutions, where two correspond to Alfv\'en waves and the third corresponds to the vortex solution of neutral hydrogen fluid.  Fig.~\ref{fig:helium-2} (blue asterisks) shows the damping rate of Alfv\'en waves using the two-fluid model. The maximal damping of Alfv\'en waves occurs near the ion-neutral collision frequency. The damping rates are reduced for the higher frequency waves contrary to the single-fluid approach (blue solid line), where the higher frequency waves show stronger damping.

Recently, \citet{soler15a} investigated Alfv\'en wave damping as a function of height in a simplified chromospheric model based on the semi-empirical  model F of \citet{fontenla93}. They found that there is a critical interval of wavelengths for which impulsively excited Alfv\'enwaves are overdamped as a result of the strong ion-neutral dissipation. All the energy stored in the impulsively generated perturbations with wavelengths belonging to the critical interval is necessarily dissipated in the vicinity of the location of the impulsive driver instead of begin transported away by propagating waves. Strong plasma heating might therefore be produced by these overdamped waves.  Equivalently, for periodically driven Alfv\'en waves there is an optimal frequency for which the damping is most effective. \citet{soler15a} concluded that ion-neutral collisions heating is much more important than Ohmic heating in the upper chromosphere for wave frequencies near the optimal frequency. \citet{soler15a} compared their results  for two sets of collision cross sections, namely those of the classic hard-sphere model and those based on recent quantum-mechanical computations \citep{Vranjes2013}, and found important differences between the results for the two sets of cross sections. For instance, the optimal frequency for the damping varies from 1~Hz to $10^2$~Hz for the hard-sphere cross sections, and from $10^2$~Hz to $10^4$~Hz for the quantum-mechanical cross sections, which points out the importance of using accurate transport coefficients that govern basic collisional phenomena in the plasma.

\begin{figure}
\center
  \includegraphics[width=0.75\columnwidth]{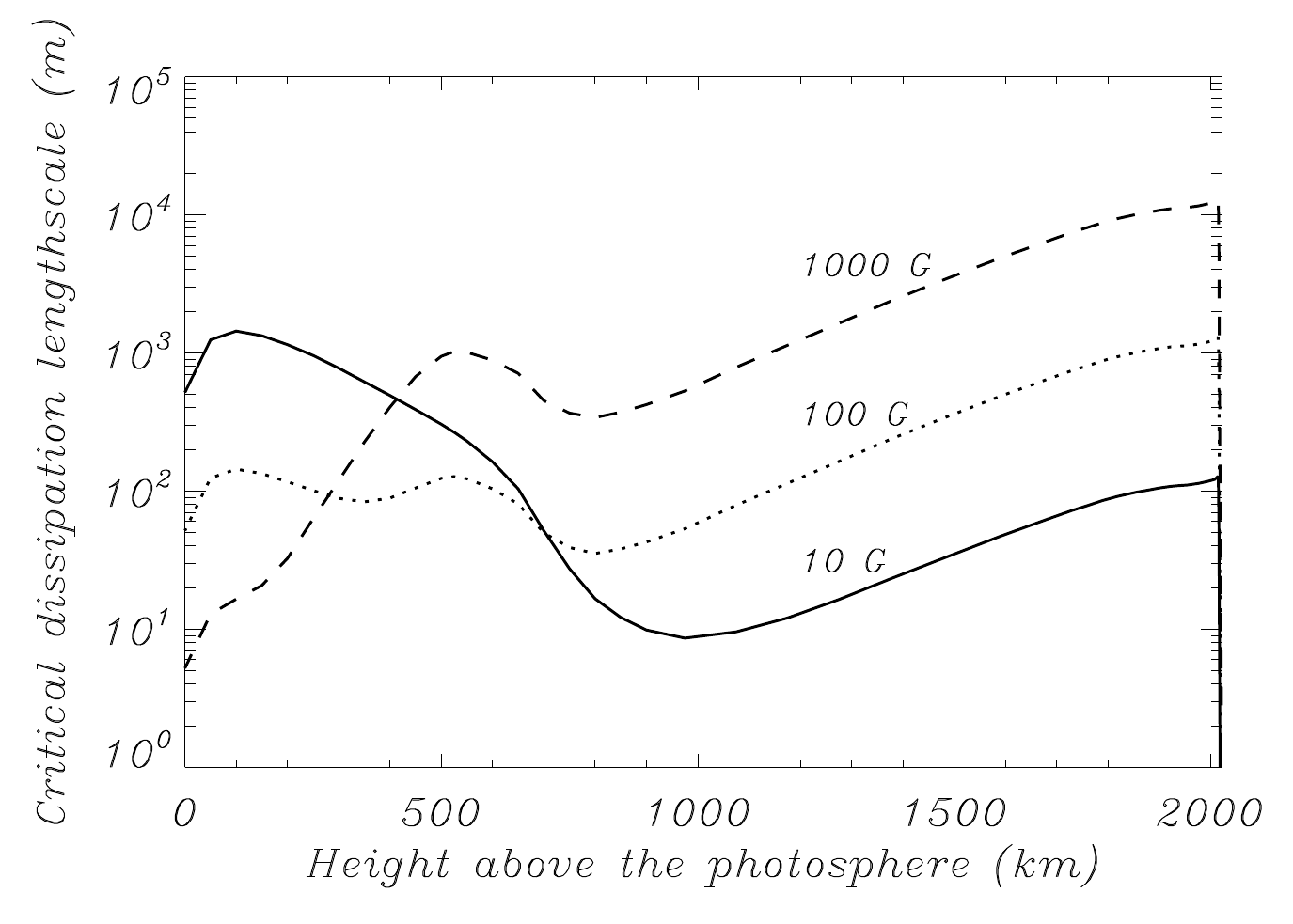}
\caption{Critical dissipation lengthscale for Alfv\'en waves as a function of height above the solar photosphere according to the chromospheric model F of \citet{fontenla93}. The various linestyles correspond to a magnetic field strength of 10~G (solid), 100~G (dotted), and 1000G (dashed). Adapted from \citet{soler15b}.}
\label{fig:soler15b}
\end{figure}

Subsequently, \citet{soler15b} used a more complete  description of the chromospheric physics to further investigate the dissipation mechanisms responsible for the overdamping of chromospheric waves and the corresponding spatial scales of the  perturbations. \citet{soler15b} found that the critical dissipation lengthscale for  waves depends strongly on the magnetic field strength. For realistic field strength in the chromosphere, the Alfv\'en wave critical dissipation lengthscale  ranges from 10~m to 1~km (see Fig.~\ref{fig:soler15b}). The mechanism responsible for the critical damping of Alfv\'en waves is Ohmic diffusion at low heights in the chromosphere and  ambipolar diffusion at medium/large heights in the chromosphere. It is found that viscosity plays no important role in the damping of chromospheric Alfv\'en waves. The results by \citet{soler15b} suggest that the spatial scales at which strong Alfv\'en wave heating may work in the chromosphere are unresolved by current observational instruments.

A step beyond the simplified analytical description in studies of dissipation and heating by Alfv\'en waves was reported recently by \citet{shelyag16}. The authors studied numerically the non-linear propagation of waves in three-dimensional strongly stratified solar flux tubes, including the effects of the mode transformation and dissipation by neutrals in a single-fluid approximation. It was shown that up to 80\% of the Poynting flux associated to these waves can be dissipated and converted into heat due to the effect of ambipolar diffusion providing an order of magnitude larger amount of energy to the chromosphere compared to the dissipation of stationary currents \citep{khomenko12}. Nevertheless, \citet{Arber2016} argued that heating produced by acoustic shocks is more important than that by Alfv\'en wave dissipation through ion-neutral collisions. Therefore, the question of chromospheric heating due to the ion-neutral interaction will require further studies in the future.

Most of the atoms in the chromosphere are neutral hydrogen, but a significant amount of neutral helium may also be present in the plasma with a particular temperature. The neutral helium atoms may enhance the damping of MHD waves due to the collisions with ions. \citet{Zaqarashvili2011a} considered three-fluid MHD approximation, where one component is electron-proton-singly ionized helium and other two components are the neutral hydrogen and neutral helium atoms. They derived the dispersion relation of linear Alfv\'en waves in isothermal and homogeneous plasma, which has a following form:
$$
\omega^4+i\left ({{\alpha_{\mathrm {H}}}\over {\rho_{\mathrm {H}}}}+{{\alpha_{\mathrm {He}}}\over {\rho_{\mathrm {He}}}}+{{\alpha_{\mathrm {H}}+\alpha_{\mathrm {He}}}\over {\rho_{\mathrm {i}}}}\right )\omega^3+\left (k^2_{\mathrm {z}}v^2_{\mathrm {A}}+{{\alpha_{\mathrm {H}}\alpha_{\mathrm {He}}\rho_{\mathrm {0}}}\over {\rho_{\mathrm {H}}\rho_{\mathrm {He}}\rho_{\mathrm {i}}}}\right )\omega^2-
$$
\begin{equation}
ik^2_{\mathrm {z}}v^2_{\mathrm {A}}\left ({{\alpha_{\mathrm {H}}}\over {\rho_{\mathrm {H}}}}+{{\alpha_{\mathrm {He}}}\over {\rho_{\mathrm {He}}}}\right )\omega+{{\alpha_{\mathrm {H}}\alpha_{\mathrm {He}}}\over {\rho_{\mathrm {H}}\rho_{\mathrm {He}}}}k^2_zv^2_A=0,
\label{eq:three-fluid}
\end{equation}
where $\rho_{\mathrm {He}}$ is the density of neutral helium atoms, $\rho_{\mathrm {i}}$ is the density of ions (protons+singly ionized helium), $\rho_{\mathrm {0}}=\rho_{\mathrm {H}}+\rho_{\mathrm {He}}+\rho_{\mathrm {i}}$ is the total density of all fluids, $v_{\mathrm {A}}=B_0/\sqrt{\mu \rho_{\mathrm {i}}}$ is the Alfv\'en speed of electron-proton-singly ionized helium fluid, $\alpha_{\mathrm {H}}$ is the coefficient of friction between neutral hydrogen atoms and ions (protons and singly ionized helium) and $\alpha_{\mathrm {He}}$ is the coefficient of friction between neutral helium atoms and ions (protons and singly ionized helium). The dispersion relation has four different roots: two complex solutions, which correspond to Alfv\'en waves damped by ion-neutral collision, and two purely imaginary solutions, which correspond to damped vortex solutions of neutral hydrogen and neutral helium fluids. Fig.~\ref{fig:helium-2} (red asterisks) shows the damping rate of Alfv\'en waves obtained by three-fluid MHD model when neutral helium atoms are also included. \citet{Zaqarashvili11b} concluded that the presence of neutral helium may significantly enhance the damping of Alfv\'en waves compared to the damping due to neutral hydrogen at certain values of plasma temperature (10 000-40 000 K) and ionization. Therefore, the height dependence of the ionization degrees of hydrogen and helium may influence the damping rate of Alfv\'en waves.

Recently, \citet{zaqarashvili2013a} studied the effect of neutral helium on the damping of torsional Alfv\'en waves in stratified, partially ionized plasma of the solar chromosphere. They considered a magnetic flux tube, which is expanded up to 1000 km height and then becomes vertical owing to merging with neighboring tubes. Consecutive derivation of single-fluid MHD equations resulted in a new Cowling diffusion coefficient in the presence of
neutral helium, which has a following form
\begin{equation}\label{eqs3}
\eta_c={B^2_0\over {\mu}}{{\alpha_{He}\xi^2_{H} +\alpha_{H}\xi^2_{He}+\alpha_{HeH}(\xi_{H}+\xi_{He})^2}\over {\alpha_{H}\alpha_{He}+\alpha_{H}\alpha_{HeH}+\alpha_{He}\alpha_{HeH}}},
\end{equation}
where $\xi_H=\rho_H/\rho$, $\xi_{He}=\rho_{He}/\rho$, while $\alpha_{{\mathrm {H}}}=\alpha_{{\mathrm {H^+H}}}+\alpha_{{\mathrm {He^+H}}}$ and $\alpha_{{\mathrm {He}}}=\alpha_{{\mathrm {H^+He}}}+\alpha_{{\mathrm {He^+He}}}$ are friction coefficients between collisions of neutral hydrogen and neutral helium atoms with ions. Based on the analytical study, \citet{zaqarashvili2013a} concluded that shorter-period ($<$ 5 s) torsional Alfv\'en waves, which are generated at the photosphere, may damp quickly in the chromospheric network due to ion-neutral collisions, but the longer-period ($>$ 5 s) waves may not reach the transition region as they become evanescent at lower heights due to stratification. But the authors also noted that the torsional Alfv\'en waves with all periods may penetrate the corona if they are excited in the higher part of the chromosphere probably due to magnetic reconnection. This point needs further clarification.

\subsubsection{Formation of spicules by ion-neutral collisions}

Spicules were discovered almost 130 years ago, but they still remain one of the most mysterious
phenomena in the solar atmosphere. Spicules are usually detected in chromospheric H$\alpha$, D$_3$ and Ca II H lines as thin and elongated structures in the solar limb \citep{sterling2000}. Spicule formation mechanisms can be formally divided into three different groups: pulses \citep{suematsu1982,murawski2010}, Alfv\'en waves \citep{hollweg1982,kudoh1999} and p-mode leakage \citep{depontieu2004}.

\citet{haerendel1992} suggested that damping of upwardly propagating Alfv\'en waves due to collisions between ions and neutrals provide average net force, which may act against gravity and lift up transition region. The net force can be calculated from MHD equations using WKB approach \citep{dePontieu1998}
\begin{equation}
F_z=\rho_n v^2_{1y}{{\omega^2}\over {2 \nu_{ni} v_A}},
\end{equation}
where $\rho_n$ is the density of neutrals, $v_{1y}$ is the perturbation of Alfv\'en waves, $\omega$ is the wave frequency, $\nu_{ni}$ is the neutral-ion collision frequency and $v_A$ is the Alfv\'en speed. \citet{dePontieu1998} showed that Alfv\'en waves with frequencies of 0.2-0.6 Hz may produce spicules.

However, \citet{james2002} numerically solved a set of fully nonlinear, dissipative 1.5D MHD equations
with continuous sinusoidal driver. They found that spicule formation is primarily caused by the impact of a series of slow shocks generated by the continuous interaction between the upward propagating driven wave train and the downward propagating train of waves created by reflection off the transition region. At lower frequencies, the heating due to ion-neutral damping was found to provide only a small benefit due to the increased thermal pressure gradient. At higher frequencies, whilst the heating effect becomes stronger, the much reduced wave amplitude reaching the transition region hinders spicule formation. \citet{james03} and \citet{erdelyi2004} came to essentially the same conclusions. The results of these works suggest that ion-neutral damping may not support spicules as described by \citet{haerendel1992}. However, the authors concluded that the effect is highly sensitive to the level of ionisation and therefore to the energy balance. Including the effects of thermal conduction and radiation may well lead to different results and hence the Alfv\'en wave damping due to ion-neutral collisions may still play a role in the formation of spicules. This point is opened for further discussion.

\subsubsection{Farley-Buneman instability in chromospheric plasma}

Motion of neutral atoms across the magnetic field in partially ionized plasmas may lead to Farley-Buneman instability (FBI) which is well studied in the Earth's ionosphere \citep{farley1963,buneman1963}. When electrons are strongly magnetized then the collisional drag of ions due to the neutral flow may generate electric currents which lead to the instability in some circumstances. \citet{fontenla2005} suggested that upward propagating fast magnetoacoustic waves could be unstable to FBI in partially ionized plasmas of the solar chromosphere, which may heat the ambient medium. The instability threshold can be expressed by the condition:
\begin{equation}
U_{thr}=C_s(1+\psi_{\perp}),
\end{equation}
where $U_{thr}$ is the threshold plasma flow, $C_s$ is the ion-acoustic speed and
\begin{equation}
\psi_{\perp}={{\nu_{ea}\nu_{pa}}\over {\Omega_e\Omega_p}}
\end{equation}
is the parameter, which controls the particle magnetization. \citet{fontenla2005} showed that the FBI would be triggered in the mid- and upper-chromosphere by wave velocity amplitudes larger or equal to the ion-acoustic speed, which is below but close to the adiabatic sound speed. \citet{fontenla2008} set up similar considerations, but instead of fast MHD waves they suggested that convective overshoot motions are drivers of the FBI, which provide enough energy to account for the upper chromospheric radiative losses in the quiet-Sun internetwork and network lanes. Recently, \citet{madsen2014} used a multi-component approach to FBI and showed that the instability may be triggered by velocities as low as 4 km s$^{-1}$, which is well below the neutral acoustic speed.

\begin{figure}
{
\includegraphics[width=6cm]{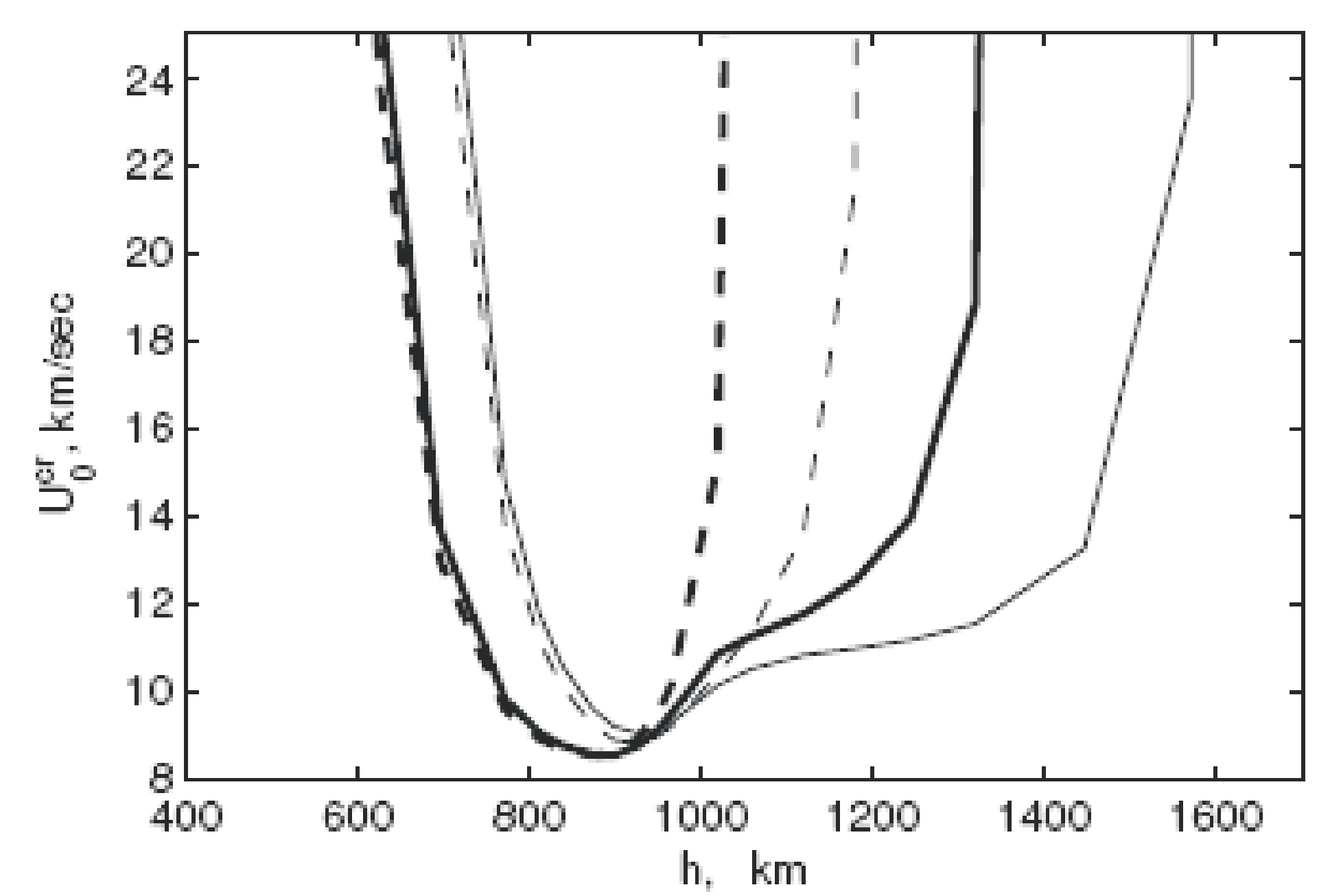}
\includegraphics[width=6cm]{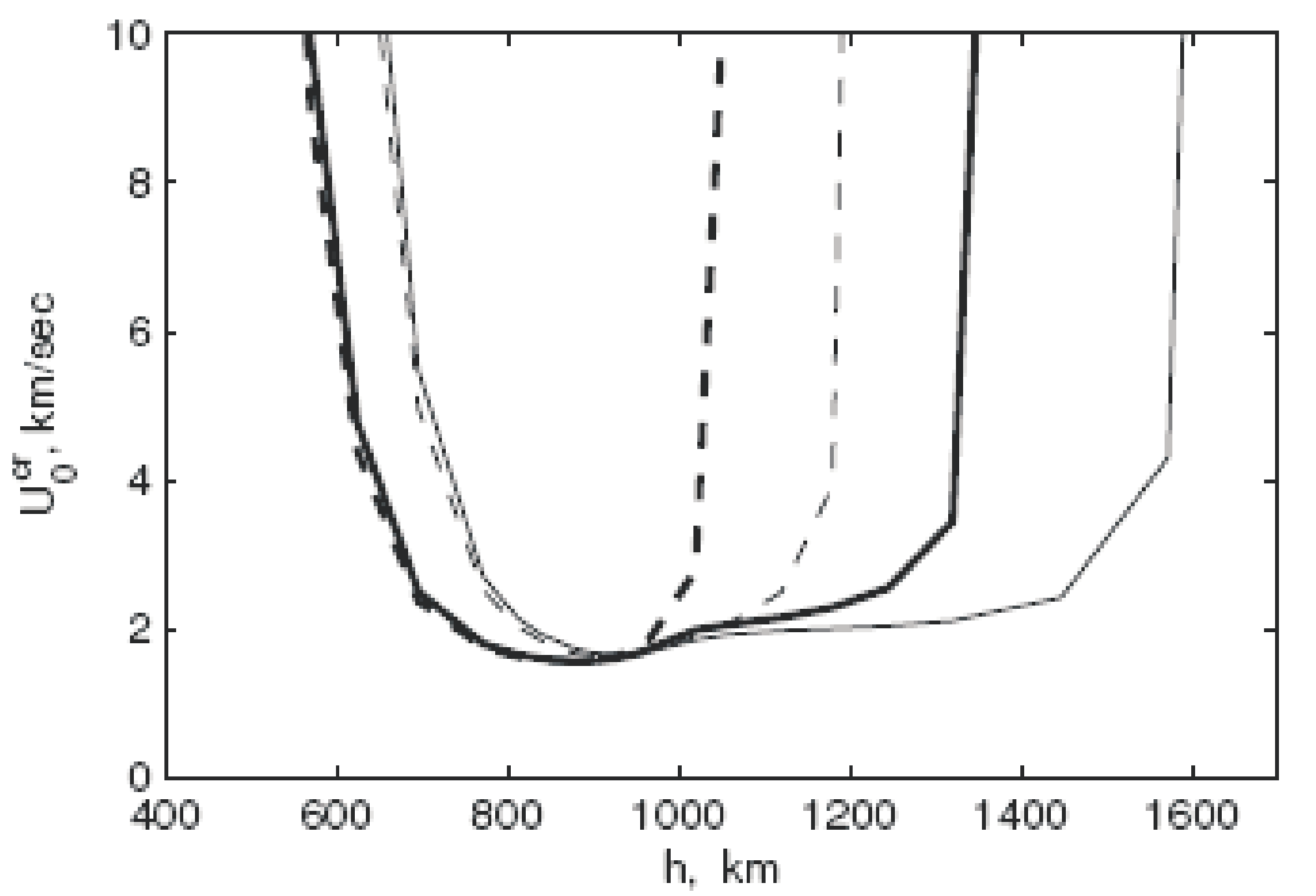}
}
\caption{Dependence of the threshold value of the velocity $U^{cr}_{0}$ on height with (solid lines) and without (dashed lines) Coulomb collisions, for $B$ = 30 G (thin lines) and for $B$ = 60 G (thick lines). The left panel corresponds to the protons and the right panel to ions with $m_i = 30 m_p$. Adapted from \citet{gogoberidze2009}.
}
\label{fig:gogoberidze2009}
\end{figure}

\citet{gogoberidze2009,gogoberidze2014} studied the FBI taking into account Coulomb collisions and particle thermal effects, respectively. Figure~\ref{fig:gogoberidze2009} shows the dependence of critical velocity on height with and without Coulomb collisions. It is seen that Coulomb collisions reduce the threshold value of the critical velocity. However, \citet{gogoberidze2009} concluded that the FBI is a less efficient heating mechanism than the collisional dissipation of cross-field currents that drive the instability. This conclusion concerns both the lower chromosphere, where the threshold velocity is decreased by heavy ions, and the middle/upper chromosphere, where the threshold velocity is decreased by the Coulomb collisions. However, local development of FBI in the presence of strong cross-field currents and/or strong small-scale magnetic fields can not be excluded. In such cases, FBI should produce locally small-scale, 10-100 cm, density irregularities in the solar chromosphere. These irregularities can cause scintillations of radio waves with similar wave lengths and provide a tool for remote chromospheric sensing.

\subsubsection{Magnetic reconnection in partially ionized chromospheric plasma}

\citet{shibata2007} discovered chromospheric anemone jets with the Solar Optical Telescope (SOT) onboard the Hinode spacecraft, which are typically $2$--$5$~Mm long with mean speed $10$--$20$~km\,s$^{-1}$. The anemone jets are probably connected to small scale magnetic reconnection in the solar chromosphere. Therefore, their discovery revived interests in chromospheric reconnection. Reconnection is also supposed to be one of the possible mechanisms of spicule formation. As the chromospheric plasma is partially ionized, it is desired to consider magnetic reconnection in the presence of neutral atoms.

\begin{figure}
\center
{
\includegraphics[width=10cm]{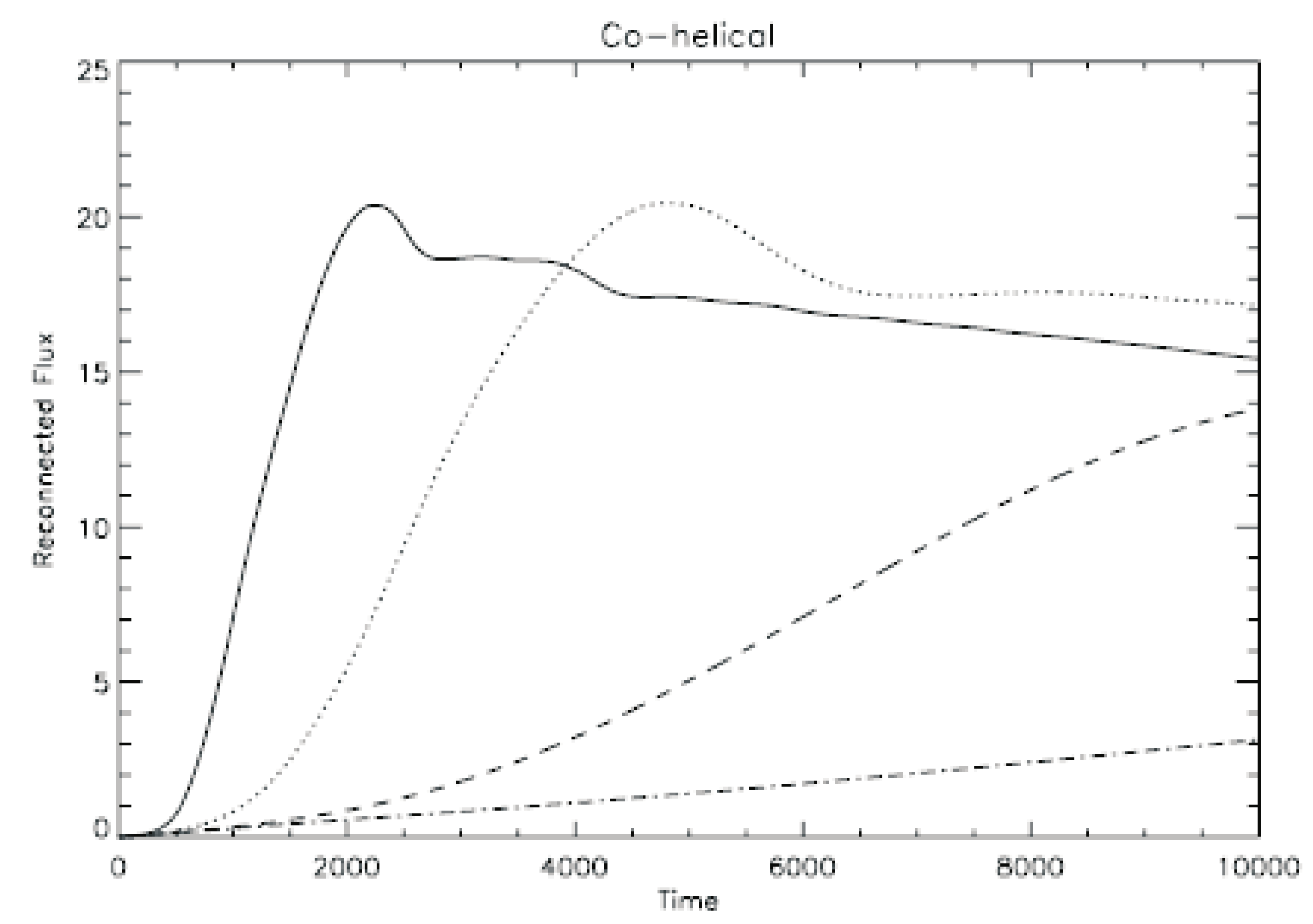}
}
\caption{Effect of neutral-hydrogen to proton density ratio on the reconnected
magnetic flux for two co-helical current loops. (solid) $\rho_n/\rho_p$ = 1,
(dotted) $\rho_n/\rho_p$ = 10, (dashed) $\rho_n/\rho_p$ = 100, and (dot-dashed) $\rho_n/\rho_p$ =
1000. Adapted from \citet{smith2008}.
}
\label{fig:smith2008}
\end{figure}

\citet{smith2008} performed 2.5D numerical simulations of coalescing current loops using a two-fluid model of partially ionized plasmas, where one fluid is neutral hydrogen atoms and second one is charged particles (protons+electrons). Figure~\ref{fig:smith2008} shows total reconnected magnetic flux as a function of time for various neutral-hydrogen to proton density ratios. The slope of each line is a direct measure of the magnetic reconnection rate. It is seen from this figure that the rates of magnetic reconnection strongly depend on the neutral-hydrogen to proton density ratio, $\rho_n/\rho_p$: the rate of magnetic reconnection for the ratio of $\rho_n/\rho_p$ = 1 (upper chromosphere) is twenty times faster than for $\rho_n/\rho_p$ = 1000 (lower chromosphere). This important result implies that observed anemone jets associated with fast magnetic reconnection tend to occur in the upper chromosphere. Comparisons of two-fluid and single-fluid simulations show significant differences in the measured rates of magnetic reconnection, particularly for the density ratios representative of the lower chromosphere. On the other hand, \citet{smith2008} showed that the rate of magnetic reconnection does not significantly depend on recombination/ionization effects and the value of ion-neutral collision frequency.

\begin{figure}
\center
{
\includegraphics[width=10cm]{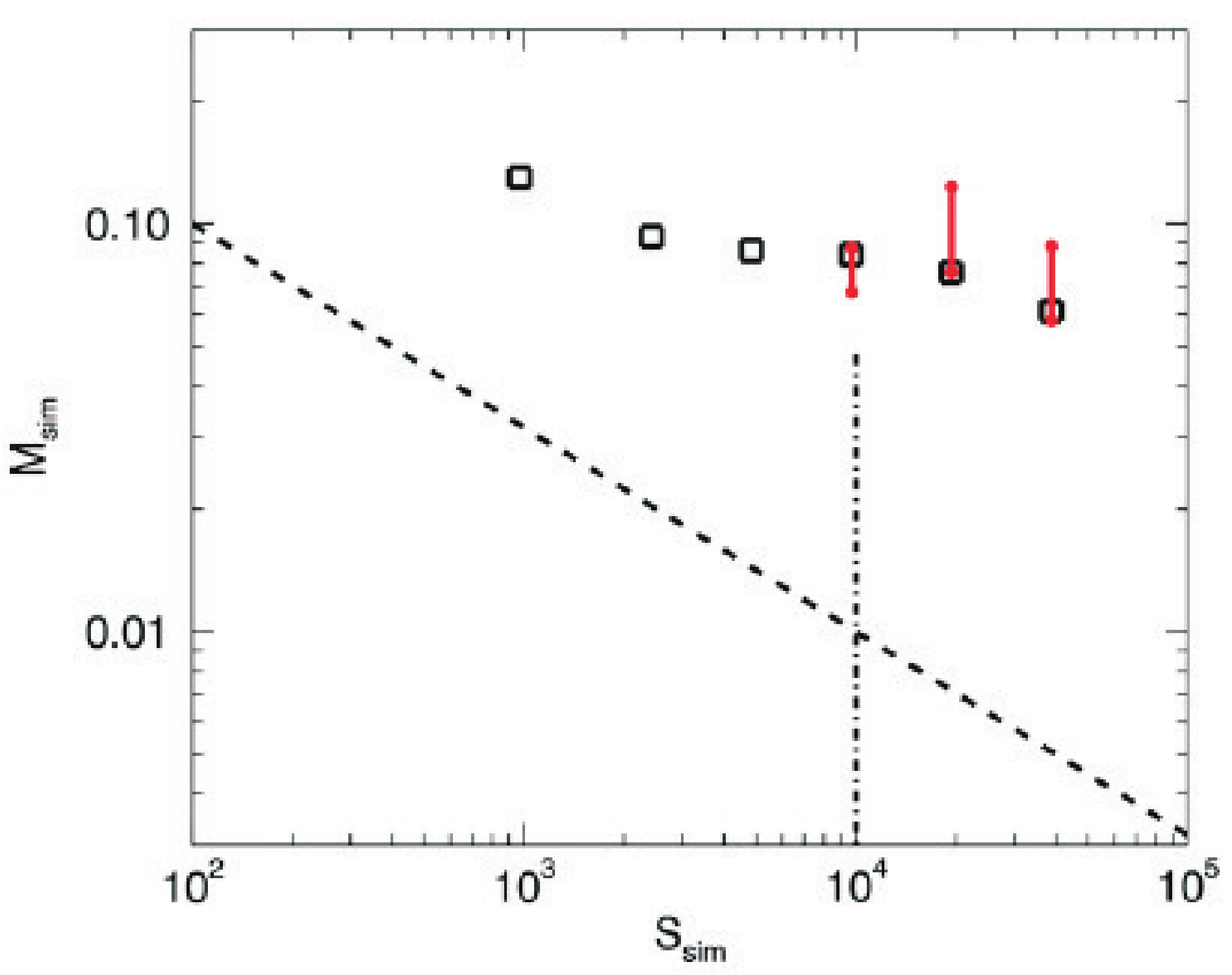}
}
\caption{Normalized magnetic reconnection rate $M_{sim}$ for six simulations with
different Lundquist number ($S_{sim}$). The squares show the reconnection rate taken
at a time in each simulation when the length of the current sheet has reached
$L_{sim}$ = 2.75 $L_0$, and in all simulations is before the onset of the plasmoid
instability. The red lines show the range in reconnection rate taken again at
later times in the three plasmoid-unstable simulations, after the plasmoids are
formed. The dashed line is the Sweet-Parker scaling law $M\sim 1/S$. The
dot-dashed line shows the separation between plasmoid-stable and plasmoid-unstable
regimes for these multi-fluid simulations. Adapted from Leake et al. (2012).}
\label{fig:leake20121}
\end{figure}

\citet{leake2012} presented the self-consistent multi-fluid simulations of magnetic reconnection in a
chromospheric weakly ionized plasma. They simulated two-dimensional magnetic reconnection in a Harris current sheet with a numerical model which includes ion–-neutral scattering collisions, ionization, recombination, optically thin radiative loss, collisional heating, and thermal conduction. They found that in the resulting tearing mode reconnection the neutral and ion fluids become decoupled upstream from the reconnection site, creating an excess of ions in the reconnection region and therefore an ionization imbalance. Ion recombination in the reconnection region, combined with Alfv\'enic outflows, quickly removes ions from the reconnection site, leading to a fast reconnection rate independent of Lundquist number, $S=\mu v_A L/\eta$. Figure~\ref{fig:leake20121} shows the dependence of the normalized magnetic reconnection rate on Lundquist number. It is seen that the consideration of neutrals significantly enhances the reconnection rate for all values of Lundquist number as compared to the single fluid Sweet-Parker reconnection rate.

\begin{figure}
\center
{
\includegraphics[width=10cm]{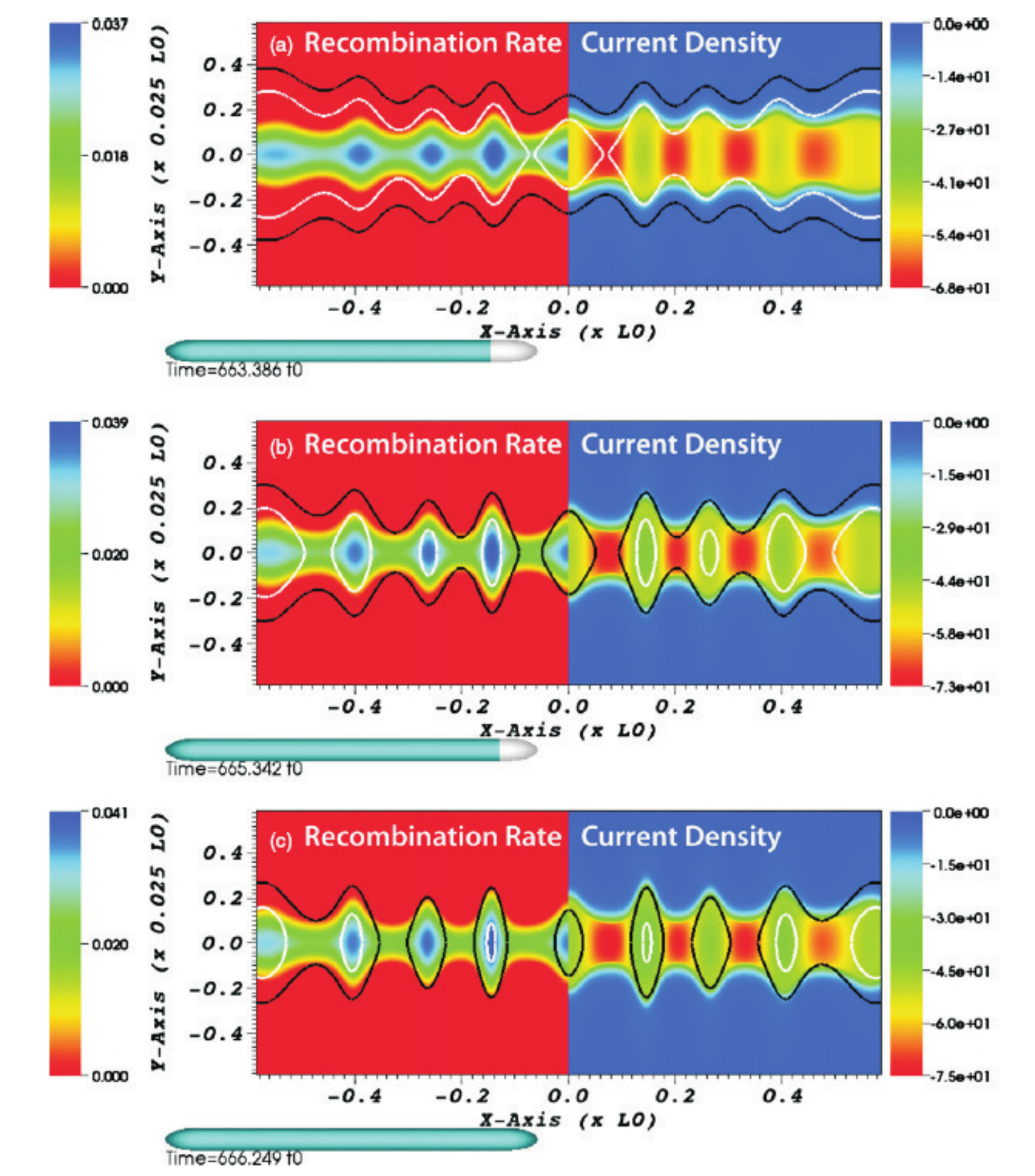}
}
\caption{Plasmoid formation and evolution: recombination rate $\Gamma^{rec}_i L^3_0 t_0$ (left) and current density $j\mu L_0/B_0$ (right), and two contour levels of $A_z$ of -0.0681 $B_0 L_0$
(white) and -0.0687 $B_0 L_0$ (black), at three different times in the simulation where $\eta=0.5 \times 10^{-5} \eta_0$. Adapted from \citet{leake2012}.}
\label{fig:leake20122}
\end{figure}

In addition, \citet{leake2012} found that the non-equilibrium partial ionization effects lead to the onset of the nonlinear secondary tearing instability (known as the plasmoid instability) at lower values of the Lundquist number than has been found in fully ionized plasmas (Figure~\ref{fig:leake20122}). The increase in the ion and electron densities in the plasmoids led to an increase in the recombination rate that allowed further contraction of the magnetic islands
on timescales comparable to the advection time of islands out of the current sheet.

Recently, \citet{murphy2015} performed 2.5D simulations of asymmetric reconnection
in weakly ionized, reacting plasmas where the magnetic field strengths, ion and neutral
densities, and temperatures were different in each upstream region. The plasma and
neutral components were evolved separately to allow non-equilibrium ionization. The asymmetric
reconnection in the chromosphere may occur when newly emerged flux interacts with
pre-existing, overlying flux. During simulations of asymmetric reconnection, the ion and neutral flows remained decoupled, but the decoupling was asymmetric. In the case of magnetic asymmetry, there was net neutral flow through the current sheet from the weak magnetic field upstream region into the strong field region, which resulted from a large scale neutral pressure gradient and imperfect coupling between ions and neutrals along the inflow direction. Similarly, the greater Lorentz force acting on the ions from the strong field upstream region led to the ions pulling the X-point into the weak field upstream region. These effects were not present during symmetric simulations. An observational consequence of these neutral flows through the current sheet is that most of the neutrals swept along with the outflow are more likely to have originated from the weak magnetic field upstream region. This may be especially important if there are different elemental abundances in each upstream region. \citet{murphy2015} also studied the Hall effect, which led to the development of a characteristic quadrupole magnetic field modified by asymmetry, but they did not find the X-point geometry expected during Hall reconnection. All simulations showed the development of plasmoids after an initial laminar phase.

The results presented in this subsection demonstrate that neutral atoms should be considered to describe magnetic reconnection in the solar chromosphere.

\subsubsection{Kelvin-Helmholtz instability in solar chromospheric jets}

The chromosphere is a highly inhomogeneous layer of the solar atmosphere, populated by a wide range of 
dynamical,  jet-like features such as type I and II spicules at the solar limb and fibrils, mottles, Rapid Redshifted and Blueshifted Excursions (REs) on the disk \citep[see the review by][]{2012SSRv..169..181T}.
These small-scale plasma structures are observed near the network boundaries in strong chromospheric spectral lines such as H$\alpha$, Ca {\sc{ii}} H \& K, and the Ca {\sc{ii}} infrared (IR) triplet. 
Some of these structures, in particular,  REs (on-disk counterparts of type II spicules)
which are absorption features detected in the blue and red wings of chromospheric lines \citep{2008ApJ...679L.167L, 2009ApJ...705..272R,2015ApJ...802...26K},
are characterised by very fast upflow velocities ($\mathrm{\sim 50-150~km~s^{-1}}$) and short lifetimes ($\sim$20-60~s).

It is suggested that the short lifetime of the spicular jets in the chromospheric lines 
may be the result of their fast heating to transition region (TR) or even coronal temperatures. \cite{2011Sci...331...55D} 
have provided evidence that TR and coronal brightenings in AIA passbands are occurring co-spatially and 
co-temporally with chromospheric REs. Furthermore, \cite{2012SoPh..280..425V} showed that the coronal hole large 
spicules observed with HINODE in the Ca ii H line are appearing in the TR O v 629.76 {\AA} line observed with the SUMER instrument on-board SOHO. 
Recently, \cite{2014ApJ...792L..15P} studied the thermal 
evolution of type II spicules using combined observations with the HINODE and Interface Region Imaging Spectrograph (IRIS) satellites and  
showed that the fading of spicules from the chromospheric Ca {\sc{ii}} H line is caused by rapid heating of the upward moving spicular plasma to higher temperatures. 
More recently, \cite{2016ApJ...820..124H} found a statistically significant match between automatically 
detected heating signatures in the corona, as observed in the AIA pass-bands, and quiet-Sun REs, with a minimum of 6\% 
of the detections at  $\sim\mathrm{10^6~K}$ (AIA Fe IX 171 {\AA}) being attributable to REs.

Despite a wealth of observations, the heating mechanism associated with chromospheric jets remains a mystery as far as 
heating timescales estimated with various dissipation mechanisms such as diffusivity, thermal conduction, radiative losses
are much longer than the dynamic timescales of observed chromospheric jets. 
Recent theoretical studies suggest that the Kelvin-Helmholtz Instability (KHI) could have a significant effect on the dynamics of chromospheric jets 
\citep{2011AIPC.1356..106Z,2015ApJ...802...26K, zaqarashvili14,2015ApJ...813..123Z}.
Mass flows in the chromospheric fine structures can create velocity discontinuities between the surface of the jets and surrounding media, 
which may trigger the KHI in some circumstances depending on the directions of flows and magnetic fields.

\begin{figure*}[t]
\begin{center}
\includegraphics[width=12.0cm]{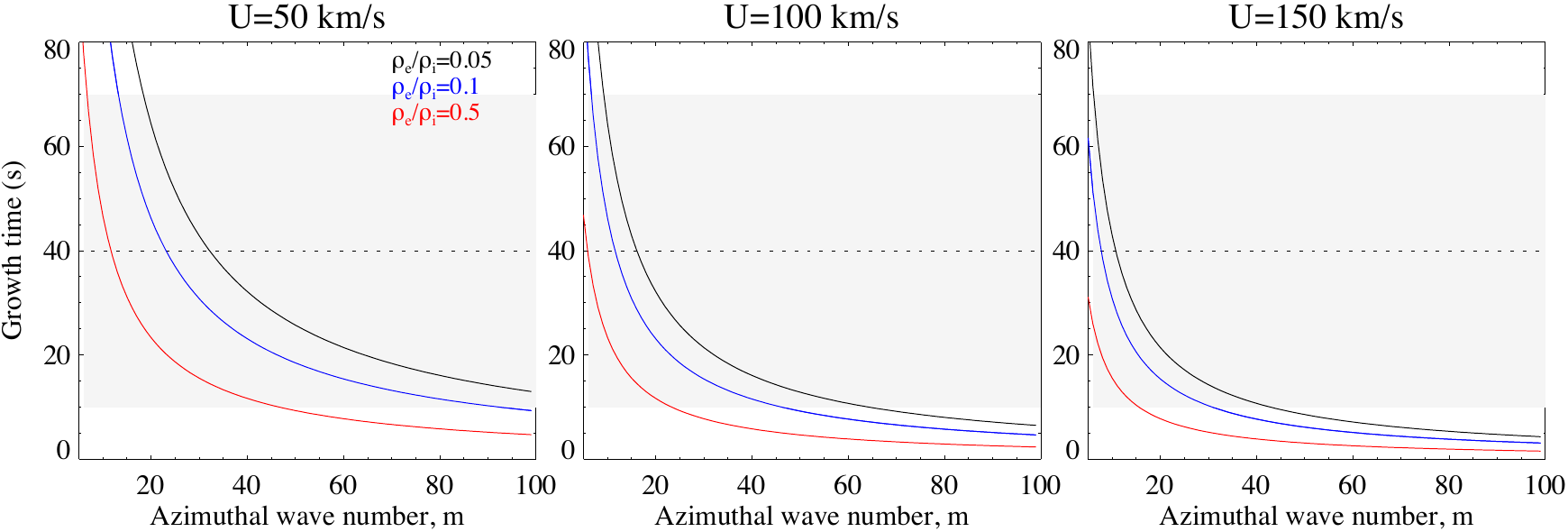}
\end{center}
\caption{KHI growth times as a function of $m$ for the twisted jets with different density ratios outside and inside the jet, $\rho_e/\rho_i,$ and flow speed $U$ from \cite{Kuridze2016}.
The grey-shaded areas indicate the typical range of lifetime ($\mathrm{\sim10 - 70~s}$) with the horizontal dotted lines indicate the average lifetime of REs and type II spicules.}
\label{fig3ku}
\end{figure*}

\cite{Kuridze2016} have investigated the dynamics and stability of quiet-Sun chromospheric jets observed at disk center 
using data obtained by the high resolution CRisp Imaging SpectroPolarimeter instrument on the Swedish 1-m Solar Telescope \citep{2003SPIE.4853..341S}.
The data revealed REs appearing as high speed jets in the wings of the H$\alpha$ line, are characterized by short lifetimes and rapid fading without any descending behavior.
To study the theoretical aspects of the stability of observed REs appearing as high speed upflows,
they modelled chromospheric jets as twisted magnetic flux tubes moving along their axis, 
and used the ideal magnetohydrodynamic approximation to derive the governing dispersion equation. 
Analytical solutions of the dispersion equation indicate that this type of jet is unstable to Kelvin-Helmholtz instability (KHI), with a very short (few seconds) instability growth time (Figure~\ref{fig3ku}). 

\begin{figure*}[t]
\begin{center}
\includegraphics[width=12.0cm]{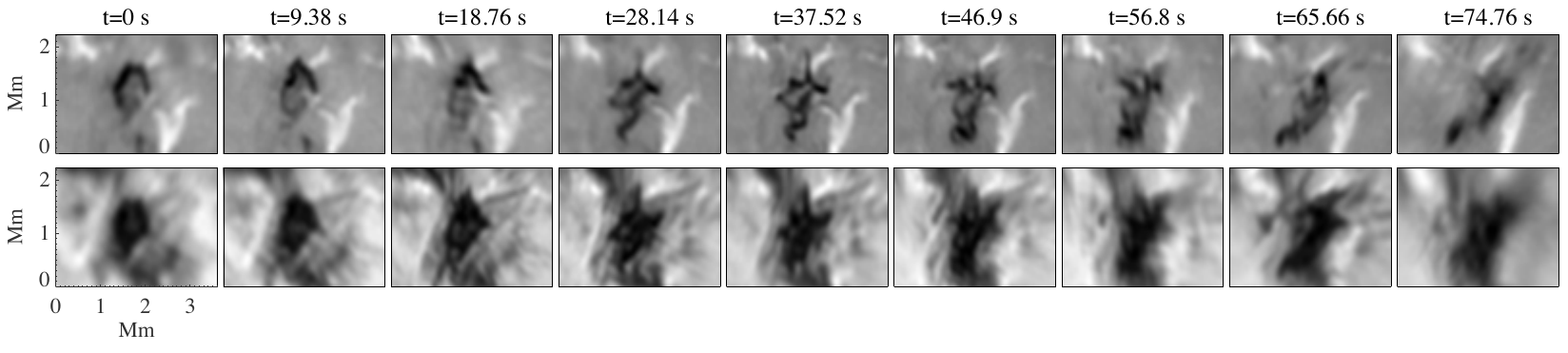}
\end{center}
\caption{Sequence of frames showing the temporal evolution of the jet ejected along the LOS
in $\mathrm{H\alpha-0.906~{\AA} (top) ~and~H\alpha - 0.543~{\AA}~of~H\alpha}$ (bottom) images from \cite{Kuridze2016}. Initially the jet has circular top, however, it 
rapidly develops azimuthal nodes around its boundary.}
\label{fig6ku}
\end{figure*}

\cite{Kuridze2016} have also detected a larger H$\alpha$ jet, 
which appears as a circular shaped absorption feature in the blue wing of H$\alpha$ (Figure~\ref{fig6ku}). 
Its radius ($\mathrm{\sim400~km}$) and LOS velocity 
($\mathrm{\sim34~km~s^{-1}}$) suggest that this jet could be the on-disk counterpart of 
a macrospicule or H$\alpha$ surge frequently observed at the solar limb. The morphology of the jet suggests that the plasma flow was oriented along the LOS.
Figure~\ref{fig6ku} shows that the structure develops azimuthal nodes over timescales of tens of seconds around its boundary. 
From a visual inspection they estimated $\mathrm{m\sim8}$ projected vortex-likes flows at around $\mathrm{t=37.52~s}$ (Figure~\ref{fig6ku}) and 
obtained growth time $\mathrm{\sim 52 - 145~s}$,  
which appears to be consistent with the timescale ($\sim$37 s) for the structure to develop the nodes in the image sequence presented in Figure~\ref{fig6ku}. 

To investigate non-linear, non-adiabatic effects \cite{Kuridze2016} have provided estimates of characteristic heating times for various
dissipation processes through the analysis of the energy equation in the partially ionized plasma.
The analysis shows that the ion-neutral collisions could be the most important 
process for the heating of the KH vortices and consequently the structure itself. For REs we estimate timescales of the heating due to the ion-neutral collisions, 
for KHI vortices with a high order of azimuthal wavenumbers, to be $\sim$20 - 130 s, comparable to the lifetimes of these chromospheric jets.

\subsubsection{Prominence equilibrium in partially ionized plasmas}
\label{prom}

Quiescent solar prominences are clouds of cool and dense plasma suspended against gravity by forces thought to be of magnetic origin. They form over a wide range of latitudes on the Sun \citep{mackay15}, but always along inversion polarity lines. Since the temperature of prominences is typically of the order of 10$^4$ K, the prominence plasma is only partially ionized. The exact ionization degree of prominences is unknown and the reported ratio of electron density to neutral hydrogen density \citep{patso02, heinzel15} covers about two orders of magnitude (0.1-10). Partial ionization brings the presence of neutrals in addition to ions and electrons, thus collisions between the different species are possible and the effects on the prominence equilibrium should be considered.

On the other hand, small amplitude oscillations in prominences and filaments are a commonly observed phenomenon \citep{oliver02}. These oscillations appear to be of local nature, and observational evidence reveals that, once excited, are damped in short spatial and temporal scales. Commonly, these oscillations have been interpreted in terms of linear magnetohydrodynamic (MHD) waves and to explain the damping different mechanisms (thermal, resonant damping in non-uniform media, partial ionization, etc.) have been invoked. The relevance of each mechanism can be assessed by comparing the spatial and time scales produced by each of them with those obtained from observations.  An extensive review on small amplitude oscillations in prominences can be found in \citet{aob12}.

One of the key problems in prominence physics is how these cool and dense structures are supported in the hotter and lighter solar corona. Usually, prominences have been considered as composed of fully ionized plasma and it has been assumed that its support is provided by the magnetic field \citep{mackay10, vialengvold}. However, the consideration of prominences as composed by partially ionized plasmas and their support has received slight attention. In order to take into account the downward plasma motions observed in prominences, \citet{mercier77} analysed the relative motion between charged and neutral species taking into account collisions between these species and found that the downward diffusion velocity of neutrals with respect to charged particles was about $6$ m/s which is much less than the global downward velocity (about $1$ km/s) in prominences. Therefore, they concluded that this effect was negligible and it can not explain the mass loss in these structures. Then, from the observational point of view the motions of both species could be considered identical. Building up on a model by  \citet{sakai84}, \citet{bakhareva92} studied the dynamic regimes of a prominence using a Kippenhahn-Schl\"{u}ter \citep{ks57} configuration and considering the prominence plasma as partially ionized. First of all, they consider a generalized Ohm's law taking into account the presence of neutrals and, then, they derived an induction equation which contains further terms coming from the non-stationarity of the problem and the presence of neutrals. Next, they assumed time dependent velocity and magnetic fields, and described the prominence dynamics in terms of a dimensionless function $a(t)$ which characterizes the degree of current-sheet compression. After perturbing the Kippenhahn-Schl\"{u}ter equilibrium, the resulting equations indicated that the equilibrium is unstable and that three different regimes can be studied. In the first dynamical regime, the non equilibrium is correlated with the process of material density oscillations growing up in the prominence, and if the relative density of neutrals decreases, the period of oscillations decreases too, and the growth time tends to infinity; in the second case, the density decreases aperiodically although the basic features of the previous regime remain; finally, in the third case, a process of slow condensation towards equilibrium density values is possible. In summary, the model shows that partially ionized plasma causes instability of the considered Kippenhahn-Schl\"{u}ter configuration, and that the excited oscillations of density, magnetic field and material velocity are able to destroy the prominence. The cause of this instability seems to be the inability of the magnetic field to support the plasma neutral component which starts to fall down, then, through ion-neutral collisions all the plasma starts to move which means that an additional current appears in the filament which, at the same time, affects the Lorentz force.

One of the typical features of solar prominences is its filamentary structure \citep{dunn60, engvold14}. The nature of these thin, long and inclined filament threads is still a matter of debate, and if one assumes that these threads delineate the magnetic field, then, the supporting mechanism against gravity remains to be understood. \citet{pecseli2000} proposed a model to explain the support of prominence threads based on the levitation of prominence plasma produced by weakly damped MHD waves in almost vertical flux tubes. The wave damping produced by ion-neutral collisions provides a wave pressure and levitation enough to support prominence plasma in the low corona. In the model, they consider that only the Lorentz force is at work and assuming that it fluctuates in time, they calculated the time average. The magnetic field is assumed to be composed of a background field, $B_0$, and a fluctuating part, $\tilde B_1$, and the time average of the Lorentz force is given by:
\begin{equation}
\bar F = \frac{1}{2\mu}k_{I}\vert B_1\vert^2 e^{-2 k_{I}z} \hat z
\label{Geq:pecselita}
\end{equation}
where $k_{I}$ is the imaginary part of the wavenumber. From a physical point of view, the wave produces a force on the medium due to the loss of wave momentum and, as a consequence, the wave is damped. Using the two fluid approach, assuming incompressibility and taking into account resistivity, the dispersion relation for this case is given by,
\begin{equation}
\omega^2 = k^2 v_a^2+k^2 \frac{\nu_{in}}{\mu \sigma}\left(1 -i \frac{\omega}{\nu_{in}}-\frac{\nu_{ni}}{\nu_{ni}-i \omega}\right)-i \omega \nu_{in}+ i \omega \frac{\nu_{in} \nu_{ni}}{\nu_{ni}-i \omega}
\label{Geq:pecselidr}
\end{equation}
where $v_a$ is the Alfv\'en velocity. Two limiting cases can be considered: (1) $\sigma \rightarrow \infty$ but $\nu_{in} \neq 0$, and (2) $\sigma$ finite and $\nu_{in} = 0$

For infinite conductivity, the dispersion relation becomes,
\begin{equation}
\omega^2 = \frac{k^2 v_a^2}{1+ \frac{\rho_n}{\rho_0} \frac{1 + i \tau_1}{1 + \tau_1^2}}
\label{Geq:pecselidrc1}
\end{equation}
where $\tau_1= \frac{\omega}{\nu_{ni}}$, and ion-neutral collisions will damp Alfv\'en waves. Considering spatial damping, and in the limit of weak damping, $k_{I} = \frac{\omega^2 \rho_n}{2 \nu_{ni} v_a (\rho_0 + \rho–n)}$, when  a real frequency is considered. For finite conductivity and the collisionless case, the dispersion relation is given by,
\begin{equation}
\omega^2 = v_a^2 k^2-i \frac{\omega}{\sigma \mu}k^2
\label{Geq:pecselidrc2}
\end{equation}
and, in the case of spatial damping, $k_{I} = \frac{\omega^2}{2 \sigma \mu v_a^3}$. Then, using typical parameters for prominence conditions, a comparison between the gravitational force and the time averaged Lorentz force was done. Numerical estimates suggest that the force exerted by Alfv\'en waves at frequencies $\omega \geq 2$ rad/s could possibly dominate over gravity. Therefore, spatial damping of Alfv\'en waves, produced by ion-neutral collisions, could give a non-negligible contribution to a force density coming from weakly damped Alfv\'en waves which could help to sustain the prominence mass.

\begin{figure}
\centering{
		  \resizebox{9cm}{!} {\includegraphics{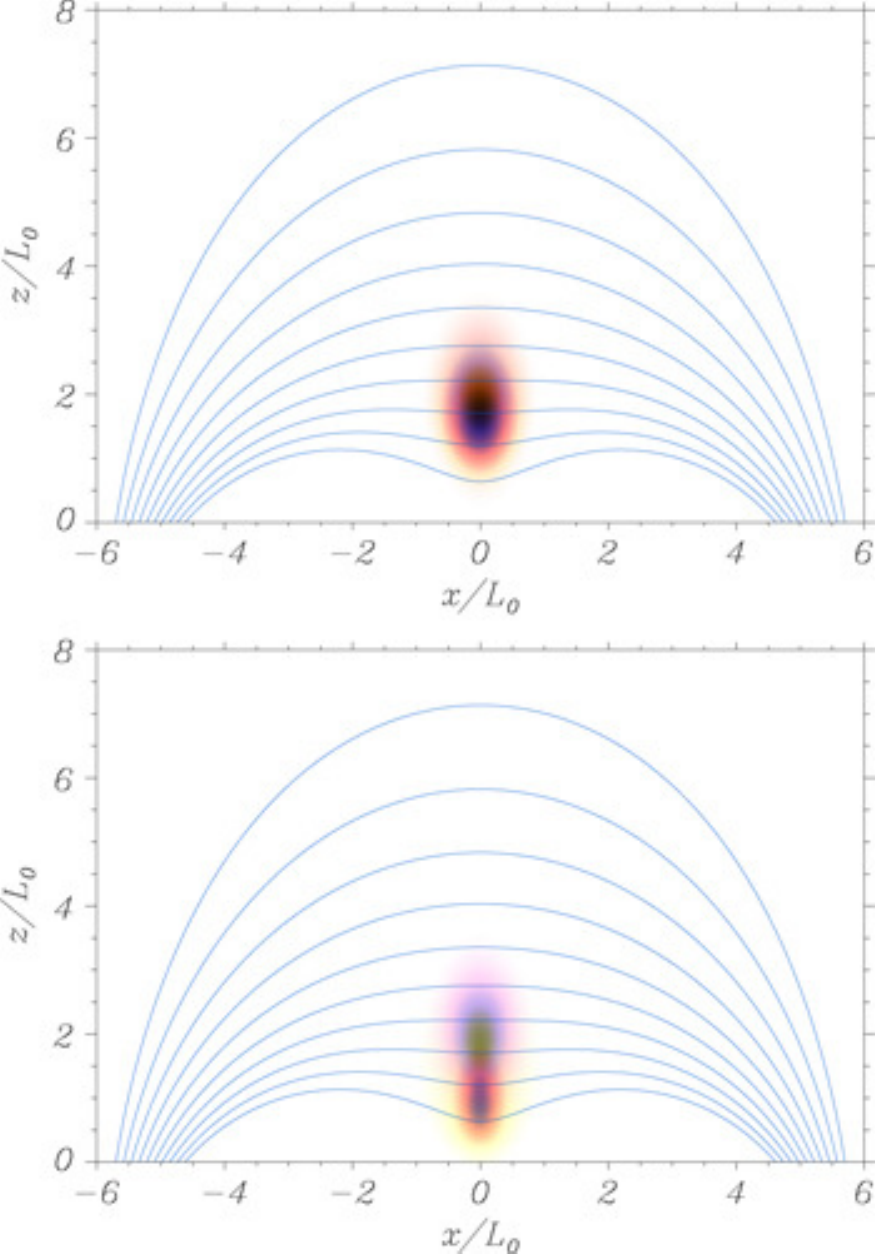}}}
\caption{Magnetic field lines (blue lines) and densities of ion-electron fluid (blue-green colours) and neutrals (yellow-pink colours) at t=4.8 \ min. Top panel: simulation corresponding to a collisional frequency $\nu_{ni}$ = 8.5\ Hz. Bottom panel: simulation corresponding to a collisional frequency $\nu_{ni}$ = 0.8\ Hz (See movie).  Adapted from \citet{terradas15}.}
\label{Gfig1}
\end{figure}

As stated before, \citet{bakhareva92} concluded that the draining of neutrals was responsible for the destabilization of the dynamical prominence model considered, since they are not attached to the magnetic field. This problem of how neutrals are supported in prominences is a matter of great interest. \citet{terradas15} have used the most simple version of two-fluid equations, including ion-neutral collisions and charge exchange collisions, to study the temporal behaviour of a prominence plasma located inside the solar corona. In the model, the prominence is represented by a large density enhancement, with respect to the background, composed by $75\%$ of neutrals and $25\%$ of ions, and the magnetic configuration is a quadrupolar configuration with dips \citep{terradas13}. The system of time dependent nonlinear equations together with boundary conditions have been solved in two dimensions and Figure \ref{Gfig1} shows the two dimensional distribution of both plasmas after $t = 4.8$ \ min. for two different collisional frequencies between neutrals and ions, $\nu_{ni}$, and in the absence of charge exchange collisions. In the top panel, we can observe that both fluids, ion-electron and neutrals, are essentially superimposed and that the spatial shape of the densities is very similar to that of the beginning of the simulation which means that mass redistribution is very small. On the contrary, in the bottom panel we observe that for a smaller collisional frequency neutrals fall, quickly diffusing through the magnetic field, therefore, neutrals can not be supported. Next, considering the top panel situation, and allowing the system to evolve longer, up to $10$ hr, the changes at the prominence core are very small since neutrals move downwards with an almost negligible velocity. This means that a partially ionized prominence plasma can be efficiently supported when the coupling between ions and neutrals is very strong, and we have a single fluid behaviour at the prominence body. In spite of the fact that the drift velocity is very small, the friction coefficient is large which means that the frictional force, oriented upwards because the drift velocity is positive, is enough to counterbalance the effect of gravity on neutrals. Considering the momentum equation for neutrals, the drift velocity in the $z$-direction can be approximated by $v_{Dz} \approx g/\nu_{ni}$, in agreement with \citet{gilbert02, gilbert11}, which means that for strong coupling the vertical component of the drift velocity is small, which is also in agreement with \citet{mercier77}. On the other hand, from the momentum equation for the ion-electron fluid it can be seen that the ionized fluid is supported by the magnetic field, but the inclusion of a neutral component produces the need to increase the deformation of the magnetic field in order to increase the restoring force. This result is only valid for strong coupling between the two fluids. Finally, including the effect of charge exchange collisions means that friction is increased (of the order of three times with respect to momentum-transfer collisions) and, as a consequence, neutrals downwards velocity suffers a further reduction which helps to sustain the prominence. In conclusion, the draining of neutrals is not an important issue for prominence evolution when a strong coupling between fluids is present.

\subsubsection{Damping of oscillations in partially ionized prominence plasmas}
\label{Gsec:damping}

Observational studies have allowed reliable values for the damping time of small amplitude prominence/filament oscillations to be obtained \citep{Molowny-Horas99, Terradas02, lin04}.  The values obtained are usually between 1 and 4 times the corresponding period, and large regions of prominences/filaments display similar damping times. Several studies have considered the damping of MHD waves in partially ionized structures of the solar atmosphere \citep{depontieu01, james03, khodachenko04,leake05, soler13} and, since prominences can be considered as partially ionized plasmas, a possible mechanism to damp prominence oscillations could be ion-neutral collisions. In this sense, \citet{khodachenko04, khodachenko06} made a qualitative study of the damping of MHD waves in a partially ionized plasma considering as damping mechanisms viscosity, collisional friction and thermal conduction. From this study, they concluded that collisional friction is the dominant damping mechanism for MHD waves in prominences.  On the other hand, flows in prominences are routinely observed in $H_\alpha$, UV and EUV lines \citep{labrosse10}.  In $H_\alpha$ quiescent filaments, the observed velocities range from $5$ to $20$ km s$^{-1}$ \citep{zirker98, lin03} and, because of physical conditions in filament plasma, they seem to be field-aligned. Furthermore, observations made with Hinode/SOT by \citet{okamoto07} reported the presence of synchronous vertical oscillatory motions in the threads of an active region prominence, together with the presence of flows along the same threads.  However, in limb prominences different kinds of flows are observed and, for instance, observations made by \citet{berger08} with Hinode/SOT have revealed a complex dynamics with vertical downflows and upflows.

In the following, assuming as general background model an unbounded, flowing, magnetized and partially ionized prominence plasma, we describe several quantitative studies which point out the role that could be played by ion-neutral collisions in the damping of MHD waves in prominence plasmas.

\subsubsection{Basic Equilibrium and Governing Equations}
\label{Gsec:be}

\citet{forteza07} derived the full set of MHD equations for a partially ionized, single-fluid hydrogen plasma (see sect.~\ref{Bsec:2-fluid}) which were applied to the study of the time damping of linear, adiabatic fast and slow magnetoacoustic waves in an unbounded prominence medium. This study was later extended to the non-adiabatic case, including thermal conduction by neutrals and electrons and radiative losses \citep{forteza08}. Next, \citet{barcelo11} considered a homogeneous, unbounded and flowing partially ionised hydrogen plasma characterized by a plasma density,
$\rho_0$, temperature, $T_0$, and number densities of neutrals, $n_{n}$, ions, $n_{i}$, and electrons $n_{e}$, with
$n_{e}=n_{i}$.  Thus, the gas pressure is $p_0 = (2 n_{i} + n_{n}) k_{\rm B} T_0$, where $k_{ B}$ is Boltzmann's constant.
The relative densities of neutrals, $\xi_{n}$, and ions, $\xi_{i}$, are given by
\begin{equation}
 \xi_\mathrm{n} = \frac{n_\mathrm{n}}{n_\mathrm{i} + n_\mathrm{n}}, \qquad \xi_{i} = \frac{n_\mathrm{i}}{n_\mathrm{i} + n_\mathrm{n}},
\end{equation}
where we have neglected the contribution of electrons.  We can now define a quantity, $\tilde \mu$,  given by,
\begin{equation}
 \tilde{\mu} = \frac{1}{1+\xi_\mathrm{i}}  \label{Geq:mu},
\end{equation}
which gives us information
about the plasma degree of ionisation. Following Eq.~(\ref{Geq:mu}), for a fully ionised
plasma $\tilde{\mu} = 0.5$, while for a neutral plasma $\tilde{\mu}
= 1$. This medium is threaded by a uniform magnetic field along the $x$-direction, with a field-aligned background flow. The equilibrium magnitudes of the
medium are given by
$$
  p_0 =\mathrm{const.}, \ \ \rho_0 = \mathrm{const.}, \ \ T_0 = \mathrm{const.},
  $$
  $${\bf B_0} =
  B_{0}\hat x, \ \ {\bf v}_0 =  v_{0}\hat x,
$$
where $B_{0}$ and $v_{0}$ are constants, and the effect of gravity
has been ignored. Since a medium with physical
properties akin to those of a quiescent solar prominence was considered, the density is
$\rho_\mathrm{0} = 5 \times 10^{-11}$  kg/m$^{3}$, the  temperature
$T_\mathrm{0} = 8 000$ \ K, the magnetic field $\vert \bf
B_\mathrm{0} \vert$= 10 \ G, and a
field-aligned flow with $v_\mathrm{0} = 15$ km/s, subalfv{\'e}nic but slightly supersonic, simulating the typical flows observed in the spines of quiescent filaments was included.

Using this background model, the dispersion relation for linear MHD waves was obtained by considering small perturbations from equilibrium,
linearising the single fluid basic equations \citep{forteza07, barcelo11}, and performing a Fourier analysis in terms of
plane waves, assuming that perturbations behave as $
f_1(\vect{r},t)=fe^{i(\omega t-\vect{k}\cdot\vect{r})}$. With no loss of generality, the
wavevector $\vect{k}$ lies in the $xz$-plane
($\vect{k}=k_x\vect{\hat{x}}+k_z\vect{\hat{z}}$),
and the wave frequency, $\omega$, could be obtained from
\begin{equation} \label{Geq:freq}
\omega = \Omega + k_{x} v_{0}
\end{equation}
$\Omega$ being the wave frequency in absence of
flow.

Following this procedure, the dispersion relation for Alfv\'en waves is given by,
\begin{eqnarray}
    \Omega^{2} -i \Omega k^{2}( \eta_\mathrm{C} \cos^{2}
     \theta + \eta \sin^{2} \theta)-v_\mathrm{a}^{2} k^{2} \cos^{2} \theta =
     0,
 \label{Geq:disp_alf1}
\end{eqnarray}
where $\theta$ is the angle between the wavenumber vector and the magnetic field, $v_a$ is the Alfv\'en speed, while $\eta$ and $\eta_\mathrm{C}$ are Spitzer's and Cowling's resistivities, respectively. Eq.~(\ref{Geq:disp_alf1}) points out that the time damping of Alfv\'en waves must be dominated by resistive effects.

The corresponding dispersion relation for thermal and magnetoacoustic waves is

\begin{eqnarray}
    (\Omega^{2} -k^{2} \Lambda^{2}) (ik^{2} \eta_\mathrm{C} \Omega-\Omega^{2})+k^{2} v_\mathrm{a}^{2}(\Omega^{2} -k_{x}^{2} \Lambda^{2}
    )+ \nonumber \\
    +i k^{2} k_{z}^{2}v_\mathrm{a}^{2} \Lambda^{2} \Xi \rho_{0} \Omega=
    0,
 \label{Geq:rdmag}
\end{eqnarray}
where $\Lambda^{2}$ is the non-adiabatic sound speed squared \citep{forteza08, soler08, soler10a} defined as
\begin{eqnarray}
    \Lambda^{2} = \frac{c_\mathrm{s}^2}{\gamma}\Big[\frac{(\gamma -1)(\kappa_{e\parallel}k^2_x+ \kappa_n k^2+\omega_T-\omega_{\rho})
     +i \gamma \Omega}{(\gamma-1)(\kappa_{e\parallel}k^2_x+\kappa_n k^2+\omega_T)+i \Omega}\Big], \label{Geq:nass}
    \end{eqnarray}
    with
   \begin{eqnarray}
   \kappa_{e\parallel} & = &\frac{T_0}{p_0}\kappa_{e\parallel}, \ \  \kappa_{n}  = \frac{T_0}{p_0}\kappa_{n} \nonumber \\
  \omega_{\rho}& = & \frac{\rho_0}{p_0}\rho_0 L_{\rho}, \ \  \omega_T = \frac{\rho_0}{p_0} T_0 L_{T} \nonumber
    \end{eqnarray}
and where the effects of optically thin radiative
losses, thermal conduction by electrons and neutrals, and a constant
heating per unit volume are included, and $\Xi$ is,
\begin{equation}
 \Xi = \frac{ \xi_\mathrm{n}^{2}\xi_\mathrm{i}}{(1+\xi_\mathrm{i}) \alpha_{n}},
\end{equation}
with $\alpha_{n}$ a friction coefficient \citep{braginski65,
khodachenko04, leake05}.  Depending on the value given to $\tilde \mu$ and to both
Spitzer's and Cowling's resistivities,
one may have different types of plasmas, and that of interest in this case is a partially ionized plasma characterized by $0.5 < \tilde \mu < 1$, $\eta \neq \eta_\mathrm{C}$ and $\Xi \neq 0$. Other important parameters are the numerical value and behaviour of the sound
$(c_\mathrm{s})$ and Alfv{\'e}n $(v_\mathrm{a})$ speeds. In the case of
 density and magnetic field assumed, the Alfv{\'e}n speed has a
constant numerical value of $126.15$ \ km/s. However, since the
sound speed depends on gas pressure, which is a function of the
number densities of ions and neutrals, its numerical value is not
constant but depends on the ionisation fraction considered.  For a
fully ionised plasma, the sound speed is $14.84$ \ km/s, while for a
partially ionised plasma with $\tilde \mu = 0.95$ its value
decreases to $10.76$ \ km/s. This variation in the sound speed can
be important since depending on the flow speed and ionisation
fraction chosen, the flow speed could be greater than, smaller than,
or equal to the sound speed, which affects the direction of
propagation of slow and thermal waves (Carbonell et al.  2009).

%%%%%%%%%%%%%%%%%%%%%%%%%%%%%
\subsubsection{Temporal damping of Alfv\'en waves}
 \label{Gsec:alfw}
%%%%%%%%%%%%%%%%%%%%%%%%%%%%%%

The solutions to the dispersion relation~(\ref{Geq:disp_alf1})
written in terms of $\omega$, frequency measured by an observer external to the flow, are,
\begin{eqnarray}
\omega_\mathrm{hfb} & = & k_x v_0 +  \nonumber \\ & + & \frac{\sqrt{4 v_\mathrm{a}^{2} k^2 \cos^2 \theta - (\eta_C \cos^2 \theta + \eta \sin^2 \theta)^2 k^4}}{2} + \nonumber \\
& + & \frac{i k^2 (\eta_C \cos^2 \theta + \eta \sin^2 \theta)}{2},  \label{Geq:aw1}
\end{eqnarray}
and
\begin{eqnarray}
\omega_\mathrm{lfb}&  = & k_x v_0 - \nonumber\\ & - & \frac{\sqrt{4 v_\mathrm{a}^{2} k^2 \cos^2 \theta - (\eta_C \cos^2 \theta + \eta \sin^2 \theta)^2 k^4}}{2} + \nonumber \\
& + & \frac{i k^2 (\eta_C \cos^2 \theta + \eta \sin^2 \theta)}{2}, \label{Geq:aw2}
\end{eqnarray}
where hfb and lfb mean high frequency and low frequency branch, respectively.

The real part of the frequency for the low frequency branch becomes zero for,
\begin{eqnarray}
k_\mathrm{lfb} = \frac{2 \sqrt{ v_\mathrm{a}^2 -v_0^2} \cos \theta}{\eta_C \cos^2 \theta + \eta \sin^2 \theta}.  \label{Geq:k1}
\end{eqnarray}
Since in this case $c_s < v_0 <  v_\mathrm{a}$, when $k < k_{lfb}$  the real part of the frequency is negative and the low frequency Alfv\'en wave propagates towards the negative part of the x-axis;  when $k = k_{lfb}$,  the low frequency Alfv\'en wave becomes non propagating, and when $k > k_{lfb}$, the real part of the frequency is positive and the low frequency Alfv\'en wave propagates towards the positive part of x-axis, with the reversal of the behaviour occurring at $k = k_{lfb}$.
Furthermore, the square root in Eqs.~(\ref{Geq:aw1}) and (\ref{Geq:aw2})  becomes zero when $k = k_c$,
\begin{equation}
k_\mathrm{c} = \frac{2 v_\mathrm{a} \cos \theta}{\eta_C \cos^2 \theta + \eta \sin^2 \theta}.
\label{Geq:kc}
\end{equation}
with $k_c > k_{lfb}$. For $k = k_c$, both real parts of the high and low frequency branches become $k_x v_0$, however, for $k > k_c$, while the real part of both Alfv\'en waves is still given by  $k_x v_0$, the imaginary parts of the frequency for both branches become,
\begin{eqnarray}
\omega_\mathrm{i,hfb} & = &   \frac{ k^2 (\eta_C \cos^2 \theta + \eta \sin^2 \theta)}{2} + \nonumber \\
& + & \frac{\sqrt{(\eta_C \cos^2 \theta + \eta \sin^2 \theta)^2 k^4 - 4 v_\mathrm{a}^{2} k^2 \cos^2 \theta}}{2}.
\end{eqnarray}
and
\begin{eqnarray}
\omega_\mathrm{i,lfb} & = &  \frac{ k^2 (\eta_C \cos^2 \theta + \eta \sin^2 \theta)}{2}  - \nonumber \\
& - &  \frac{\sqrt{(\eta_C \cos^2 \theta + \eta \sin^2 \theta)^2 k^4 - 4 v_\mathrm{a}^{2} k^2 \cos^2 \theta}}{2},
\end{eqnarray}
Therefore, for $k > k_c$, we end up with two Alfv\'en waves propagating in the same direction, with the same real frequency ($k_x v_0$) but with different imaginary parts of the frequency i. e. having different damping times.

 \begin{figure}
	  \centering{
		  \resizebox{8cm}{!} {\includegraphics{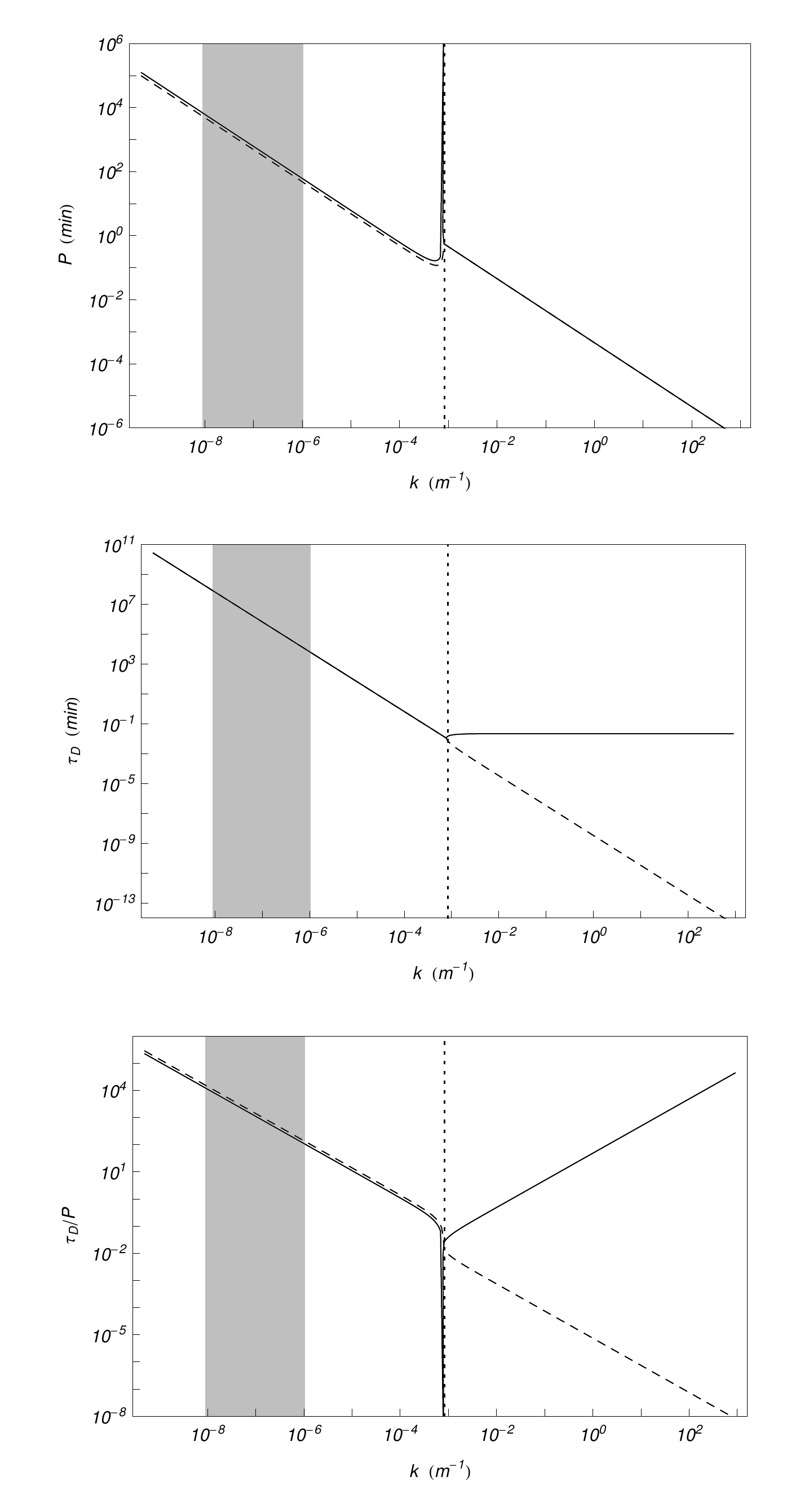}}}
		   \vspace{-3mm}
		   \caption{Period, damping time, and ratio of the damping time to period versus the wavenumber
for the long (solid) and short (dashed) period Alfv\'en waves in a PIP  with
$\tilde{\mu}=0.8$. The background flow speed is $15$ \ km/s. In all the panels, the shaded region
corresponds to the interval of observed wavelengths in prominence
oscillations. Adapted from \citet{barcelo11}.}     \label{Gfig2}
\end{figure}

Figure~\ref{Gfig2} displays the period ($P=
\frac{2 \pi}{\omega_\mathrm{r}}$), damping time ($\tau_\mathrm{D} =
\frac{1}{\omega_\mathrm{i}}$) and the ratio of damping time to period ($\tau_D/P$) versus the wavenumber and confirms the above analytical results. It shows the unfolding in period due to the flow and the behaviour of the period, for both waves, around the critical wavenumbers. For wavenumbers larger than $k_c$, both waves display the same period. Also, in the case of the damping time it can be seen that once the critical wavenumber $k_c$ is attained we are left with two different damping times for wavenumbers larger than $k_c$. One damping time is almost constant while the other decreases in a continuous way. Of course, the ratio $\tau_D/P$ has also two branches which behave quite differently for wavenumbers larger than $k_c$. Furthermore, we can also observe that  the critical wavenumbers, $k_{lfb}$ and $k_c$, are outside the region of observed wavelengths in prominence oscillations. Regarding the damping time of Alfv\'en waves, Figure~\ref{Gfig2} clearly shows that,  within the range of observed wavelengths, it is very large and quite incompatible with the up to now reported damping times from observations. The same happens with the ratio between the damping time and the period. This clearly points out that, at least within the range of observed wavelengths in prominence oscillations, the ion-neutral collisions mechanism can not provide with an explanation for the temporal damping of Alfv\'en waves in partially ionised prominence plasmas.

%%%%%%%%%%%%%%%%%%%%%%%%%%%%%
\subsubsection{Temporal damping of magnetoacoustic waves}
 \label{Gsec:magw}
%%%%%%%%%%%%%%%%%%%%%%%%%%%%%%

  \begin{figure}
	  \centering{
		  \resizebox{8cm}{!} {\includegraphics{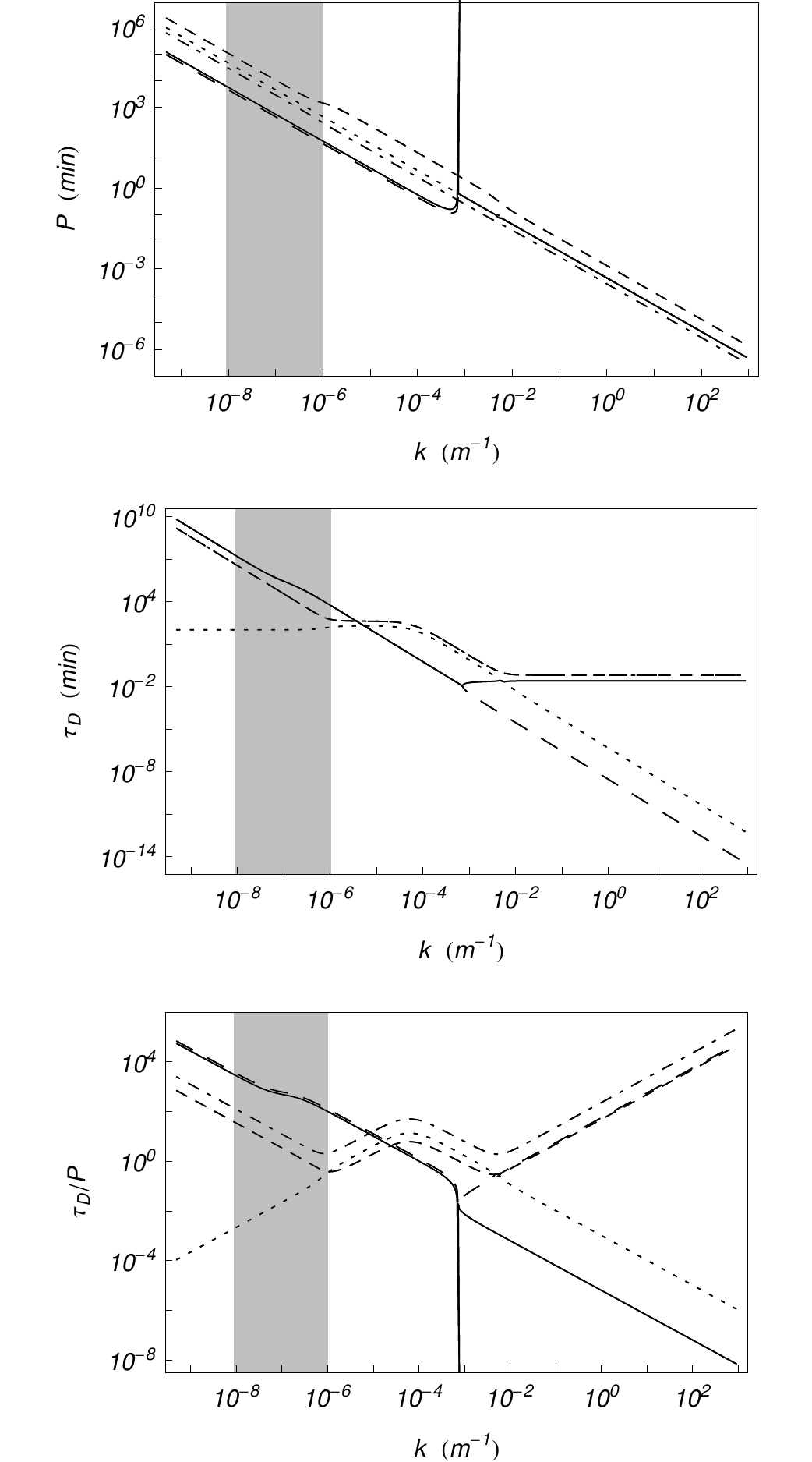}}}
		   \vspace{-3mm}
		   \caption{Period, damping time, and ratio of the damping time to period versus the wavenumber
for magnetoacoustic waves in a PIP  with
$\tilde{\mu}=0.8$. Slow waves (short-dashed and dot-dashed lines), fast waves (continuous and long-dashed lines), thermal wave (dotted line).  The location of the critical wavenumbers is defined by the presence of peaks and bifurcations in the plots. The background flow speed is $15$ km/s and the shaded region
corresponds to the interval of observed wavelengths in prominence
oscillations. Adapted from \citet{barcelo11}.}      \label{Gfig3}
\end{figure}

In this case, the obtained dispersion relation (Eq.~\ref{Geq:rdmag}) couples fast, slow and thermal waves. This dispersion relation is a fifth degree polynomial in the frequency providing two fast, two slow and one thermal wave. For non-parallel propagation, the dispersion relation can be solved numerically and Figure~\ref{Gfig3} displays the period, damping time and the ratio of damping time to period versus the wavenumber for magnetoacoustic and thermal waves. The results suggest that the behaviour of fast waves is similar to that of Alfv\'en waves (see \ref{Gsec:alfw}), although in a partially ionized plasma the presence of Cowling's resistivity results in the critical wavenumbers for fast waves being displaced towards small wavenumbers with respect to the case of a fully ionized and resistive plasma. On the other hand, the effect of the coupling is reflected in the behaviour of the damping times for fast and slow waves. At small wavenumbers, the damping time of fast waves is affected by thermal effects and additional distortions, with respect to the uncoupled case, appear for slow waves at large wavenumbers. On the contrary, the thermal wave is not affected by the coupling. Regarding the damping ratio, $\tau_D/P$, Figure~\ref{Gfig3} shows that, within the region of interest, it is very large for fast waves and very small for the thermal wave, while for slow waves, for wavenumbers between $10^{-7}$ and $10^{-6}$ m$^{-1}$, it is close to damping ratios found in observations. However, this ratio correspond to very long period oscillations having a more or less similar damping time.

Another important parameter is the ionisation fraction $\tilde{\mu}$. Varying this parameter, going from an almost fully ionised to an almost neutral prominence, the main effects appear in fast and Alfv\'en waves since they are influenced by Cowling's resistivity which depends on the ionisation fraction. The modification of Cowling's resistivity modifies the location of critical wavenumbers appearing in fast and Alfv\'en waves. Furthermore, our results suggest that in flowing partially ionised prominence plasmas, and within the range of observed wavelengths in prominence oscillations,  resistive effects, dominated by Cowling's resistivity, are not enough efficient to damp MHD waves responsible for these oscillations. However, the efficiency of ion-neutral collisions could be improved if the commonly assumed values for the characteristic prominence parameters (density, magnetic field, ionisation fraction) were different and as a result the numerical value of Cowling's resistivity was increased. This points out the need for an accurate and proper determination of these characteristic magnitudes.

Finally, for a particular flow speed we may have that at two additional wavenumbers the period of the high-period branch of slow waves becomes infinite, which is due to the coincidence between the numerical values of flow speed and the real part of the non-adiabatic sound speed. Since the non-adiabatic sound speed is determined by plasma physical conditions, these conditions become of paramount importance.

%%%%%%%%%%%%%%%%%%%%%%%%%%%%%
\subsubsection{Effect of Helium}
 \label{Gsec:He}
%%%%%%%%%%%%%%%%%%%%%%%%%%%%%%

Along above studies, a hydrogen plasma has been considered, however, $90\%$ of the prominence chemical composition is hydrogen while the remaining $10\%$ is helium. Therefore, it is of great interest to know the effect of the presence of helium on the behaviour of magnetohydrodynamic waves in a partially ionized plasma with prominence physical properties. This study has been done by \citet{soler10a} in an unflowing prominence plasma like that considered in previous paragraphs, but composed of hydrogen and helium. The species present in the medium are electrons, protons, neutral hydrogen, neutral helium (HeI), and singly ionized helium (HeII), while the presence of He III is negligible \citep{gl09}. Under such conditions the basic MHD equations for a non-adiabatic, partially ionized, unflowing single-fluid plasma have been generalized.

\begin{figure*}[!t]
\includegraphics[width=0.5\textwidth]{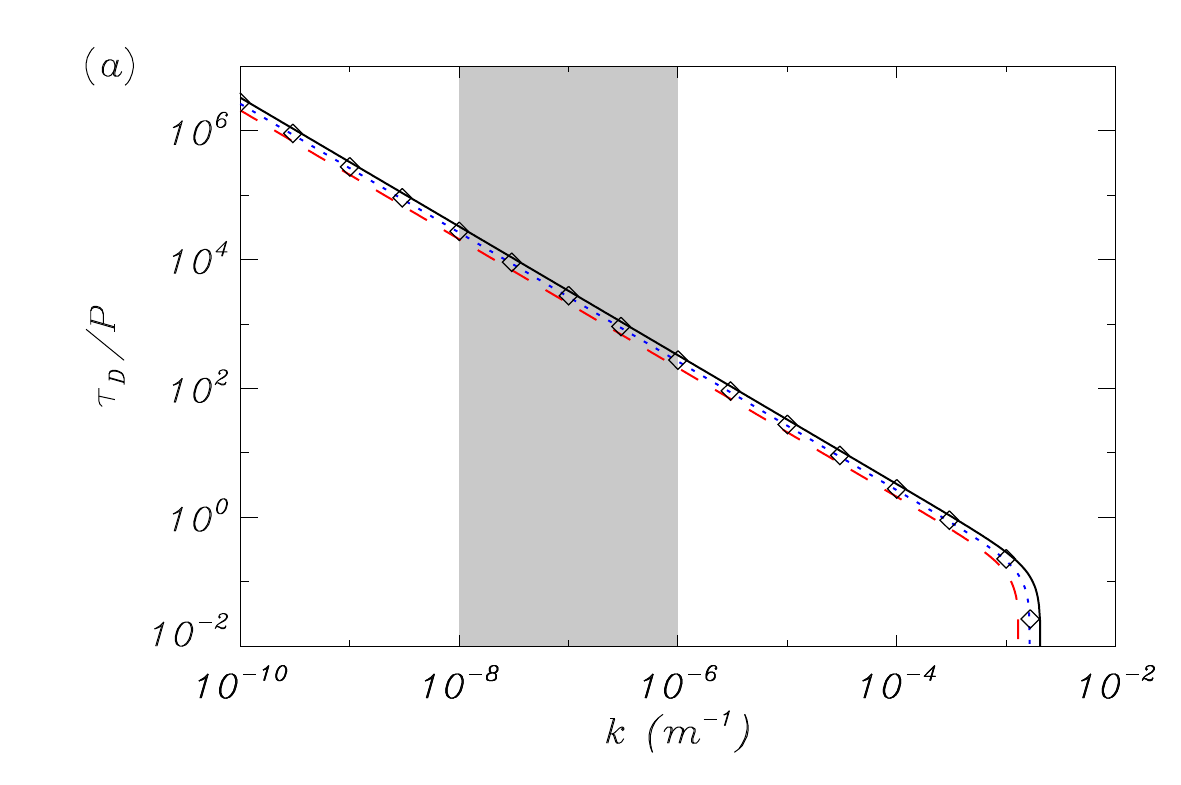}
\includegraphics[width=0.5\textwidth]{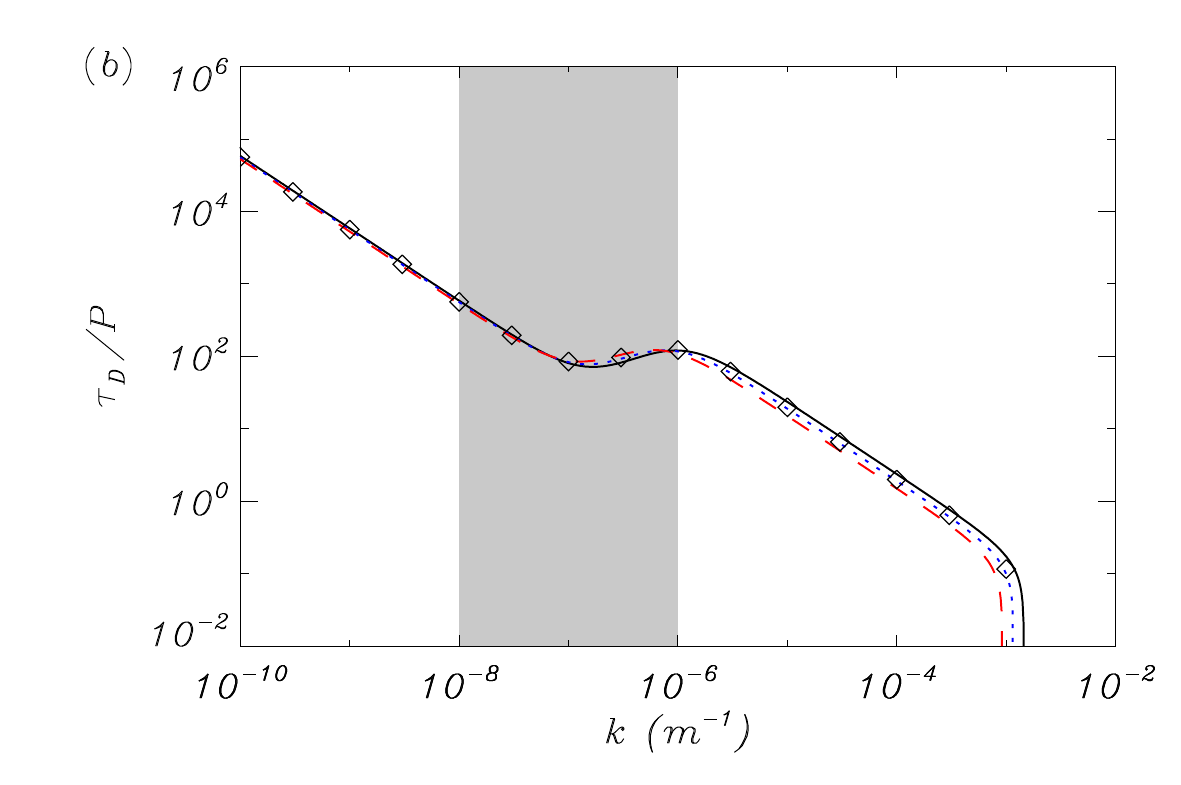}\\
\includegraphics[width=0.5\textwidth]{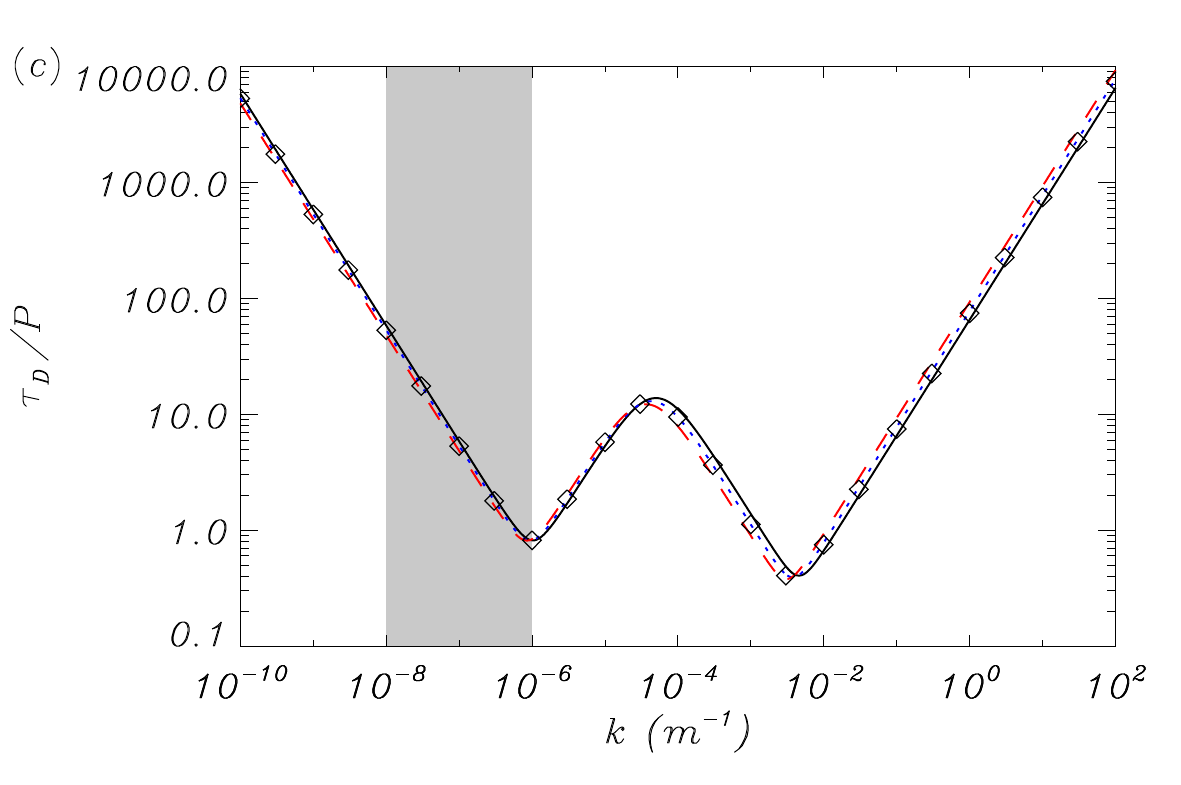}
\includegraphics[width=0.5\textwidth]{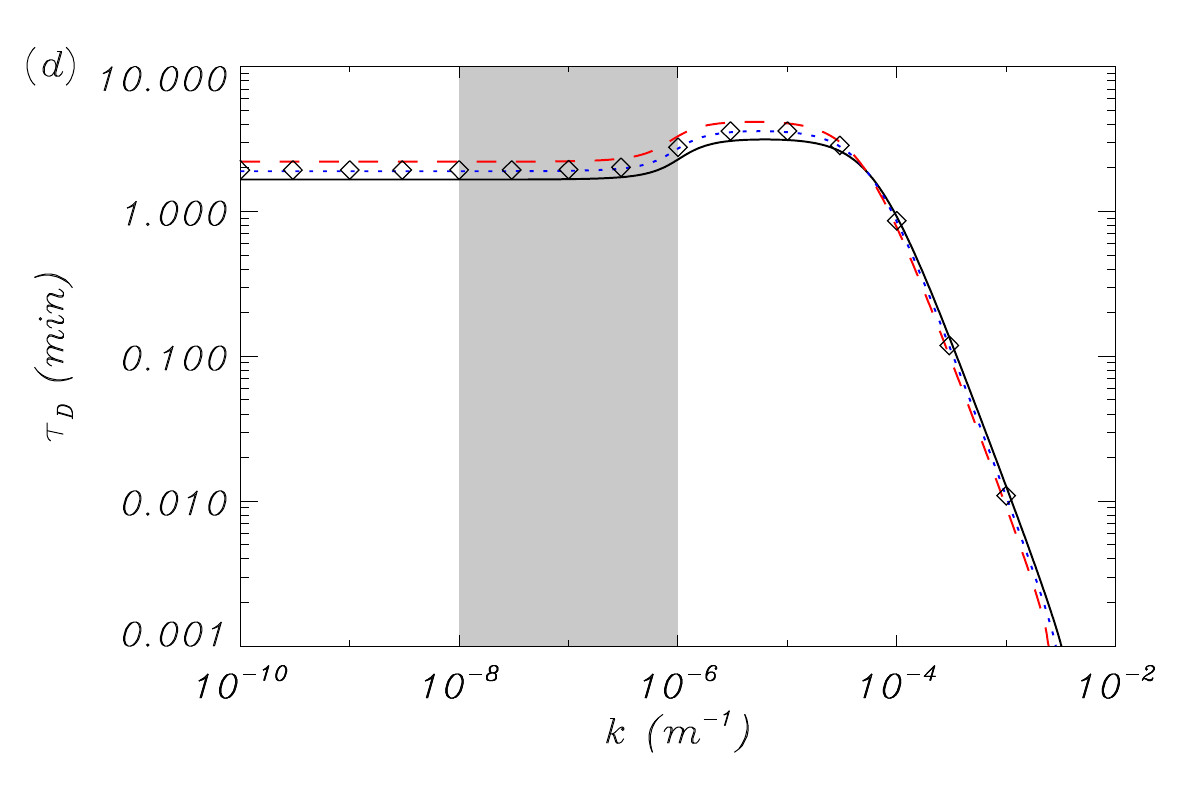}
\caption{($a$)--($c$): Ratio of the damping time to the period, $\tau_D/P$, versus the wavenumber, $k$, corresponding to the Alfv\'en wave,  fast wave, and  slow wave, respectively.
($d$) Damping time, $\td$, of the thermal wave versus the wavenumber, $k$. The different linestyles represent: $\xi_{{\rm HeI}} = 0\%$ (solid line), $\xi_{{\rm HeI}} = 10\%$ (dotted line), and $\xi_{{\rm HeI}} = 20\%$ (dashed line). In all computations, $\mutilde_{\rm H} = 0.8$ and $\delta_{\rm He} = 0.1$. The results for $\xi_{{\rm HeI}} = 10\%$ and $\delta_{\rm He} = 0.5$ are plotted by means of symbols for comparison. The shaded regions correspond to the range of typically observed wavelengths of prominence oscillations.  In all the figures shown, the angle, $\theta$, between the wavevector and the x-axis is $\pi/4$. Adapted from \citet{soler10a}.}
\label{Gfig4}
\end{figure*}

The hydrogen ionization degree is characterized by $\mutilde_{\rm H}$ which varies between $0.5$, for fully ionized hydrogen, and $1$ for fully neutral hydrogen. The helium ionization degree is characterized by $\delta_{\rm He} = \frac{\xi_{{\rm HeII}}}{\xi_{{\rm HeI}}}$, where $\xi_{{\rm HeII}}$ and $\xi_{{\rm HeI}}$ denote the relative densities of single ionized and neutral helium, respectively. Figure~\ref{Gfig4} displays $\tau_D/P$ as a function of the wavenumber $k$ for the Alfv\'en, fast, and slow waves, and the results corresponding to several helium abundances are compared for hydrogen and helium ionization degrees of $\mutilde_{\rm H} = 0.8$ and $\delta_{\rm He}=0.1$, respectively. It is observed that the presence of helium has a minor effect on the results. In the case of Alfv\'en and fast waves (Fig.~\ref{Gfig4}a,b), a critical wavenumber, $k_{\rm c}$ occurs (See
Eq.~(\ref{Geq:kc})) and, since Cowling's diffusivity $\eta_{\rm C}$ is larger in the presence of helium because of additional collisions of neutral and singly ionized helium species, $k_{\rm c}$ is shifted toward slightly lower values  than when only hydrogen is considered, so the larger $\xi_{{\rm HeI}}$, the smaller $k_{\rm c}$.  In the case of the slow wave (Fig.~\ref{Gfig4}c), the maximum and the right hand side minimum of $\tau_D/P$ are also slightly shifted toward lower values of $k$. Previous results from \citet{carbonell04} and \citet{forteza08} suggest that thermal conduction is responsible for these maximum and minimum of $\tau_D/P$. The additional contribution of neutral helium atoms to thermal conduction produces this displacement of the curve of $\tau_D/P$. In the case of Alfv\'en and fast waves, this effect is not very important.
Finally, the thermal mode has been considered. Since it is a purely damped, non-pr\-o\-pa\-ga\-ting disturbance, only the plot of the damping time, $\tau_D$, as a function of $k$ for $\mutilde_{\rm H} = 0.8$ and $\delta_{\rm He}=0.1$ is shown (Fig.~\ref{Gfig4}d). We observe that the effect of helium is different in two ranges of $k$. For $k > 10^{-4}$~m$^{-1}$, thermal conduction is the dominant damping mechanism, so the larger the amount of helium, the smaller $\tau_D$ because of the enhanced thermal conduction by neutral helium atoms. On the other hand, radiative losses are more relevant for $k < 10^{-4}$~m$^{-1}$. In this region, the thermal mode damping time grows as the helium abundance increases.  Since these variations in the damping time are very small, we again conclude that the damping time obtained in the absence of helium does not significantly change when helium is taken into account. In summary, this study points out that the consideration of neutral or single ionized helium in partially ionized prominence plasmas does not modify the behaviour of MHD waves found by \citet{forteza07, forteza08}.
However, \citet{Zaqarashvili11b} used a multifluid approach to study the damping of Alfv\'en waves in an isothermal and homogeneous partially ionized plasma when helium is included. They considered a three-fluid MHD approximation, where one component was electron-proton-singly ionized helium, and the other two components were neutral hydrogen and helium atoms. The results indicate that the presence of neutral helium enhances in a significant way the damping of Alfv\'en waves for a certain range of plasma temperature (10 000 - 40 000 K). Then, in the case of prominences, the damping of Alfv\'en waves in the prominence-corona transition region could be significantly enhanced while it is not affected in the prominence core, which agrees with the results by \citet{soler10a} using the single-fluid approximation.

%%%%%%%%%%%%%%%%%%%%%%%%%%%%%%%%%%%%%%%%%%%%%%%%%%%
\subsubsection{Spatial damping of Magnetohydrodynamic waves}
\label{Gsec:spa}
%%%%%%%%%%%%%%%%%%%%%%%%%%%%%%%%%%%%%%%%%%%%%%%%%%%

\citet{Terradas02} analyzed small
amplitude oscillations in a polar crown prominence and reported the presence of a plane propagating wave as well as an standing
wave.  In the case of the propagating wave, which was interpreted
as a slow MHD wave, the amplitude of the oscillations spatially decreased in a substantial way after a distance of $2-5 \times
10^4$ km from the location where wave motion was being generated.  This distance could be considered as a typical spatial
damping length, $L_\mathrm{d}$, of the oscillations.

Using the dispersion relations for Alfv\'en and magnetoacoustic-thermal waves given by Eqs.~(\ref{Geq:disp_alf1}) and (\ref{Geq:rdmag}), respectively, \citet{carbonell10} studied the spatial damping of MHD waves in a flowing partially ionized prominence plasma. The wavelength of the waves is given
by $\lambda = \frac{2 \pi}{k_\mathrm{r}}$, the damping length by $L_\mathrm{d} = \frac{1}{k_\mathrm{i}}$ and the damping length per wavelength is
${L_\mathrm{d}}/{\lambda}$, with $k = k_r + i k_i$.

\begin{figure}[!t]
\includegraphics[width=0.5\textwidth]{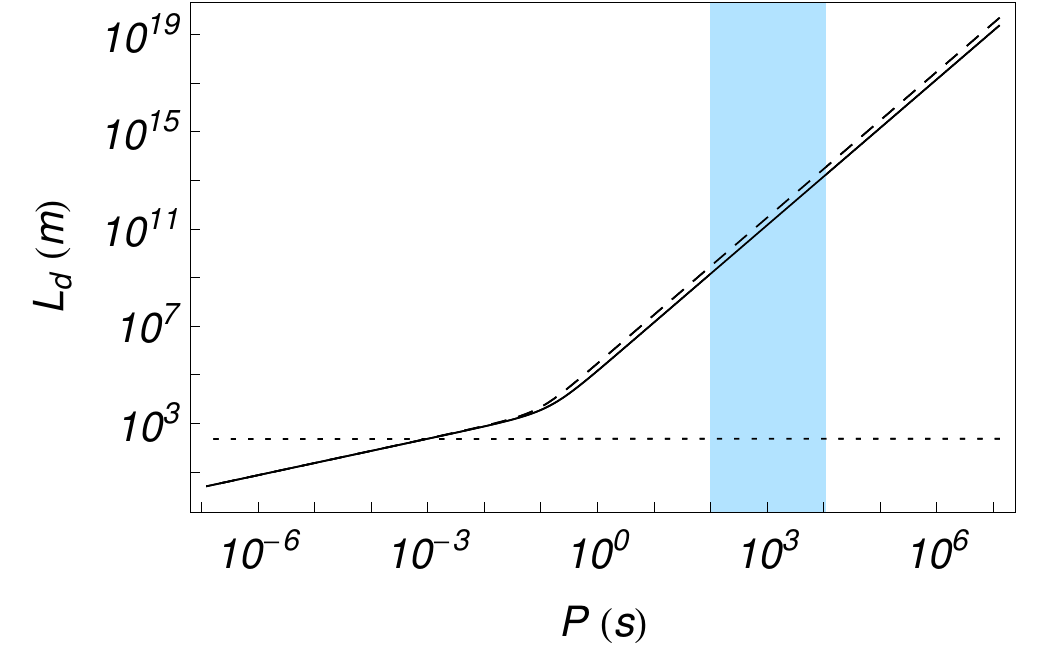}
\includegraphics[width=0.5\textwidth]{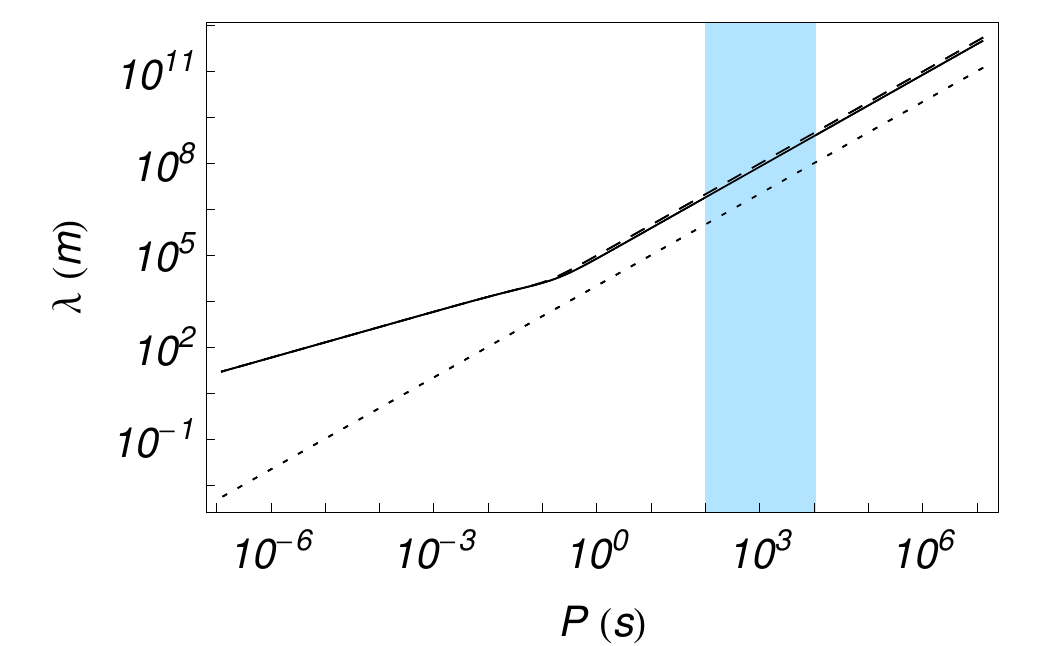}\\
\begin{minipage}{0.5\textwidth}
\includegraphics[width=\textwidth]{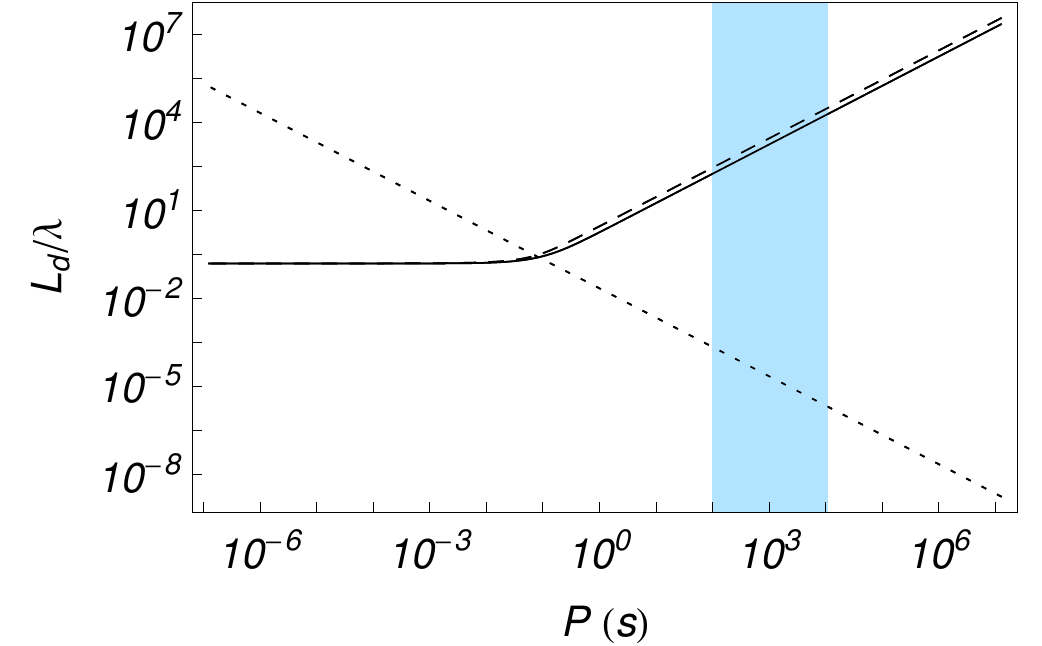}
\end{minipage}
\hspace{0.1cm}
\begin{minipage}{0.45\textwidth}
\caption{Damping length, wavelength, and ratio of the damping length to the wavelength versus period for the three (solid, dashed, dotted) Alfv\'en waves in a partially ionized plasma with an ionization degree $\tilde{\mu}=0.8$ and with a background flow of $10$ \ km s$^{-1}$. In all the panels, the shaded region corresponds to the interval of observed periods in prominence oscillations.  Adapted from \citet{carbonell10}}
\label{Gfig5}
\end{minipage}
\end{figure}

\begin{figure}[h]
 \includegraphics[width=0.5\textwidth]{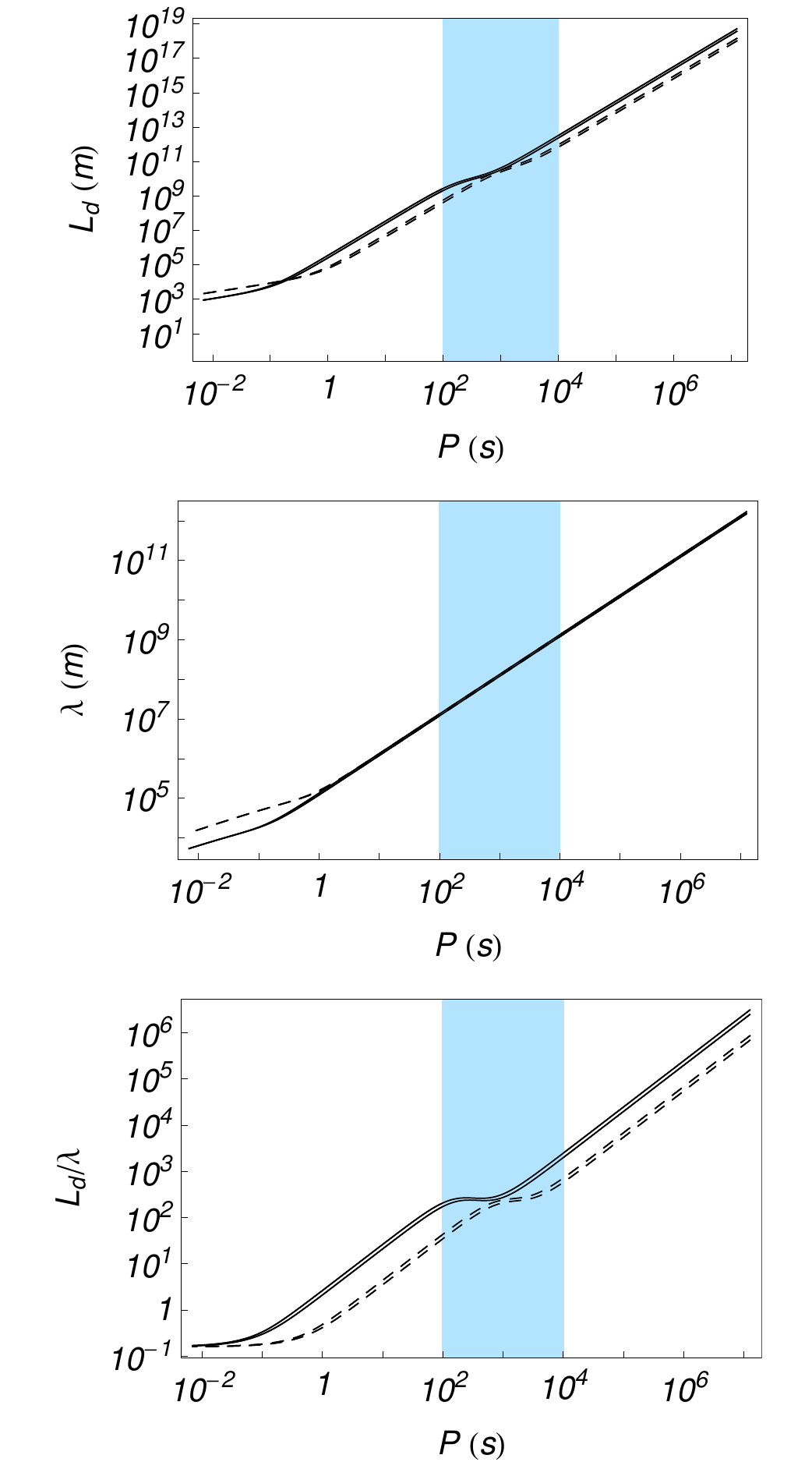}
\includegraphics[width=0.5\textwidth]{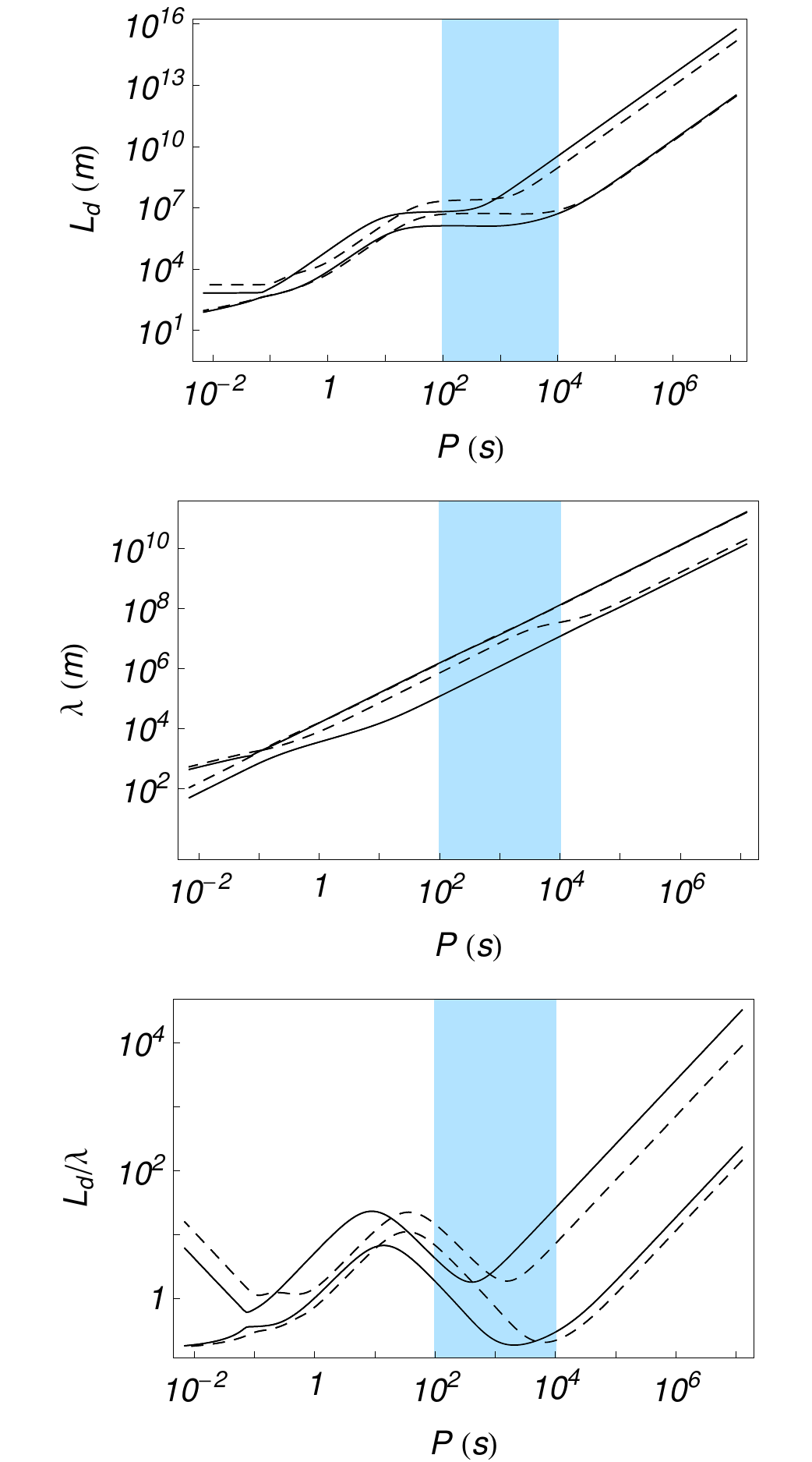}
		   \caption{Damping length, wavelength, and ratio of the damping length to the wavelength versus period
for the non-adiabatic fast (left panels), slow (right panels) waves in a partially ionized plasma with an ionization degree
 $\tilde{\mu}=0.8$ (solid) and $\tilde{\mu}=0.95$ (dashed). The flow speed is $10$ \ km s$^{-1}$. Adapted from \citet{carbonell10}}   \label{Gfig6}
 \end{figure}

In the case of Alfv\'en waves, three propagating Alfv\'en waves are obtained, and Figure~\ref{Gfig5} shows the numerical solution of dispersion relation (\ref{Geq:disp_alf1}).   For all the interval of periods considered, a strongly damped additional Alfv\'en wave appears, while on the contrary, the other two Alfv\'en waves are very efficiently damped for periods below $1$ s. However, within the interval of periods typically observed in prominence oscillations these waves are only efficiently attenuated when almost neutral plasmas are considered.

When Equation~(\ref{Geq:rdmag}) is expanded, it becomes a seventh
degree polynomial in the wavenumber $k$, whose solutions are three propagating fast waves, two slow waves and two thermal waves. Figure~\ref{Gfig6} displays the
behaviour of the damping length, wavelength and ratio of damping length versus wavelength for fast and slow  waves. The damping length of a fast wave in a partially ionized plasma is strongly diminished by neutral's
thermal conduction for periods between $0.01$ and $100$ s, and, at the same time, the radiative plateau present in fully ionized ideal plasmas  \citep{carbonell06} almost disappears.
The behaviour of slow waves is not so strongly modified as for fast waves, although thermal conduction
by neutrals also diminishes the damping length for periods below $10$ s, and a short radiative plateau still remains for
periods between $10$ and $1000$ s \citep{carbonell06}.  Finally, thermal waves are only slightly modified although the effect of partial
ionization is to increase the damping length of these waves, just the opposite to what happens with the other waves. Also, in the presence of flow, wavelengths and damping lengths are modified, and since for slow waves sound speed and observed flow speeds are comparable this means that the change in wavelength
and damping length are important, leading to an improvement in the efficiency of the damping.  Moreover, the maximum of efficiency
is displaced towards long periods when the ionization decreases, and for ionization fractions from $0.8$ to $0.95$ it is clearly
located within the range of periods typically observed in prominence oscillations with a value of $L_\mathrm{d}/\lambda$ smaller
than $1$.  This means that for a typical period of $10^{3}$ s, the damping length would be between $10^{2}$ and $10^{3}$ km,
the wavelength around $10^{3}$ km and, as a consequence, in a distance smaller than a wavelength the slow wave would be strongly attenuated. In conclusion, the joint effect of non-adiabaticity, flows and partial ionization allows spatially damping of slow waves in an efficient
way within the interval of periods typically observed in prominences.

\subsubsection{Wave heating and energy balance in partially ionised prominences}

The energy balance in solar prominences, and the understanding of the processes involved with heating and cooling of the plasma, are difficult problems that are intimately linked to the prominence formation and structure \citep[see][]{gilbert2015}. Incident radiation is generally accepted as the dominant prominence heating mechanism. However, when the balance between incident  radiation and cooling is taken into account, the obtained radiative-equilibrium temperatures are lower than what is typically inferred from observations.  An additional, non-negligible source of heating seems to be necessary to raise the prominence temperature up to the expected values \citep{labrosse10,Heinzel2010,Heinzel2012}. Dissipation of magnetohydrodyamic (MHD) wave energy in the partially ionised prominence plasma is proposed as another possible source of prominence heating. Recently, \citet{soler2016} explored the role of Alfv\'en waves in prominence heating. They used a slab model with a transverse magnetic field to represent a solar prominence embedded in the corona and modeled the prominence medium as a three-fluid plasma composed of a charged ion-electron single fluid and two neutral fluids comprised of neutral hydrogen and neutral helium. These three fluids exchange momentum because of particle collisions, which cause the damping of the waves and dissipation of wave energy.  \citet{soler2016} consistently computed the  plasma heating rate  and compared it with the prominence radiative losses. They concluded that wave heating is efficient for waves with periods shorter than 100~s and may be efficient enough to compensate for a fraction of the radiated energy. \citet{soler2016}  estimated the  volumetric  heating integrated over the range of periods between 0.1~s and 100~s  to be as large as 10\% of the bulk radiative energy of the cool prominence plasma and it can possibly account for the additional heating necessary to explain the observed prominence core temperatures \citep{Heinzel2010}. Thus, partial ionisation effects are also shown to be an important actor concerning the energy balance in prominences.

\subsubsection{Kelvin-Helmholtz, Rayleigh-Taylor and dissipative instabilities}

Observations of quiescent prominences have revealed the presence of flows that show a turbulent behavior  \citep[see][]{berger10,ryutova10}. The Kelvin-Helmhotz instability (KHI) has been proposed as a feasible mechanism that can contribute to the development of turbulence in prominences \citep{ryutova10}. In solar coronal plasmas, this instability has also been observed in coronal mass ejections \citep{foullon11,ofman11,mostl13} and coronal streamers \citep{feng13}.

The KHI is well studied in hydrodynamics and it arises at the interface of two fluid layers that move with different speeds. Flows are
generally uniform in both layers, but strong velocity shear arises near the interface, which consequently becomes unstable to spiral-like perturbations. Inclusion of magnetic field obviously modifies the thresholds and growth rates of KHI. It has been shown that the flow-aligned magnetic field suppresses the KHI for sub--Alfv\'enic flows of fully ionized plasmas due to magnetic tension force \citep[see, e.g.,][]{chandrasekhar61}. The flow velocities that are  measured  in quiescent prominences are typically lower than 30~km~s$^{-1}$ \citep[e.g.,][]{zirker98,berger10}, while the expected Alfv\'en velocity in  prominence threads is of the order of 100~km~s$^{-1}$. Therefore, the onset of the KHI in quiescent prominences seems unlikely according to the classical criterion as the mass flows are presumably field-aligned in prominence threads.

\begin{figure}
  \includegraphics[width=0.49\columnwidth]{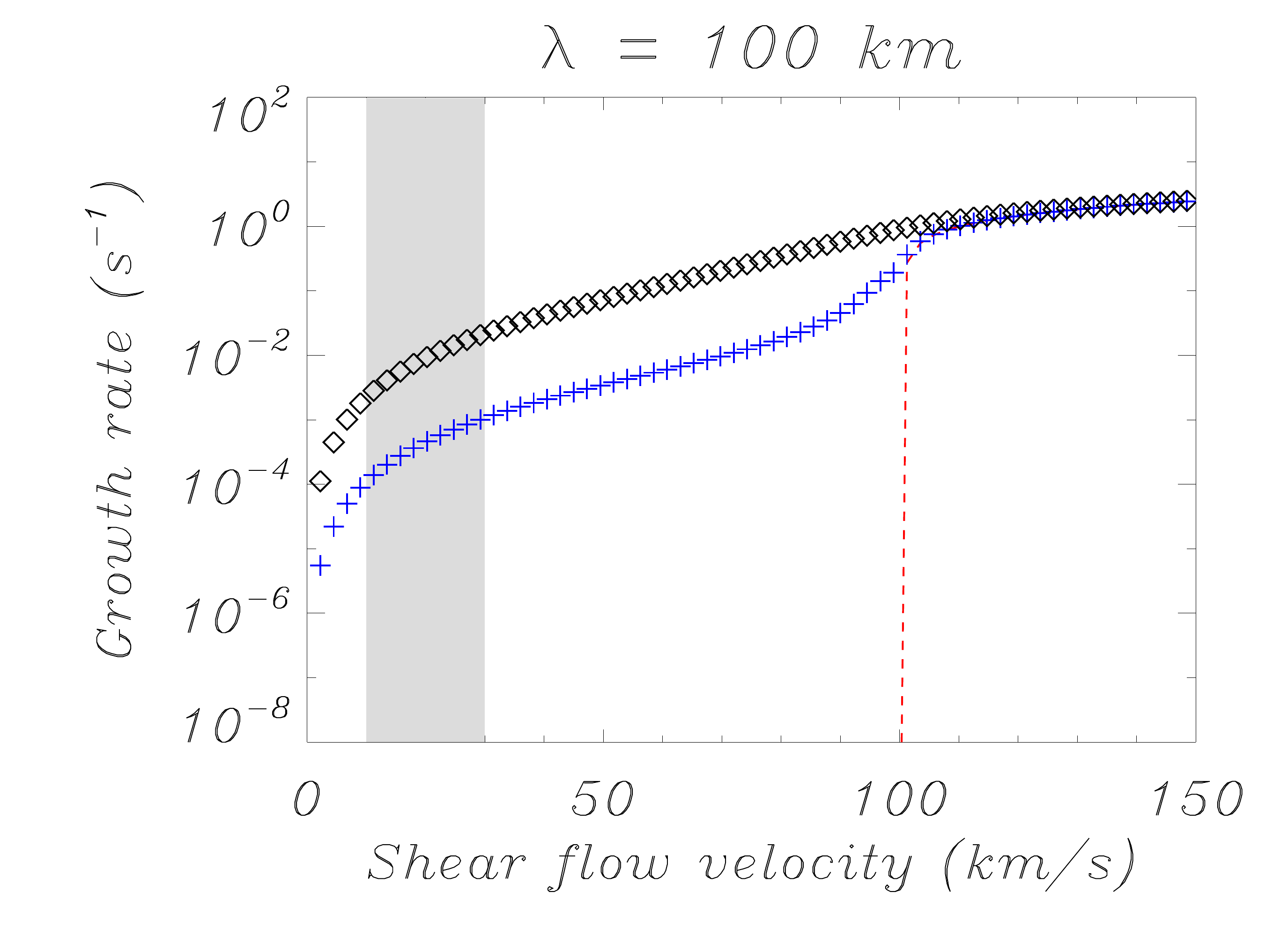}
   \includegraphics[width=0.49\columnwidth]{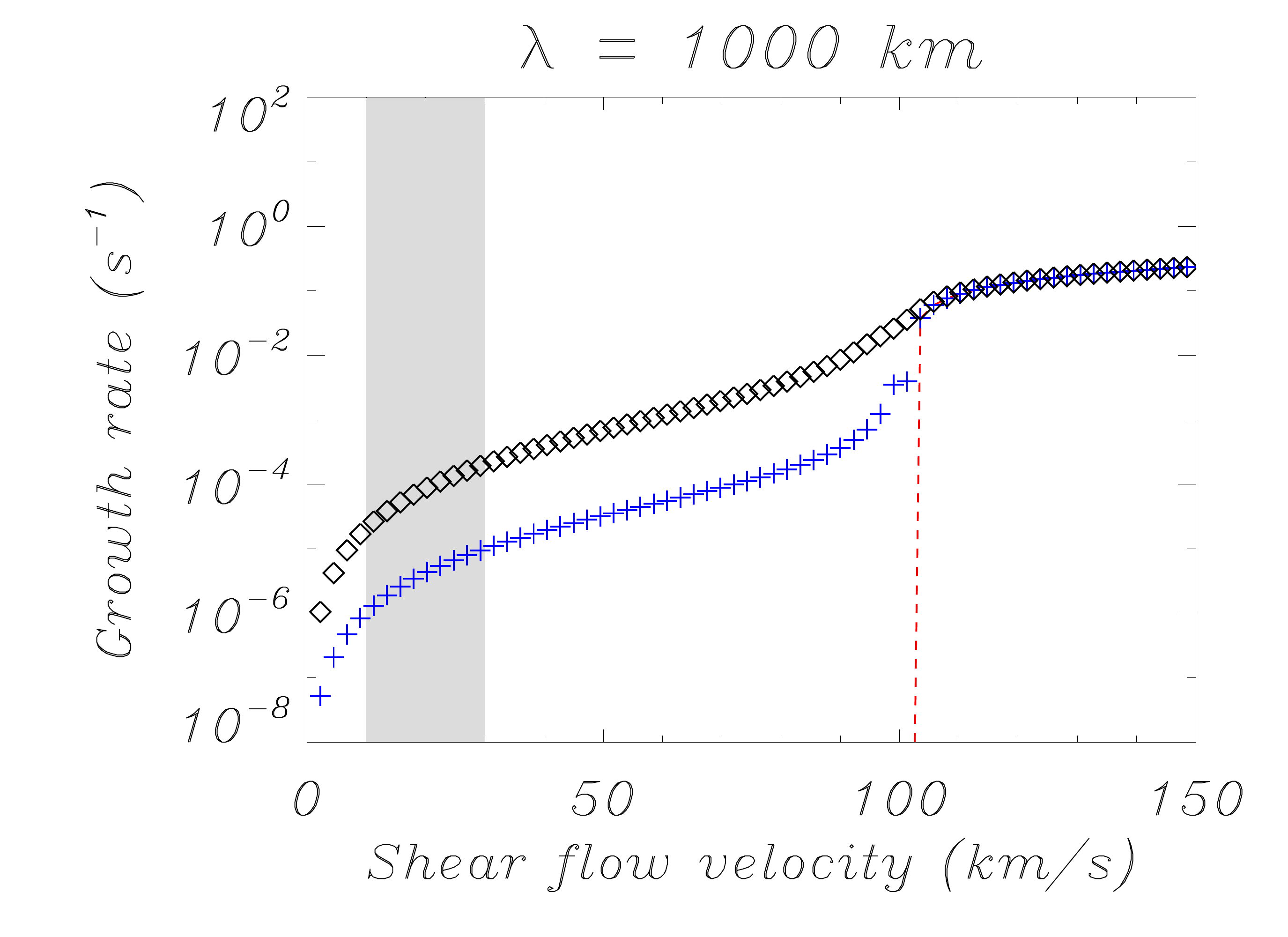}
\caption{Growth rate of the KHI linear stage as a function of the shear flow velocity in a prominence thread model for two different values of the perturbation wavelength: 100~km (left) and 1000~km (right). The red dashed lines, blue crosses, and black diamonds correspond to fully ionized plasma, partially ionized plasma, and weakly ionized plasma, respectively. The shaded area denotes the region of typically observed flow velocities in quiescent prominences. Adapted from \citet{martinez15}.}
\label{fig:martinez15}
\end{figure}

However, the prominence plasma is not fully ionized but partially ionized. The existence of a neutral component in the plasma may invalidate the classical criterion for the onset of the KHI and may allow its development for sub-Alfv\'enic field-aligned flows. The linear stage of the KHI in partially ionized  plasmas was studied by \cite{watson04,soler12KHI,martinez15}. These studies showed that partially ionized plasmas can develop the KHI even when the velocity of the shear flow is sub-Alfv\'enic. The physical reason for this result is that neutrals largely ignore the stabilizing effect of the magnetic field and only feel its influence indirectly through collisions with ions. Although the ion-neutral coupling exerts a significant influence, so that increasing the ion-neutral collision frequency reduces the growth rate of the instability, it is not possible to completely suppress the onset of the KHI originated in the neutral component (see Fig.~\ref{fig:martinez15}). Because of the imperfect coupling between ions and neutrals that naturally occur in real prominence plasmas, the onset and development of the KHI appears to be unavoidable in such structures even for slow flow velocities. These theoretical results support the explanation of the origin of the observed turbulent flows in prominences in terms of the sub-Alfv\'enic KHI in partially ionized plasmas.
The KHI can be also developed for sub-Alfv\'enic motion of fully ionized plasma if the magnetic field has small component perpendicular to the flow direction \citep[see, e.g.,][]{chandrasekhar61}. This situation may arise if the prominence threads are modeled as twisted flux tubes moving along their axis. In this case, the threshold of KHI is reduced significantly, which allows to sub-Alfv\'enic flows to be unstable \citep{zaqarashvili10b, zaqarashvili14}. Obviously, consideration of neutral particles may enhance the growth rate of KHI in twisted tubes. However, it is not known which mechanism is dominant for occurrence of KHI in sub-Alfv\'enic flows. This point clearly needs further discussion.

Another instability of interest, which has been mostly studied in the context of fully ionised plasmas, is the Rayleigh-Taylor instability (RTI). \citet{diaz12} considered an equilibrium configuration made of two compressible and magnetized plasmas composed of ions, electrons and neutrals separated by a contact surface, and in the presence of gravity. Then, using the two-fluid description, \citet{diaz12} studied how the classical criterion for the onset and growing rate of the RTI is modified in a partially ionized plasma. It was found that compressibility and ion-neutral collisions reduce the linear growth rate, but do not affect the critical threshold of the onset of the RTI. In particular, due to ion-neutral collisions, the growth rate can be decreased by an order of magnitude as compared with the corresponding value for the collisionless case. Prominences are cool and dense structures embedded in the solar corona, supported against gravity by its magnetic field, which are likely to develop RTI. On the other hand, prominences are made of partially ionized plasma, therefore, \citet{diaz12} applied the above study to a plasma with physical properties akin to those of prominences. Using typical parameters for prominence threads as well as coronal values, the resultant time for the RTI instability was of the order of $30$ min, which is of the same order as the observed lifetime of the prominence threads. This results greatly differs from the classical result, of the order of $1$ min, unable to explain the reported lifetime of threads.

Simulations of the non-linear phase of the RTI at the PCTR considering partial ionization of the  prominence plasma in the single-fluid formulation were reported by  \cite{Khomenko+etal2014}. These simulations included the ambipolar term in the generalized Ohm's law, as the dominant effect. The ambipolar term is larger in the regions with low density, i.e. regions with, generally, low momentum $\rho \vec{u}$. Therefore, the impact of such regions into the overall flow dynamics was found to be not large. A similar conclusions were reached in the simulations of multi-fluid turbulence in molecular clouds \citep{Downes11, Downes12}. However, statistically, simulations with/without ambipolar term develop flows that are statistically different at small scales. In agreement with the linear theory, non-linear simulations demonstrate that the introduction of the ambipolar diffusion removes the cut-off wavelength for the growth rate of the instability and allows the small scales to develop. As the non-linear development of the instability is such that small scales merge and give rise to larger scales \citep{jun95, stone07,  hillier12}, the larger growth rate of small scales leads to larger extreme velocities in the ambipolar case \cite{Khomenko+etal2014}. In addition, some 30\% larger temperatures at the PCTR were found in the simulations with ambipolar term, as a result of the Joule dissipation. A significant drift momentum, defined as $\vec{p}_D=\sqrt(\rho_i\rho_n)(\vec{u}_i-\vec{u}_n)$ is present at the PCTR \cite{Khomenko+etal2014}.

Finally, another example of instability is the dissipative instability which arises at the interface between two media and is related with the phenomenon of negative energy waves. This instability appears for flows whose speeds are below the KH threshold value. Usually, the interface between two media allows the propagation of two modes travelling in opposite directions, however, for flow speeds larger than a critical value, the propagation direction of the two waves is the same. Then, the wave having the smaller phase speed is a negative energy wave which means that dissipation produces an amplification of the wave amplitude, which leads to an instability, while at the same time wave energy decreases. In this case, the dissipative mechanisms working in the two regions amplify this negative energy mode leading to dissipative instability. \citet{ballai15} have studied the dissipative instability at the interface between two magnetized media, representing the corona and the partially ionised prominence plasma, in the incompressible limit. The dissipative mechanisms at work in the coronal and prominence media are viscosity and Cowling's resistivity, respectively. Using the limit of weak damping, a dispersion relation for Alfv\'enic waves propagating along the interface was derived. The imaginary part of it describes the evolution of the instability, and the results show that while the forward propagating wave is always stable, for the backward propagating wave there is a threshold of the flow, below the KH threshold, for which the wave becomes unstable. This analysis also shows that partial ionisation has a stabilising effect on the interface, for any degree of ionisation, and that the unstable behaviour is due to the viscosity of the coronal plasma.

\subsubsection{Resonant Absorption in partially ionized plasmas}

The process of resonant absorption, caused by plasma inhomogeneity across the magnetic field, has important implications in the behavior and energy transport of transverse MHD waves in magnetic flux tubes of the solar atmosphere \citep[see the review by][]{goossens11}. Heating of coronal loops involving the process of resonant absorption was first suggested by \citet{ionson78} and has been extensively investigated in the literature afterward. Resonant absorption produces a radial flux of transverse wave energy toward the nonuniform boundary of the flux tube, where the energy is absorbed into the continuous Alfv\'en spectrum \citep[see, e.g.,][]{tataronis75,poedts89,goossens13}. Phase mixing of the continuum Alfv\'en modes causes the energy to cascade from large spatial scales to small spatial scales \citep[see, e.g.,][]{heyvaerts83,cally91,soler15}. Dissipation of MHD wave energy by, e.g., magnetic resistivity and/or viscosity becomes efficient when the generated spatial scales are sufficiently small. Resonant absorption of wave energy and phase mixing occur simultaneously and  are  intimately linked as both processes are caused by  inhomogeneities across the magnetic field. Such inhomogeneities naturally occur in magnetic flux tubes of the solar atmosphere, including solar prominence threads.  Recent observations by \citet{okamoto15} and \citet{antolin15} claim to provide direct evidence of the process of resonant absorption of transverse waves taking place in solar prominence threads.

The theory of resonant waves in the solar atmosphere has mainly been studied assuming fully ionized plasma \citep[see][]{goossens11}. Early attempts to study resonant waves in partially ionized plasmas were performed by \citet{soler09b,soler11}. These authors used the single-fluid approximation to investigate  resonant Alfv\'enic waves in a model of a partially ionized thread of a solar prominence. \citet{soler09b,soler11} found that the process of resonant absorption of wave energy is not altered by partial ionization in the single-fluid approximation. The results in  partially ionized flux tubes concerning the damping of transverse waves and the energy flux toward the resonance location are the same as in fully ionized flux tubes when total density is equal in both cases. The single-fluid approximation breaks down when small length scales approaching the ion-neutral collision length are involved. Resonant wave perturbations can develop very small length scales in the vicinity of the resonance position owing to phase mixing \citep[see, e.g.,][]{tirry96,ruderman99,vasquez05,terradas06}.  This led \citet{soler12a}  to analytically study resonant  waves in partially ionized plasmas using the multi-fluid treatment. They found that ion-neutral collisions generate a dissipative layer around the location of the resonance in a similar way as magnetic resistivity and viscosity do in fully ionized plasmas \citep[see, e.g.,][]{hollweg88,poedts90,sakurai91}. The conserved quantity at the resonance and the jump of the perturbations across the dissipative layer in the multi-fluid treatment are the same as in fully ionized plasmas \citep[see][]{goossens95}. In the case of Alfv\'enic waves in partially ionized flux tubes,  \citet{soler12a} found the resonant absorption rate is inversely proportional to the wave frequency, while the ion-neutral collisions damping rate is inversely proportional to the square of the frequency. For the observed wave frequencies in the solar atmosphere, the resonant absorption rate should be faster than the ion-neutral collisions damping rate. Hence, the resonant absorption process should not be altered in agreement with the single-fluid results. However, for wave frequencies of the order of the ion-neutral collision frequency the  damping rate due to ion-neutral collisions could be faster, thus effectively suppressing the resonant energy transfer toward to nonuniform boundary of the tube. Numerical simulations beyond analytic studies should be used in the future to determine the actual impact of partial ionization on the time dependent process of resonant absorption and on the deposition of resonant wave energy into the plasma.

\subsubsection{Coronal rain}

Coronal rain are dense condensations with chromospheric to transition region temperatures falling down in the much hotter corona.  In most theoretical studies, these condensations  have been considered to be made of fully ionized plasma \citep{oliver2014}, however, these cold blobs are made of ionized and neutral material which must be strongly coupled, since they fall together, and they can be used as a tool to explore the interaction between neutral and ionized plasma. To describe the temporal evolution of a partially ionized hydrogen blob which falls under the action of gravity, pressure and friction forces, \citet{oliver16} considered a two-fluid description for charged and neutral fractions. After assuming a static atmosphere, at t=0 a dense blob, composed of neutral and charged material, was injected and, using a one dimensional numerical simulation, its temporal evolution, along a vertical path, was investigated. The temporal evolution is described by a system of partial differential equations together with a set of imposed boundary conditions \citep{oliver16}, and the blob was left to evolve under the joint action of gravity, pressure and friction forces between ions and neutrals. The results indicate that the falling material displays two different phases: In the first, the blob accelerates while, in the following phase, it maintains a practically constant speed, the duration of the first phase being longer for denser blobs. Furthermore, the results also indicate that blob's dynamics is determined by its mass while the ionization degree is irrelevant, and that a correlation exists between the blob maximum speed and its initial density ratio with respect to the corona. On the other hand, the interplay of forces acting on the neutral and charged fractions of the blob is such that both fractions are subject to the same acceleration regardless of their respective densities. Finally, it must be taken into account that coronal rain condensations fall following a curved path, then, it becomes important to determine if the coupling, between the charged and neutral fraction, is strong enough to impose neutrals to follow the magnetic field lines or not. Therefore, for a full understanding of the temporal behaviour of partially ionized coronal rain blobs, multidimensional numerical simulations are needed.

\subsubsection{Multi-fluid modelling of high-frequency waves in the partially ionised solar plasma}

The single-fluid MHD approximation with a generalised Ohm's Law provides an accurate description of low-frequency waves in the solar plasma even when the plasma is only partially ionised. However, when the frequency of the waves is of the same order of or higher than the frequencies of the individual collisions between different species, the single-fluid model is not applicable and more complicated approaches are needed. Recently, \citet{MST16,MST17} theoretically studied high-frequency waves in a plasma composed of hydrogen and helium and went beyond the usual single-fluid description by considering a multi-fluid approach in which the various components of the plasma are treated as separate fluids. In their model, electron inertia was neglected, which allowed them to obtain an expression for the electric field that includes the effects of Hall's current and magnetic resistivity. In addition, they took into account the friction due to collisions between different species.

First of all, \citet{MST16} considered the case of full ionisation, so that the background hydrogen-helium plasma was composed of three distinct ionic species, namely protons, HeII, and HeIII. Through the analysis of the wave dispersion relations and temporal numerical simulations, it was found that at high frequencies ions of different species are not as strongly coupled as in the low-frequency limit. Hence, different ions cannot be treated as a single fluid. In addition, elastic collisions between the distinct ionised species are not negligible for high-frequency waves and an appreciable damping and energy dissipation is obtained. Furthermore, Coulomb collisions between ions are able to remove the cyclotron resonances and the strict cutoff regions, which are present when those collisions are not taken into account.

Subsequently, \citet{MST17} extended their previous model by allowing the presence of neutrals. Hence, \citet{MST17} used a five-fluid model with three ionised (protons, HeII, and HeIII) and two neutral (H and HeI) components. They discussed the effect of momentum transfer collisions on the ion-cyclotron resonances and compared the importance of magnetic resistivity, ion-neutral and ion-ion collisions on the wave damping at various frequency ranges. Three specific environments were explored: the higher chromosphere (largely ionised), a solar prominence (partially ionised), and the lower chromosphere (weakly ionised). The investigation and comparison of environments with such different degrees of ionisation led to a comprehensive understanding of the influence of neutral species on the propagation of high-frequency waves in the small-amplitude regime. This improved our knowledge of the physics involved in the damping and energy dissipation and paved the way for future studies that should focus on the full nonlinear evolution of the waves and the associated plasma heating. 

\newpage

\section{Planetary Ionospheres}
\label{Fsec:Ionosphere}

Planetary ionospheres are created when an extended neutral atmosphere is subjected to sources of ionization.  Accordingly, the planet's upper atmosphere consists of a neutral gas co-existing with an embedded ionized component which typically represents only a small fraction of the total gas density.  In this manner, planetary upper atmospheres/ionospheres are quintessential examples of partially ionized plasmas.

Planetary ionospheres vary considerably among the planets in our solar system.  Furthermore, they can be expected to differ considerably among the ionospheres of as yet undiscovered extra-solar planetary systems.  The main characteristics of planetary ionospheres depend on a given planet's atmospheric constituents, characteristics of the ionizing radiation, intrinsic magnetic fields (if any), gravitation field, proximity to the stellar ionization source and stellar wind, the planet's rotation rate and the orientation of its rotation axis, etc.  In this section, rather than review all planetary ionospheres, we consider only the ionospheres of the terrestrial planet, Earth, and those of the extended gaseous atmospheres of the outer planets, Jupiter and Saturn. 

Among the many reviews of planetary ionospheres, we direct the reader in particular to \citet{bauer1973}, \citet{cravens1997}, \citet{schunknagy2004}, \citet{mendillo2002}, and \citet{nagy2008}.  \citet{leake2014} compares the partially ionized gases of the Sun's chromosphere with the Earth's upper atmosphere/ionosphere.

\subsection{Earth's Ionosphere: Introduction}

The Earth's outer atmosphere is partially ionized due to three main ionizing sources:  (1) photo-ionization due to impinging EUV and soft X-ray energy from the Sun, (2) impact ionization at high latitudes of  energetic, precipitating particles such as those associated with the visible aurora, and (3) impact ionization due to meteor ablation which creates a small but important population of metallic ions.  Charge exchange between neutrals and ions is another process that influences the distribution of ion species in the Earth's ionosphere.

As the ionosphere and upper atmosphere together form a partially ionized plasma, their characteristics are influenced by ion-neutral coupling, displaying an important dynamic interplay between upper atmospheric motions, or winds, and plasma drifts.  Because the earth's strong, dipole geomagnetic field influences the charged particle dynamics, the ionospheric plasma supports an intrinsic system of electrical currents which depend on the local conductivities and collision frequencies, varying with altitude and latitude.  Furthermore, as the magnetic field threads the partially ionized plasma to form the magnetosphere at much higher altitudes, it ``connects'' or couples the ionosphere/upper atmosphere to the magnetospheric plasma and solar wind and their sources of energy and momentum, including large electric fields, field-aligned currents, and highly variable particle precipitation.

The term ``ionosphere'' refers to the ionized component of the Earth's upper atmosphere (approximately 90 to 1000 km) and the term ``thermosphere'' refers to upper atmosphere neutral gases in the altitude region of approximately 90-600 km in which the neutral temperature is significantly increased.  The ``mesosphere'' is that part of the upper atmosphere between the stratosphere and lower ionosphere, essentially between 40-90 km.  Collectively, these regions are often referred to as the ionosphere-thermosphere-mesosphere or ``ITM'' system.  

An outline of this section is as follows:  After a description of the basic properties of the earth's ionosphere, we discuss how electrons and ions are magnetized at different altitudes because their collision frequencies with the neutral gas are substantially different.  We then describe how the resulting Pedersen and Hall conductivities enable a global system of electrical currents to flow in the lower ionosphere followed by a discussion of the dynamics of the coupled neutral and ionized gases, magnetospheric coupling, and the creation of enhanced conductivity regions associated with the aurora.  The discussion here follows the review article of \citet{pfaff2012}.  Other references that provide general descriptions of the earth's ionosphere and upper atmosphere include the following monographs:  \citet{hargreaves1992}, \citet{kelley2009}, \citet{prolss2004}, \citet{risbeth1969}, and \citet{schunknagy2004}, among others.

\subsubsection{Basic properties of the Ionosphere}

The ionospheric environment is complex due to the many processes (chemical, dynamical, and electrodynamical) acting within the region and due to its interaction with external processes such as those related to coupling with the magnetosphere above and tropospheric forcing below.  Distinct regions form within the ionosphere primarily as a function of altitude because the composition of the atmosphere varies with height as do the depths of the penetrating EUV and soft X-ray radiation.  The ionosphere also varies considerably with latitude, since the high latitude gases are connected to the magnetosphere via highly inclined magnetic field lines of force whereas the mid and low latitude ionosphere exists on closed magnetic field lines.  Accordingly, these regions are dominated by physical processes which also vary with altitude and latitude, including ion-neutral coupling, the creation of ionospheric currents, and the motions of plasma due to drifts associated with external and internal electric fields as well as due to forcing by neutral winds.  Furthermore, we note that the physical processes operating on and within the ionosphere often have very different spatial and temporal scales.  

\begin{figure}
	  \centering{
		  \resizebox{8cm}{!} {\includegraphics{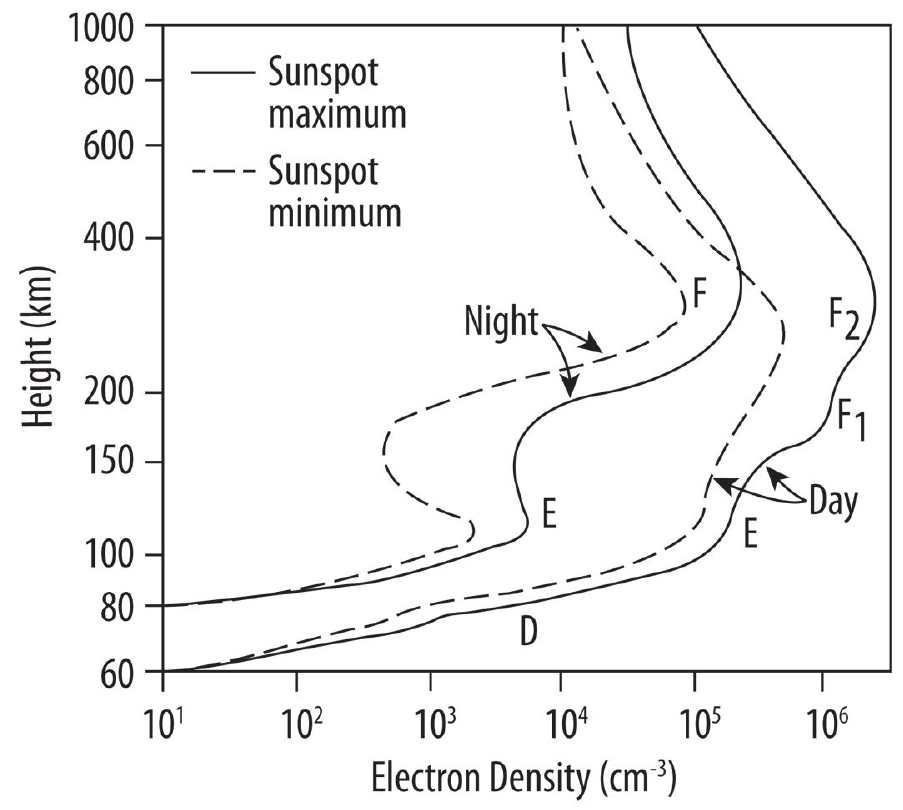}}}
		   \vspace{-3mm}
		   \caption{Representative plasma density profiles for day and night at solar minimum and maximum conditions at a mid-latitude location [Hargreaves, 1992].  }      \label{Pfaff_1}
\end{figure}

The three main layers of the Earth's ionosphere are characterized by its different plasma density regions, namely the D region (60-90 km altitude), E region (90-150 km altitude), and F region (above 150 km altitude), as shown in Figure~\ref{Pfaff_1}.  This figure illustrates how the ionospheric density varies between day and night and also how it varies between sunspot minimum and maximum, since the solar cycle governs the strength of the EUV ionizing source.  During both the daytime and nighttime, the electron density is generally largest at the higher altitudes, 250-450 km or the F region, which is divided into two different regions (F1 and F2 layers) during the day.  The E region densities are also largest during the daytime, yet diminish considerably during the night as the ion recombination rates in this region are rapid.  The lowest altitude region, the D region, has a complex chemistry of positive and negative ions including water cluster ions.  It has a much reduced plasma density which is even smaller at night.

\begin{figure}
	  \centering{
		  \resizebox{8cm}{!} {\includegraphics{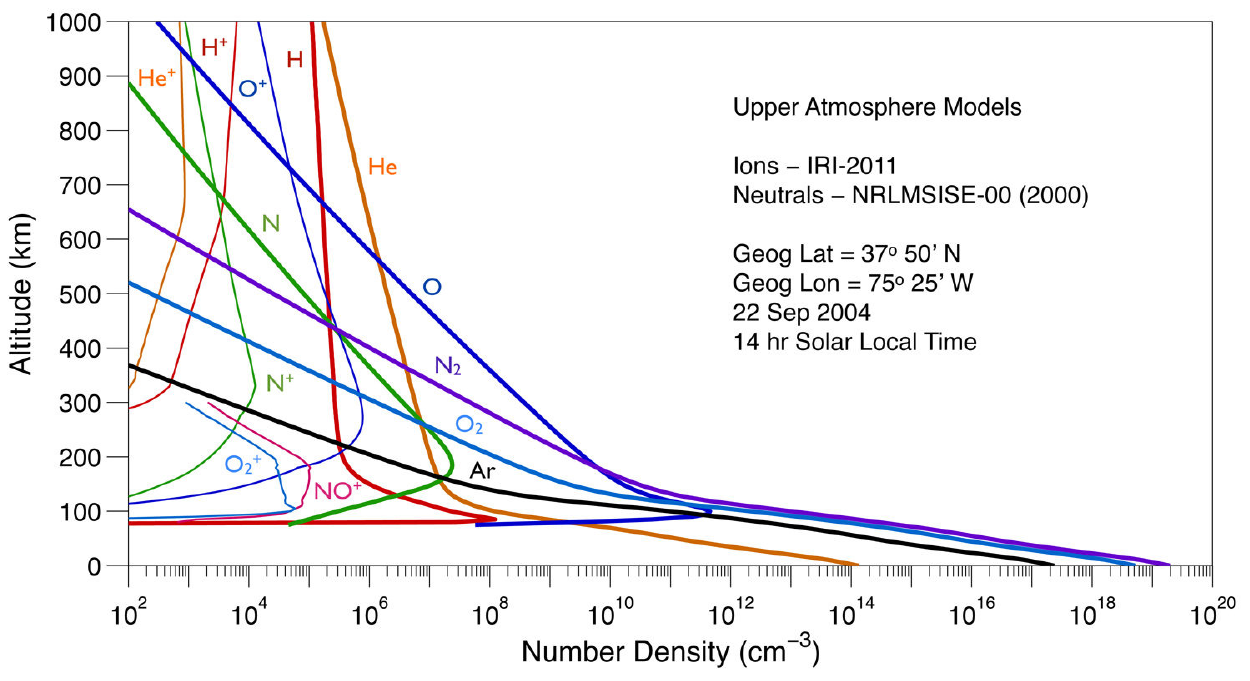}}}
		   \vspace{-3mm}
		   \caption{Model distributions of neutral and ion constituents as a function of altitude for a mid-latitude location (Wallops Island, Va., USA) at 14 hr Solar Local Time \citep{pfaff2012}.  Ion distributions below 80 km are not provided by the model.}      \label{Pfaff_2}
\end{figure}

To illustrate how the earth's atmospheric constituents vary with altitude and provide a basis for the creation of ionosphere constituents via EUV and soft X-ray radiation, Figure~\ref{Pfaff_2} provides NRLMSISE-2000 model \citep{picone2002} distributions of the neutral atomic and molecular species as a function of altitude for a mid-latitude location at 14hr LT \citep{pfaff2012}.  The plot shows how the familiar nitrogen, oxygen, and argon atmospheric constituents at the earth's surface decrease exponentially with altitude until about 100 km, at which time the neutral atmosphere changes markedly in character.  Above this altitude, the atmosphere changes from well mixed to free molecular flow, the temperature increases, the exponential decay of the total number density becomes less steep, and a variety of other atmospheric gas constituents emerge, as shown in the figure.  

The upper atmosphere is subject to EUV radiation which serves two very important functions:  First, it dissociates the diatomic oxygen molecules and releases heat, thereby expanding the upper atmosphere.  Second, it ionizes a very small fraction of the upper atmosphere via photo-ionization.  The main processes are diatomic oxygen dissociating to two oxygen atoms plus energy or:
\begin{equation}
O_2 + h\nu  \rightarrow  O + O + energy
\label{1}
\end{equation}					
and the photo-ionization of the oxygen atoms:

\begin{equation}
O + h\nu  \rightarrow  O^{+} + e^{-}	
\label{2}					
\end{equation}
$N_2$ then charge exchanges with $O^+$ to form $NO^{+}$ + N while $O_2$ charge exchanges with $O^+$ to form $O_2^{+}$ + O, both within the lower portions of the ionosphere.  These molecular ions are the main ionospheric ions below about 180 km and are extremely important as they recombine very quickly, which means that the lower ionosphere essentially vanishes at night whereas the upper ionosphere (F region) generally remains throughout the evening.  The bottomside ledge of the ionosphere is, hence, much higher in altitude at night.  This characteristic of the earth's nighttime ionosphere has a variety of important consequences ranging from effects on radio wave propagation to instabilities which form on the bottomside ledge of the low latitude ionosphere just after sunset.  

To complete the picture of the ionized component of the Earth's upper atmosphere, we comment briefly on metallic ions which result from the influx of meteoroids which laden the earth with 108 kg/yr of dust, grains, and atoms released via ablation \citep{ceplecha1992}.  The main source of metallic ions in the ionosphere is impact ionization caused by meteors striking the atmosphere with sufficient kinetic energy to remove electrons from the neutral particles.  Typical metallic ions in the upper atmosphere include $Fe^{+}$, $Mg^{+}$, and $Na^{+}$.  Metallic ions may also be created as meteoritic atoms undergo charge exchange and also via photo-ionization.  Metallic ions recombine very slowly and, in general, form layers near the lower edge of the ionosphere (near 100 km) due to wind shear processes.  They may also be transported over long distances due to upper atmospheric winds and electric fields. 

\begin{figure}
	  \centering{
		  \resizebox{8cm}{!} {\includegraphics{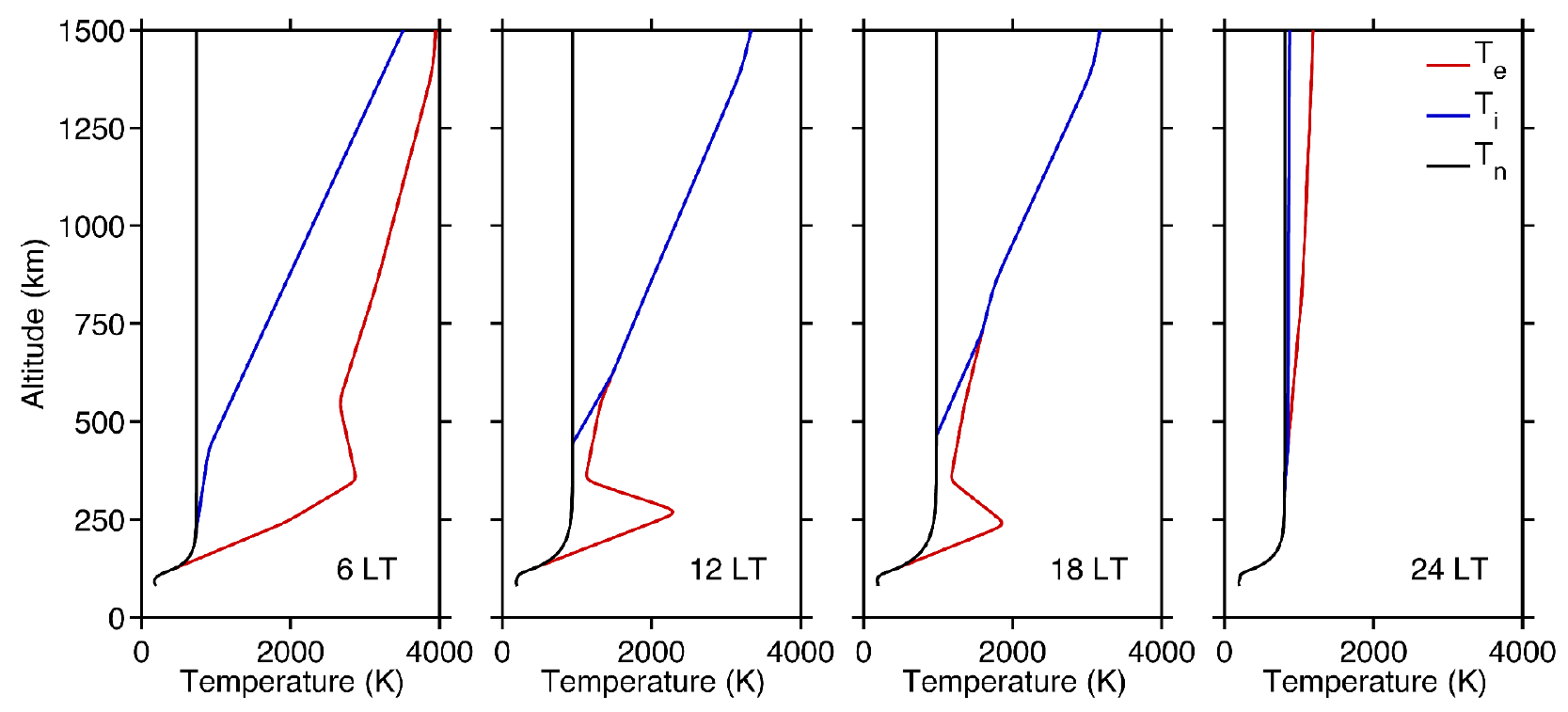}}}
		   \vspace{-3mm}
		   \caption{ IRI model calculations showing the plasma and neutral temperatures at the magnetic equator at four different local times \citep{pfaff2012}.}      \label{Pfaff_3}
\end{figure}

The temperature of the various constituents of the upper atmosphere changes significantly with altitude and local time, as shown in Figure~\ref{Pfaff_3} \citep{pfaff2012}.  This figure shows typical plasma ($T_e$ and $T_i$) and neutral temperatures ($T_n$) based on the IRI model for 6LT, 12LT, 18LT, and 24LT at the equator. The greatest variation in temperature is with the electrons, whose average temperature is roughly 200 K at 100 km yet increases to 4000 K at 1500 km at 6LT.  When neutral constituents such as oxygen are photo-ionized, the newly freed electrons that are emitted typically have energies of several eV, much higher than that of the newly created ions.  This is due to the significantly lower mass of the electron compared to the ion and the conservation of momentum.  The higher electron energy eventually dissipates via collisions with the neutral gas as well as with ions via electron-ion collisions.  

As shown in the leftmost panel in Figure~\ref{Pfaff_3} corresponding to 6AM local time, the electron temperature, $T_e$, is highest in the early morning, as the neutral atmosphere has yet to warm and is still contracted from the nightside.  Hence, the newly formed electrons have less neutral atmosphere with which to collide, particularly at the higher altitudes.  During the day, the neutral atmosphere expands and the electron temperature is decreased, equilibrating with the ion temperature, Ti, at the higher altitudes.  While photo-ionization continues during the day within the lower portion of the ionosphere, the electron temperature maintains a daytime enhancement near 250 km simply due to the fact that the neutral atmosphere density falls off with increasing altitude.  At night, there are no newly formed electrons and ions and hence the temperatures of the neutrals, ions, and electrons converge to similar values via collisions.  At the higher latitudes, as discussed further on below, the plasma and neutral temperatures can increase locally due to Joule heating, precipitating particles, and plasma instabilities.

\subsubsection{Ion-Neutral Coupling and Magnetization of Charged Particles}

As the ions and neutrals are immersed in a common volume in the ionosphere, the two gases are coupled via collisions, with the coupling efficiency dependent on the gas densities and temperatures.  A critical parameter within any partially ionized plasma in the presence of a magnetic field is under what conditions are the charged particles ``magnetized'' --  in other words, at what altitudes are they able to execute their gyrations about the ambient magnetic field without their motions being substantially altered by collisions with neutral particles.

Charged particles in the ionosphere are compelled by the Lorentz force to gyrate about magnetic field lines at their cyclotron frequencies, $\Omega_j$, given by

\begin{equation}
\Omega_j  =  \frac{e \vert B \vert}{m_j}
\label{3}
\end{equation}
where e is the electric charge, $\vert B \vert$ is the magnetic field strength, and $m_j$ is the mass of the species which is designated by j.  For $O^{+}$ ions and a magnetic field strength of 0.5 Gauss, the gyrofrequency is 48 Hz, whereas this frequency is 1.4 MHz for electrons for the same magnetic field.

\begin{figure}
	  \centering{
		  \resizebox{8cm}{!} {\includegraphics{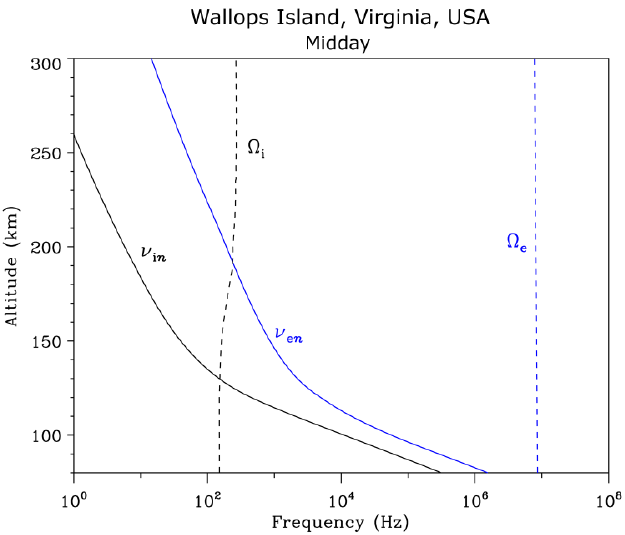}}}
		   \vspace{-3mm}
		   \caption{ Altitude profiles of the electron-neutral collision frequency and the ion-neutral collision frequency.  Profiles of the mean ion gyrofrequency and electron gyrofrequency are also shown \citep{pfaff2012}.}      \label{Pfaff_4}
\end{figure}

As shown in Figure~\ref{Pfaff_4}, the gyrofrequencies for the ions and electron species are plotted versus altitude for a mid-latitude location (Wallops Island, USA) where the ambient magnetic field on the ground is approximately 0.5 Gauss.  Here, we use the mean mass of the ions which accounts for the slight variation of the ion gyro frequency near 180 km altitude as the dominant ion switches from molecular ($O2^{+}$ and $NO^{+}$) to atomic oxygen ($O^{+}$) above this altitude.  

Since the ionospheric gases are dense and only weakly ionized, collisions with neutrals significantly influence the individual and bulk motions of the plasma.  Because of these collisional effects, the statistical collective behavior of the particles is defined using the equations of hydrodynamics.  The ion-neutral collision frequencies ($\nu_{in}$) and the electron-neutral collision frequencies ($\nu_{en}$) are also plotted versus altitude in Figure~\ref{Pfaff_4} using standard collision operators, in this case, those provided by \citet{banks1973}.

Notice immediately in Figure~\ref{Pfaff_4} that the electron gyrofrequency, $\Omega_e$, is larger than the electron-neutral collision frequency throughout the ionosphere.  Hence, we say that the electrons are``magnetized'' throughout the region, since the electrons are able to execute their gyromotions about the magnetic field lines of force and consequently undergo unimpeded E x B drifts across magnetic field lines.  It is only at the very lowest altitudes, below 70 km, where the electron-neutral collision frequency succeeds in significantly altering the normal electron gyrations.

The ions, on the other hand, are collision dominated below about 130 km where $\nu_{in} \sim \Omega_i$.  At lower altitudes, the ion motions are severely restrained by the neutral atmosphere to the extent that their motions become essentially those of the neutral gas or winds.  Importantly, since the electrons are magnetized, the lower ionosphere region between 100-120 km, where $\nu_{in}/\Omega_i > 1$ and $\nu_{en}/\Omega_e < 1$, constitutes the conducting or ``dynamo'' region where strong electrical currents flow, as discussed in the next section. 

Above 130 km, the ion gyro frequency is greater than the ion-neutral collision frequency although the effect of ion-neutral collisions is still significant.  At higher altitudes ( $>$ 200 km) both the electrons and ions are considered magnetized and both execute $\vec E \times \vec B$ drifts.  However, even at these and higher altitudes, the ions and neutrals still ``collide'' to an extent where their motions influence each other, as we shall see further on below.

\subsubsection{Mobility and Conductivity Tensors and Dynamo Currents}

Because the plasma distribution function in the presence of a magnetic field is highly anisotropic, we define a mobility tensor, $\underline{ \underline{\mu}}$, which relates the velocity (V) and force (F) vectors:

\begin{equation}
\vec V  = \underline{\underline{ \mu}} \cdot \vec F
\label{4}
\end{equation}			  			
where 

\[\underline{\underline{ \mu}} = \left( \begin{array}{ccc}
\mu_1 & \mu_2 & 0 \\
-\mu_2 & \mu_1 & 0 \\
0 & 0 & \mu_0 \end{array} \right) \]
This tensor has been defined in a geometry such that the magnetic field, $\vec B$, is parallel to the z-axis.  For single particle motion subject to an applied force, we have the following mobilities:

$$
\mu_{0j} =  	1/m \nu_{jn} 
$$
for  parallel mobility,

$$\mu_{1j} =  	(1/m \nu_{jn})(\nu_{jn})^2/(\Omega_j^2 + \nu_{jn}^2)$$     	
for Pedersen mobility, and

$$\mu_{2j} =  	(1/m \nu_{jn})(\nu_{jn}\Omega_j)/(\Omega_j^2 + \nu_{jn}^2) $$
for  Hall mobility. The mobilities are higher at lower latitudes due to the weaker magnetic field strength, since they are roughly proportional to $1/\Omega_i^2$.

We now define a conductivity tensor, $\underline{\underline{\sigma}}$, that relates the electric field, $\vec E$, and current density, $\vec J$:

\begin{equation}
\vec J  =  \underline{\underline{\sigma}} \cdot \vec E
\label{5}
\end{equation} 						 
which can also be expressed in terms of the difference of the ion and electron motions:

\begin{equation}
\vec J  =  N_e (\vec V_i - \vec V_e)
\label{6}
\end{equation}				
The conductivity may be expressed in terms of the mobilities such that $\underline{\sigma}  =  N_e^2(\mu_i + \mu_e)$.  The components of the conductivity are thus:

$$\sigma_0  =  N_e^2(\mu_{0e} + \mu_{0i})$$   		
for parallel conductivity,

$$\sigma_1  =  N_e^2(\mu_{1e} + \mu_{1i})$$ 		
for Pedersen conductivity, and

$$\sigma_2  =  N_e^2(\mu_{2e} - \mu_{2i}) $$
for  Hall conductivity

\begin{figure}
	  \centering{
		  \resizebox{8cm}{!} {\includegraphics{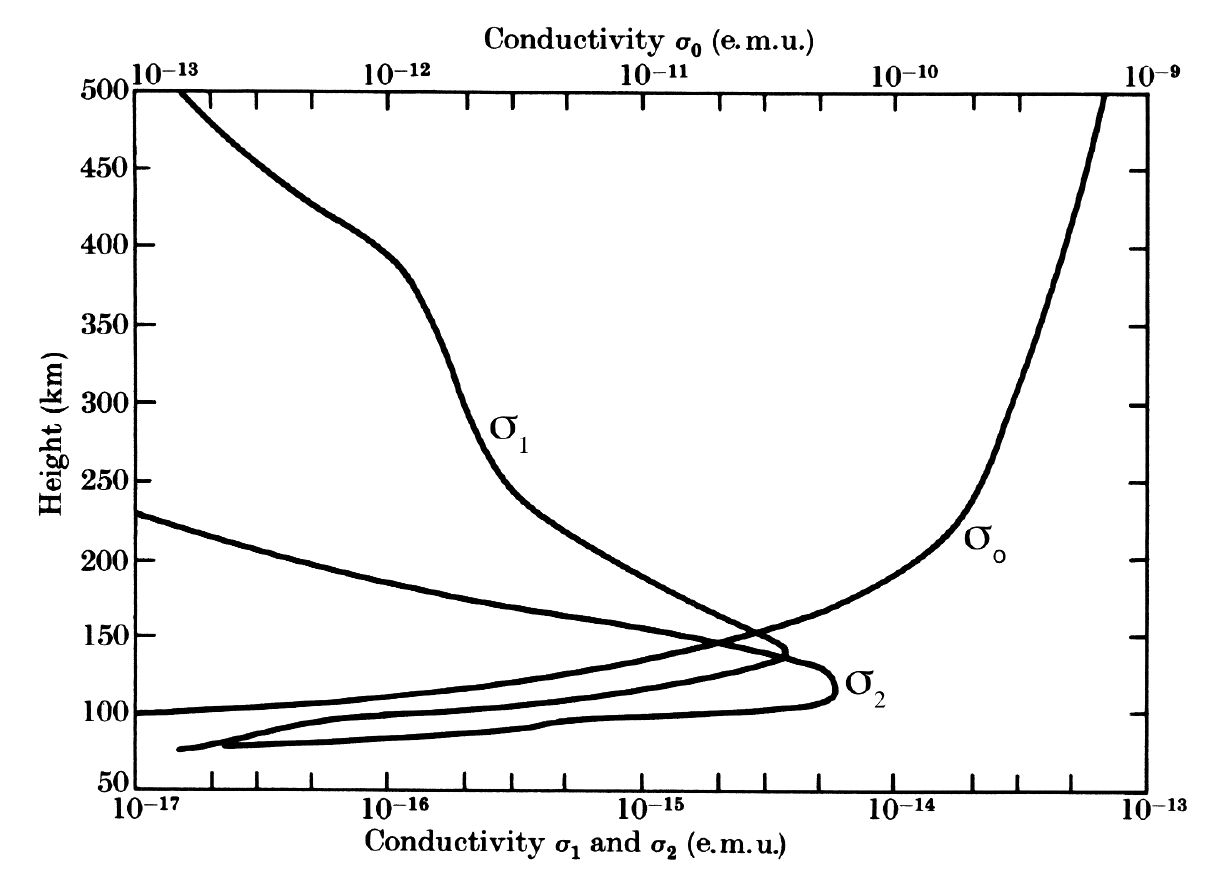}}}
		   \vspace{-3mm}
		   \caption{Ionospheric conductivities for daytime, mid-latitude conditions \citep{akasofu1972}.  (See text for explanation of symbols).}      \label{Pfaff_5}
\end{figure}

Clearly, the ionospheric conductivity depends linearly on the ambient plasma density and couples the neutral and ionized gases via the anisotropic mobilities as organized by the magnetic field.  In Figure~\ref{Pfaff_5}, typical parallel ($\sigma_0$), Pedersen ($\sigma_1$), and Hall ($\sigma_2$) conductivities are plotted versus altitude for daytime, mid-latitude conditions \citep{akasofu1972}.  Notice how the Hall conductivity peaks at a lower altitude than the Pedersen conductivity where it is responsible for the global dynamo current system.  The parallel conductivity is orders of magnitude larger than that of the Hall and Pedersen conductivity (see scale at top of figure).

Equation~(\ref{5}) may now be expressed more fully:

\begin{equation}
\vec J  =  \sigma_1\vec E^{'}_{\perp} + \sigma_2 \hat {\bf b}^{'} \times \vec E^{'}_{\perp} + \sigma_0 \vec E_{\parallel}
\label{7}
\end{equation}				
where $\hat {\bf b}^{'}$ is the unit vector along the magnetic field direction and $\vec E_{\perp}$ is the perpendicular electric field in the earth-fixed frame, such that $\vec E^{'}_{\perp}  =  \vec E_{\perp} + \vec U \times \vec B$.  This expression shows that for an electric field, $\vec E$, perpendicular to the magnetic field, $\vec B$, the current will have two components: one along $\vec E$ but perpendicular to $\vec B$ (Pedersen) and one perpendicular to both $\vec E$ and $\vec B$ (Hall).  Their magnitudes differ by $\nu_{in}/\Omega_i$ and the angle, $\alpha$, between them is given by $\alpha  =  arc tan (\Omega_i/\nu_{in})$.

\begin{figure}
	  \centering{
		  \resizebox{8cm}{!} {\includegraphics{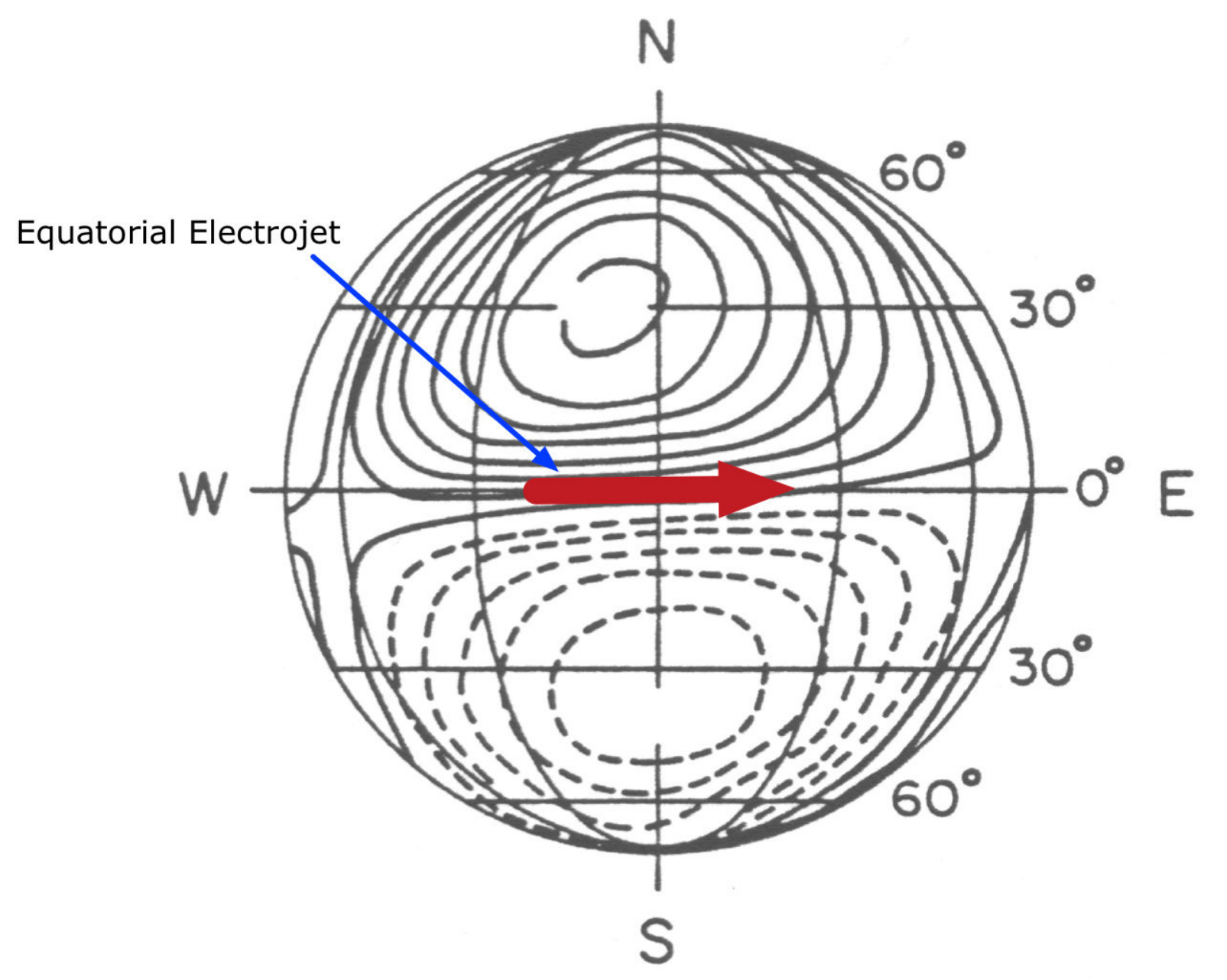}}}
		   \vspace{-3mm}
		   \caption{Current patterns set up by the atmospheric dynamo for the daytime hemisphere \citep{matsushita1965}.}      \label{Pfaff_6}
\end{figure}

As the solar EUV heats the neutral upper atmosphere, the neutral gases generally expand away from the sub-solar point, creating a system of upper atmosphere tidal motions or winds.  The neutral motions drag ions across magnetic field lines via ion-neutral collisions, creating a very small charge displacement with the electrons that subsequently sets up relatively weak, ionospheric electric fields.  As discussed above, at the lower altitudes ($<$ 130 km), the ion-neutral collision frequencies become so large that the ions are collision dominated whereas the electrons, due to their much smaller mass, are not restricted and hence, remain magnetized. Due to the differential drifts of the ions and electrons at these low altitudes, a system of global scale, horizontal current patterns develops, in which internal, polarization electric fields are generated to restrict the current flow to be non-divergent.  This worldwide system of currents is called the ``solar quiet'' (or sq) dynamo which form current loops at mid-latitudes which close at the magnetic equator, as shown in Figure~\ref{Pfaff_6} \citep{matsushita1965}. 

The strongest current driven by the atmospheric dynamo is the equatorial electrojet.  This current system exists as a result of both the dynamo action at all latitudes and the enhanced conductivity set up by the horizontal magnetic field geometry at the magnetic dip equator.  At this location, in the altitude range of roughly 95 - 115 km, significant zonal (east-west) electron drifts are set up in the following way.  Vertical $\vec E \times \vec B$ drifts due to the ambient eastward electric fields (driven by the solar forcing discussed above) combined with the north-south magnetic field are inhibited for ions because the ion-neutral collisions are sufficiently large such that $\vec V_i = 0$.  Hence, only the electrons are magnetized and the resulting vertical charge separation between the electrons and ions establishes a vertical polarization field, $\vec E_p$.  This polarization electric field, in turn, creates a horizontal $\vec E_p \times \vec B$ drift which reinforces the small, existing dynamo current and thus creates the strong electrojet current.  The equatorial electrojet is strongest within a degree of the magnetic equator, where the magnetic field vector is precisely horizontal, thus enabling the vertical, polarization electric field described above to be set up.  This current can also be explained in terms of the Cowling conductivity, which is an enhancement, by a factor of $1+ \sigma_2^2/\sigma_1^2$, of the Pedersen conductivity at the magnetic equator \citep{kelley2009}. 

At night, the zonal electric field reverses direction and thus the polarization electric field is downward. The resulting $\vec E \times \vec B$ drift again enhances the electrojet current since the zonal current is also of opposite direction at night. The magnitude of the nighttime current densities is much lower than during the daytime due to the greatly reduced ambient electron density. 

In contrast to the low latitudes where the magnetic field is nearly horizontal, at high latitudes the magnetic field lines are highly inclined and connect the ionosphere to the large amplitude magnetospheric electric fields.  These DC electric fields that map down from the magnetosphere set the high-latitude plasma in motion via $\vec E \times \vec B$ drifts.  At lower altitudes (below 200 km), however, the ion motion is appreciably affected by ion-neutral collisions such that the ion plasma drift is noticeably slowed and undergoes a change in direction.  The electrons, on the other hand, remain fully magnetized and continue to execute $\vec E \times \vec B$ drifts at altitudes as low as 70 - 80 km. Below these altitudes, the electron-neutral collision frequency becomes significant when compared to the electron gyrofrequency, as discussed earlier. 

\begin{figure}
	  \centering{
		  \resizebox{8cm}{!} {\includegraphics{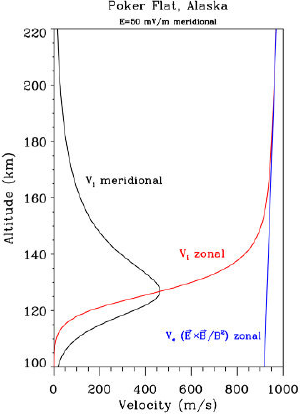}}}
		   \vspace{-3mm}
		   \caption{Ion drift velocity perpendicular to the magnetic field in the zonal (red) and meridional (black) directions, and the $\vec E \times \vec  B$ drift in the zonal direction (blue), as a function of altitude for a constant DC electric field of 50 mV/m applied in the meridional direction.  The variations of the ion drift component with altitude are due to ion-neutral collisions, as explained in the text.  The slight decrease of the $\vec E \times \vec B$ magnitude at lower altitudes is due to the corresponding small increase of the magnetic field strength closer to the earth \citep{pfaff2012}.}      \label{Pfaff_7}
\end{figure}

The ion and electron plasma drifts, calculated versus altitude for a 50 mV/m poleward (meridional) DC electric field at Poker Flat, Alaska ($\vert \vec B \vert$ = 0.54 G), are plotted in Figure~\ref{Pfaff_7} \citep{pfaff2012}.  Ions and electrons both undergo an $\vec E \times \vec B$ drift of approximately 950 m/s at 220 km.  However, at lower altitudes, where the ion-neutral collisions become increasingly larger, the ion drift begins to both slow and change its orientation towards the direction of the DC electric field.  At about 125 km altitude, where the Pedersen mobility peaks, the ion drift components are half the values of those of the $\vec E \times \vec B$ components and are equally divided between flow along and perpendicular to the electric field direction.  As the ion-neutral collisions slow the ion drifts, momentum is transferred to the neutral gas, as discussed in the next section. 

\subsubsection{Coupled Neutral and Plasma Dynamics at Low, Middle, and High Latitudes}

The coupled neutral and ionized gases within the Earth's ionosphere undergo a variety of circulation patterns and motions.  The main dynamics at low and mid latitudes are driven by neutral motions or winds which collide with ions, setting them in motion.  In addition, relatively small amplitude (0.5-5 mV/m) intrinsic DC electric fields are set up within the mid and low latitude ionosphere which subsequently drive plasma motions via $\vec E \times  \vec B$ drifts.  At high altitudes, on the other hand, magnetospheric electric fields with relatively large amplitudes (10-100 mV/m or greater) map down along the conducting field lines and set the plasma in motion.  In this case, the plasma, in general, drives the neutral winds.  The interplay between the motions of the two gas populations is quite complex, given the very different time constants to set the gases in motion which depend on the ion-neutral collision frequencies which vary significantly with altitude. 

At low and mid-latitudes, the Earth's upper atmosphere has a global circulation of neutral winds created by pressure gradients from both solar and auroral heating.  There is also forcing by tidal energy from the troposphere and mesosphere below.  The global thermospheric winds tend to blow horizontally from the subsolar heated region around the Earth to the coldest region on the nightside.  As the wind develops, Coriolis forces act to deflect the flow.  The horizontal wind speeds vary depending on the geomagnetic conditions.  For quiet geomagnetic conditions the speeds typically range from 100 to 300 m/s at altitudes of several hundred km.  Furthermore, because of the inclined angle of the magnetic field, the horizontal neutral winds are effective in transporting plasma to higher or lower altitudes.

\begin{figure}
	  \centering{
		  \resizebox{8cm}{!} {\includegraphics{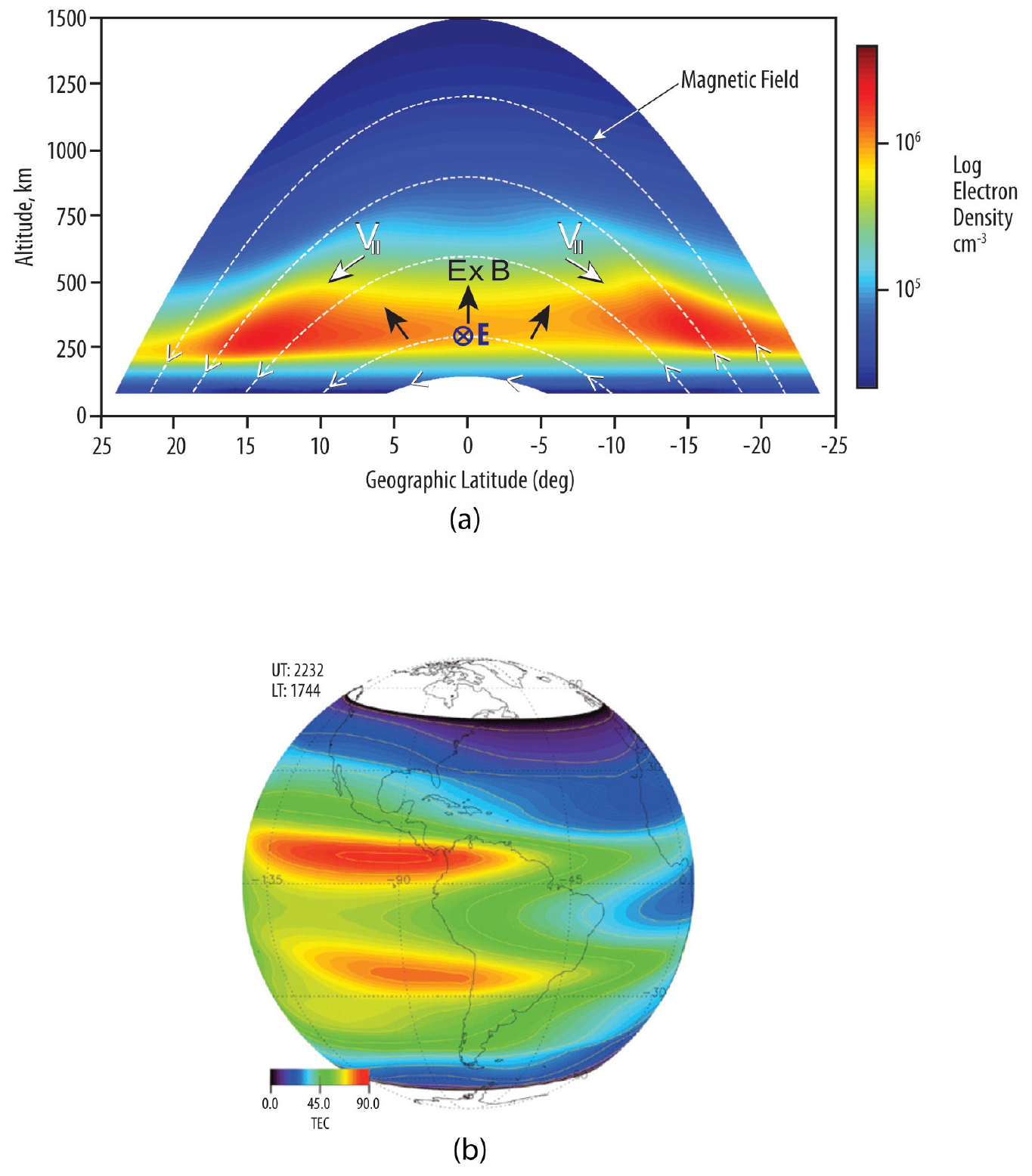}}}
		   \vspace{-3mm}
		   \caption{(a) SAMI2 model calculations versus latitude and altitude of the plasma density for 1500 SLT at 205o East.  The upward E x B drift at the magnetic equator is driven by the eastward electric field, and there is subsequent flow downward along the magnetic field lines off the equator. (b) SAMI2 model calculations of the total electron content (TEC) at 22:32 UT, showing the equatorial anomaly enhancement in the late afternoon [Figure courtesy J. Huba].}
		   \label{Pfaff_8}
\end{figure}

Neutral winds move the conducting plasma of the ionosphere across geomagnetic field lines at low and middle latitudes on the Earth's day side, driving an atmospheric dynamo and equatorial electrojet discussed earlier.  As part of this process, the winds also set up a global system of small (typically 0.5-2 mV/m) DC electric fields to ensure that the divergence of the currents is zero.  These fields drive plasma motions, such as shown in Figure~\ref{Pfaff_8}(a) in which a small zonal eastward electric field at the magnetic equator drives the plasma upwards which then slides back down along the magnetic fields on either side of the equator, creating enhanced plasma density at off-equator latitudes during the daytime.  This process is called the ``fountain effect'' \citep{hargreaves1992} and is an excellent example of how intrinsic electric fields, driven by neutral winds, alter the ambient plasma density distribution on a global scale, as shown in the plasma density model data in Figure~\ref{Pfaff_8}(b).  Indeed, this figure demonstrates how the plasma density on the dayside does not peak at the sub-solar point, which might otherwise be expected based solely on the maximum intensity of incoming solar EUV radiation.

\begin{figure}
	  \centering{
		  \resizebox{8cm}{!} {\includegraphics{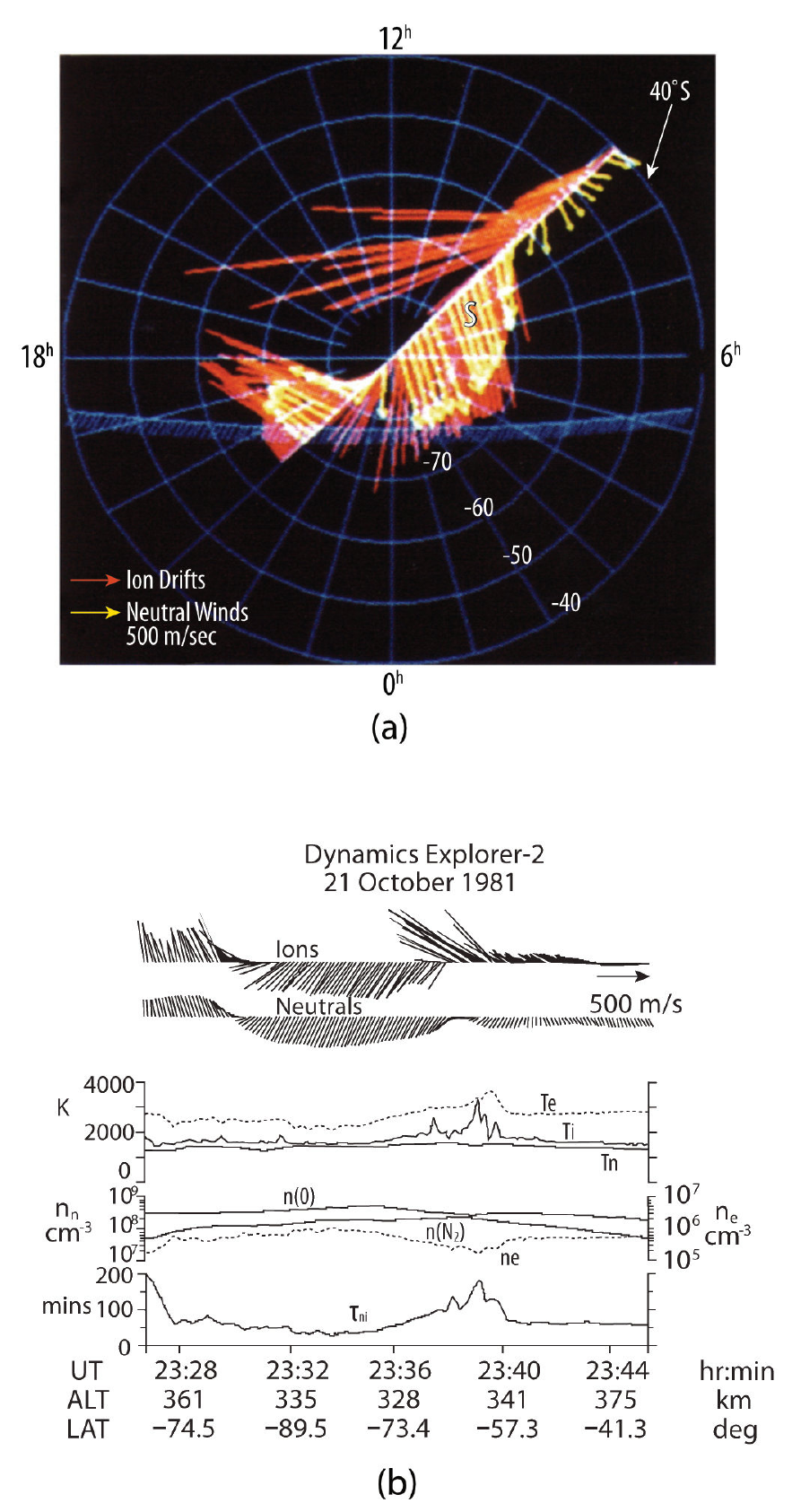}}}
		   \vspace{-3mm}
		   \caption{(a) The neutral wind, and ion drift vectors measured by DE-2 on a perigee pass over the southern polar region, plotted in geographic polar coordinates.  The neutral winds are coded by the yellow arrows, and the ion drifts are coded by the red bars.  The curved line represents the location of the solar terminator (90° solar zenith angle) \citep{killeen1984}.  
(b) Geophysical quantities measured along the track of DE 2 during the orbit shown in (a).  The ion drifts and the neutral winds are shown in the top two traces plotted against time, altitude, and latitude of the spacecraft.  The second panel shows the electron, ion, and neutral temperatures measured along the track, and the third panel shows the atomic oxygen and molecular nitrogen number densities (left-hand scale) and the electron density (right-hand scale).  The bottom trace shows the ion-neutral coupling time constant measured along the track, as discussed in \citet{killeen1984}.}
		   \label{Pfaff_9}
\end{figure}

In the polar regions, on the other hand, it is the strongly drifting ions, moving in response to the imposed higher amplitude magnetospheric electric fields, that ``drag'' the neutrals and thus generate neutral winds with speeds exceeding 1 km/s in the high-latitude F-region thermosphere.  Such $\vec E \times  \vec B$ drifts drive the high latitude ionosphere with two-celled ``convection'' patterns that characterize the polar cap ionosphere.  In this manner, the associated upper atmosphere wind system would exhibit corresponding patterns, at least initially.  An example of the ion drifts and winds measured by the Dynamics Explorer-2 satellite traversing the high latitude region is shown in Figure~\ref{Pfaff_9}(a) \citep{killeen1984}.  

In cases where the relative motions of the neutral and plasma gas populations are not the same, frictional or ``Joule'' heating occurs. The Joule heating rate per unit volume, $Q_j$ , may be represented as: 

\begin{equation}
        Q_j  =  \vec J \cdot \vec E^{'}  =  \sigma_p \vec E^{'2}
        \label{8}
        \end{equation}
where $\vec E^{'} = (\vec E + \vec U \times \vec B)$ and $\sigma_p$ is the Pedersen conductivity \citep{thayer2000}.  An example of Joule heating measurements by satellite probes is shown in Figure~\ref{Pfaff_9}(b) \citep{killeen1984} for the same data shown in Figure~\ref{Pfaff_9}(a).  This is a fundamental physical process inherent to differential velocities of ion and neutral gas populations within partially ionized plasmas.

Note that whereas the ionized gas motions are governed by the ambient magnetic field, the neutral gases are not.  Thus, when the magnetospheric electric fields are shut off or change direction, the winds that were initially driven by the $\vec E \times \vec B$ plasma drifts continue moving, and may proceed equatorward from the polar cap.  The coupling of the neutral wind and plasma drifts underscores how the upper atmosphere/ionosphere of the earth must be understood as a dynamic system, in which the coupling efficiencies and time constants strongly depend on the ion-neutral collision frequencies and hence, altitude.

\subsubsection{Response to energetic particles at high latitudes}

In addition to responding to strong, imposed electric fields, the high latitude ionosphere also responds to vast swaths of precipitating energetic particles accelerated in the magnetosphere, with typical incoming energy spectra ranging from tens of eV to several keV (e.g., energetic electrons associated with the visual aurora).  These energetic particles are responsible for subsequent increases in localized conductivity (discussed below), heating, and field-aligned currents.  The field-aligned currents are associated with incoming electromagnetic energy flux into the region manifested in the form of Poynting flux and Alfv\'en waves \citep{kelley2009}.  The coupling between the solar wind and the magnetosphere-ionosphere (M-I) system is important in determining the global energy budget in the high latitude ionosphere/thermosphere system in which energy is input from the magnetosphere in the form of both electromagnetic fields and particle kinetic energy flux. 

\begin{figure}
	  \centering{
		  \resizebox{8cm}{!} {\includegraphics{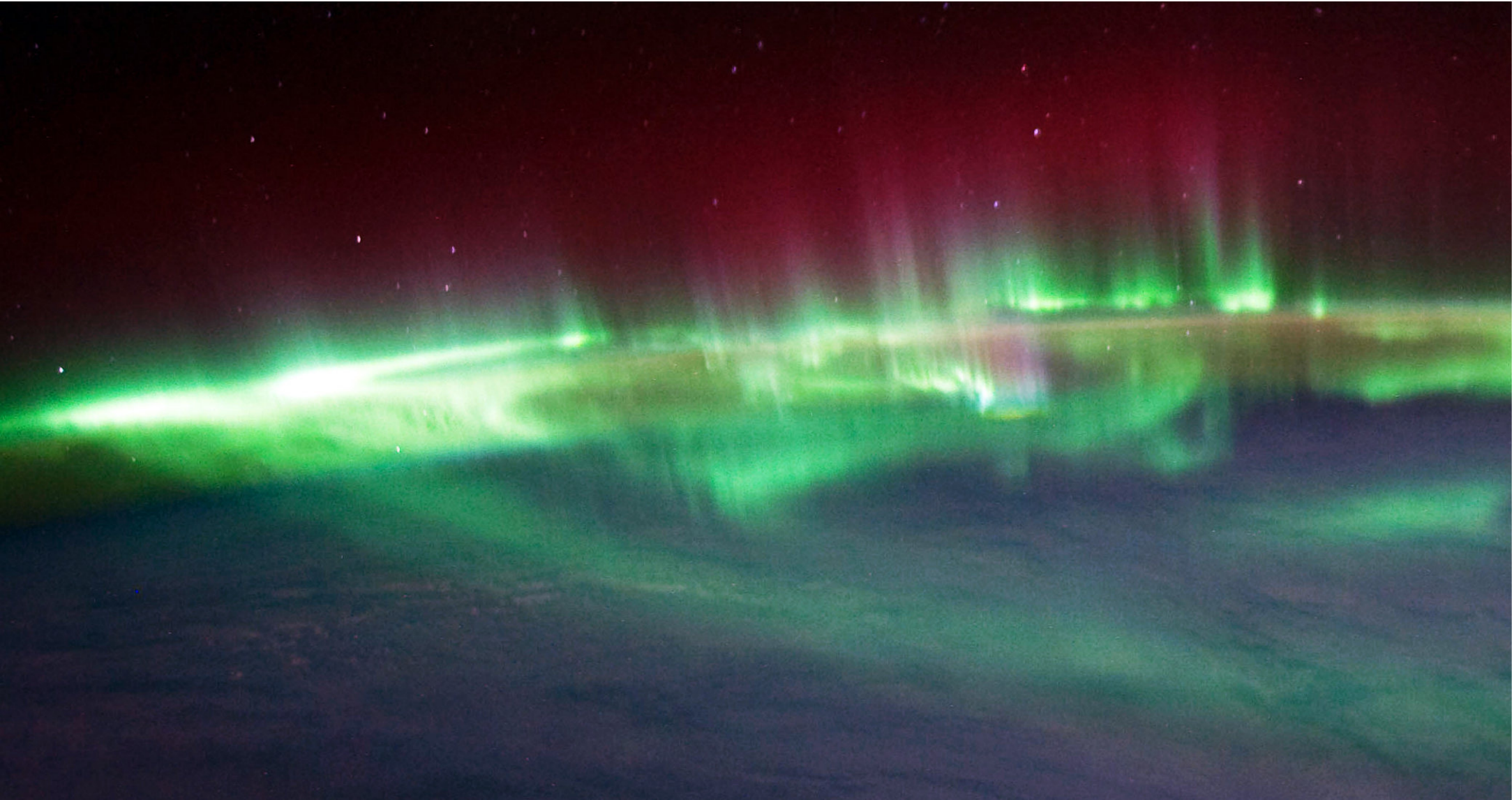}}}
		   \caption{View of the aurora photographed from the International Space Station on March 4, 2012, 17:20:33 UT on a pass over the South Indian Ocean.  Figure courtesy NASA.}
		   \label{Pfaff_10}
\end{figure}

Figure~\ref{Pfaff_10} shows a photograph of the aurora from the International Space Station in which the auroral optical emissions vary as a function of altitude as the incoming energetic electrons interact with the neutral atmosphere at different altitudes, depending on their energy.  Note that at the higher altitudes ($\sim$130-250 km) the aurora is red/pink as the incoming electrons with lower energy interact with atomic oxygen yet is green below these altitudes corresponding to the incoming electrons with higher energy (typically keV) interacting, again, with oxygen.

The influx of precipitating energetic particles into the high latitude upper atmosphere/ionosphere system represents a source of thermal plasma density (i.e., increased conductivity) that results when the energetic electrons are ``braked'' by the neutral atmosphere.  In such cases, incoming beams of energetic particles (mostly electrons) interact with the upper atmosphere which undergo impact ionization to create thermal plasma and heat the ambient gases.  The precipitating electrons are a source of field-aligned currents whose circuit is ultimately completed through the lower ionosphere where the Hall and Pedersen mobilities enable the currents to close.  

\begin{figure}
	  \centering{
		  \resizebox{8cm}{!} {\includegraphics{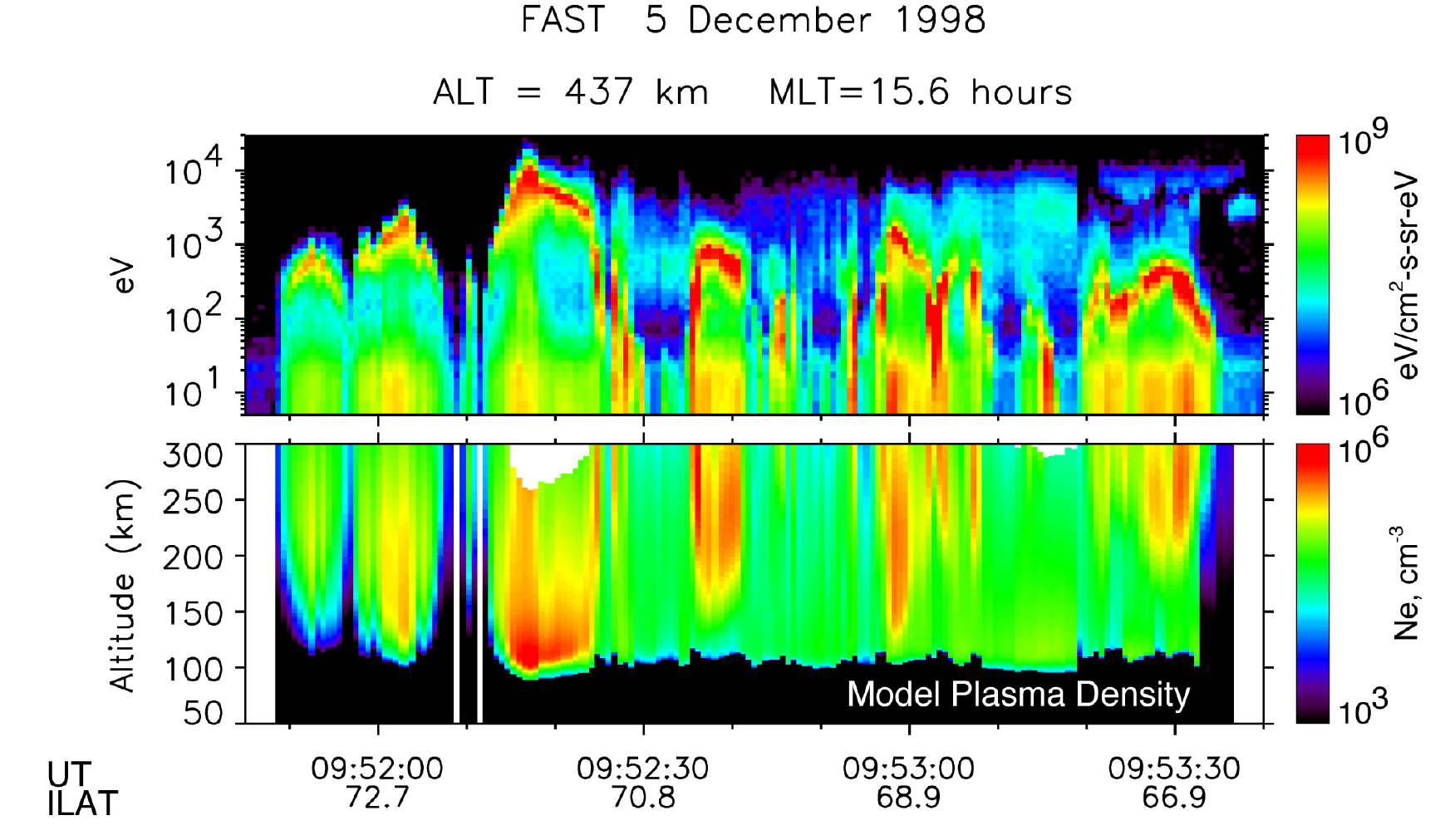}}}
		   \vspace{-3mm}
		   \caption{Observations gathered on the FAST satellite near 437 km are shown in the top panel and reveal narrow beams of downgoing, energetic electron spectra.  The bottom panel shows the corresponding model thermal plasma created by these electrons [Pfaff, 2012].}
		   \label{Pfaff_11}
\end{figure}

To illustrate how the incoming energetic electrons create thermal plasma, Figure~\ref{Pfaff_11} \citep{pfaff2012} shows a pass of incoming keV electrons from a perigee pass of the FAST satellite in the upper panel.  Each of these groups of energetic electrons corresponds to a spatial extension of energetic electrons through which the satellite has traversed and which would constitute an auroral arc formation where they interact with the ambient neutral gases at lower altitudes.  The lower panel shows the thermal panel produced by these electrons, using the model of \citet{strickland1993}.  Notice the ``clumps'' of enhanced thermal plasma (i.e., enhanced conductivity patches) whose altitude depends on the energy of the incoming electrons in the upper panel.  This illustrates how ``patchy'' partially ionized plasmas are formed in the high latitude regions of the earth's upper atmosphere and may be common in astrophysical plasmas where electron beams with sufficient energy to ionize the neutral gas are present.

\subsubsection{Response of the Ionosphere to Disturbances/Magnetic storms}

Geomagnetic storms can occur in response to enhanced solar activity that interacts with the Earth i.e., sudden changes in the solar wind dynamic pressure at the magnetopause caused by the passage of a coronal mass ejection. The interaction causes a temporary disturbance in the magnetosphere which leads to an expansion of plasma convection and particle precipitation patterns, stronger electric fields, and intensified precipitation \citep{dremukhina1999}. These changes are accompanied by substantial increases in the Joule and particle heating rates and the electrojet currents.  The ionosphere receives a large amount of energy during storms and substorms, and large storms can significantly modify the density, composition, and circulation of the ionosphere on a global scale.  Accordingly, neutral flows in the lower ionosphere increase at higher latitudes during storms as well.  Substorms correspond to the explosive release of energy in the auroral region near magnetic midnight in which energy is released from the magnetotail and injected into the high latitude ionosphere, often causing discrete auroral arcs.  Eventually, the disturbances associated with substorms typically encompass the entire high-latitude region.

\begin{figure}
	  \centering{
		  \resizebox{8cm}{!} {\includegraphics{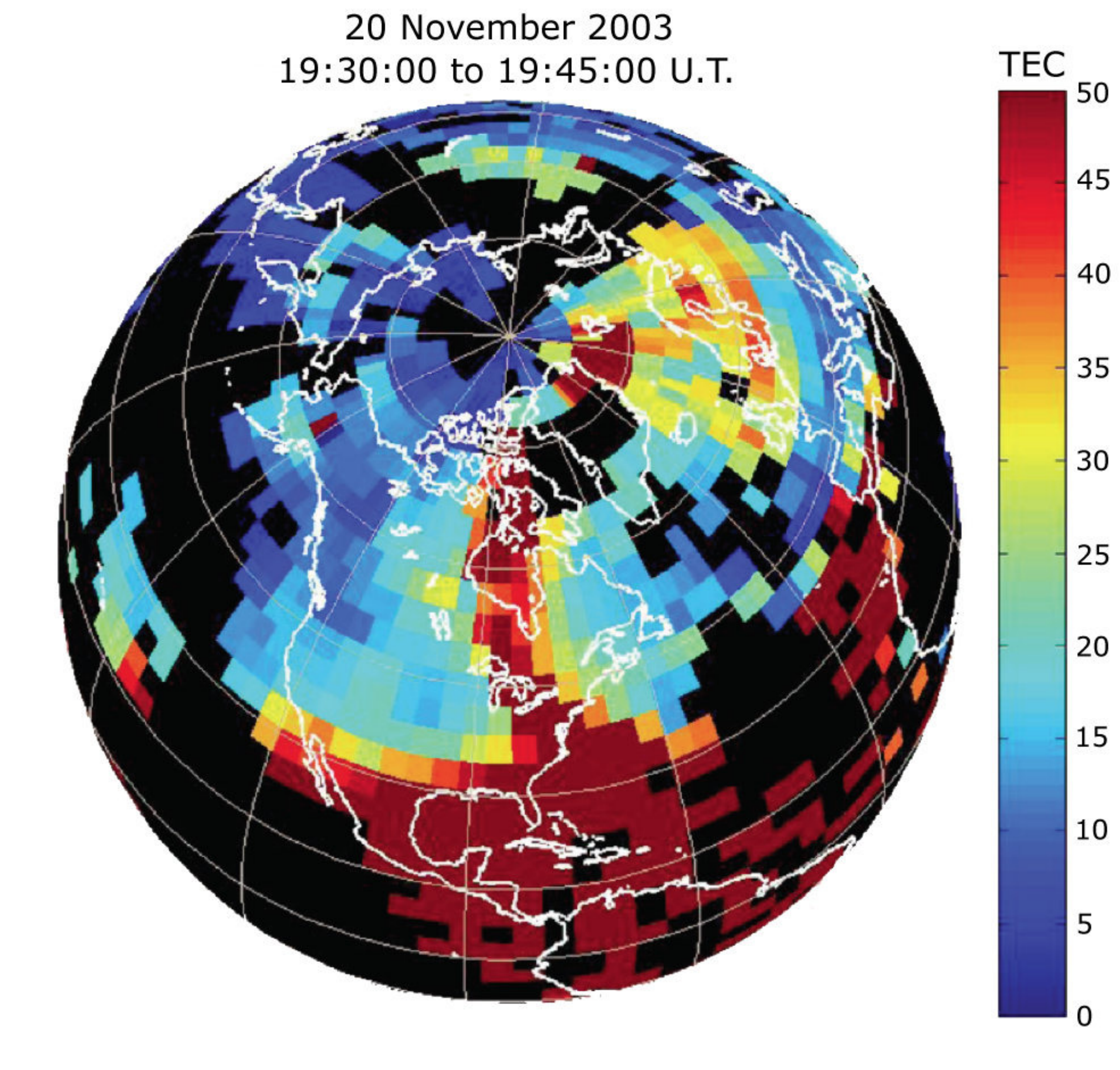}}}
		   \vspace{-3mm}
		   \caption{ Storm-enhanced plasma density (SED) signatures in total electron content (TEC) observed on November 20, 2003.  These are believed to be connected to plasmasphere erosion and driven by sub-auroral electric fields from the inner magnetosphere.  Strong plasma density gradients are observed by a network of ground-based GPS observatories (Figure courtesy of A. Coster).}
		   \label{Pfaff_12}
\end{figure}

The low and mid-latitude ionosphere is also highly subject to magnetic disturbances and storms and reveals a complex response based on penetration electric fields that are applied from the magnetosphere response to the increased solar wind pressure and electric fields associated with the storm.  These processes include the abrupt transport of ionospheric plasma to very high altitudes at mid- and low- latitudes.  An example of a storm-enhanced plasma density feature is shown in Figure~\ref{Pfaff_12} (see also \citet{foster2005}) and demonstrates how the ionospheric plasma enhancements may be confined within narrow longitude regions based on the magnetic storm input.  The neutral gases also respond to this sudden increases of energy.  Indeed, the entire upper atmosphere  --  both neutral and ionized components  --  respond in dramatic fashion to externally imposed disturbances associated with magnetic storms.

In summary, the most accessible partially ionized plasma available for detailed study, the earth's ionosphere/thermosphere system is particularly rich in processes due to the many sources of energy, momentum, and ionization that drive and maintain the coupled system as well as due to the presence of the dipole magnetic field that permeates and helps organize the system.  Global patterns of electric fields, currents, plasma density, neutral winds and other parameters are set up which display variations as a function of altitude, latitude, local time, and solar activity.  Detailed measurements with satellite, sounding rocket, and remote sensing instruments, as well as advanced modeling techniques, reveal fundamental processes important not only for understanding important physics of planetary ionospheres but also for comparisons with partially ionized plasmas associated with solar and astrophysical gases.  Indeed, the coupled neutral and ionized gases that exist within the Earth's upper atmosphere yield a highly complex system with a wide variety of physical processes that are among the most captivating and important topics in space research.

%%%%%%%%%%%%%%%%%%%%%%%%%%%%%%%%%%%%%%%%
\subsection{Jupiter's and Saturn's thermosphere and ionospheres}
\label{Fsubsec: Jupiter's and Saturn's thermosphere and ionospheres }

\paragraph{Thermosphere Parameters}

\begin{table}
\begin{center}
\caption{Key Thermosphere Parameters for Earth, Jupiter and Saturn}
\label{tab:1}    
\begin{tabular}{p{4.6cm}p{1.8cm}p{1.8cm}p{1.8cm}}
\hline\noalign{\medskip}
Parameters (units) & Earth\quad &
 Jupiter & Saturn  \\
\hline\noalign{\medskip}
Homopause temperature (K) &$\;\;200  $&$\;\; 200 $&$\;\;
 140 $\\
\hline\noalign{\medskip}
Homopause pressure ($ 10^{-6}$ bar)&$\;\;1.0$&$\;\;2$&$\;\;0.1$ \\
\hline\noalign{\medskip}
Homopause density  ($10^{19}\; m^{-3}$)&$\;\;3.7$&$\;\;
7.3$ &$\;\;
0.52$ \\
\hline\noalign{\medskip}
Homopause scale height$^a$ ($km$)&$\;\;
6.0$&$\;\;
35.7$ &$\;\;
64.6$  \\
\hline\noalign{\medskip}
Exospheric temperature (K)&$\;\;1000$&$\;\;
940$&$\;\;
420$\\
\hline\noalign{\medskip}
Exospheric scale height$^b$ ($km$)&$\;\;
52.6$
 &$\;\;
335.6$ &$\;\;
387.0$  \\
\hline\noalign{\medskip}
Critical density ($10^{13}\,m^{-3}$) &$\;\;
10$ &$\;\;
2.5$&$\;\;
2.5 $ \\
\hline\noalign{\medskip}
Critical pressure ($10^{-12}$ bar)&$\;\;4$&$\;\;
1$&$\;\;
1$ \\
\hline
\end{tabular}
\end{center}
\vspace{-0.2cm}
\noindent $^a$ Scale height for N$_2$ for Earth and H$_2$ for Jupiter and Saturn. \\
 \noindent $^b$ Scale height for $O$ for Earth and $H$ for Jupiter and Saturn.
\end{table}

The thermosphere is the uppermost region of a planet's neutral atmosphere. It is characterised as a region in which the temperature steadily increases with altitude until a maximum (exospheric) limit is reached. Mean free path lengths for thermospheric species are long - sometimes up to hundreds of kilometres - and the mixing of the atmosphere by convection is almost non-existent. The level at which convective mixing is no longer important is known as the homopause. Above the homopause, atmospheric atomic and molecular species settle out diffusively; each species has its own scale height, $H_S$, given by:

\begin{equation}
\label{eq:scaleheight}
H_S=k\,T/g\,m_S
\end{equation}
where $k$ is Boltzman's constant, $T$ is the thermospheric temperature, $g$ the acceleration due to gravity and $m_S$ is the atomic or molecular weight of species $S$. At higher altitudes, the thermosphere merges into the exosphere. The base of the exosphere, the exobase, is characterised by a critical density, $N_C(S)$, at which the horizontal mean free path of the main thermospheric species, $S$, is equal to the scale height. $N_C(S)$, is given by:
\begin{equation}
\label{eq:scaleheight1}
N_C(S)=\frac{1}{\pi\,d^2_S\,H_S}
\end{equation}
 where $d_S$ is the diameter of species $S$. Some approximate values for key parameters for the Earth, Jupiter and Saturn are compared in Table~\ref{tab:1} \citep{Miller-2005}.
 
  \begin{figure}
\centering
\includegraphics[width=0.8\textwidth]{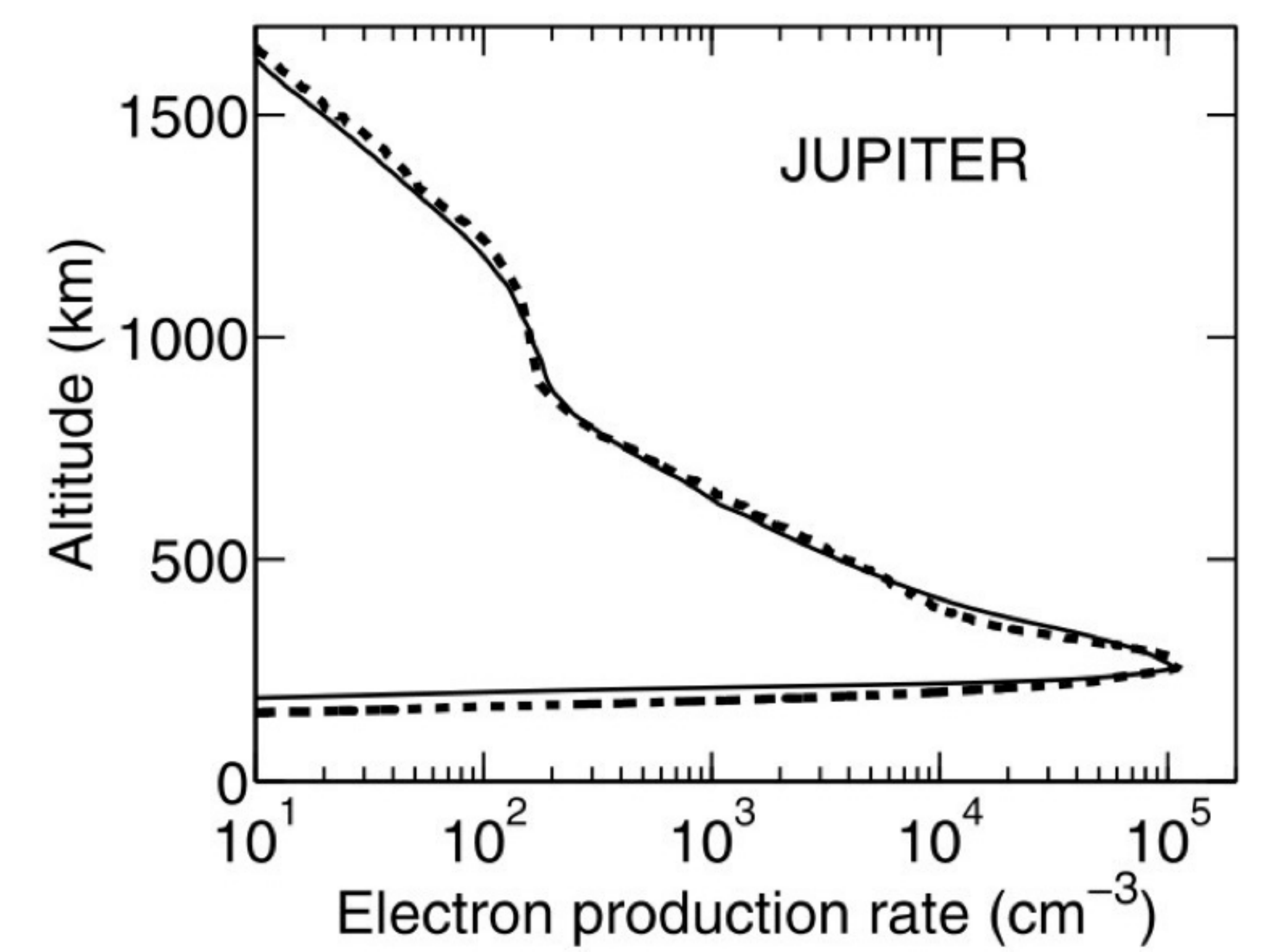}
\caption{ Profile in altitude of the electron production rate (solid line) calculated by  \citet{Galand-et-al-2011} and by \citet{Grodent-et-al-2001} for similar thermospheric conditions at Jupiter.}
\label{fig:alexeev_new}
\end{figure}

Figure~\ref{fig:alexeev_new} shows a comparison between the electron production rate, which has been calculated by \citet{Galand-et-al-2011} (solid line) and that derived by \citet{Grodent-et-al-2001} (dashed line). Both profiles in altitude of the electron production rate assuming the same triple Maxwellian distribution for the incident auroral electrons as  \citet{Grodent-et-al-2001} and similar thermospheric conditions at Jupiter. The Jupiter's ionospheric profile can be compared with number densities of the Earth's ionosphere (see Figure~\ref{Pfaff_2}) 
 
 \begin{figure}
\centering
\includegraphics[width=0.8\textwidth]{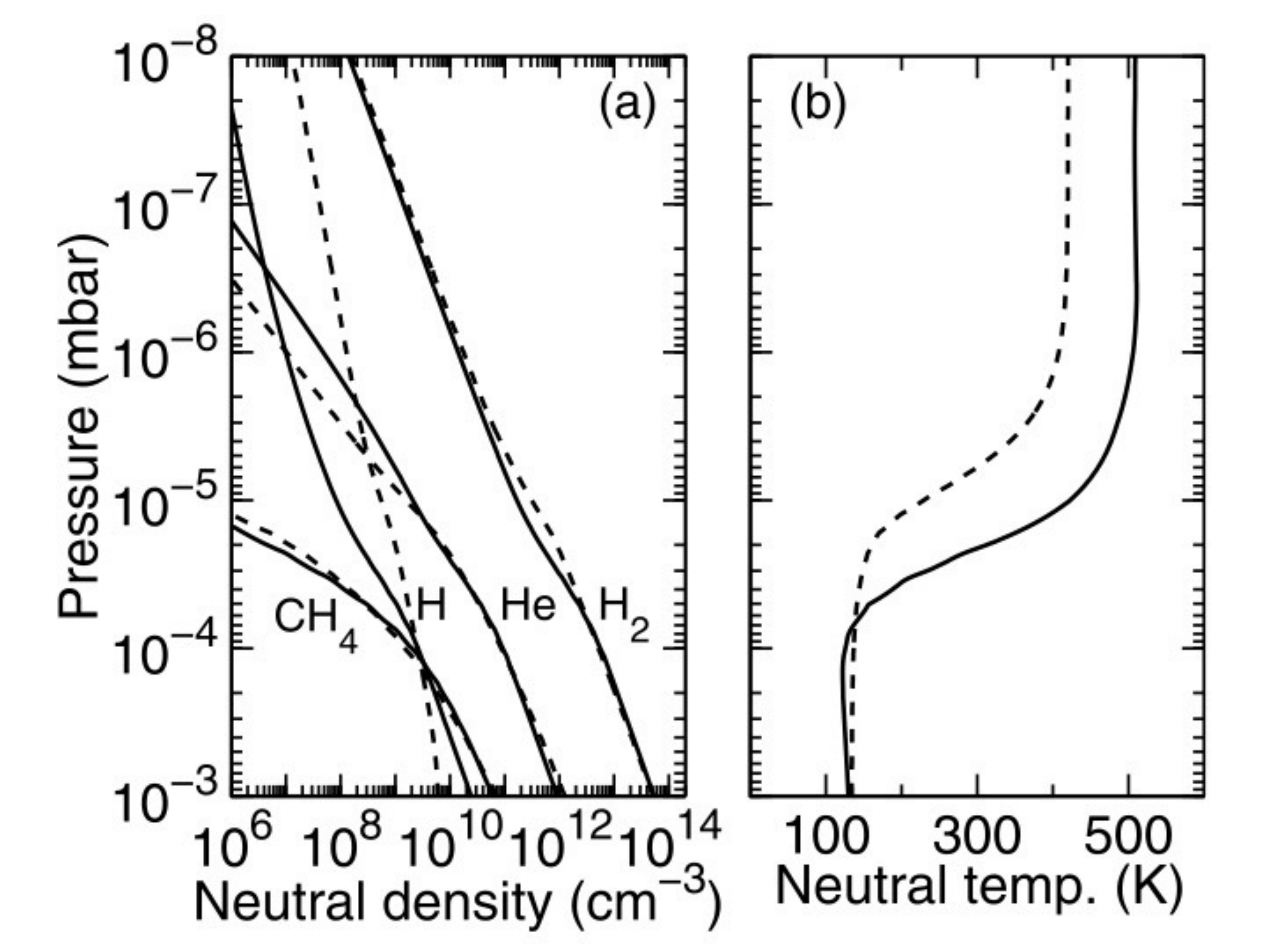}
\caption{ 
Saturn neutral atmospheric (a) density and (b) temperature profiles in altitude resulting from the 3D neutral Saturn Thermosphere Ionosphere Magnetosphere (STIM) model at 78$ ^\circ$S latitude at equinox ($T_{exo} = 510\, K$, solid lines) \citep{Galand-et-al-2011}. 
The reference profiles ($T_{exo} = 420\, K$) derived by \citet{Moses-et-al-2000} are shown as dashed lines.
 }
\label{fig:`Jup_e_profile2}
\end{figure}

The Saturn thermospheric densities and temperature extracted at a latitude of 78$ ^\circ$S at 12LT at equinox during solar minimum conditions are presented in Figure~\ref{fig:`Jup_e_profile2} \citep{Galand-et-al-2011}. Though the quantities are plotted for noon, no significant local time dependence is found, which is explained by the fast rotation of Saturn. The exospheric temperature, $T_{exo}$, is strongly correlated with the amount of Joule heating present in the high-latitude regions. The electron energy flux and convection electric field, which drive the Joule heating, are chosen such that the derived value for $T_{exo}$ is in agreement with observations from UV occultation. 

\begin{figure}
\centering
\includegraphics[width=0.8\textwidth]{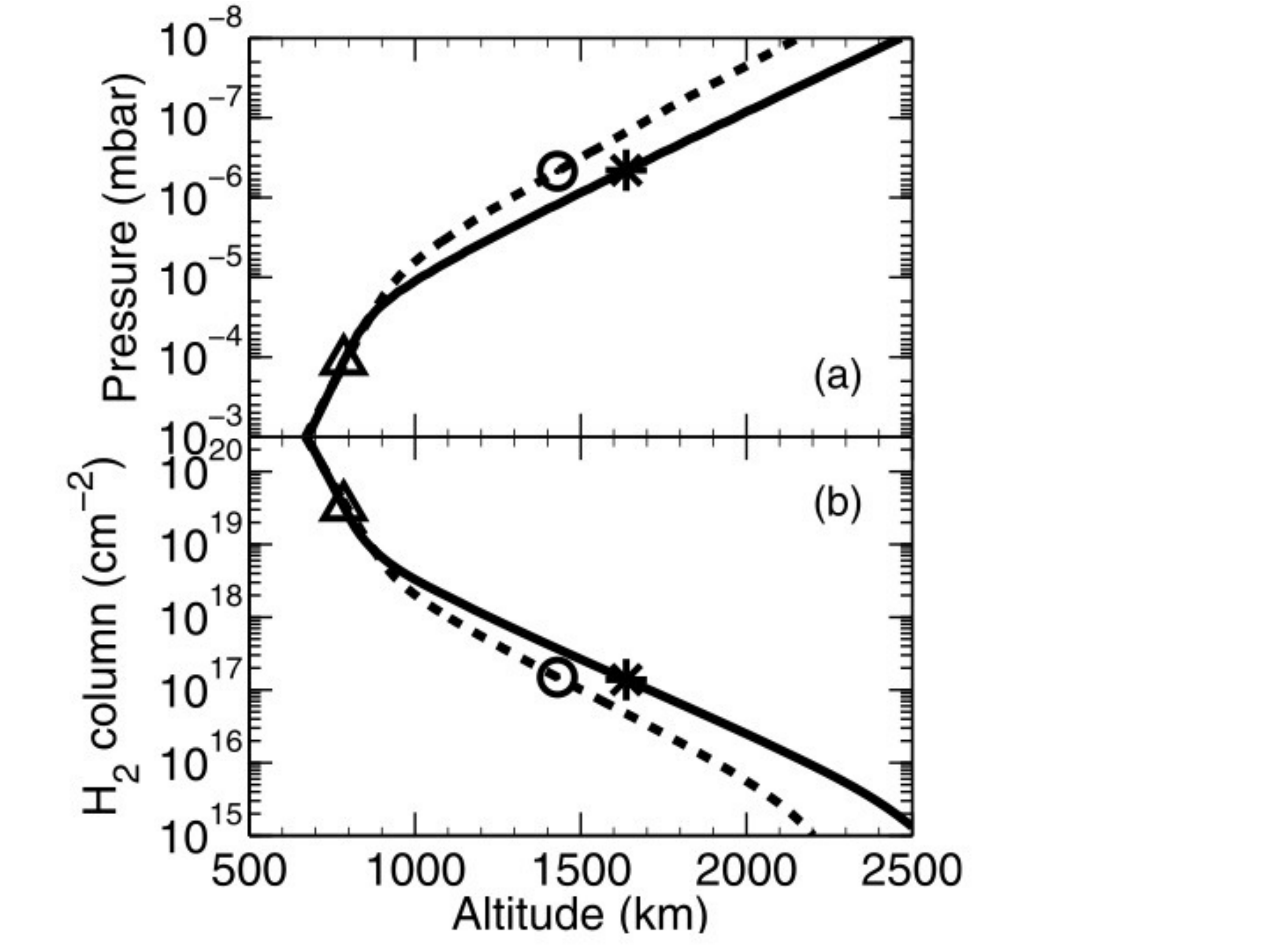}
\caption{(a) Pressure versus altitude derived from the 3D STIM atmospheric profile ($T_{exo} = 510\, K$ 
\citep{Galand-et-al-2011},
solid line) and from the reference profile by \citep{Moses-et-al-2000} ($T_{exo} = 420\, K$) (dashed line). (b) Same as Figure (a) but for the H$_2$ column density above. The markers correspond to the pressure and H$_2$ column density at the altitude of maximum energy deposition for $E_{m} = 10$ keV (triangle) using the STIM atmosphere and for $E_{m} = 500$ eV using the 78$ ^\circ$S STIM atmosphere (star) and reference atmosphere (circle). For visibility reasons, the case of 10 keV using the reference atmosphere has not been plotted as it overlaps with the 10 keV case using the STIM atmosphere. The transport of auroral electrons and the column density valid along the path were calculated for the dip angle at 78$ ^\circ$S, i.e., 82$ ^\circ$. The reference altitude is taken to be the 1 bar level, and the top of the atmosphere is taken at a pressure of $4.4 \times 10^{-9}$ mbar.
 }
\label{fig:`Jup_e_profile3}
\end{figure}

In Figure~\ref{fig:`Jup_e_profile3}, the midlatitude atmospheric model of \citet{Moses-et-al-2000}, widely used by the community at high-latitudes \citep{Gerard-et-al-2004, Gustin-et-al-2009} is shown as dashed lines in Figure~\ref{fig:`Jup_e_profile2}. The associated exospheric temperature of 420 K was derived from the analysis of the Voyager 2 UVS solar ingress occultation. Plotted as a function of pressure, the H$_2$ density obtained for $T_{exo}$ = 510 K is very close to the 420 K reference model. The same applies to the minor species He and CH$_4$ at high pressures above $3 \times 10^{-5}$ mbar, keeping in mind that the mixing ratios assumed for these species at the lower boundary in the STIM model are those from \citet{Moses-et-al-2000}. At lower pressure, large differences in the H and He densities are clearly apparent between the  Saturn Thermospher-Ionosphere Model (STIM)profiles and the reference profiles. Such differences are primarily due to global dynamics. Because of the difference in temperature profiles between the 510 K STIM (solid line) and the 420 K reference (dashed line) atmospheric models, the conversion from pressure to altitude is similar at pressures larger than $2 \times 10^{-5}$ mbar, but significantly different at pressures smaller than $2 \times 10^{-5}$ mbar, as illustrated in Figure~\ref{fig:`Jup_e_profile3}a. For instance, on the one hand, 10 keV auroral electrons penetrate deeply into the atmosphere to an altitude of 790 km (triangle in Figure~\ref{fig:`Jup_e_profile3}a) corresponding to an H$_2$ column density of $3 \times 10^{-19}$ $cm^{-2}$ (triangle in Figure~\ref{fig:`Jup_e_profile3}b) and have a penetration altitude largely insensitive to the atmospheric models. On the other hand, for 500 eV electrons there is a difference of 208 km between the high-latitude STIM (stars) and the midlatitude reference (circles) models. Therefore, the conversion between the penetration altitude of auroral electrons and the mean energy of the auroral electrons using alter- native atmospheric models needs to be taken with caution when focusing on soft electron precipitation.

\begin{figure}
\centering
\includegraphics[width=0.8\textwidth]{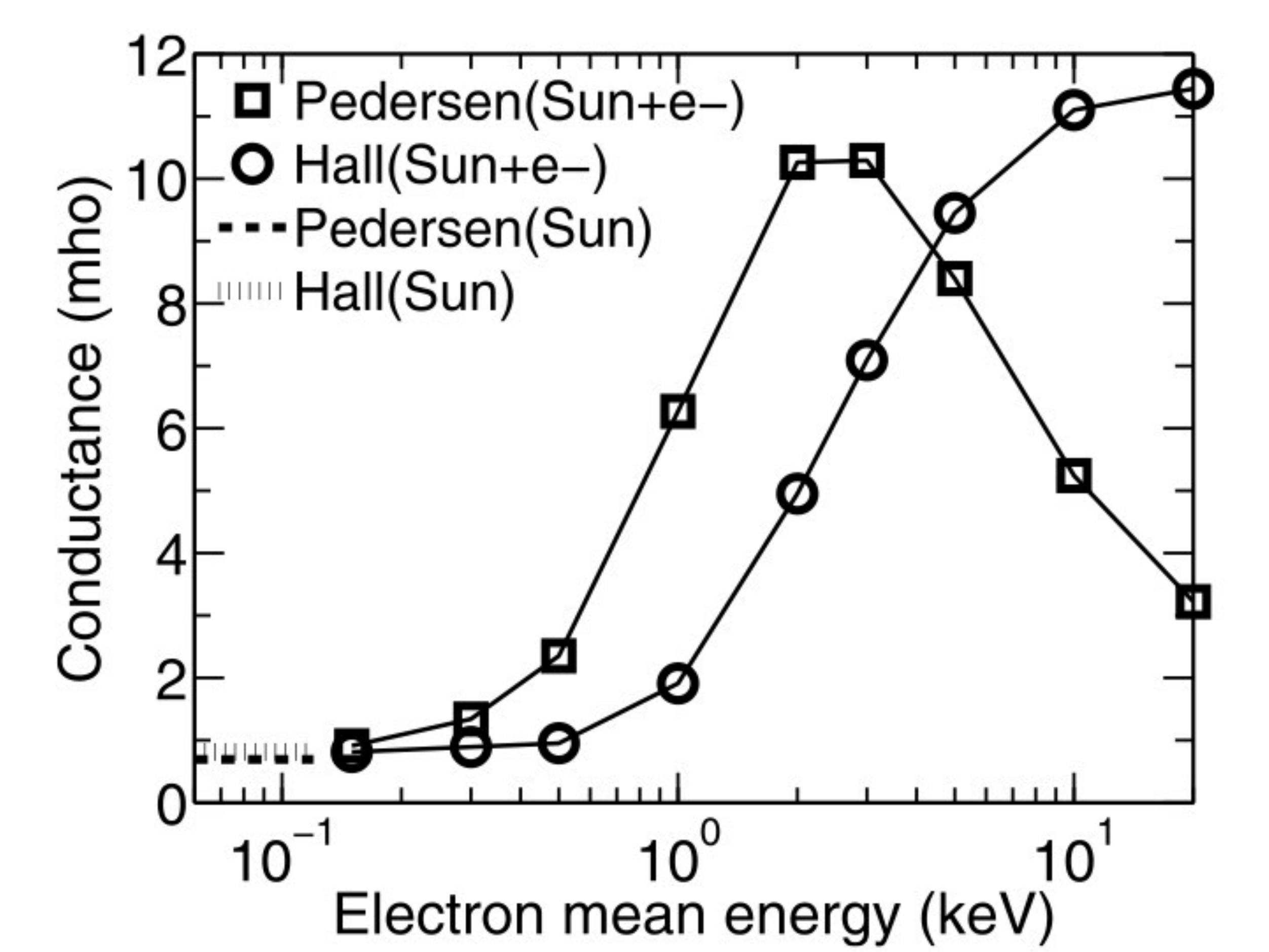}
\caption{Pedersen (squares) and Hall (circles) ionospheric conductances as a function of the mean energy of the incident auroral electrons, Em, under solar illumination angle of 78$ ^\circ$ at noon. The distribution of the incident electrons is assumed to be Maxwellian with an energy flux, $Q_0$, of $0.2\, mW \cdot m^{-2}$. The Pedersen and Hall ionospheric conductances obtained for solar illumination alone (no auroral particle precipitation) are shown with horizontal, dotted and dashed lines, respectively,
\citep{Galand-et-al-2011}.
}
\label{fig:`Jup_e_profile4}
\end{figure}

In Figure~\ref{fig:`Jup_e_profile4}, the ionospheric Pedersen conductance SP (squares) and Hall conductance SH (circles) are plotted as a function of the mean energy of the incident electrons. The energy dependence is different between Pedersen and Hall conductances, as their conductivities do not peak in the same altitude region. Auroral electrons with incident mean energy in the 2 - 3 keV range deposit their energy near the homopause, where the Pedersen conductivity peaks. As a result, the Pedersen conduc- tance reaches a maximum in the 2 - 3 keV incident energy range. The Hall conductivity peaks well below the homopause. As a consequence, the Hall conductance is increasing over the whole electron mean energy range considered here spreading from 150 eV to 20 keV. At very low electron mean energy, the auroral deposition peak altitude is located at altitudes higher than the solar deposition peak altitude, in a region less efficient for generating conductivities. As a result, the Pedersen and Hall conductances converge towards the solar values equal to 0.7 mho and 0.8 mho, respectively. At 150 eV, the Pedersen conductance is within 30\% of the solar value, while the Hall conductance whose conductivity peaks lower than the Pedersen conductivities, has already reached the solar value. 

\paragraph{Magnetospheric plasma}

The plasma features and particles beams are regulated by magnetospheric magnetic field. The general view of the Jupiter and Saturn magnetosphere is given by the magnetospheric models of
 \citet{Belenkaya-et-al-2005, Alexeev-et-al-2006, Alexeevetal2015}. Particles, momentum and energy are exchanged between the planetary upper atmosphere and magnetosphere via the ionosphere in the high latitude regions. There is a net momentum transferred from the atmosphere to the magnetosphere, while energy, for instance particle precipitation, is deposited from the magnetosphere to the atmosphere \citep[see e.g.,][]{Hill-1979, Hill-2001, Cowley-Bunce-2001}. These particles primarily originate from moons (e.g., Io and Europa at Jupiter, and Enceladus at Saturn), and to a lesser extent from the planetary atmosphere and the solar wind (e.g., polar regions at Jupiter). Some of the ions resulting from ionization of the moon$'$s gas torus are neutralized through charge exchange and leave the system; the others are picked up by the planetary magnetic field closely rotating at the planet's rotation rate and flow outward through the planetary magnetosphere \citep{Bagenal-Delamere-2011}.
      The resulting upward currents, flowing from the atmosphere to the magnetosphere, that supply the required angular momentum accelerate the particles, increasing their energy and energy flux \citep[see e.g.,][] {Ray-et-al-2010, Ray_et-al-2012a}. Particles can also precipitate as a result of wave-particle interactions (e.g. \citet{Radioti-et-al-2009}).

When the energized particles reach the high latitude upper atmosphere, they collide with the atmospheric species, depositing energy through ionization, excitation and dissociation of the neutral gas. This yields the so-called auroral emissions defined as the photomanifestation of the interaction of energetic, extra-atmospheric particles with an atmosphere (e.g. 
      \citet{Bhardwaj-Gladstone-2000, Galand-Chakrabarti-2002, Fox-et-al-2008, Slanger-et-al-2008}). 
      In the Jupiter system, atmosphere and magnetic field are not just Jupiter, but a light atmosphere and small magnetic field also exist on one of the largest moons, Ganymede. As a result of Jovian magnetospheric plasma flow past Ganymede, we have auroral particle acceleration and Ganymede$'$s auroral emission (see for example \citet{Lavrukhin2015})
      
Auroral particle degradation results in an increase in ionospheric densities and electrical conductances (e.g.  \citet{Millward-et-al-2002, angeo-26-77-2008, Galand-et-al-2011}). Ionospheric currents, which allow closure of the magnetospheric current system, are enhanced and 
induce strong Joule heating of the high-latitude thermosphere (e.g. \citet{Miller-2005, Muller-Wodarg-et-al-2012}). This high-latitude atmospheric heating is a key player in the energy balance at the giant planets (e.g.  \citet{Yelle-Miller-2004}). In other words, particle precipitation, which can be traced via auroral emissions, plays a critical role in the thermosphere-ionosphere system and its coupling to the magnetosphere \citep{Alexeevetal2014}.

 The global magnetospheric dimensional  sizes are provided in
 Table~\ref{Ftab2:magnetosphere parameters} \citep{Alexeevetal2015}.
The scaling relations allow one to adapt the magnetopause and the tail current systems
developed for the Earth's magnetosphere to the case of  Jupiter and Saturn.
However, for the
currents caused by the rapid planetary rotation (magnetodisk for Jupiter and Saturn)
 there is no analogy
in the terrestrial magnetosphere. The magnetodisk is the main source of Jupiter's magnetospheric magnetic field. Its effective magnetic moment dominates
Jupiter's dipole magnetic moment ($\sim 2.6$ times
\citet{Alexeev-Belenkaya-2005}).

The dimensions of planetary magnetospheres may vary by a factor of several hundreds (see Table~\ref{Ftab2:magnetosphere parameters}), but different planets have similarly shaped magnetopauses. The front part of the
magnetosphere coincides with the paraboloid of revolution, which has a symmetry
axis that is a line joining the planet and the Sun. The magnetic field at the
subsolar point can be determined from the balance of the solar wind plasma dynamic
pressure and the pressure of the magnetic field inside the magnetosphere. This field does not
depend on the value of the planetary dipole and is determined unambiguously by the
solar wind dynamic presssure.
 The planetary dipole determines the size of the
magnetosphere, $R_{ ss}$. For Jupiter$'$s and Saturn$'$s magnetospheres the equatorial
magnetodisks are formed by the rotational uploading of 
the satellite{'}s plasma. As a result, $R_{ ss}$ is determined not by the planetary dipole only, but by some ``effective'' dipole which is bigger than the planetary one.

\begin{table}
\begin{center}
\caption{Solar System Planetary Magnetosphere Parameters\index{magnetosphere parameters}}
\label{Ftab2:magnetosphere parameters}      
\begin{tabular}{p{5.2cm}p{1.9cm}p{0.8cm}p{1.0cm}p{1.1cm}}
\hline\noalign{\medskip}
\noalign{\smallskip}
Parameters (units) & \hspace*{0.1cm} Equation  &\hspace*{0.1cm} Earth &
\hspace*{0.1cm} Jupiter & Saturn  \\
\noalign{\smallskip}
\hline\noalign{\medskip}
Heliocentric distance (AU) &\hspace*{0.cm} $\quad r_0$ &
\hspace*{0.4cm} 1 &\hspace*{0.3cm}  5.2 &\hspace*{0.4cm}  9.5\\
\hline\noalign{\medskip}

Equatorial radius ($ R_{E}=\; 6,371\, \mbox{km} $)&\hspace*{0.4cm}$ r_P$ &
\hspace*{0.4cm}1&\hspace*{0.4cm}11.2&\hspace*{0.35cm}9.45 \\
\hline\noalign{\medskip}

Magnetic moment $( T \cdot km^3 ) $ &\hspace*{0.cm}$\; B_0\cdot r_P^3$ &
\hspace*{0.3cm}$0.008$&\hspace*{0.35cm}$150.$&\hspace*{0.4cm}$4.6$  \\
\hline\noalign{\medskip}

Dipole tilt angle (degrees)&\hspace*{0.4cm}$\psi$ &
$\quad 10.5^\circ$ &$\; \quad 10^\circ$ &$\qquad 0^\circ$  \\
\hline\noalign{\medskip}

Equatorial magnetic field ($\mu $T) &\hspace*{0.4cm} $ B_0$&
\hspace*{0.4cm} 30.&\hspace*{0.3cm} 420.
&\hspace*{0.4cm} 20. \\
\hline\noalign{\medskip}

Dipole hemisphere magnetic flux (GWb)&\hspace*{0.cm}$\; \; 2\pi B_0 r_P^2$ &
$\quad 7.7$ &$\; \; 13,450$ &$\; \quad 456.$  \\
\hline\noalign{\medskip}

Open field line magnetic flux (GWb)&\hspace*{0.15cm}$\pi B_0 r_P^2\frac {r_P}{R_{ss}}$ &
$\quad 0.42$ &$\quad 450$ &$\; \quad 11.4$  \\
\hline\noalign{\medskip}

Polar oval radius (degrees)&\hspace*{0.cm}$\sin \theta=\sqrt \frac {r_P}{R_{ss}}$
&$\quad 20^\circ$ &$\; \quad 15^\circ$ &$\qquad 13^\circ$  \\
\noalign{\smallskip}
\hline\noalign{\medskip}

Average IMF magnitude (nT) &\hspace*{0.cm}$\frac {5\sqrt{1+r_0^2}}{\sqrt{2}\cdot r_0^2}$
 &\hspace*{0.4cm}
 5 &\hspace*{0.45cm}1&\hspace*{0.4cm} 0.5 \\
 \noalign{\smallskip}
\hline\noalign{\medskip}

Nominal Parker spiral angle $\phi $ &$\tan\phi =\frac{1}{r_0}$ &
$\; \quad 45^\circ$ &$\; \quad 80^\circ$ &$\; \; \quad 85^\circ$  \\
\noalign{\smallskip}
\hline\noalign{\medskip}

Solar wind ram pressure (nPa) &\hspace*{0.2cm} $\frac {1.7}{ r_0^2}$ &
\hspace*{0.4cm}1.7&\hspace*{0.35cm}0.07&\hspace*{0.3cm}0.015 \\
\hline\noalign{\medskip}

Subsolar magnetic field magnitude (nT) &\hspace*{0.3cm}$B_{ss}=\frac {74.5}{r_0}$
 &\hspace*{0.3cm}
 74.5&\hspace*{0.45cm}14.3&\hspace*{0.4cm} 7.8 \\
\hline\noalign{\medskip}

Subsolar magnetopause distance ($R_P$)&\hspace*{0.4cm} $ R_{ss}$&
\hspace*{0.4cm}$10\; R_E$&
\hspace*{0.4cm} $70\; R_J$&\hspace*{0.3cm}$22\; R_S$  \\
\hline\noalign{\medskip}

\end{tabular}
\end{center}
\end{table}
   
\subsubsection{Energy Deposition of Precipitating Auroral Particles}
  \label{Fsubsubsec:Energy Deposition}
  
 \paragraph{Energetic Electrons}

The incident auroral electron characteristics derived from the spectroscopic analysis of the ultraviolet auroral emissions  have been summarized for the main auroral ovals of Jupiter and Saturn in \citep{Gerard-et-al-2013}.
The two sets of values in the corresponding Table of \citet{Gerard-et-al-2013} correspond to two different atmospheric models used for the analysis.  Energetic electrons interact with the atmospheric neutrals through elastic scattering and inelastic collisions, the latter including ionization, excitation, dissociation or a combination of them. Ionization yields the production of secondary electrons, which can in their turn interact with the atmosphere. Furthermore, suprathermal electrons interact with the thermal, ionospheric electrons through Coulomb collisions. This yields an increase in the ionospheric electron temperature  
\citep{Grodent-et-al-2001,Galand-et-al-2011}.

A major, additional source of energy originates in the high latitude regions, where magnetospheric currents can deposit globally several tens of TW in Saturn case, more than 50 times the absorbed solar EUV value, as thermal energy, primarily via Joule heating 
\citep{Cowley-et-al-2004a}. The assessment of Joule heating, plasma flows, and current system at high latitudes, all requires the knowledge of the ionospheric state, and in particular electrical conductances in the auroral regions 
[e.g., \citet{Cowley-et-al-2004b} and \citet{Cowley-et-al-2008}].

As a result of the interaction with the atmospheric species, the suprathermal electrons undergo degradation in energy and redistribution in pitch angle (the angle between the electron velocity and the local magnetic field). As the energy loss is a function of the electron energy, and secondary electrons are added towards lower energies, the initial electron energy distribution at the top of the atmosphere changes, as the electrons penetrate deeper in the atmosphere. The calculation of the distribution of electrons in both position and velocity space is required. Three approaches have been applied to auroral electrons at Jupiter and Saturn, all assuming steady-state conditions and the guiding center approximation 
\citep{Rees-1989}:

 a) The ``Continuous Slowing Down Approximation'' (CSDA) method assumes that the energy loss is a continuous rather than a discrete process 
 \citep{Gerard-Singh-1982, Singhal-et-al-1992, Rego-et-al-1994, Prange-et-al-1995, Dalgarno-et-al-1999}. 
The variation $dE$ in electron energy per path length $ds$ in an atmosphere composed of species $k$ with neutral density $n_k$ and energy loss $L_k$ is given by:
\begin{equation}
 \label{ion1}
\frac{dE}{ds}=-\Sigma_k n_k(s) L_k(E)
\end{equation}
The method, simple to implement, requires - in order to be able to integrate Eq. (\ref{ion1}) - that either the atmospheric composition is independent of altitude (e.g., \citet{Dalgarno-et-al-1999})
 or that atmospheric species have energy losses proportional to each other [e.g., 
 \citet{Rego-et-al-1994}]. 
 The method is limited to high energies where the assumption of a continuous loss is justified and scattering is neglected. The CSDA method allows the calculation of the profiles in altitude of ionization and excitation rates. 

 b) An alternative method is to utilize transport models based on the explicit, direct solution of the Boltzmann equation, which can use a two-stream approach (up/down)
  \citep{Waite-et-al-1983, Achilleos-et-al-1998, Grodent-et-al-2001, Gustin-et-al-2009} 
or multistream approach (more than two pitch angles considered) \citep{Kim-et-al-1992, Perry-et-al-1999, Menager-et-al-2010, Galand-et-al-2011}. 
The Boltzmann equation expresses the conservation of the number of particles in the phase space, as given by:

\begin{equation} \label{ion2}
\frac{df}{dt}+f\nabla_u\cdot\frac{F}{m}=\left(\frac{\delta f}{\delta t} \right)_{coll}+S_{ext}
\end{equation}
where $f (r, u, t)$ is the suprathermal electron distribution at position $r$, velocity $u$ and time 
$t$. The second term on the LHS takes into account the effect of any dissipative forces $F$. The first term on the RHS represents variation due to collisions and the second term is associated with external sources (e.g., photoelectrons, secondary electrons from an ion beam).

The Boltzmann equation is solved in terms of the suprathermal electron intensity 
($I_e =\frac{u^2}{m} f $), which is a measurable quantity. The phase space is usually reduced to three dimensions, path length $s$ along the magnetic field line, kinetic energy $E$, and cosine $\mu$ of the pitch angle $\theta$. Scattering is included. 
Beside ionization, excitation, and dissociation rates this method allows the calculation of thermal electron heating rates.

c) Monte Carlo simulations refer to a stochastic method based on the collision-by-collision algorithm 
\citep{angeo-26-77-2008, Gerard-et-al-2009, Tao-et-al-2011}. A large number of particles is considered and followed in the simulated atmosphere. The Monte Carlo approach avoids the use of an energy grid, which can be of great interest for problems with electron energies ranging over five orders of magnitude. Its drawback is that it is computationally expensive, since it requires a large number of particles to reduce the statistical noise. At Jupiter and Saturn, only excitation, ionization and dissociation processes have been included; thermal electron heating, which is efficient at low energies ($<1\, eV$), has not been considered. 

Suprathermal electron transport models are driven by the electron intensity at the top of the atmosphere, which is a function of energy and pitch angle. For Jupiter and Saturn, it is shown that Joule heating and frictional effects, due to ion-neutral coupling can produce large amounts of energy that may account for their high exospheric temperatures \citep{Miller-2005}. One key question for all studies of the upper atmospheres of the giant planets is that measured exospheric temperatures are several hundred degrees higher than can be produced by the effects of solar EUV heating alone \citep{Yelle-Miller-2004}.

%%%%%%%%%%%%%%%%%%
\newpage
%%%%%%%%%%%%%%%%%%
\section{Molecular clouds}
\label{mc}
%%%%%%%%%%%%%%%%%%

\subsection{Introduction} \label{Jsec:intro}

Molecular clouds (MCs) are the densest and coldest structures in the
interstellar medium (ISM) of our and other galaxies, and the sites of
most, if not all, star formation (SF) in these systems. There, the ionization
fraction is $\sim 10^{-7}$ or lower \citep[e.g.,] [Ch.\ 27] {Shu92}, and
so ambipolar diffusion (AD, often also referred to as ion-neutral drift)
has the potential to be important in various aspects of MC dynamics. In
particular, AD was central to the so-called ``standard model'' of
magnetically-regulated, AD-mediated star formation \citep[see the
reviews of] [and references therein; see also Sec.\ \ref{Jsec:std_mod}
below] {Shu+87, Mouschovias91}. It has also been considered as a
possible mechanism responsible for a number of observed properties in
the ISM and the MCs of our galaxy, and in this chapter we will review
the most frequently discussed ones among these applications. However, in
most cases, other mechanisms have also been proposed as being
responsible for those oberved properties, and so we will present those
alternative processes here as well, hopefully providing a broad and
unbiased view. Specifically, we will discuss the formation and collapse
of dense cores in MCs (Sec.\ \ref{Jsec:cores_n_collapse}), the observed
correlation (and lack thereof) between the density and magnetic fields
in the ISM, the formation of MC cores, the origin of the nearly constant
width of filamentary structures withn MCs, and the resolution of the
so-called ``magnetic braking catastrophe'' in protostellar accretion
disks. However, because most of these processes are related to the
underlying assumptions for the structure and state of MCs, we start with
a brief recount of the evolution of our view of MCs.  A more detailed
account up to the ``turbulent support'' picture can be found in
\cite{MK04}.

\subsection{The Evolving Picture of MCs} \label{Jsec:MC_pic_evol}

\subsubsection{The early scenario of collapse and its demise}

MCs have been known to contain highly supersonic nonthermal motions for
over 40 years now, after the discovery that their molecular-line
emission exhibits widths corresponding to velocities up to one order of
magnitude larger than their thermal speeds \citep{Wilson+70}.
Originally, such motions were interpreted as a possible signature of
global gravitational contraction \citep{GK74}, but it was soon argued
that, if this were the case, then the global SF rate (SFR) of The Galaxy
would be much larger than observed \citep{ZP74}. Indeed, assuming that
all molecular gas in The Galaxy (with a mass of $10^9 \Msun$, where
$\Msun \approx 2\times 10^{30}$ kg is the mass of the Sun), at a mean
number density of $n \sim 10^8 \pcm$ ($=10^2 \pcc$, corresponding to a
mass density $\rho = 3.94 \times 10^{-19} \kpcm$, for a mean particle
mass $\mu = 2.36$), is collapsing on its free-fall time ($\tff =\sqrt{3
  \pi/32 G \rho} \approx 3.35$ Myr $\approx 1.06 \times 10^{14}$ s),
then the SFR would be $\sim 300\, \Msun {\rm yr}^{-1}$. Instead, the
observed SFR is currently estimated to be roughly two orders of
magnitude smaller \citep[see, e.g.,] [for a current estimate] {CP11}.
Moreover, \cite{ZE74} suggested that, if MCs were dominated by global
collapse (or, in general, any cloud-scale velocity gradient), such that
the envelopes of clouds were infalling onto their centers, then one
should observe systematic Doppler shifts between lines produced at the
periphery of clouds and the lines produced near the center of the clouds
around the newly-formed stars, an effect that is not observed.
Therefore, \cite{ZE74} concluded that, instead of consisting of global,
cloud-scale motions, the observed supersonic linewidths corresponded to
small-scale turbulent motions.

\subsubsection{The magnetic and turbulent support scenarios}

Shortly thereafter, these supersonic motions were reinterpreted as MHD
waves \citep{AM75}, in particular as Alfv\'en waves, which were thought
to be less dissipative than the shock waves that would be produced by
hydrodynamic supersonic turbulence, and thus to allow the motions to
persist over the whole lifetmes of the clouds. At the time, these were
estimated to be of the order of a few times $10^7$ yr  (recall that 1 yr
$\approx  3.15 \times 10^7$ s) \citep{BS80}.
Moreover, the magnetic field was believed to be sufficiently strong to
prevent the global collapse of the clouds. This scenario, to which we
will refer as the ``magnetic support'' model of MCs, was reinforced by
the observation of significant magnetic fields in MCs \citep [see] [and
references therein] {MG88}, and prevailed until the late 1990s, when
numerical simulations of supersonic MHD turbulence became feasible, and
showed that MHD turbulence dissipates just as rapidly as pure
hydrodynamic turbulence \citep{ML+98, Stone+98, PN99, AV01}, except
possibly if the energy flux up and down magnetic field lines is
imbalanced \citep{Cho+02}. The supersonic nonthermal motions were then
reinterpreted as genuine supersonic MHD turbulence. Moreover, it was
also noted that giant molecular clouds (GMCs) exhibit near equipartition
between their nonthermal kinetic and gravitational energies
\citep{Larson81, Heyer+09}. This fact was then interpreted as evidence
of near virial equilibrium in the clouds between their self-gravity and
turbulence, although the latter required continuous driving in order to
be maintained \citep[see, e.g.,] [] {VS+00, MK04, ES04, BP+07, MO07}.
The prime candidate for injection of kinetic energy into the ISM and MCs
in particular was feedback from supernova explosions and HII regions
around massive stars, although other mechanisms such as
magnetorotational instabilities, protostellar outflows, massive-star
winds, the passage of spiral density waves, etc., were also considered
\citep[see the review by] [and references therein] {MK04}.  We will
refer to this view as the ``turbulent support'' model of MCs.

\subsubsection{The hierarchical gravitational collapse scenario}

However, in the last decade or so, the turbulent-support paradigm has
been questioned by a number of authors, on the following accounts.
First, it has been pointed out that the near-equipartition may be a
consequence of unimpeded gravitational collapse rather than virial
equilibrium \citep{BP+11a, Dobbs+14, Palau+15}. Second, in the
virial-equilibrium interpretation, stellar-driven turbulence would
provide the counteracting agent against the clouds' self-gravity,
through a hypothetical turbulent pressure. However, this interpretation
faces some problems: i) The near-equipartition is observed even
in clouds that have little or no ongoing star formation \citep{MT85}, so
that the source of their kinetic energy cannot be the feedback from
stellar sources. ii) There appears to be no {\it a priori} reason why the
energy injection by stellar sources would adjust itself to maintain the
clouds in near-equilibrium. Although some idealized models of energy
balance have been produced in which the feedback self-regulates to do
this \citep [e.g.,] [] {Krumholz+06, Goldbaum+11}, full numerical
simulations including ionization feedback suggest that, once the star
formation activity fully develops in the cloud, the latter is disrupted
\citep [partly evaporated, partly torn apart;] [] {VS+10, Dale+12,
  Colin+13}, at least for clouds of masses $< 10^6 \Msun$. Moreover,
\cite{Colin+13} found that the clouds appear in near equipartition {\it
  before} SF has fully developed in the clouds, and attributed the
equipartition to the clouds' gravitational contraction instead. Thus,
stellar driving of a stabilizing turbulence seems to face significant
problems. 

Another mechanism that has been advocated as the driver of MC
turbulence is a combination of Kelvin-Helmholz, Rayleigh-Taylor,
thermal, and nonlinear thin-shell instabilities in the clouds themselves
as they are assembled by the convergence of streams in the warm atomic
medium of The Galaxy \citep{KI02, AH05, Heitsch+05, VS+06, Heitsch+06,
  KH10}. However, numerical simulations of the process have repeatedly
shown that the Mach numbers produced in the clouds by this mechanism
fall short of the values observed in MCs \citep{KI02, Heitsch+05,
  VS+06}. Furthermore, simulations including self-gravity show that the
near-equipartition is only established once gravitational contraction has
engaged in the clouds, while at earlier times the turbulent kinetic
energy dominates over the gravitational energy \citep{VS+07, Colin+13}.

In consequence, a return to the globally-collapsing scenario has been
advocated by some authors, albeit with the novel feature that the
collapse should be hierarchical and chaotic rather than monolithic
\citep[e.g.,] [] {BH04, VS+07, Peretto+07, HB07, HH08, Heitsch+08,
  VS+09, GM+09, Schneider+10, BP+11a, BP+11b, Peretto+13}. That is, the
assembly of the clouds by the collision of streams of diffuse gas
produces, via the thermal instability, large masses of cold gas that
contain many Jeans masses. Also, the turbulence produced by the
collision in turn produces nonlinear density fluctuations, which have shorter
free-fall times than the average in the cloud \citep{HH08, VS+09}. Thus,
the collapse occurs in a multi-scale fashion, with the collapse of the
small-scale fluctuations preceding that of the whole cloud, as
expected from linear theory. 

Molecular clouds in this scenario are then in a hierarchical state of
collapse, and their nonthermal motions are dominated by collapses within
collapses, rather than by actual random turbulence. The latter rather
applies to the large-scale ISM, in which the MCs constitute the sites of
collapse. The MCs also contain tuurbulence, but it is relatively weak,
and not coupled to the density or size scales, its only role being to
provide the seeds for the hierarchical collapse, but not to provide any
support against the collapse. The solution to the SFR conundrum of
\cite{ZP74} is accomplished by early destruction of the clouds by the
stellar feedback, before more than $\sim 10\%$ of their mass has been
converted to stars \citep{VS+10, Dale+12, ZA+12, Colin+13, ZV14}. We
will refer to this as the ``hierarchical gravitational collapse''
scenario of MCs. At the time of this writing, the turbulent-support
scenario is still the most frequently discussed one in the literature,
although the hierarchical collapse model is also discussed, in
particular for addressing evolutionary features of MCs \cite{ZA+12, ZV14}.

On the basis of this changing picture of MCs we now describe the role
that AD has played as a proposed solution to various features and
properties of MCs, together with the alternative mechanisms that have
been proposed for accomplishing the same results, thus hoping to provide
the reader with a complete and unbiased view of the current status of
the field.

\subsection{The Formation and Collapse of Molecular Cloud Cores}
\label{Jsec:cores_n_collapse} 

As mentioned in Sec.\ \ref{Jsec:MC_pic_evol}, GMCs have typical average
mass densities $\rho \sim 4 \times 10^{19} \kpcm$ (molecule number
densities $\sim 100 \pcc$). Essentially all SF in The Galaxy occurs in
the interiors of GMCs, although not throughout their volume, but rather only
in their densest regions, called ``dense cores'', with typical
densities $\rho \gtrsim 4 \times 10^{21} \kpcm$ (number densities $n \gtrsim
10^4 \pcc$).  The origin and collapse of the dense cores is
therefore one of the key issues that needs to be accounted for by the
various models for MCs.

\subsubsection{The standard model} \label{Jsec:std_mod}

In the magnetic support model, the fundamental underlying quantity was
the ratio between the self-gravitational ($W$) and the magnetic ($\Em$)
energies of a certain parcel of the fluid. Assuming, for simplicity, a
spherical volume with a uniform density and magnetic fields inside, and
neglecting them outside, we can write\footnote{In this section, for
  consistency with the most frequent convention in the field, we use cgs
units.}
\beq
\frac {|W|} {\Em} = \frac {\eta G M^2 R^{-1}} {\frac{1}{8 \pi}\int_V
  B^2\, dV},
\label{Jeq:Em_to_W_ratio}
\eeq
where $\eta = 3/5$ for the spherical, uniform region, but in general it
is a factor of order unity accounting for the shape and mass
distribution of volume $V$. For uniform density and magnetic fields,
this then reduces to
\beq
\frac {|W|} {\Em} = \frac {6 \eta G M^2} {R^4 B^2}.
\label{Jeq:Em_to_W_ratio_v2}
\eeq
Defining the magnetic flux across a cross-sectional area $A$ across
volume $V$ as
\beq
\Phi \equiv \int_A \boldB \cdot \nhat\, d A,
\label{Jeq:mag_flux}
\eeq
where $\nhat$ is the unit vector perpendicular to $d \boldA$, eq.\
(\ref{Jeq:Em_to_W_ratio_v2}) becomes
\beq
\frac {|W|} {\Em} = \frac {6 \pi^2 \eta G M^2} {\Phi^2}
\label{Jeq:Em_to_W_ratio_v3}
\eeq
for our uniform, spherical gas cloud. When this ratio equals unity, the
cloud is in balance between its self-gravity and the magnetic pressure
gradient (with respect to the exterior), and so we define the ``critical''
value of the so-called mass-to-magnetic flux ratio as this equilibrium
value: 
\beq
\left(\frac {M} {\Phi} \right)_{\rm cr} \equiv \frac {1} {\sqrt{6 \pi^2
    \eta G}}.
\label{Jeq:mfr_cr}
\eeq
In what follows, we denote the mass-to-flux ratio (M2FR), normalized to this
critical value, by $\mu$.

An important property of the M2FR is that it does not
depend on the size of the cloud, but only on its mass and magnetic flux.
For Lagrangian fluid parcels (following the mass) and under ideal MHD
conditions, this quantity is then invariant, in particular upon
compressions, meaning that no amount of external compression may cause
the self-gravity of the cloud to dominate over the magnetic support and
induce its collapse if the cloud's M2FR is smaller than
the critical value ($\mu < 1$). These clouds are termed {\it
  magnetically subcritical}. Conversely, clouds with $\mu >1$ cannot be
supported against their self-gravity by the magnetic pressure alone.
Such clouds are called {\it magnetically supercritical}. It is important
to note that, in general, the energy balance must take into account that
the magnetic support is at least partially counteracted by the surface
magnetic term arising in the virial theorem \citep[see the discussions
in] [] {MZ92, Shu92, VS99, BP+99}. In particular, for a uniform magnetic
field, the surface term completely cancels the magnetic energy, and so
some amount of gravitational contraction that increases the field
strength inside the cloud is required before magnetic support can
operate on a subcritical cloud, but eventually it will, halting the
contraction.

In the magnetic-support model, then, clouds were assumed to be globally
magnetically subcritical, and thus absolutely supported against their
self-gravity as long as the ideal MHD regime is applicable. In order for
{\it some} fraction of the mass to undergo gravitational collapse, in
this model it is necessary for that material to lose magnetic support.
This can be accomplished by ambipolar diffusion (AD), a process by which
the neutrals, driven by self-gravity, drift away from the ions,
effectively losing magnetic flux, which remains anchored on the ions.
Eventually, the M2FR of this neutral mass may become larger than unity
and proceed to dynamical collapse. We now revisit the standard
single-flud approximation of the process, closely following the
discussion by \cite{Shu+87}, which has been employed heavily in the
description of MCs.

\subsubsection{The single-fluid approximation for ambipolar diffusion} 
\label{Jsec:single_fluid_AD}

The main process that needs to be described is the drift between
neutrals and ions in a partionally ionized plasma. To do this, we
consider the drag force between the two types of particles, given by 
\begin{equation}
\fd = \rhon \rhoi \gamma (\ui - \un),
\label{Jeq:drag_for}
\end{equation}
where $\rhoi$ and $\rhon$ are respectively the ion and neutral
densities, $\ui$ and $\un$ are their velocities, and $\gamma$ is the
drag coefficient (damping rate) associated with momentum exchange in
ion-neutral collisions.  When the collision process is dominated by the
dipole moment induced in the neutral by the passing ion, this
coefficient takes the value $\gamma \approx 3.5 \times 10^{13}$ cm$^3$
g$^{-1}$ s$^{-1}$ \citep{Draine+83}.

The ion-neutral drift is induced by the Lorentz force, and, at terminal
drift velocity, the two forces balance. We thus have, for the drift
velocity, 
\begin{equation}
\vd \equiv \ui - \un = \frac{1} {4 \pi \rhon \rhoi \gamma} \left(\nabla
  \times \BB \right)\times \BB.
\label{Jeq:drift_vel}
\end{equation}

Now, for MC conditions, the cyclotron frequency $eB/m_{\rm i}c$ of the
typical ion (an electron) is much greater than the collision frequency,
and thus the magnetic field can be considered to be frozen in the ions,
because they gyrate many times around the field lines before being
knocked off by a collision. In this case, we have the field-freezing
condition for the ions:
\begin{equation}
\frac{\partial \BB} {\partial t} + \nabla \times \left(\BB \times \ui
\right) = 0.
\label{Jeq:ff_ions}
\end{equation}
Substituting eq.\ (\ref{Jeq:drift_vel}) into eq.\ (\ref{Jeq:ff_ions}) we obtain
\begin{equation}
\frac{\partial \BB} {\partial t} + \nabla \times \left(\BB \times \un
\right) = \nabla \times \left\{\frac {\BB} {4 \pi \rhon \rhoi \gamma}
\times \left[\BB \times \left(\nabla \times \BB \right) \right]\right\}.
\label{Jeq:flux_AD}
\end{equation}
Equation (\ref{Jeq:flux_AD}) has the dimensions of a
diffusion equation, of the form
\begin{equation}
\frac{B} {\tad} \sim {\cal D} \frac{B} {L^2},
\label{Jeq:ord_mag_diff}
\end{equation}
where $\tad$ is the characteristic AD timescale, $L$ is the
characteristic length scale of the system, and ${\cal D}$ is the
associated ``diffusion coefficient''. Comparing expression
(\ref{Jeq:ord_mag_diff}) with eq.\ (\ref{Jeq:flux_AD}), we see that
\begin{equation}
{\cal D} \sim \va^2 t_{\rm ni}, 
\label{Jeq:AD_diff_coef}
\end{equation}
where
$\va = B/\sqrt{4 \pi \rhon}$ is the Alfv\'en speed in the medium
(assuming $\rhon \gg \rhoi$) and 
\begin{equation}
t_{\rm ni} \equiv (\gamma \rhoi)^{-1}
\label{Jeq:tni}
\end{equation}
is the mean collision time with an ion per neutral particle. 
On the other hand, for the AD timescale, from expression
(\ref{Jeq:ord_mag_diff}) we have 
\beq
\tad \sim L^2/{\cal D}.
\label{Jeq:tau_AD}
\eeq

It is interesting to compare the AD timescale with the dynamical time,
$\tdyn$. For a magnetized medium in which the information-transmission
speed is $\sim \va$, we have $\tdyn \sim L/\va$. In addition, if
the medium is in approximate equilibrium between magnetic support and
its self-gravity, then the dynamical time should also be of the order of
the free-fall time, so that we also have $\tdyn \sim (G \rhon)^{-1/2}$.
Combining these relations, and noting that the ion and neutral densities
are related to first order by 
\begin{equation}
\rhoi \approx C \rhon^{1/2}, 
\label{Jeq:rhoi_vs_rhon}
\end{equation}
with $C \approx 3 \times 10^{-16}$ cm$^{-3/2}$ g$^{1/2}$ \cite{Elm79}, we find
\begin{equation}
\frac {\tad} {\tdyn} \sim \frac {\gamma C} {G^{1/2}} \approx 40.
\label{Jeq:tad2tdyn}
\end{equation}
This is only an order-of-magnitude estimate, but more precise
calculations \citep[e.g.,] [] {Shu83} still give values $\sim 10$ for
the ratio $\tad/\tdyn$, implying that AD operates on timescales much
longer than the free-fall time. This, together with the fact that only
the densest material could proceed to collapse, was the reason, in the
magnetic-support model, for the low efficiency of star formation:
low-mass star-forming clouds were believed to have subcritical M2FRs in
general, and therefore, to be globally supported against collapse.
Gravitational contraction would only proceed in the densest regions
(cores) as allowed by AD on its characteristic timescale, and thus much
more slowly than in free-fall, and involving only a small fraction of
the total mass in the MCs. Massive-star-forming clouds, which were then
thought to be exceptional, were believed to be the few instances of
magnetically supercritical clouds, and undergoing unimpeded collapse.

\subsubsection{Problems with the magnetic-support model}

In recent years, it has come to be realized that there are
several problems with the magnetic-support model. We now briefly discuss
some of these caveats. A complementary discussion can be found in 
\cite{MK04}. 

\subsubsection{Discrepancy with observed timescales}
\label{Jsec:timescales}

As discussed in Sec.\ \ref{Jsec:single_fluid_AD} (cf.\ eq.\
[\ref{Jeq:tad2tdyn}]), the AD timescale was originally estimated to be
at least one order of magnitude longer than the free-fall time of dense
clumps. Since, in the magnetic-support model, this slow-contraction
stage corresponded to the prestellar stage of a dense core, this implied
that the typical duration of the ``starless'' stage of a contracting
core should be much longer (roughly by one order of magnitude) than the
duration of the proto-stellar stage (the stage where a collapsed object
--- a protostar --- has already appeared). However, systematic surveys
of dense cores \cite[e.g.,] [] {LM99, Jijina+99} were able to estimate
the ratio of the durations of these two stages by measuring the ratio of
the number of starless to protostellar cores. The observed ratios turned
out to be $\sim 0.05$--0.67, indicating that the starless stage was
actually {\it shorter} than the protostellar one, and shorter than the
prediction from AD models by factors $\sim 2$--50 \cite{LM99}.

As a solution to this conundrum, \cite{CB01} noted that the AD timescale
becomes shorter as the initial M2FR of a core is closer to the critical
value, so that the measurements of nearly critical values of the M2FR in
dense cores would imply contraction times in the prestellar cores
shorter by up to an order of magnitude than the ``classical'' AD
timescale, bringing the revised AD timescale into agreement with the
observed duration of the prestellar stage of the cores. Similarly,
\cite{FA02} considered the fluctuations in the AD timescale caused by
local fluctuations of the magnetic and density fields, concluding that
the ``effective'' AD timescale would be shortened by roughly the same
factor. 

However, since the revision to the AD timescale brings it
close to the free-fall time, effectively no delay is introduced in the
collapse timescale by the AD-mediated contraction, reducing the role of
the magnetic field to providing support to the envelopes of the dense
cores. But then, as discussed in Sec.\ \ref{Jsec:no_subcrit}, the current
state of affairs is that subcritical dense cores are not observed, and
in this case the envelopes are not supported by the magnetic field either.

\subsubsection{The mass-to-flux ratio as a boundary condition}
\label{Jsec:M2FR_as_BC}

\begin{figure}
\begin{center}
\includegraphics[scale=.4]{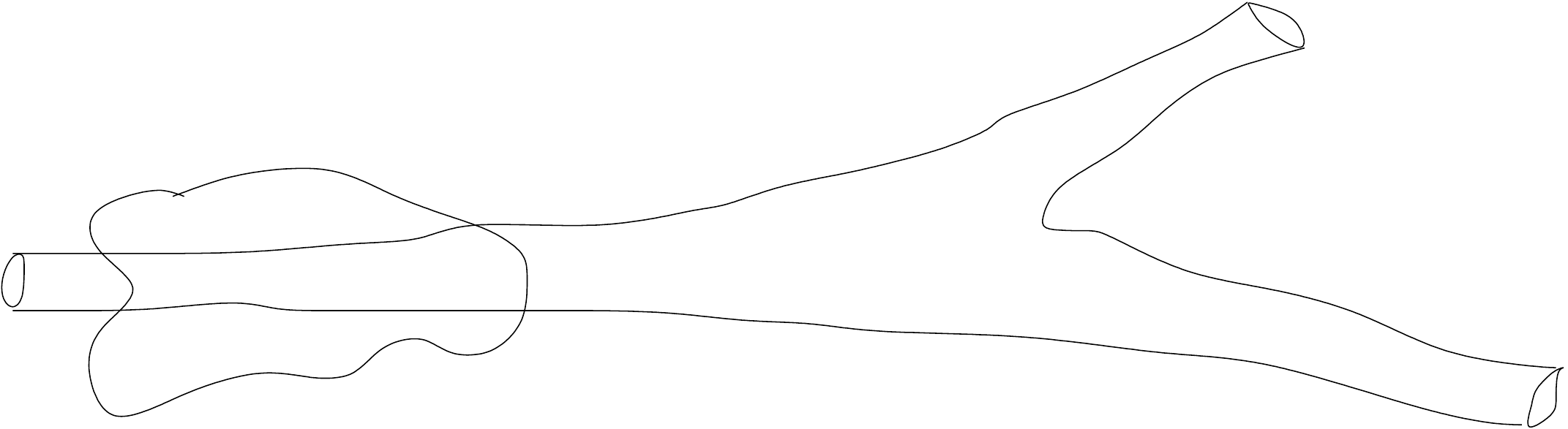}
\end{center}
\caption{Schematic illustration of a flux tube that loses its identity
  far from a cloud.}
\label{Jfig:div_flux_tubes}
\end{figure}

Another important problem is that, due to the condition $\nabla \cdot
\boldB =0$, magnetic field lines do not end, and, as a consequence, flux
tubes in general extend indefintely beyond the boundaries of any cloud
that may have been defined by some ad hoc criterion, such as, for
example, being detected with a certain tracer, or, in numerical
simulations, as a connected region above some density threshold. In
fact, the mean magnetic field in the Galactic plane is roughly azimuthal
\citep[e.g., ] [] {Han13, Beck15}, and therefore flux tubes might in
principle extend for hundreds or thousands of parsecs. In reality, flux
tubes may lose their identities as the field lines that define them
drift apart from each other, as schematically illustrated in Fig.\
\ref{Jfig:div_flux_tubes}.  Nevertheless, it is in general possible in
principle to consider a segment along a flux tube that is long enough 
to be magnetically supercritical.  This critical length is known as the
{\it accumulation length} \citep[see, e.g., ] [] {Mestel85, HBB01,
  VS+11}.  The mass-to-flux ratio of a system then becomes a boundary
condition; that is, it is determined by the size of the system
boundaries relative to the accumulation lentgh.

In early studies of AD-mediated core contraction \citep[e.g., ] []
{Mouschovias76}, the core was often assumed to be bounded by a warm,
tenuous phase that isolated the core and did not contribute to the
gravitational potential. Thus, the M2FR was well defined. However, in
recent years it has become clear that MCs and their substructures may be
accreting in general---the MCs may be accreting from their atomic
envelopes \citep[e.g., ] [] {HP99, BP+99, AH05, AH10, Heitsch+05, HH08,
  VS+06, VS+07, VS+11, Myers09, Molinari+14, Heiner+15}, while
the cores accrete from the filamentary structures in which they are
embedded \citep[e.g., ] [] {Myers09, Schneider+10, Kirk+13, Peretto+13,
  Battersby+14, GV14}. This makes it impossible to uniquely define ``the
core''.

In an attempt to resolve this ambiguity, \cite{VS+11} noted that the
relevant mass for evaluating a physically meaningful mass-to-flux ratio
is the mass that dominates the gravitational potential; that is, the
mass in the densest component, which is increasing due to accretion. For
magnetized structures accreting along field lines, the M2FR is
increasing with time, not because of AD, but through accumulation along
field lines, driven either by large-scale inertial compressions (i.e.,
larger-scale turbulence) or larger-scale gravitational instabilities in
the surroundings of the compact object under consideration.

On the other hand, it should be noted that the notion of accumulation
along field lines has been questioned by a number of authors, on the
basis that the accumulation lengths necessary for producing magnetically
supercritical molecular clouds are too long \citep[on the order of
hundreds of parsecs; see, e.g., ] [] {HBB01, VS+11}, and so turbulent
reconnection diffusion would interfere with the accumulation process
\citep{Lazarian14}, or simply the flows would not have remained coherent
over such lengths \citep[e.g., ] [] {KB15}. However, these arguments may
have overlooked the fact that the flows may orient the magnetic field
lines rather than the other way around \citep [e.g., ] [] {HP00, Han13}
and that converging flows on the scales of kiloparsecs are likely to be
set up in spiral arms. In any case, numerical simulations designed to
test the feasibility of the accumulation mechanism on the several-kpc
scale are necessary.

\subsubsection{ The gradient of mass-to-flux ratio in dense cores}
\label{Jsec:M2FR_vs_n} 

The assembly of dense structures by accumulation of material implies
another important difference with the magnetic-support model. In the
latter, the redistributon of magnetic flux due to AD implies that the
central, densest parts of the cores are those that have already lost
part of their magnetic flux to the envelope, and thus 
are expected to have a {\it larger} value of $\mu$ than the average in
their parent cloud, $\mu_0$. Instead, if the dense structures are
assembled by means of partial compression (i.e., accumulation from
distant regions along field lines), with no action of AD, the central
parts are expected to have a {\it lower} value of $\mu$ than the cloud's
average.  Specifically, the M2FR of a subregion of size $\ell$ of a
cloud of size $R$ and mean M2FR $\muo$ is expected to lie in the range

\begin{equation}
\muo \frac{\ell}{L} \le \mul \le \muo
\label{Jeq:mu_subregion}
\end{equation}

\citep[] [] {VS+05}. This range can be understood as follows: the lower
limit applies for the case when the ``core'' is actually simply a
subregion of the whole cloud of size $L$, with the same density and
magnetic field strength. Since the density and field strength are the
same, the mass of the core simply scales as $(\ell/L)^3$, while the
magnetic flux scales as $(\ell/L)^2$.  Therefore, the MFR of a subregion
of size $\ell$ scales as $(\ell/L)$. Of course, this lower-limit
extreme, corresponding to the case of a ``core'' of the same density and
field strength as the whole cloud, is an idealization, since
observationally such a structure cannot be distinguished from its parent
cloud. Nevertheless, as soon as some compression has taken place, the
core will be observationally distinguishable from the cloud (for
example, by using a tracer that is only excited at the core's density),
and the measurement of the MFR {\it in the core} will be bounded from
below by this limit.

On the other hand, the upper limit corresponds simply to the case where
the entire cloud of size $L$ has been compressed isotropically to a size
$\ell$, since in this case both the mass and the magnetic flux are
conserved, and so is the MFR. Any intermediate case has an intermediate
value of the M2FR given by relation (\ref{Jeq:mu_subregion}). 

The ratio of the central-to-envelope M2FR for four cores was measured by
\cite[] [] {Crutcher+09}, and found to be $<1$, although the
uncertainties were large, and the result was challenged by \cite[] []
{MT10}. Nevertheless, a subsequent Bayesian study by \cite[] []
{Crutcher+10} seemed to confirm the result, suggesting that the
accumulation scenario for cores may indeed be at work in those cores.
These results suggest that the M2FR gradient in dense cores and their
envelopes is at least partially due to mass accumulation by turbulent or
gravitational compression and fragmentation, and not due to the action
of AD (at least in the early stages of the process).

\subsubsection{The observed lack of magnetically subcritical
molecular clouds and cores} \label{Jsec:no_subcrit}

Another reason why the magnetic-support model has lost predominance in
recent years is that detailed statistical analyses of the magnetic field
strengths in clouds and cores, derived from Zeeman observations
\citep{Crutcher+10, Crutcher12}, have shown that a) for densities $n
\gtrsim 300 \pcc$, $B$ scales as $\sim n^{0.65}$, in good agreement with
models of clouds with magnetic fields too weak to support them, so that
they are contracting gravitationally, and dragging the field along with
them because of flux freezing. b) There are almost no observations of clouds
and/or cores that are clearly magnetically subcritical,
suggesting that the clouds start out being already supercritical.

In addition, today we understand that most stars, in particular low-mass
stars, form in clusters together with the high-mass stars \cite{LL03}.
This is in contradiction with the magnetic-support scenario, because
there it was postulated that star formation is bimodal, with low-mass
stars forming in magnetically subcritical clouds, and massive stars and
clusters forming in supercritical clouds. The current evidence suggests
instead that most low-mass stars form in the same environment as massive
stars and clusters, i.e., in supercritical clouds, negating
the bimodal picture of the magnetic-support scenario.

\subsection{(De)correlation between magnetic and density fields}
\label{Jsec:B-rho_corr}

\begin{figure}
\begin{center}
\includegraphics[scale=.25]{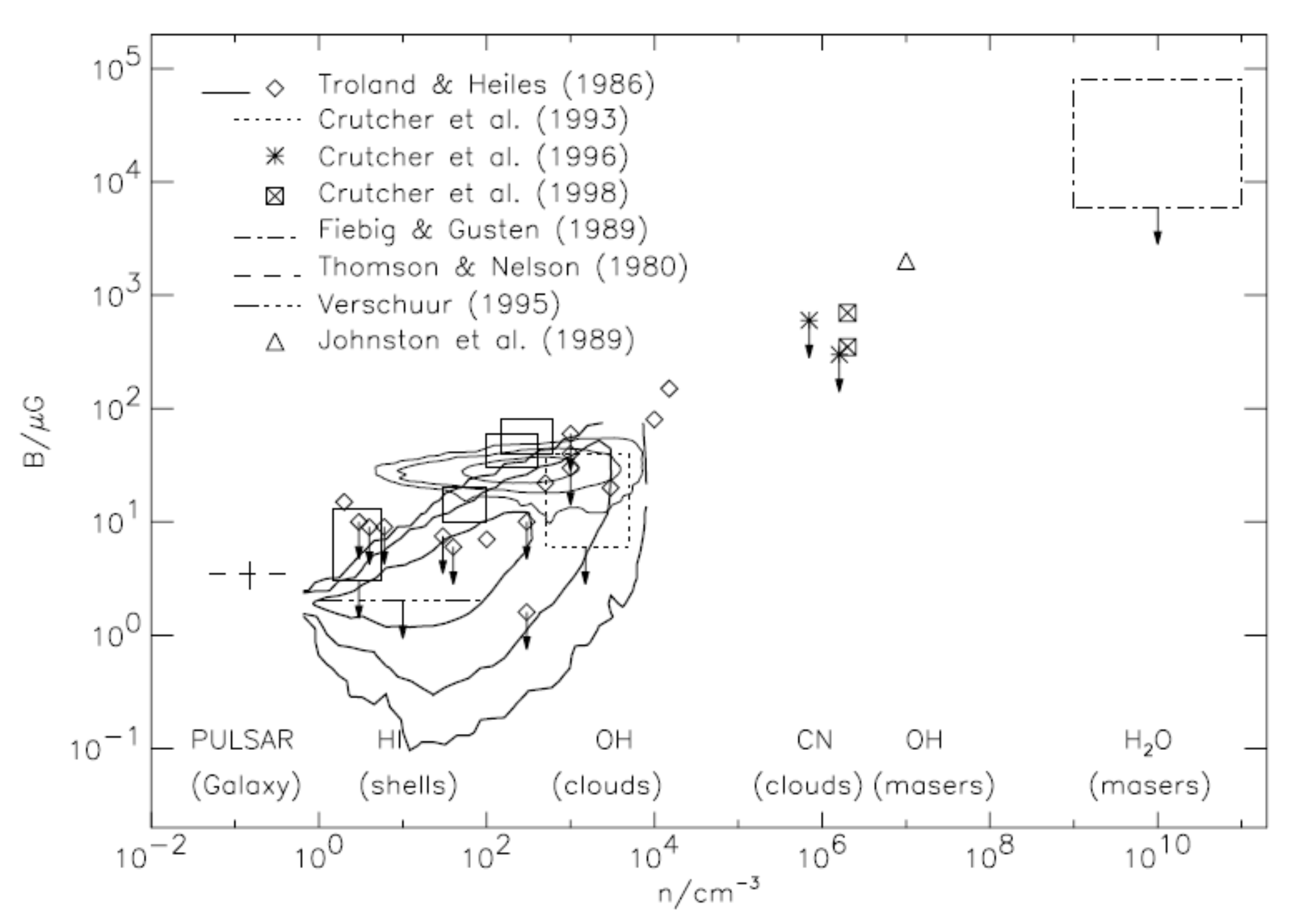}
\includegraphics[scale=.25]{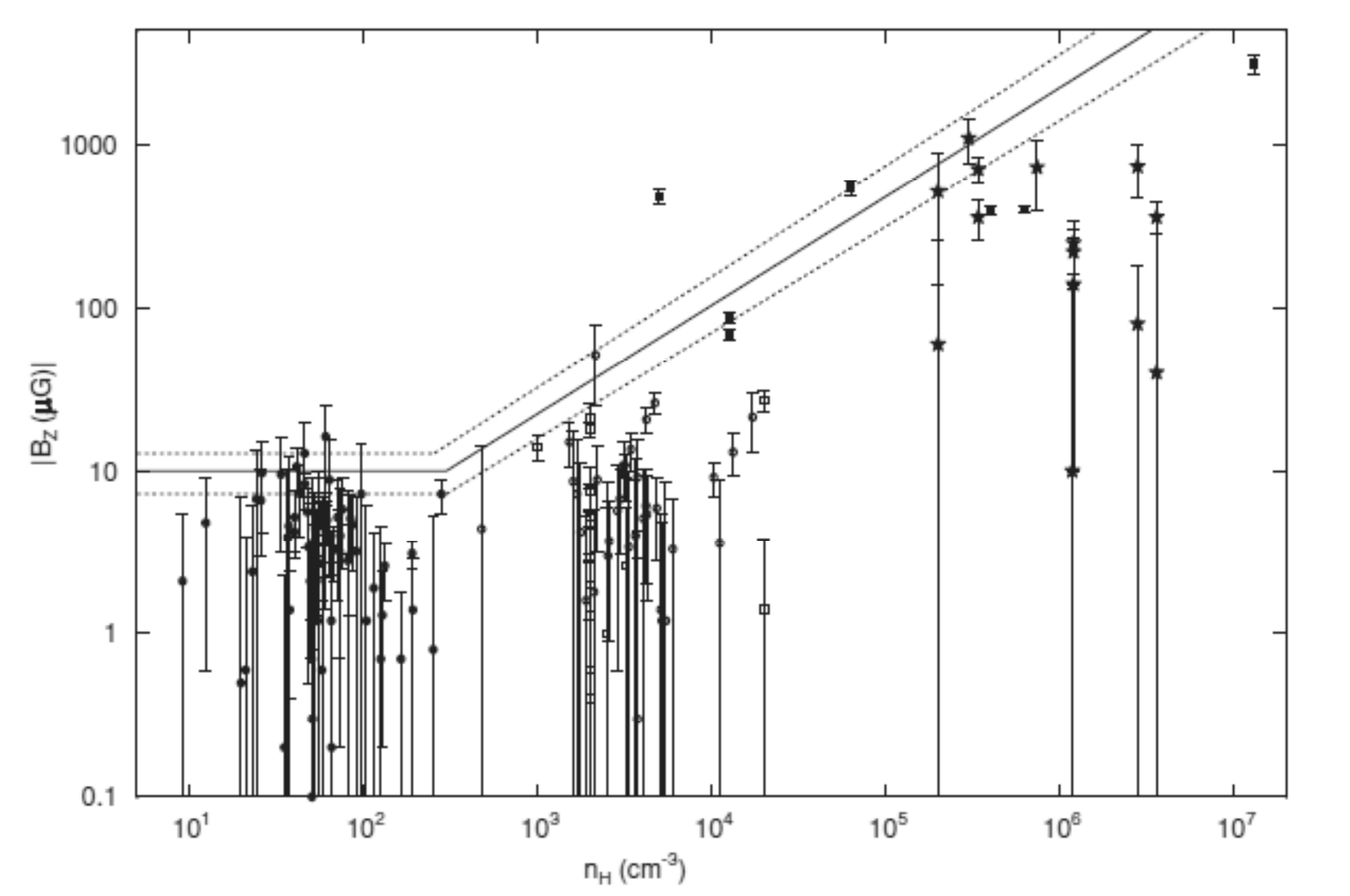}
\end{center}
\caption{{\it Left Panel:} $B$-$n$ correlation in a numerical simulation
  of super-Alfv\'enic turbulence and in several observational surveys
  \citep[from] [] {PN99}. {\it Right panel:} Summary of Zeeman
  determinations of the magnetic field strength {\it versus} density
  $n$, as compiled by \cite{Crutcher+10}.}
\label{Jfig:B-n_corr_PN99}
\end{figure}

In a supersonic, turbulent MHD flow, such as that generally assumed to
permeate MCs, magnetic fluctuations as well as density and velocity
fluctuations are present. Since the field may be capable of providing
support against the self-gravity of the density fluctuations, it is
important to determine the correlation (if any) between the
density and the magnetic fluctuations induced by the turbulence. This
problem was first investigated by several numerical studies \cite [] []
{Passot+95, PN99, Ostriker+99, Ostriker+01}. In general, it was found
that the magnetic field exhibited a {\it range} of values at a given
density, with indications that, at low densities, the magnetic field
tended to approach a constant, minimum value, while at large densities,
it tended to increase with increasing density. Moreover, the
distribution of points in the $B$-$n$ diagram was seen to have a large
scatter, so that a full range of values of the field strength was
observed at every value of the density (Fig.\ \ref{Jfig:B-n_corr_PN99},
left panel). This result is consistent with the observed $B$-$n$
correlation from Zeeman splitting in the diffuse atomic medium and
molecular gas, as reported, for example, by \cite[] [] {Crutcher+10} (Fig.\
\ref{Jfig:B-n_corr_PN99}, {\it right panel}).

Several mechanisms have been proposed to explain this behavior.
\cite{Heitsch+04} suggested that turbulence enhances the diffusion of
the ratio $B/\rho$ through an enhancement of ion-neutral drift at small
scales, referring to this mechanism as {\it turbulent ambipolar
  diffusion}. More recently, \cite[] [see also the review by
\cite{Lazarian14}] {Lazarian+12} have proposed that the decorrelation
between the magnetic and density fields at low densities, as well as
the absence of magnetically subcritical dense cores, is due to {\it
turbulent reconnection diffusion}, a process consisting in the
enhancement of magnetic dissipation by the enhanced turbulent
dissipation (reconnection) induced by turbulence, which brings
magnetic field lines in close contact, fostering reconnection and thus
dissipating magnetic energy at sites of high densities. It is
important to emphasize that turbulent reconnection implies actual
dissipation of magnetic energy through Ohmic resistivity, while
turbulent ambipolar diffusion implies only a spatial redistribution of the
magnetic flux within the core, without actually dissipating
magnetic energy. This happens because, as the neutrals percolate
through the ions, mass drifts towards the core's center without
dragging the flux with it. Thus, effectively the magnetic flux drifts
outward of the core.

Both of the mechanisms proposed above are based on some form of
diffusion of the magnetic flux. It is important to remark, however, that
it is {\it not} necessary to have some form of diffusion in order to
accomplish a decorrelation of the magnetic and the density fields. in
fact, it is common to encounter the erroneous notion that under ideal
MHD conditions the magnetic field should be correlated with density.
However, this is by no means so, and a lack of correlation is expected
in the ideal MHD case as well, because of the possibility of free flow
along field lines. This was shown by \cite[] [hereafter PV03] {PV03}, who
investigated this problem analytically in the isothermal case, by
decomposing the flow into nonlinear, so-called ``simple'' waves
\citep[e.g.,][] {LL59, Mann95}, which are the nonlinear extensions of
the well known linear MHD waves, and have the same three well-known
modes: fast, slow, and Alfv\'en.

For illustrative purposes, note that
compressions along the magnetic field lines are one instance of the {\it
slow} mode, while compressions perpendicular to the field lines (i.e.,
{\it magnetosonic} waves) are an instance of the {\it fast} mode.

PV03 concluded that each of the modes is characterized by a
different scaling between the magnetic pressure ($\propto B^2$) and the
density, as follows:
\begin{eqnarray}
B^2 &\propto c_1 - \beta \rho  & \hbox{~~~~~slow,} \label{Jeq:slow_B_rho}\\
B^2 &\propto \rho^{2}       & \hbox{~~~~~fast,} \label{Jeq:fast_B_rho}\\
B^2 &\propto \rho^{\gamam}  & \hbox{~~~~~Alfv\'en,} \label{Jeq:Alfv_B_rho}
\end{eqnarray}
where $c_1 > 0$ is a constant, and $\gamam$ is a parameter that can take
values in the range (1/2,2) depending on the Alfv\'enic Mach number
\citep[see also][]{MZ95}. Note that eq.\ (\ref{Jeq:slow_B_rho}) implies
that for $\rho > c_1/\beta$ the slow mode disappears \citep{Mann95}, so
that only the fast and Alfv\'en modes remain. Conversely, note that, at
low density, the magnetic pressure due to the fast and Alfv\'en modes
becomes negligible in comparison with that due to the slow mode, which
approaches a constant. This implies that a log-log plot of $B$ vs.\
$\rho$ will exhibit an essentially constant value of $B$ at very small
values of the density. In other words, {\it at
low values of the density, the domination of the slow mode implies that
the magnetic field exhibits essentially no correlation with the density.}

\begin{figure}
\begin{center}
\includegraphics[scale=.3]{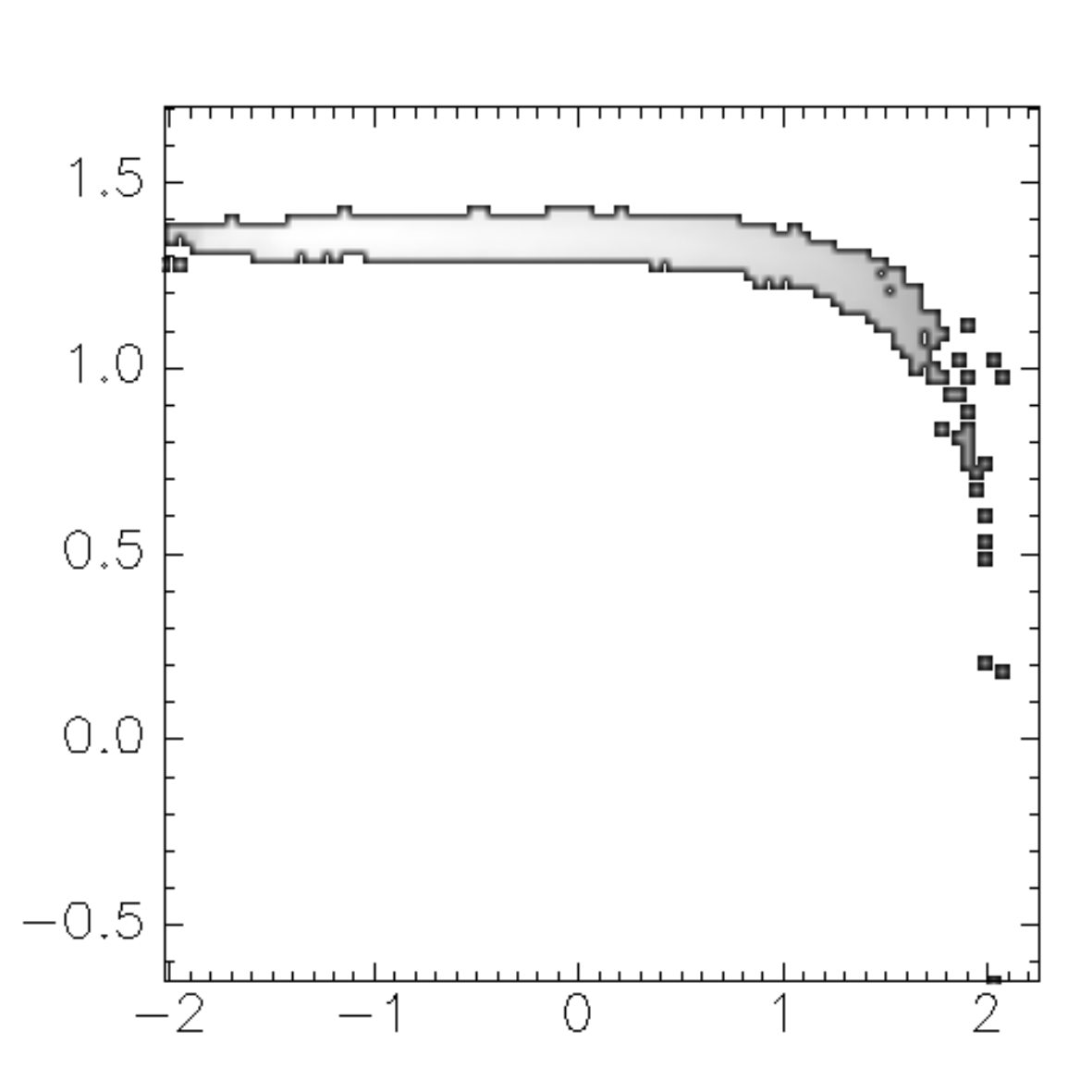}
\includegraphics[scale=.3]{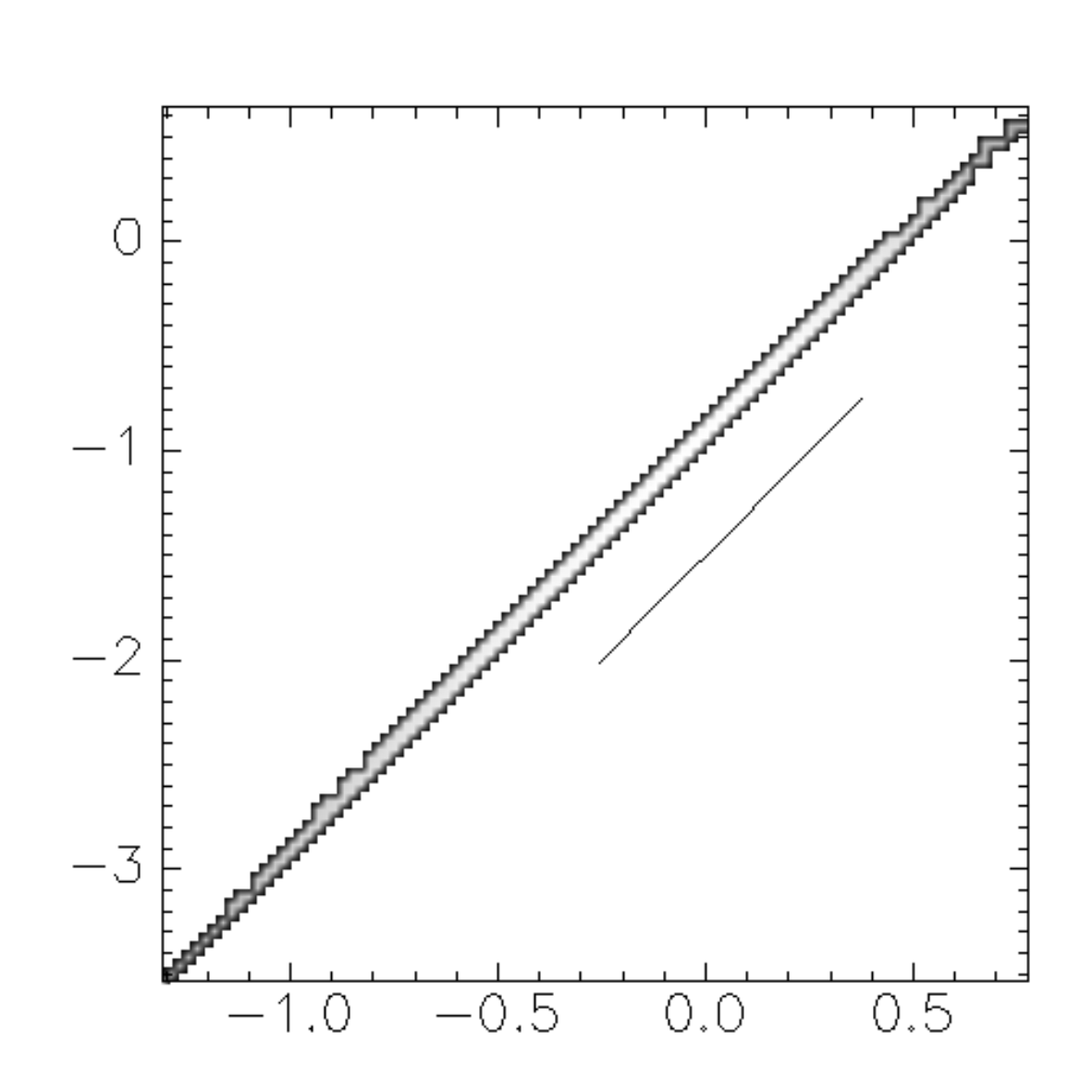}
\includegraphics[scale=.3]{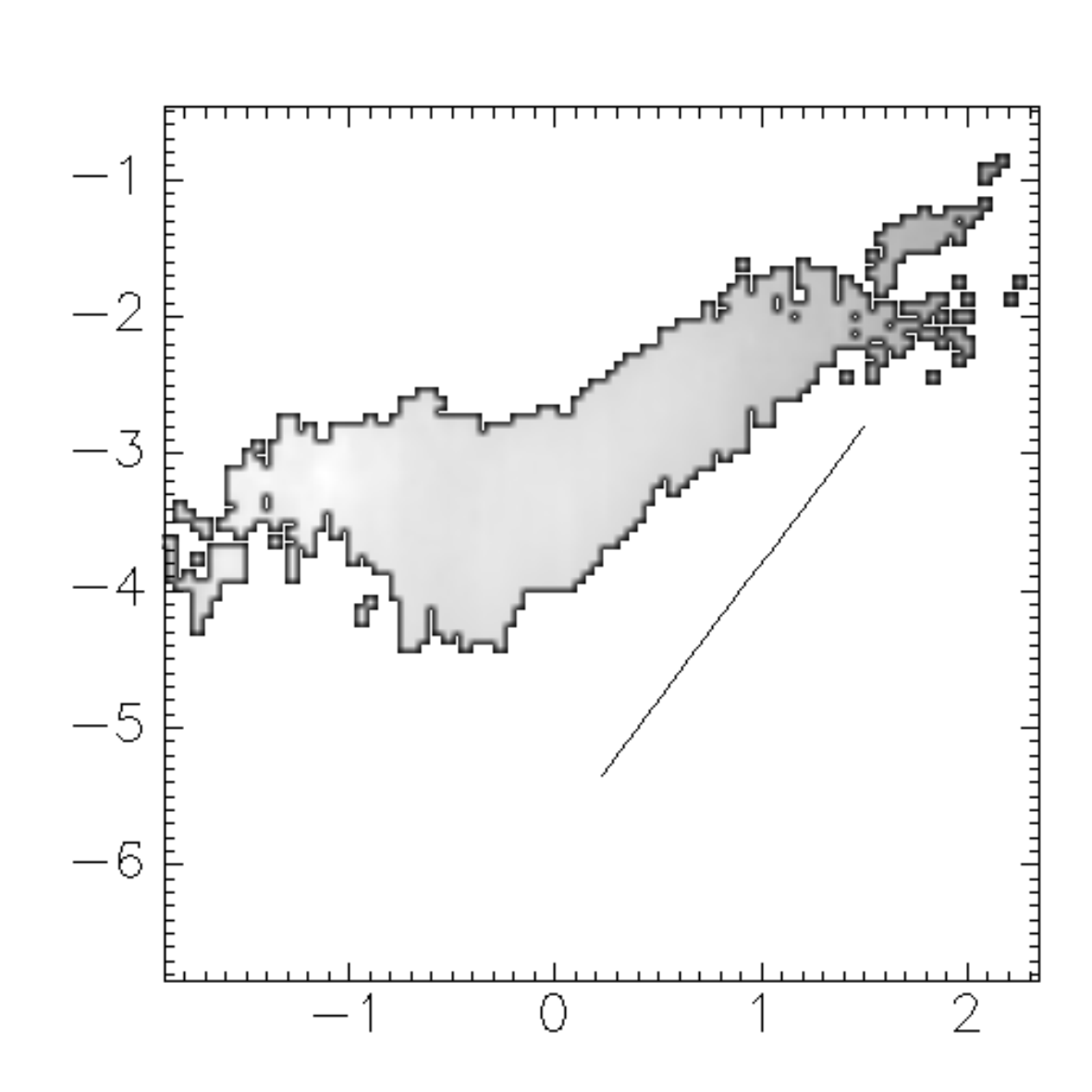}
\end{center}
\caption{Two-dimensional histograms of the grid cells in numerical
simulations in the $\ln B^2$ (vertical axis)-$\ln \rho$(horizontal axis)
space. The gray scale indicates the density of points in this
space. {\it Left panel:} A slab-geometry numerical simulation by PV03
dominated by the slow mode, exhibiting the behavior indicated by eq.\
(\ref{Jeq:slow_B_rho}). {\it Middle panel:} Same as the {\it left} panel,
but for a numerical simulation dominated by the fast mode, exhibiting
the behavior indicated by eq.\ (\ref{Jeq:fast_B_rho}). {\it Right panel:}
Two-dimensional histogram of the grid cells in the $\ln B^2$-$\ln n$
space from a numerical simulation by PV03 in which both the slow and the
fast modes are active. At low densities, the slow mode causes a
density-independent magnetic field strength, while at higher densities,
the fast mode produces a positive correlation. The straight-line segment
has a slope of 2.}
\label{Jfig:PV03_f1-2}
\end{figure}

PV03 also tested these results numerically, 
by isolating, or nearly isolating, the three different wave modes.  The
{\it left} panel of Fig.\ \ref{Jfig:PV03_f1-2} shows the distribution of
points in the $\ln B^2$-$\ln \rho$ space for a simulation dominated by
the slow mode, exhibiting the behavior outlined above, corresponding to
eq.\ (\ref{Jeq:slow_B_rho}). In contrast, the {\it middle} panel of
Fig.\ \ref{Jfig:PV03_f1-2} shows the distribution of points in the same
space for a simulation dominated by the fast mode, exhibiting the
behavior indicated by eq.\ (\ref{Jeq:fast_B_rho}). Finally, the {\it
  right} panel of this figure shows a simulation where both modes are
active.

The most important conclusion from eqs.\
(\ref{Jeq:slow_B_rho})--(\ref{Jeq:Alfv_B_rho}) is that, in a turbulent
flow in which all modes are active, the net, average scaling of the
magnetic field with the density will arise from the combined effect of
the various modes. Moreover, since at low densities the values of $B$
produced by the fast and Alfv\'en modes are also small, while the field
strengths produced by the slow mode remain roughly constant, the field
fluctuations will be dominated by the latter mode at low densities, and
a roughly density-independent field strength is expected. Conversely, at
high densities, the slow mode disappears, while the contribution from
the fast and Alfv\'en modes will dominate, producing a field strength
that increases with increasing density.  Finally, because each mode
produces a different dependence of the magnetic field strength with the
density, we expect that the instantaneous value of the density at a
certain location in physical space is not enough to determine the value
of the magnetic field strength there. Instead, this value depends on the
{\it history of modes} of the nonlinear waves that have passed through
that location, naturally implying that, within a large cloud, a large
scatter in the measured values of the magnetic field is expected. The
expected net scaling of the field strength with the density is
illustrated in the {\it right panel} of Fig.\
\ref{Jfig:PV03_f1-2}. These results are in qualitative agreement
with detailed statistical analyses of the magnetic field distribution in
the ISM \citep{Crutcher+10}, as illustrated in the {\it right} panel of
Fig.\ \ref{Jfig:B-n_corr_PN99}.

In conclusion, the observed constancy of the mean magnetic field
strength at low densities may be due to either the superposition of the
various modes of nonlinear MHD waves in the turbulent ISM, and/or to the
diffusion of magnetic flux by AD, and/or to the dissipation of the
magnetic energy by turbulent reconnection. Certainly, there is no
shortage of available mechanisms for producing it!

\subsection{Density profiles of filaments in MCs}

More recently, AD has been invoked to explain the radial density
profiles of filamentary structures in MCs. In recent years, observations
with the {\it Herschel} Space Observatory have revealed that the denser
regions of MCs are characterized by having a strongly filamentary
structure, and that the filaments host most
of the dense cores in MCs \cite[e.g.] [] {Andre+10, Molinari+10}. Since
then, molecular-line observations have additionally provided information
on the kinematic structure of the filaments \cite[e.g.,] []
{Schneider+10, Kirk+13, Peretto+13, Battersby+14}, suggesting that there
is flow {\it onto} the filaments, mainly perpendicular to their axes,
from their surroundings, as well as flow {\it along} the filaments, onto
the cores they contain.

One important feature of the filaments is their radial column density
structure, which, when averaged along the length of the filaments,
exhibits a Plummer-like profile of the form \cite{Arzoumanian+11}
\begin{equation}
  \Sigma_P(R) = \frac{A_p \rho_c R_c}
                  {\left[1+\left(R/R_c\right)^2\right]^{\frac{(p-1)}  {2}}},
                  \label{Jeq:sigma}
\end{equation}
where $R$ is the distance perpendicular to the filament's long axis, $p$
is the (negative) logarithmic slope of the volume density profile at
large $R$, and $A_p = \int^\infty_{-\infty} ( 1+u^2 )^{-p/2} du$.  Also,
\cite{Arzoumanian+11} found that the filaments are well fitted for
values of $p$ in the range $1.5<p<2.5$, and a characteristic Gaussian
full width at half maximum (FWHM) of $0.1$ pc, which corresponds to $R_c
\approx 0.03$ pc, roughly independently of the central column density.
Therefore, those authors suggested that this width is probably a
universal property of MC filaments.

The value of the slope, $p \sim 2$, is significantly different from that
expected for a self-gravitating, isothermal, hydrostatic cylinder
extending to infinity, but with finite linear density (mass per unit
length), for which the theoretically predicted value is $p=4$
\cite{Ostriker64}. This value of the slope, and the near-independence of
the characteristic width with the central column density of the
filaments, have prompted the producton of several analytical models. In
particular, it has been found by \cite{FM12} that truncated hydrostatic
equilibrium solutions, subjected to a finite, external pressure, change
slope as the external pressure is varied. Specifically, they found that
the slope is $\sim 2$ when the ratio of the central to the external
pressure is $\sim 10$. On the other hand, \cite{Heitsch13} has taken
into account the possibility of accretion onto the filaments and
considered magnetized cases \cite[se allso] [] {FP00}, finding that
accreting models produce profile slopes consistent with observations
only in the case of low-mass filaments, or when the magnetic field
scales weakly with density, for example, as $B\propto n^{1/2}$.

More directly related to the subject of this chapter, \cite{HA13} have
proposed that the near column density independence of the central width
may be a result of the central region being in virial equilibrium, so
that the turbulence in that region (driven by accretion onto the
filament) exerts a pressure that balances the ram-pressure exerted by
the accretion. Since in this case the rate at which the internal
turbulence dissipates is crucial for determining the stationary
turbulent Mach number, they tested two different timescales of
dissipation: the standard turbulent cascade timescale, given simply by
the turbulent crossing time across scale $\ell$, $\tturb =
\ell/\sigma_{\ell}$, where $\sigma_\ell$ is the turbulent velocity
dispersion at scale $\ell$, and the AD dissipation timescale of Alfv\'en
waves, given by \cite{KP69}
\beq
\tdad = \frac {2 \gamma \rhoi} {\va^2 (2 \pi/\lambda)^2},
\eeq
where $\gamma$ is the damping rate introduced in eq.\
(\ref{Jeq:drag_for}), $\rhoi$ is the ion density, given by eq.\
(\ref{Jeq:rhoi_vs_rhon}), $\va$ is the Alfv\'en speed, and $\lambda$ is
the wavelength, assumed to be of the order of the radius of the
filament, which is the relevant scale at which energy dissipation is
considered. These authors find that the filament's radius $\rf$ depends
on the central density $\rhoc$ as $\rf \sim \rhoc^{-1}$ in the case of
the standard turbulent dissipation, while, for dissipation dominated by
AD, they find $\rf \sim \rhoc^{-0.2}$. Thus, they conclude that
dissipation due to AD allows a weak enough dependence of the central
width on the density as to be in agreement with the observed near
constancy of the central width in MC filaments.

However, it should be pointed out that molecular-line studies of
filaments often do not recover the same characteristic width as the
Herschel observations, showing instead a large scatter of values
\cite[e.g.,] [] {Panopoulou+14}, or much larger widths \cite[e.g.,] []
{Furuya+14}, suggesting that the 0.1-pc characteristic width of the
Herschel data may result from a bias introduced by the analysis
procedure \cite{Heitsch13}. 

Moreover, numerical simulations not including magnetic fields nor AD
\cite[e.g.,] [] {GV14, Federrath15}, as well as simulations including
the magnetic field but not AD \cite [e.g.,] [] {Federrath15, Kirk+15},
also exhibit widths close to the observed values. In particular, 
\cite{Federrath15} has interpreted the the width of the filaments as a
manifestation of the turbulent ``sonic scale'', the scale at which the
turbulent velocity dispersion coincides with the sound speed. In
addition, he also interpreted the radial density profile as a
consequence of the scaling of the post-shock density with the post-shock
thickness in the collision of two planar shocks, thus providing an
AD-unrelated interpretation of the filament nealy-constant thickness.

One can conclude from the above discussion that there exist several
possible explanations for the radial column density structure of MC
filaments, with AD being just one of them, but at this point it has not been
conclusively determined which mechanism is mainly responsible for it.

\subsection{Avoiding the magnetic braking catastrophe}
\label{Jsec:mag_brak}

During the collapse of a region of a MC, conservation of angular
momentum would seem to imply the existence of a strong ``angular
momentum problem'', which has been known for a long time
\cite{Hoyle45}, and according to which, the increase in any initial
rotation rate present in the cloud would generate such strong
centrifugal forces that the cloud would be disrupted and produce
fragments with masses too small compared to actual stellar
masses. This is, however, contrary to the observation that the
specific angular momentum of cloud substructures seems to decrease
with their size \cite[see, e.g.,] [and references therein]
{Bodenheimer95}. This implies that, somehow, angular momentum must be
transferred outward from the collapsing material during the collapse and
fragmentation of a cloud.

One possible mechanism to avoid the angular momentum problem is the
so-called ``magnetic braking'' \cite{Mouschovias77}, by which a
rotating protostar permeated by a magnetic field would drag the field
lines, causing them to rotate at the same angular velocity, and thus
accelerating material outwards, transferring the angular momentum to
large radii from the protostar. However, numerical as well as
analytical simulations have shown that this effect is too strong under
ideal MHD conditions \cite{KK02, Allen+03, Galli+06}, causing a
``catastrophic'' magnetic braking, which removes too much angular
momentum from the protostellar disk. This reulst in the formation of a
``pseudo-disk'' which, in spite of having a flattened geometry (because
of contraction preferentially along field lines), does not have a
Keplerian rotation regime, and so it is not rotationally supported.
This is in contradiction with the Keplerian regime often observed in
protostellar disks \cite[see, e.g.,] [and references therein]
{Mundy+00}, although the question of whether this regime is the norm
or not is far from settled \cite[see, e.g., the review by] [] {Li+14}.

This problem persists even under the presently accepted weaker typical
values of the magnetic field strength in MCs and their cores (cf.\ Sec.\
\ref{Jsec:no_subcrit}), according to which the typical core has a
normalized M2FR $\mu \sim$ a few. This implies that, although the field is not
strong enough to prevent collapse of the cores, it is in general still
strong enough to cause catastrophic magnetic braking in the ideal MHD
regime \cite[e.g.,] [] {Li+11}.

AD, as well as Ohmic dissipation and the Hall effect, have been proposed
as mechanisms capable of preventing catastrophic magnetic braking in the
cores, with Ohmic dissipation being the best candidate so far because it
really reduces the magnetic flux, while AD seems to only redistribute
it, allowing it to pile up in a small circumstellar region where magnetic
braking is still too strong \cite[] [see also the review by
\cite{Li+14}] {DB10, Li+11, Dapp+12}.

However, other resolutions have been proposed for the magneting braking
catastrophe, involving turbulence, either by itself \cite{Seifried+12},
or in conjunction with turbulent reconnection \cite{Santos+12}. The
former study suggested that the whole magnetic braking catastrophe
arises from an unrealistic choice of initial conditions, which neglected
the presence of turbulent motions, and presented simulations of
gravitational collapse of magnetized cores in which rotationally
supported disks readily form. Those authors argued that the formation of
Keplerian disks in this case is due to the absence of coherent
rotational motions at different scales together with the transport of
angular momentum by shearing turbulent motions. This is in qualitative
agreement with the result by \cite{JK04} that the specific angular
momentum in nonmagnetic turbulence simulations decreases with size, in
agreement with observations, in spite of not allowing for magnetic
breaking. These authors attributed this result to gravitational torques,
but it is quite likely that hydrodynamic torques are at play as well. In
addition to this, \cite{Santos+12} also considered turbulent
reconnection, pointing out that dissipation of magnetic energy by
reconnection is greatly enhanced in the presence of turbulence, allowing
magnetic flux to be transported to the outskirts of the disk on
timescales comparable to the collapse, and thus the formation of a
rotationally supported disk of dimensions up to $\sim 100$ pc.

\subsection{Conclusions} \label{Jsec:concls}

In this chapter we have reviewed several applications of AD in the ISM
and star formation, from the correlation of the magnetic field with the
gas density, through the formation of dense cores and their
subsequent gravitational collapse, to the formation of rotationally
supported accretion disks around protostars. In each case, we also
discussed alternative mechanisms that have been proposed to accomplish
the same effects without resorting to AD.

To place the discussion in context, we started with a brief historical
recount of how star formation models have evolved in the last 40 years
or so, emphasizing the diminishing role played by the magnetic field in
the process, as recent observations suggest that MCs and their cores are
in general magnetically supercritical. This implies that the field is
not strong enough to support the cores against collapse, although the
typical values suggest it is still strong enough to cause catastrophic
magnetic braking in the ideal MHD regime. In the absence of magnetic
support, current theories consider that MCs are either supported by
turbulent pressure, or else that they are undergoing global,
hierarchical collapse. In the latter case, the role of turbulence is
reduced to producing density fluctuations that can collapse earlier than
the whole cloud.

We next reviewed the standard single-fluid formulation for AD,
introducing its characteristic diffusivity and timescale, and then
proceeded to the applicactions. First was the ``standard'' model of
magnetic support and AD-mediated collapse of dense cores in MCs,
presenting the standard results, and then noting the objections to this
model. One very important problem is the fact that the M2FR of a cloud
is a rather subjective and ill-defined quantity, depending strongly on
the choice of boundary conditions, and becoming a variable of the
problem when accretion is considered. Under these considerations, and
with the revised values of the M2FR in cores, the effective AD timescale
becomes of the order of the free-fall time, and thus the role of AD
is significantly weakened.

The second application is the observed lack of correlation of the
magnetic field strength with the gas density at low densities ($n
\lesssim 100 \pcc$). This has been attributed to the diffusion of the
magnetic flux caused by turbulent AD and/or reconnection, although it
has also been suggested that it may be a consequence of the different
scaling of the magnetic field with the density for different modes of
MHD waves.

A third application is the possibility that AD may set the observed
near-universal width of the filamentary structures in
MCs. The specific suggestion is that this width is controlled by a
balance between the ram pressure exerted on the filament by material
accreting onto it (perpendicularly to its axis) and the pressure from
the turbulence driven by the accretion itself in the central parts of the
filament. For this balance to produce filament widths consistent with
observations, the dissipation of the turbulence should be dominated by
AD. However, other mechanisms not involving AD have also been
proposed, and nonmagnetic simulations (where dissipation cannot be
dominated by AD) exhibit similar widths, so there is no conclusive
evidence that AD-dominated turbulence dissipation is the controlling
mechanism of the filaments' width.

Finally, we discussed the magnetic braking catastrophe, which consists
of excessive removal of angular momentum during core collapse, which
prevents the formation of rotationally-supported Keplerian disks in the
ideal MHD regime. AD diffusion has been proposed as a possible mechanism
to remove magnetic flux from the forming disk, allowing the retention of
sufficient angular momentum, but numerical simulations suggest that in
general the effect is not sufficient for the task, and it is currently
accepted that Ohmic dissipation is a better candidate. However, it has
also been proposed that the ``catastrophe'' rather arises from the
choice of unrealistically smooth non-turbulent initial conditions, and
numerical simulations with turbulent initial conditions seem to be able
to readily form Keplerian disks. This would be due to the nonlinear
exchange of angular momentum between fluid parcels in this regime.

It can be concluded that AD has seen a large number of intended
applications in various important aspects of the ISM and star formation,
but in all cases, there exist alternative possible mechanisms that do
not rely on AD to accomplish the same tasks. The remaining challenge
lies in being able to discern between the various candidate mechanisms
and find the truly dominant mechanism in each case.

%%%%%%%%%%%%%%%%%%%%%%%%%%%%%%%
\newpage
\section{Exoplanets}
\label{Fsec:Exoplanets}
%%%%%%%%%%%%%%%%%%%%%%%%%%%%%%%

  \subsection{Introduction}
  \label{Fsubsec:Introduction: Exoplanet -- discovery status}
The study of planets beyond the Solar system, exoplanets, is one of the fastest growing fields in present day space science. On one hand, it is driven by the continuously growing number of discovered exoplanets and the goal to find a potentially habitable world similar to our Earth. On the other hand, the specific physical conditions expected on many exoplanets, untypical for the relatively well investigated Solar system planets, inspire researches to investigate them deeply for better understanding their nature. 
Almost two decades after the discovery of 51 Peg b, the first Jupiter-type gas giant outside our solar system, more than 3000 exoplanets, including about 600 multiple systems, have been detected mainly by space- and ground-based photometric transit surveys ($\sim 2600$) and ground-based radial velocity ($\sim 600$) measurements (http://exoplanetarchive.ipac.caltech.edu; /http://exoplanet.eu/catalog/). In addition, about $>2500$ "\emph{planet candidates}" have been found by the Kepler mission
\citep{Batalha-etal-2013, Vanderburg-etal-2015}. The constantly growing number of discovered exoplanets and accumulation of data regarding their physical and orbital characteristics provide an empirical background for more detailed investigation of general principles and major trends in formation and evolution of the planetary systems, including the potential habitability aspect of the terrestrial-type planets.

\begin{figure}[b]
\begin{center}
\includegraphics[width=3.4in]{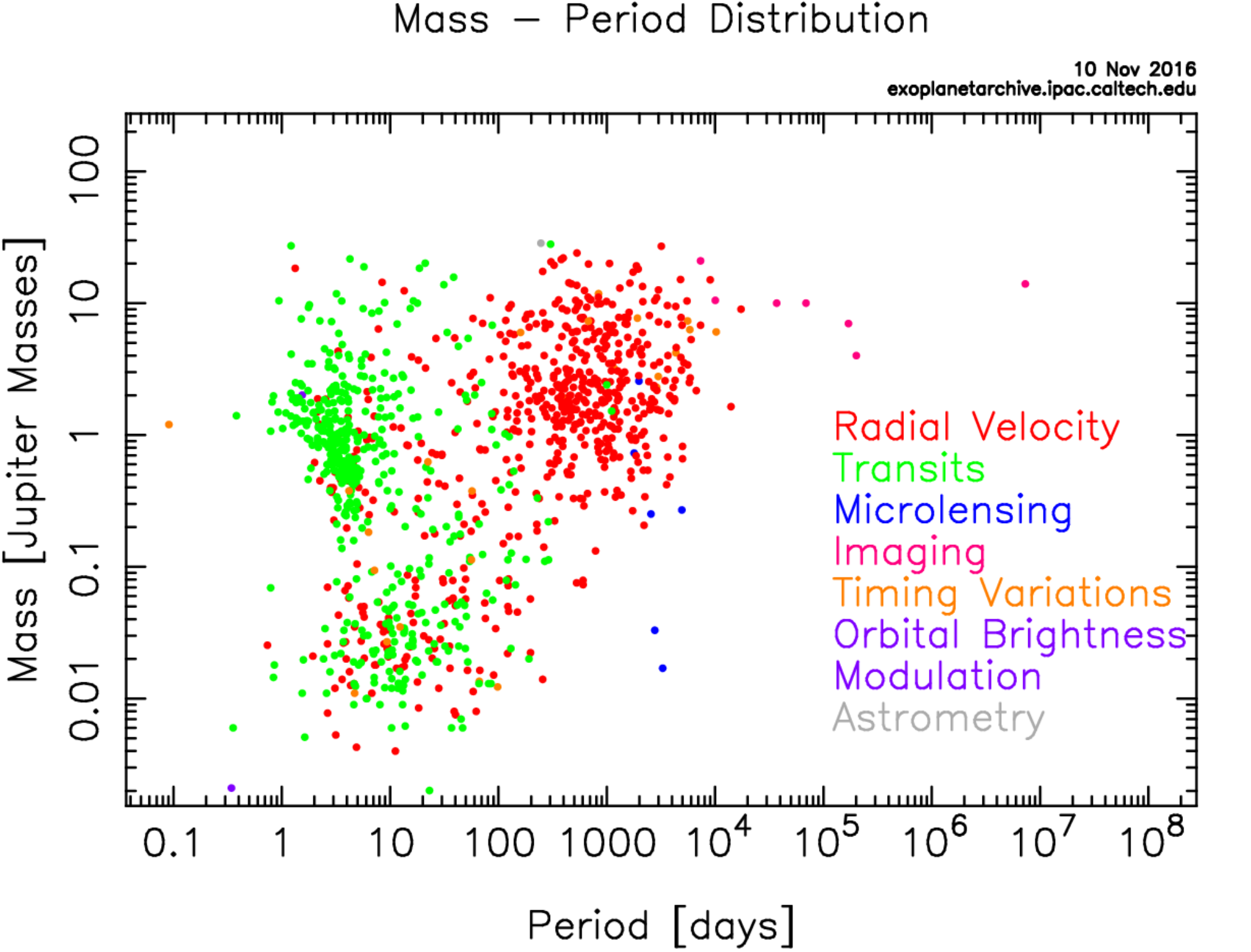}
\caption{Mass-Orbital period distribution of confirmed exoplanets (http://exoplanetarchive.ipac.caltech.edu/exoplanetplots/)}
\label{fig_1}
\end{center}
\end{figure}

Although most of the exoplanets discovered so far are thought to be gas or ice giants, like Jupiter or Neptune, some potentially rocky planets have been identified around M stars \citep{Bonfils-etal-2013}. The discovery of these planets became possible due to several international ground-based transit search projects, as well as the {\emph{COROT}} and {\emph{Kepler}} space observatories. More than a half of known exoplanets have orbits around their host stars shorter than 0.6 AU (Figure~\ref{fig_1}). By this, an evident maximum in the orbital distribution of exoplanets takes place in the vicinity of 0.05 AU, with two well pronounced major populations there with orbital periods $P < 30$ days corresponding to the giant type planets ($0.2 M_{\rm J} < M_{\rm p} < 8 M_{\rm J}$), so called {\emph{Hot Jupiters}} (HJs), and less massive ($0.008 M_{\rm J} < M_{\rm p} < 0.08 M_{\rm J}$), {\emph{Neptune-}} and {\emph{super-Earth}} type planets. Here $M_{\rm J}$ stays for the mass of Jupiter. Altogether the HJs constitute about 16$\%$ of the total number of known exoplanets.

The detection of exoplanets at orbital distances $\leq 0.05$ AU rises questions regarding their upper atmosphere structure, the planet interaction with the extreme stellar radiation and plasma environment \citep{Yelle-2004, Holmstrom-etal-2008, Ekenback-etal-2010, Kislyakova-etal-2013, Kislyakova-etal-2014}, the role of possible magnetospheres for atmospheric protection \citep{Khodachenko2007a, Khodachenko-etal-2007b, Khodachenko-etal-2012}, destructive tidal forces between the host star and the planet \citep{Koskinen-etal-2010, Koskinen-etal-2013, Trammell-etal-2011, Trammell-etal-2014}, as well as the stability of atmospheres against different erosion and mass loss processes \citep{Guillot-etal-1996}. The study of close-orbit exoplanets and their survival in hard stellar radiation and plasma conditions also helps to understand how terrestrial planets and their atmospheres, including early Venus, Earth and Mars evolved during the active early evolution phase of their host stars.

\citet{Lammer-etal-2003} were the first to show that a hydrogen-rich thermosphere of a HJ at close orbital distance will be heated to several thousand Kelvin such that hydrostatic conditions will no longer be valid and the thermosphere will dynamically expand \citep{Lammer-etal-2013}. The stellar X-ray/EUV (XUV) radiation energy deposition results in heating, ionization, and consequent expansion of atmospheres of the close-orbit exoplanets. These appear the major driving factors for the mass loss of a planet \citep{Lammer-etal-2003, Yelle-2004, Erkaev-etal-2005, Tian-etal-2005, Garcia-Munoz-2007, Penz-etal-2008, Guo-2011, Guo-2013, Koskinen-etal-2010, Koskinen-etal-2013, Lammer-etal-2013}. Applied hydrodynamic (HD) models by \citet{Yelle-2004},
\citet{Tian-etal-2005}, \citet{Garcia-Munoz-2007}, \citet{Penz-etal-2008}, \citet{Guo-2011}, \citet{Guo-2013}, \citet{Shaikhislamov-etal-2014}, and \citet{Khodachenko-etal-2015} as well as the quasi-empirical modeling by \citet{Koskinen-etal-2010}, also indicate that close-in exoplanets experience extreme heating by stellar XUV radiation, which results in an expanding supersonic planetary wind up to Roche lobe with mass loss rates $\sim 10^{7} - 10^{8}$ kg$\cdot$s$^{-1}$. Such loss rates are also supported by {\emph{Hubble Space Telescope (HST)}}/Space Telescope Imaging Spectrograph (STIS) observations \citep{Vidal-Madjar-etal-2003, Vidal-Madjar-etal-2004}, which detected a $15\% \pm 4\%$ intensity drop in the high-velocity part of the stellar Ly$\alpha$ line during the planet (HD 209458b) transit, which was interpreted as a signature of hot neutral hydrogen atoms in the expanding planetary atmosphere.

The problem of upper atmospheric erosion of close-orbit exoplanets and their mass loss is closely connected with the study of the whole complex of stellar-planetary interactions, including consideration of the influences of intensive stellar radiation and plasma flows (e.g., stellar winds and coronal mass ejections) on the planetary plasma and atmosphere environments. The following processes have to be considered in that respect in their mutual relation and influence.
\begin{enumerate}
\item {The heating of the planetary upper atmosphere by the stellar XUV radiation results in its expansion, which under certain conditions could be so large that the majority of light atmospheric constituents overcome the gravitational binding and escape from the planet in the form of a HD wind \citep{Yelle-2004, Tian-etal-2005, Koskinen-etal-2010, Koskinen-etal-2013, Erkaev-etal-2013, Shaikhislamov-etal-2014}. This contributes to the so-called {\emph{thermal}} mass loss process of atmospheric material.
}
\item {Simultaneously with the direct radiative heating of the upper atmosphere, the processes of ionization and recombination as well as production of energetic neutral atoms by sputtering and various photochemical and charge exchange reactions take place \citep{Yelle-2004, Lammer-etal-2009, Lammer-etal-2013, Shematovich-2012, Guo-2011, Guo-2013}. Such processes result in the formation of extended (in some cases) coronas around planets, filled with hot neutral atoms.
}
\item {The expanding, XUV-heated and photochemically energized, upper planetary atmosphere and the hot neutral corona may appear in a direct contact with the plasma flow of stellar wind and/or coronal mass ejections with the consequent loss due to an ion pickup mechanism. That contributes to the {\emph{non-thermal}} mass loss process of the atmosphere \citep{Lichtenegger-etal-2009, Khodachenko-etal-2007b, Kislyakova-etal-2013, Kislyakova-etal-2014}.
}
\end{enumerate}

\subsection{Magnetospheric protection of close-orbit exoplanets}
\label{Fsubsec:Magnetospheric protection of close-orbit exoplanets}

The planetary intrinsic magnetic field appears a crucial factor that influences both, {\emph{thermal}} and {\emph{non-thermal}} types of the mass loss of close-orbit exoplanets and protects the planetary upper atmospheric environment. The protective role of planetary magnetic field has two major aspects. First, the large-scale magnetic fields and electric currents, related to the planetary magnetism, form the planetary magnetosphere, which acts as a barrier for the upcoming stellar wind. Magnetosphere protects the ionosphere and upper atmosphere of a planet against direct impact of stellar plasmas and energetic particles, thus reducing the {\emph{non-thermal}} mass loss \citep{Khodachenko2007a, Khodachenko-etal-2007b, Lammer-etal-2007}. Second, the internal magnetic field of the planetary magnetosphere strongly affects the {\emph{thermal}} mass loss by influencing the streaming of the expanding planetary wind plasma and its interaction with the stellar wind \citep{Adams-2011, Trammell-etal-2011, Trammell-etal-2014, Owen-Adams-2014, Khodachenko-etal-2015}. Therefore, the processes of material escape, planetary mass loss, and formation of the planetary magnetosphere  have to be considered jointly in a self-consistent way in their mutual interconnection. The expanding partially ionized plasma of the planetary wind interacts with intrinsic magnetic dipole field and appears a strong driver in formation and shaping of exoplanetary magnetosphere, which in its turn influences the overall mass loss of a planet.

\subsubsection{Planetary magnetism}
\label{Fsubsec:Planetary magnetism}

The intrinsic magnetic field of a planet, which influences the character of the magnetospheric obstacle, as well as the dynamics of the inner magnetospheric plasma, is generated by a magnetic dynamo. The existence and efficiency of the dynamo are closely related to the type of planet and its interior structure. Not all planets have intrinsic magnetic fields, or in other words, -- the efficiently operating dynamos. Planetary magnetic dynamo requires the presence of an electrically conducting region (i.e. a liquid outer core for terrestrial planets, or a layer of electrically conducting liquid hydrogen for gas giants) with non-uniform flows organized in a certain manner, which create a self-sustaining magnetic field. According to dynamo theory, this flow should be convective in nature \citep{Stevenson-1983}. Therefore, convection can be regarded as a necessary requirement for a planetary magnetic field \citep{stevenson2003}.

Limitations of the existing observational techniques make direct measurements of the magnetic fields of exoplanets impossible. At the same time, a rough estimation of an intrinsic planetary magnetic dipole moment $\mathcal{M}$ can be obtained by simple scaling laws derived by the comparison of different contributions in the governing equations of planetary magnetic dynamo theory \citep{Farrell-etal-1999, Sanchez-Lavega-2004, Griesseier-etal-2004, Christensen-2010}. Most of these scaling laws reveal a connection between the intrinsic magnetic field and the rotation of a planet. \citet{Griesseier-etal-2004} estimated the intrinsic planetary magnetic dipole moments of exoplanets and corresponding sizes of their magnetospheres using the following scaling laws for $\mathcal{M}$:
\begin{equation}
\begin{array}{ll}
\mathcal{M} \propto \rho_c^{1/2} \omega_{\rm p} r_c^4 & \quad \mbox{\emph{Busse} 1976,\ }\quad
\\
\mathcal{M} \propto \rho_c^{1/2} \omega_{\rm p}^{1/2} r_c^3 \sigma^{-1/2} & \quad
\mbox{\emph{Stevenson} 1983,\ }\quad
\\
\mathcal{M} \propto \rho_c^{1/2} \omega_{\rm p}^{3/4} r_c^{7/2} \sigma^{-1/4} & \quad
\mbox{\emph{Mizutani et al.} 1992,\ }\quad
\\
\mathcal{M} \propto \rho_c^{1/2} \omega_{\rm p} r_c^{7/2} & \quad \mbox{\emph{Sano} 1993.\ }\quad
  \end{array}
  \label{eq-1}
\end{equation}
Here $r_c$ is the radius of the dynamo region (also called the core radius), and $\omega_{\rm p}$ is the cyclic frequency of a planet rotation around its axis. The internal properties of a planet, such as the mass density and the conductivity of the dynamo region are denoted by $\rho_c$ and $\sigma$, respectively (for details of the model parameters estimation see \citet{Griesseier-etal-2004}). More recently, \citet{Reiners-Christensen-2010}, based on scaling properties of convection-driven dynamos \citep{Christensen-Aubert-2006}, calculated the evolution of average magnetic fields of HJs and found that (a) extrasolar gas giants may start their evolution with rather high intrinsic magnetic fields, which then decrease during the planet life time, and (b) the planetary magnetic moment may be independent of planetary rotation.

The equations (\ref{eq-1}) provide a range $\mathcal{M}_{\rm min} - \mathcal{M}_{\rm max}$ of reasonable planetary magnetic moment values. In spite of being different in details, all these models yield an increase of $\mathcal{M}$ with an increasing planetary angular velocity $\omega_{\rm p}$. In that respect it is necessary to take into account the fact, that close-orbit exoplanets very likely are tidally locked to their host stars. The angular rotation of a tidally locked planet is synchronized with its orbital revolution so, that $\omega_{\rm p}$ is equal to the orbital cyclic frequency $\Omega$ determined by Kepler's law. The time scale for tidal locking $\tau_{\rm sync}$ depends on the planetary structure, orbital distance to the host star, and the stellar mass \citep{Showman-Guillot-2002}. By this, the planets for which $\tau_{\rm sync} \leq 0.1$ Gyr, can be assumed to be tidally locked, since the age of a planet is at least an order of magnitude longer. On the other hand, the planets with $\tau_{\rm sync} \geq 10$ Gyr are almost certainly tidally unlocked. The influence of tidal locking on the value of an expected planetary magnetic dipole was studied for different planets (giants and terrestrial-type) in \citet{Griesseier-etal-2004, Griesseier-etal-2007a, Griesseier-etal-2007b}. It was shown, that the magnetic moments of slowly rotating tidally locked exoplanets usually are much smaller than those for similar, but freely rotating tidally unlocked planets. In Table~\ref{Table-sw-mdip-Rs} possible ranges of planetary magnetic dipole moments defined by (\ref{eq-1}) for a Jupiter-type ($M_{\rm p} = M_{\rm J}$; $R_{\rm p} = R_{\rm J}$) exoplanet orbiting around a solar-analogue G-type star ($M_* = M_{\rm Sun}$, $R_* = R_{\rm Sun}$) are provided. The values of $\mathcal{M}$ are scaled in units of the present time Jupiter magnetic moment $\mathcal{M}_{\rm J} = 1.56 \times 10^{27}$ A m$^2$.
\begin{table}
\caption{
The stellar wind parameters for a solar-analogue G-type star ($M_* = M_{\rm Sun}$, $R_* = R_{\rm Sun}$, age $t_* \sim 4$ Gyr) and the ranges of $\mathcal{M}$ given by (\ref{eq-1}) as well as corresponding $R_{\rm s} = R_{\rm s}^{(dip)}$ provided by (\ref{eq-Rs}) for a Jupiter-type exoplanet
at different orbital distances $d$. The stellar wind velocity includes also the contribution of the Keplerian planetary orbital velocity $V_K$ $^1$: Tidally locked. $^2$: Possible tidally locked. $^3$: Not tidally locked.}
\label{Table-sw-mdip-Rs}
\begin{center}
\begin{tabular}[t]{lcccc}
\hline\noalign{\smallskip}
Orbital distance $d$ & $n_{\rm sw}$  & $\tilde{v}_{\rm sw}$ & $\mathcal{M}$ & $R_{\rm s}^{(dip)} (\mathcal{M}_{\rm min})$~/~$R_{\rm s}^{(dip)} (\mathcal{M}_{\rm max})$ \\
{[AU]}               & {[cm$^{-3}$]} & {[km/s]}             & {[$\mathcal{M}_{\rm J}$]} & {[$R_{\rm p}$]} \\
\noalign{\smallskip}\hline\noalign{\smallskip}
$0.045^1$            & $9.1e3$ & $210$ & $0.12 ... 0.3$ & $4.3 ... 6.2$ \\
$0.1^2$              & $1.2e3$ & $260$ & $0.04 ... 1.0$ & $3.8... 12$   \\
$0.3^3$              & $92$    & $340$ & $1.0 ... 1.0$  & $15 ... 15$    \\
\noalign{\smallskip}\hline
\end{tabular}
\end{center}
\end{table}
\subsubsection{Magnetic shielding}
\label{Fsubsec:Magnetic shielding}

For the efficient magnetic shielding (i.e. magnetospheric protection) of a planet against the stellar wind, the size of its magnetosphere characterized by the magnetopause stand-off distance $R_{\rm s}$ should be much larger than the height of the exobase. By this, the value of $R_{\rm s}$ is determined from the balance between the stellar wind ram pressure and the planetary magnetic field pressure at the sub-stellar point \citep{Griesseier-etal-2004, Khodachenko2007a}. In most studies, the investigation of exoplanetary magnetospheric protection is performed within a highly simplifying assumption of a planetary {\emph{dipole-dominated}} magnetosphere. This means that only the intrinsic magnetic dipole moment of an exoplanet $\mathcal{M}$ and the corresponding magnetopause electric currents (i.e., "screened magnetic dipole" case) are considered as the major magnetosphere forming factors. In this case, i.e. assuming $B(r) \propto \mathcal{M} / r^3$, the value of $R_{\rm s}$ is defined by the following expression:
\begin{equation}
  R_{\rm s} \equiv R_{\rm s}^{(dip)} = \left[ \frac{\mu_{\rm 0} f_{\rm 0}^2 {\mathcal{M}}^2}{8 \pi^2 \rho_{\rm sw} \tilde{v}_{\rm sw}^2} \right]^{1/6},
   \label{eq-Rs}
\end{equation}
where $\mu_{\rm 0}$ is the diamagnetic permeability of free space, $f_{\rm 0} \approx 1.22$ is a form-factor of the magnetosphere caused by the account of the magnetopause electric currents, $\rho_{\rm sw} = n_{\rm sw} m$ is the mass density of the stellar wind, and $\tilde{v}_{\rm sw}$ is the relative velocity of the stellar wind plasma which includes also the Keplerian planetary orbital motion velocity $V_K$.

The stellar wind parameters appear to be the important factors which influence the planetary magnetosphere size (e.g., see (\ref{eq-Rs})). Recently, there have been important developments towards indirect detections of stellar winds through their interactions with the surrounding interstellar medium. In particular, the stellar mass loss rates and related stellar wind parameters have been estimated by observing astrospheric absorption features of several nearby G- and K- stars. Comparison of the measured absorption to that calculated by hydrodynamic codes made it possible to perform empirical estimation of the evolution of the stellar mass loss rate as a function of stellar age \citep{Wood-etal-2002, Wood-etal-2005} and to conclude about the dependence of $n_{\rm sw}$ and $v_{\rm sw}$ on the age of the stellar system. In particular, the younger solar-type G- stars appeared to have much denser and faster stellar winds as compared to the present Sun. Combining the stellar mass loss measurements of \citet{Wood-etal-2005} with the results of \citet{Newkirk-1980} for the age-dependence of stellar wind velocity, \citet{Griesseier-etal-2007a} proposed a method for calculation of stellar wind density $n_{\rm sw}$ and velocity $v_{\rm sw}$ at a given orbital location of an exoplanet $d$ for a given mass $M_*$, radius $R_*$ and age $t_*$ of star. As an example, the values of the stellar wind plasma parameters for a solar-analog G-type star ($M_* = M_{\rm Sun}$, age $t_* \sim 4$ Gyr) at orbital distances of 0.045 AU, 0.1 AU, and 0.3 AU are given in Table~\ref{Table-sw-mdip-Rs}.
For the tidally locked close-orbit exoplanets with weak magnetic moments exposed to a dense and/or fast stellar wind plasma flows, (\ref{eq-Rs}) yields rather small values for sizes of dipole-dominated magnetospheres, $R_{\rm s} = R_{\rm s}^{(dip)}$, compressed by the stellar wind plasma flow (see Table~\ref{Table-sw-mdip-Rs}), which in the most extreme cases of collision with CMEs may even shrink down to the planetary surface, i.e. $R_{\rm s} \rightarrow R_{\rm p}$. Such an approach to estimation of the magnetosphere size, based on (\ref{eq-Rs}), resulted in a commonly accepted conclusion, that in order to have an efficient magnetic shield, a planet needs a strong intrinsic magnetic dipole $\mathcal{M}$.

\citet{Khodachenko-etal-2007b} studied the mass loss of the Hot Jupiter HD 209458b due to the ion pick-up mechanism caused by stellar CMEs, colliding with the planet. In spite of the sporadic character of the CME-planetary collisions, in the case of a moderately active host star of HD 209458b, it has been shown that the integral action of the stellar CME impacts over the exoplanet's lifetime can produce a significant effect on the planetary mass loss. The estimates of the \emph{non-thermal} mass loss of the tidally locked, and therefore, weakly magnetically protected hot giant HD 209458b due to the stellar wind ion pick-up lead to significant and sometimes unrealistic values -- up to several tens of planetary masses $M_{\rm p}$ lost during the planet life time \citep{Khodachenko-etal-2007b}. In view of the fact that multiple close-in giant exoplanets, comparable in mass and size with the Solar System Jupiter exist, and that it is unlikely that all of them began their life as ten times, or even more massive objects, one may conclude that additional factors and processes have to be taken into consideration in order to explain the protection of close-in exoplanets against of destructive {\emph{non-thermal}} mass loss. Regarding that problem, \citet{Khodachenko-etal-2012} proposed a more complete model of the magnetosphere of a giant-type exoplanet, which due to consequent account of the specifics of the close-orbit exoplanets (addressed in the following subsections) provides under similar conditions larger sizes for the planetary magnetospheric obstacles, than those given by the simple screened magnetic dipole model, traditionally considered in such cases.
\subsection{Magnetodisk-dominated magnetosphere of a Hot Jupiter}

The investigation of exoplanetary magnetospheres and their role in evolution of planetary systems forms a new and fast developing branch. Magnetosphere of a close-orbit exoplanet is a complex object, whose formation depends on different external and internal factors. These factors may be subdivided on two basic groups: (a) {\emph{stellar factors}}, e.g., stellar radiation, stellar wind plasma flow, stellar magnetic field and (b) \emph{planetary factors}, e.g., type of planet, orbital characteristics, escaping material flow, and planetary magnetic field. The structure of an exoplanetary magnetosphere depends also on the speed regime of the stellar wind and planet interaction \citep{Erkaev-etal-2005, Ip-etal-2004, Matsakos-etal-2015, Strugarek-etal-2015}. In particular, for an exoplanet at sufficiently large orbital distance when the stellar wind is super-sonic and super-Alfv\'{e}nic, i.e. when the ram pressure of the stellar wind dominates the magnetic pressure, an Earth$/$Jupiter-type magnetosphere with a bow shock, magnetopause, and magnetotail, is formed. At the same time, in the case of an extremely close-orbit location of an exoplanet (e.g., $d < 0.03$ AU for the Sun analogue star), where the stellar wind is still under acceleration and remains sub-magnetosonic and sub-Alfv\'{e}nic \citep{Ip-etal-2004, Preusse-etal-2005, Strugarek-etal-2015}, an Alfv\'{e}nic wing-type magnetosphere without a shock in the upstream region is formed \citep{Woodward-McKenzie-1999, Strugarek-etal-2015}. The character of the stellar wind impact on the planetary nearby plasma environment and inner atmosphere is different for the super- and sub- Alfv\'{e}nic types of the magnetosphere and in each particular planet case it has to be properly taken into account. In the present paper, further on, we do not consider the Alfv\'{e}nic wing-type magnetospheres, aiming at moderately short orbit giant planets near solar-type stars, under the conditions of a  super-Alfv\'{e}nic stellar wind flow, i.e., with the magnetospheres having in a general case a bow shock, a magnetopause, a magnetotail, similar to the case of the solar system Jupiter.

\begin{figure}[b]
\begin{center}
 \includegraphics[width=4.5in]{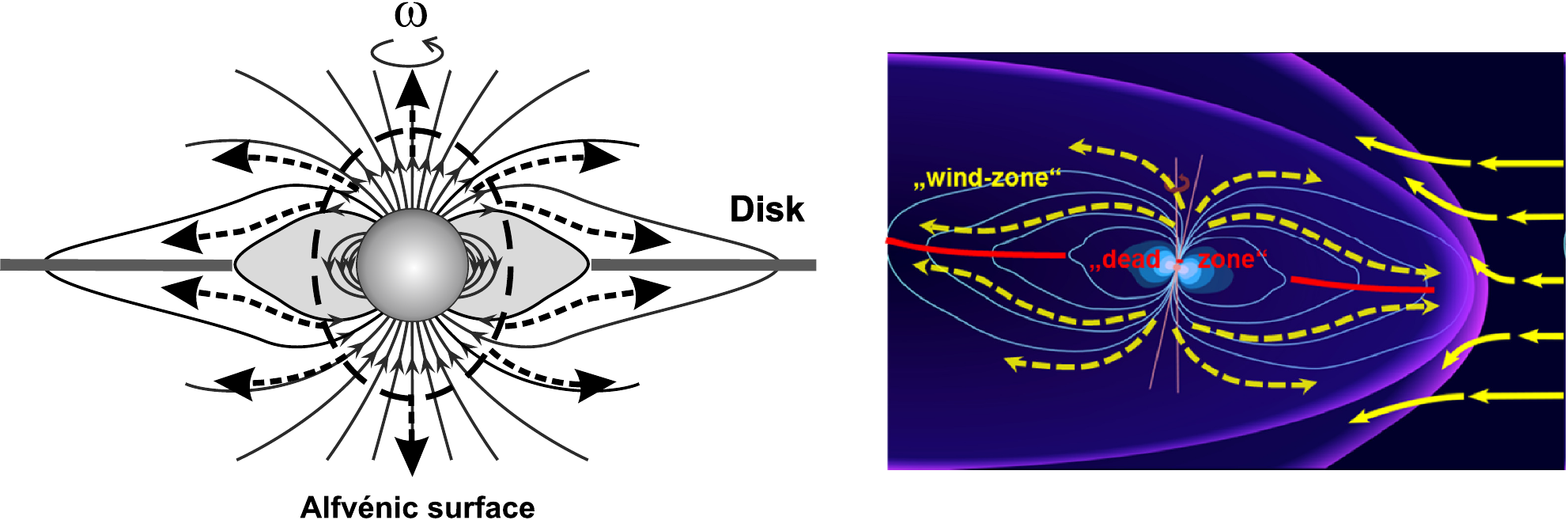}
 \caption{Schematic view of the magnetodisk formation mechanisms and structuring of HJ's magnetosphere (adapted from \citet{Khodachenko-etal-2012}).}
   \label{fig-disk}
\end{center}
\end{figure}

To explain an obvious survival and, therefore, sufficient magnetospheric protection of close-orbit giant exoplanets under the extreme conditions of their host stars \citet{Khodachenko-etal-2012} proposed a generic model of an exoplanetary magnetosphere. A key element in the proposed approach consists of the account of the expanding and escaping dynamical gas layer of the upper atmosphere of planet heated and ionized by the stellar XUV radiation \citep{Johansson-etal-2009, Koskinen-etal-2010, Koskinen-etal-2013, Garcia-Munoz-2007, Guo-2011, Guo-2013, Shaikhislamov-etal-2014}. The interaction of the outflowing atmospheric partially ionized plasma with the rotating planetary magnetic dipole field leads to the development of a current-carrying magnetodisk surrounding the exoplanet. In spite of the generality of the proposed modelling approach, so far it has been applied mainly to HJs -- the close-orbit giant exoplanets, as the material escape processes and related magnetic topology effects are is expected to be the most pronounced there. According to \citet{Khodachenko-etal-2012}, the magnetodisk of a HJ can be formed by two different mechanisms, acting simultaneously (see Figure~\ref{fig-disk}): 1) the outflow of the thermally escaping planetary plasma wind, heated and ionized by the stellar radiation, -- the so-called {\emph{thermally-driven}} magnetodisk formation mechanism and 2) the centrifugal acceleration of plasma by the rotating planetary magnetic field in the co-rotation region, with subsequent release of material beyond the centrifugal Alfv\'{e}nic cylinder, -- the so-called {\emph{rotation-driven}} or "sling" \citep{Alexeev-etal-2003, Alexeev-Belenkaya-2005} mechanism.
The inner edge of the magnetodisk is located at the surface where the energy density of plasma (including its kinetic and thermal energies) becomes equal to the energy density of the planetary magnetic field. In the case of a well established and dominating motion of the plasma, like that realized in the co-rotation regime under the conditions of a {\emph{rotation-driven}} magnetodisk formation \citep{Khodachenko-etal-2012}, this boundary surface, from which the magnetodisk builds up, is the Alfv\'{e}nic surface ($r=R_{\rm A}$) where the plasma ram pressure gets equal to the magnetic pressure, or the Alfv\'{e}n Mach number becomes $M_{\rm A}^2 =1$. In the case of a {\emph{thermally-driven}} magnetodisk formation, realized during the partially ionized planetary atmospheric material outflow accelerated by the thermal pressure gradient forces under the conditions of a slowly rotating tidally locked planet, the ruts of magnetodisk are located at the surface where the plasma thermal pressure is equal to the magnetic field energy density, i.e. in the region of plasma $\beta = 1$. Beyond the above mentioned boundary surfaces (depending on particular mechanism of the magnetodisk formation) the expanding plasma is not guided any more by the dipole magnetic field. It deforms the field lines leading to creation of a current-carrying magnetodisk (see Figure~\ref{fig-disk}), which in turn crucially changes the topology of the HJ's middle and outer magnetosphere.
Two major regions with different topology of the magnetic field \citep{Mestel-1968} can be distinguished in the magnetosphere of a HJ, driven by the escaping plasma flow. The first region corresponds to the inner magnetosphere, or so-called {\emph{dead zone}}, filled with closed dipole-type magnetic field lines. The magnetic force in the {\emph{dead zone}} is strong enough to lock plasma with the planet. In the second region, so-called {\emph{wind zone}}, the expanding plasma drags and opens the magnetic field lines, leading to the formation of a thin current-carrying magnetodisk (see Figure~\ref{fig-disk}). The field of the magnetodisk can exceed significantly the dipole field. Altogether, this leads to the development of a new type of a escaping-wind-driven, magnetodisk-dominated magnetosphere of an exoplanet, which has no analogs among the solar system planets \citep{Khodachenko-etal-2012}.

A simultaneous self-consistent description of both, the {\emph{rotation-driven}} and {\emph{thermally-driven}}, of the magnetodisk formation is a complex physical problem. So far, these processes are treated separately. It is especially possible in the case of close-orbit tidally locked exoplanets, which are subject to strong radiative energy deposition, whereas the rotational effects are usually much weaker, as the planetary rotation is synchronized with the orbital revolution \citep{Khodachenko-etal-2012, Trammell-etal-2011, Trammell-etal-2014, Owen-Adams-2014, Shaikhislamov-etal-2014, Khodachenko-etal-2015}. In this case, the radial expansion of hot planetary plasma dominates the co-rotation in the inner magnetosphere. Nowadays, besides a semi-analytical qualitative treatment of origin and interconnection of the inner (dipole dominated) and outer (magnetodisk-dominated) parts of the HJ's magnetosphere \citep{Khodachenko-etal-2012}, also laboratory experiments on simulation of the magnetodisk formation under the conditions of outflowing plasma in the background magnetic dipole field have been performed \citep{Antonov-etal-2013}.
\subsection{Simulation of Hot Jupiter's \emph{thermally-driven} magnetodisk}

Different modeling approaches and approximations for the simulation of expanding exoplanetary atmospheres and related thermal mass losses, proposed so far, have addressed various important factors, such as atmospheric composition and photochemistry, boundary conditions, the spectrum of ionizing radiation and distribution of primary energy input, radiative and adiabatic cooling processes, as well as tidal and centrifugal forces. A number of self-consistent 1D HD models have been developed \citep{Yelle-2004, Garcia-Munoz-2007, Koskinen-etal-2010, Koskinen-etal-2013, Shaikhislamov-etal-2014} that do not rely anymore on artificial simplifications and idealizations such as isothermal flow, and specific boundary conditions to initiate the wind. However, another crucial factor -- the intrinsic planetary magnetic field -- was either not considered at all, or included in a non-self-consistent way (e.g., by prescribing specific magnetic configurations). To include the magnetic field, one has to employ instead of the widely used 1D spherically symmetric non-magnetized HD model, at least an axisymmetric 2D magnetohydrodynamics (MHD) approach.

In the first attempts to incorporate planetary magnetic field, a specific topology of the HJ's inner magnetosphere with the predefined {\emph{wind}} and {\emph{dead zones}} was assumed, and the material escape of the planetary atmosphere was investigated in a semi-analytic way \citep{Trammell-etal-2011}. \citet{Adams-2011} considered outflows from close-in gas giants in the regime, where the flow is controlled by a static magnetic field. The isothermal Parker solution was constructed along the open magnetic field lines, which gives a faster supersonic transition \citep{Adams-2011} than in the spherical expansion cases. More self-consistent treatments based on 2D MHD codes has been recently performed by \citet{Owen-Adams-2014} and \citet{Trammell-etal-2014} in which {\emph{dead}} and {\emph{wind zones}} have been shown to form in the expanding planetary wind. However, the thermosphere heating and the MHD flow initiated close to the planetary surface were simulated with rather simplified models, assuming a mono-energetic XUV flux, homogenous (e.g., \citet{Trammell-etal-2014}) or empirically estimated gas temperature, and variable boundary conditions at the planet surface. As a result, the obtained estimations for the magnetic field, at which the escaping planetary wind of the particulary considered HD 209458b is significantly suppressed, vary in different papers by more than an order of magnitude from less than 0.3 G in \citet{Owen-Adams-2014} up to more than 3 G in \citet{Trammell-etal-2014}. At the same time, the major conclusion of these more advanced simulations stays the same -- the exoplanetary {\emph{thermal}} mass loss can be significantly suppressed by the presence of magnetic field, as compared to the quasi-spherical HD outflows, that often were used in the modeling of the planetary wind and related mass loss. Moreover, despite the recognition of importance of the {\emph{dead}} and {\emph{wind zones}} in the context of the planetary magnetosphere topology and related atmospheric mass loss, another important structure - the magnetodisk - which is closely associated with these regions and influences the global size of the magnetosphere \citep{Khodachenko-etal-2012}, has not been sufficiently modelled and investigated in these papers.

\subsubsection{MHD model of the planetary wind}

It should be emphasized that the processes of material escape and planetary magnetosphere formation have to be considered jointly in a self-consistent way in their mutual relation and influence. In that respect, the self-consistent modeling of HJ's planetary wind, driven by the stellar radiative heating in the presence of an intrinsic magnetic field, still remains an actual task. The most recent results on that issue are provided in \citet{Khodachenko-etal-2015}, where a self-consistent 2D MHD model of an expanding HJ's upper atmosphere in the presence of the planetary intrinsic magnetic dipole field is presented. For the sake of definiteness, typical physical parameters of a modeled HJ were taken as those of HD 209458b (i.e., $r_{\rm p} = 1.38 R_{\rm J}$ , $M_{\rm p} = 0.71 M_{\rm J}$ ) orbiting a Sun-like G-star ($M_{*} = 1.148 M_{\rm Sun}$, age $\sim 4$ Gyr) at a distance of 0.047 AU. A purely hydrogen atmosphere of a HJ composed in the general case of molecular and atomic hydrogen as well as corresponding ions and electrons was considered. The 2D MHD model by \citet{Khodachenko-etal-2015} generalizes the 1D HD model developed in \citet{Shaikhislamov-etal-2014} to the case of a magnetized planet. It includes in a self-consistent way (a) a realistic solar-type spectrum of the XUV radiation (to calculate correctly the intensity and column density distribution of radiative energy input); (b) basic hydrogen (photo)chemistry for the appropriate account of atmosphere composition and $H3+$ cooling; (c) stellar-planetary gravitational and rotational forces. The model operates also with such boundary conditions, that does not influence the numerical solution \citep{Shaikhislamov-etal-2014}. The basic equations of the model are those of the standard HD (continuity, momentum, and energy balance equations), supplemented by the equations for the magnetic field, hydrogen ionization, by XUV and electron collision, and recombination, as well as the equations for most important chemistry (altogether 16 reactions) and radiation diffusion in a partially ionized plasma \citep{Shaikhislamov-etal-2014, Khodachenko-etal-2015}. The model includes the energy- and temperature- dependent reactions between $H$, $H^+$, $H_2$, $H_2^+$, $H_3^+$, and $e$ components. The photo ionization rate is split between $H$ and $H_2$ species,correspondingly to their cross sections.

To simulate the expanding upper atmosphere of a HJ, the multi-fluid MHD model by \citet{Khodachenko-etal-2015} includes the following species: $H$, $H^+$, $H_2$, $H_2^+$, $H_3^+$. A pure molecular hydrogen atmosphere of an HD209458b analogue in a barometric equilibrium with the base (inner boundary) temperature of $1000$ K was taken as the initial state for the simulation of the expanding planetary wind driven by the stellar XUV radiative heating. The planetary wind plasma was treated as a quasi-neutral fluid in thermal equilibrium with $T_e = T_i$. As it was shown in \citet{Shaikhislamov-etal-2014}, \citet{Khodachenko-etal-2015}, as well as in \citep{Yelle-2004} and \citet{Koskinen-etal-2010, Koskinen-etal-2013}, the components  $H_2$, $H_2^+$, $H_3^+$ can exist around an HD 209458b analogue planet only at very low heights $< 0.1 r_{\rm p}$ in a relatively dense thermosphere. The reason for the inclusion of the hydrogen chemistry, besides the need to account for the IR cooling produced by $H_3^+$ molecule, is that a significant part of stellar XUV energy is absorbed at heights where molecular hydrogen dominates, as compared to the atomic one. In the upper layers of the expanding planetary wind the main processes, responsible for the transformation between neutral and ionized hydrogen particles of the planetary origin are photo-ionization, electron impact ionization and dielectronic recombination \citep{Shaikhislamov-etal-2014, Khodachenko-etal-2015}. The radiative heating term in the energy equation of the model \citep{Shaikhislamov-etal-2014, Khodachenko-etal-2015} is derived by integration of the stellar XUV spectrum. For the solar type host star of HD209458b, the spectrum of the Sun is used as a proxy.

Another kind of important interaction between the considered particle populations is resonant charge-exchange collisions. Indeed, charge-exchange has the cross-section of about $\sigma_{exc} = 6 \times 10^{-15}$cm$^2$ at low energies, which is an order of magnitude larger than the elastic collision cross-section. When planetary atoms and protons slip relative each other, because they have different thermal pressure profiles and protons feel electron pressure while atoms do not, the charge-exchange between them leads to velocity and temperature interchange. \citep{Shaikhislamov-etal-2014, Khodachenko-etal-2015} describe this process in the momentum equation for planetary protons with a collision rate $n_H \sigma_{exc} V$, where the interaction velocity $V$ depends in general on thermal and relative velocities of the interacting fluids (i.e., protons and neutral atoms). For the densities above $10^5$cm$^{-3}$ and other parameters typical for HD209458b, the charge-exchange between planetary atoms and protons ensures that they move practically together with a relatively small slippage and equal temperatures, constituting a hydrodynamic planetary wind in thermal equilibrium. The model also includes atomic-atomic and proton-atomic elastic collisions with the cross-sections of $\sim 10^{-16}$cm$^2$  which, however, are relatively unimportant. For the typical temperatures $\leq 10^4 $K realized in the upper atmosphere of a HJ  \citep{Koskinen-etal-2010, Koskinen-etal-2013, Shaikhislamov-etal-2014}, the proton-proton Coulomb collision cross section is well above $10^{-13}$cm$^2$. Therefore, as has been also shown in \citet{Garcia-Munoz-2007, Trammell-etal-2011, Koskinen-etal-2010, Koskinen-etal-2013}, and \citet{Guo-2011}, the atomic, charge-exchange, and Coulomb collisions in the most of the modeled regions of a HJ's upper atmosphere are efficient enough to ensure the "nonslippage" approximation for neutral hydrogen, ions, and electrons such that the assumptions of $\textbf{V}_{\rm n} = \textbf{V}_{\rm i} = \textbf{V}_{\rm e} = \textbf{V}$ as well as temperature equilibrium $T_{\rm n} = T_{\rm i} = T_{\rm e} = T$ hold true. Here the indexes $n$, $i$, and $e$ denote the neutral, ionized and electron components of the partially ionized atmospheric gas, respectively. The "nonslippage" assumption may, however, be wrong in the relatively narrow inner (close to the planet) regions of the expanding partially ionized planetary wind filled with a sufficiently strong magnetic field. In these regions the effects of relative motion of spices and related energy dissipation processes become important. This constitutes an as yet unstudied so far aspect of the HJ's magnetospheric physics which requires special detailed investigation. The momentum transfer between similar particles due to thermal motion in the fluid approach is described by viscosity. Its dimensionless value can be estimated as relation of mean-free path to the system size. For the planetary protons experiencing the Coulomb collisions at a temperature of $10^4$K and density of $\sim (10^6 - 10^5)$cm$^{-3}$ the dimensionless viscosity is about $\sim 10^{-5}$.

Because of the extremely large system size, the magnetic Reynolds number, $Re_m$, of the problem is everywhere exceedingly large, except for a region very close to the planet where the particle density sharply increases. Thus, the dynamics of the magnetic field in the most of the modelled HJ's magnetosphere should be dissipation-less. This is achieved in the numerical code by taking sufficiently high, though finite, conductivity, which corresponds to $Re_m \geq 10^5$. Close to the planet surface where electron-atom collisions dominate as compared to the electron-ion collisions, the corresponding expressions from \citet{braginski65} are used to calculate the electric conductivity.

The application of HD/MHD approaches in the planetary (HJ's) aeronomy can be justified only below the exobase, which corresponds to the region where the Knudsen number $Kn = \bar{l} / \Delta =1$, where $\bar{l}$ is the mean free path of particles and $\Delta$, a typical scale of flow. However, in the regions far from the planet, $\Delta$ may be taken to be of the order of the planetary radius $R_{\rm p}$, and the condition $Kn \sim 1$ is realized approximately in the layers with density $\sim 10^4$ cm$^{-3}$. That remains well above the height range ($\sim 10 R_{\rm p}$) considered in the modeling simulations. Close to the planet surface, the barometric height $H$, which is smaller than $R_{\rm p}$, has to be taken as the typical scale $\Delta$. Because of much larger densities there, the mean free path $\bar{l}$ also becomes smaller, and the condition $Kn < 1$ holds true, making the fluid approach fairly valid.

\subsubsection{Planetary wind propulsion force}

In the first studies of the planetary atmosphere expansion aimed at estimating the material \emph{thermal} escape rate and related mass loss, an isothermal Parker-type solution was imposed with an a priori prescribed boundary condition at the planetary inner atmosphere (or surface) related to the XUV heating of a fixed, predefined thin layer \citep{Watson-etal-1981}. However, it was shown in \citet{Tian-etal-2005} that in a thin-layer-heating approximation the expanding atmosphere mass loss rate strongly depends on the altitude of the layer where the radiation energy
is deposited. Moreover, as it was suggested already by Parker, the realistic solution that corresponds to an outflowing (expanding) planetary wind regime requires an additional volume heating in order to compensate for the adiabatic cooling of the expanding plasma envelope. This was also confirmed in the follow-up numerical modeling studies. For example, in the case of the solar wind, an additional acceleration factor is called for at a few solar radii. This additional driver is attributed to the dissipation of Alfv\'{e}n waves \citep{Usmanov-etal-2011}. For an outflowing atmosphere of a HJ, one of the possible candidates for the role of such a distributed accelerator is the absorption of XUV, which takes place everywhere atomic or molecular hydrogen is present. However, it turns out that at distances of about several planetary radii, where the additional forcing of the
expanding planetary atmospheric material is needed to keep it in the outflowing regime (i.e., to escape the gravity tension), the gas becomes highly ionized, and the absorption of XUV decreases. Nevertheless, the same fact of the increasing ionization degree, which reduces the efficiency of XUV heating, contributes to the additional acceleration of the wind and appears as an additional booster for the planetary plasma wind. That is because
the thermal pressure $p = (n_{\rm a} + n_{\rm i} + n_{\rm e}) kT$ increases (up to two times) at certain intervals of heights (in the ionization region) because of the contribution of the electron and ion pressure, whereas the density of material $\approx (n_{\rm a} + n_{\rm i}) m_{\rm i}$ remains practically unchanged. The created pressure gradient provides an additional driving force for the expansion of ionized atmospheric material. In the ionization region, the expanding planetary wind experiences an additional acceleration and reaches velocities that sustain the continuous outflow regime.

\begin{figure}[b]
\begin{center}
 \includegraphics[width=3.4in]{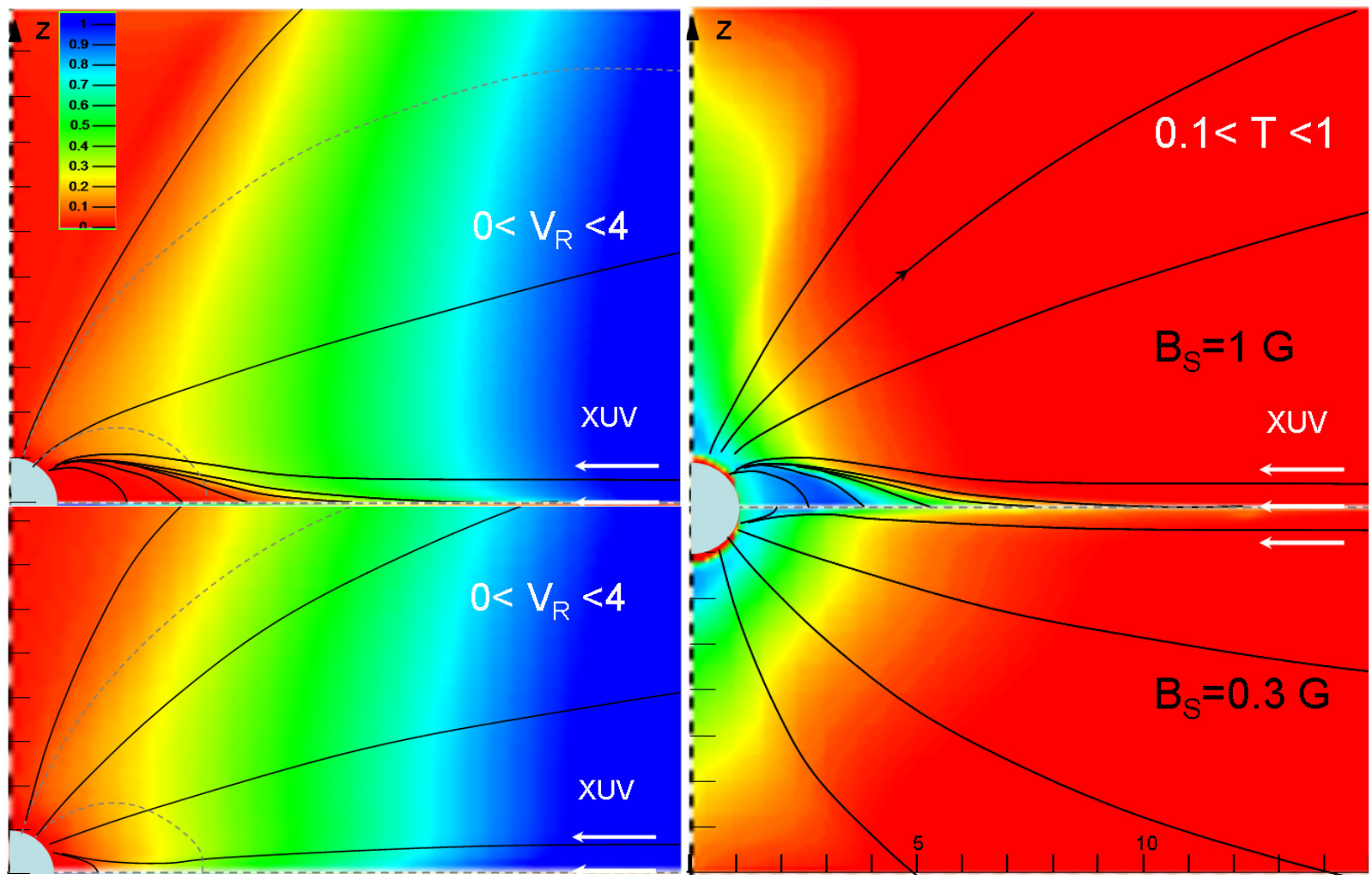}
 \caption{HJ's wind radial velocity, temperature and magnetic field structure simulated for a magnetized analog of HD209458b with equatorial surface field $B_{\rm s}=1$G (top panels) and $B_{\rm s}=0.3$G (bottom panels).  Variation ranges of the simulated physical quantities are indicated in the top of the corresponding panels. The color scale between the normalized minimal and maximal values of a simulated physical quantity is presented in the upper left panel. All values below or above the corresponding minimal and maximal values, are plotted as uniformly red or blue, respectively.
Magnetic field is shown with black solid lines (adapted from \citet{Khodachenko-etal-2015}).}
   \label{fig-num-solution}
\end{center}
\end{figure}

\subsubsection{Magnetic field and plasma flow structure in HJ's inner magnetosphere}

The self-consistent MHD model of the expanding magnetized HJ's wind developed in \citet{Khodachenko-etal-2015} for the first time enabled the detailed and realistic simulation of the whole complex of the physical parameters of the planetary inner magnetosphere, including also the formation of a {\emph{thermally-driven}} magnetodisk and the quantifying of the related magnetically controlled planetary thermal mass loss. To characterize the intrinsic planetary magnetic dipole field, its surface value at equatorial plane $B_{\rm s}$ is used. The major physical parameters and results of the simulations are scaled in units of the characteristic values of the problem, defined as follows: for temperature, $\tilde{T}_{\rm c} = 10^4$ K; for speed, $V_{\rm Ti}(T_{\rm c}) = \tilde{V}_{\rm c} = 9.1$ km $\cdot$ s $^{-1}$; for distance, the planet radius $R_{\rm p}$; for time, the typical flow time $\tilde{t}_{\rm c} = R_{\rm p} / \tilde{V}_{\rm c} \approx 3$ hr.

In line with qualitative expectations, the intrinsic planetary magnetic dipole field influences the dynamics of the escaping plasma and structures the planetary wind. In particular, in the case of a sufficiently small magnetic field, no {\emph{dead zone}} is formed. This happens if the gas thermal and ram pressures exceed everywhere the magnetic pressure, so that the latter cannot stop the developing flow. For the conditions of an HD 209458b analog
planet such a situation takes place at $B_{\rm s} < 0.1$ G. The resulting spatial structure of the magnetic field and planetary wind is close to that in the case of radially stretched field lines following the streamlines realized with a broadly distributed azimuthal current $J_{\varphi} \sim (\sin \theta) / R^2$ centered on the equatorial plane.

\begin{figure}[b]
\begin{center}
 \includegraphics[width=3.4in]{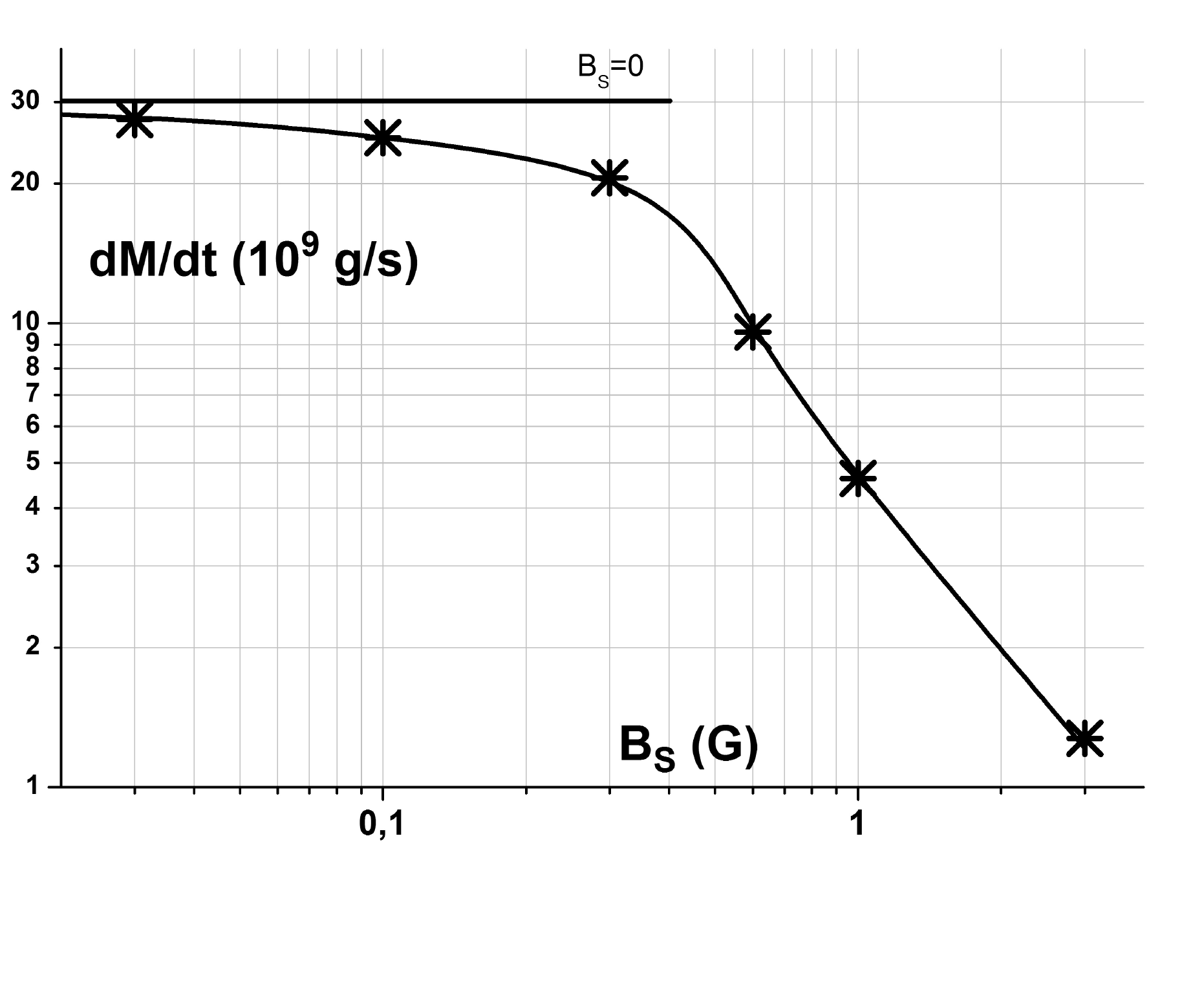}
 \caption{Time-averaged mass-loss rate of an HD 209458b analog planet as a function of the planetary equatorial surface magnetic field value $B_{\rm s}$. The horizontal line indicates the mass-loss rate without magnetic field (adapted from \citet{Khodachenko-etal-2015}).}
   \label{fig-massloss}
\end{center}
\end{figure}

Figure~\ref{fig-num-solution} demonstrates the solution obtained with $B_{\rm s}=0.3$G and $B_{\rm s}=1$G. In these cases the planetary wind is influenced by the magnetic field, and a {\emph{dead zone}} is formed. The effect of magnetic field is visible close to polar regions, where the flow is unable to fully stretch the dipole field lines near the planet and it follows along them. Due to the larger divergence of the material flow controlled by the magnetic dipole field and the corresponding faster adiabatic cooling, as well as the suppression of the zonal flow by the magnetic and tidal forces, the bulk velocity in polar regions is significantly diminished and does not reach supersonic values. This effect was predicted in \citet{Shaikhislamov-etal-2014}. The magnetic field lines passing near the {\emph{dead zone}} are stretched strongly in the vicinity of the equatorial plane and have the shape typical for the current disk. The regions of increased temperature above the poles and in the {\emph{dead zone}} where magnetic field and tidal force suppress the plasma outflow are clearly visible in the right panels of Figure~\ref{fig-num-solution}.

Formation of an extended and thin current sheet in the equatorial plane admits that the reconnection of reversed magnetic field lines might be triggered. According to \citet{Khodachenko-etal-2015}, the {\emph{thermally-driven}} magnetodisk shows a cyclic behavior, composed of consequent phases of the disk formation and magnetic flux accumulation followed by its short explosive relaxation during the magnetic reconnection with the ejection of a ring-type plasmoid.

\subsubsection{Magnetic control of HJ's mass loss}

The reconnection events in the magnetodisk strongly modulate in time the simulated instant mass-loss rate of a HJ measured through the outer boundary of the modelling box. With the increase of the surface magnetic field, the quasi-periodic mass-loss pulses become more pronounced in amplitude and more frequent in time \citep{Khodachenko-etal-2015}. The time-averaged dependence of the mass-loss rate on planetary surface magnetic field is presented in Figure~\ref{fig-massloss}.

As can be seen in Figure~\ref{fig-massloss}, a sharp decrease of the mass-loss rate takes place for the magnetic field values beyond $B_{\rm s} \sim 0.3$G, and at $B_{\rm s}=1$G the averaged mass-loss rate becomes about an order of magnitude smaller than that without the magnetic field. This effect may be explained by the increase of polar- and {\emph{dead zone}} areas of the suppressed material outflow with the increasing planetary magnetic field at the expense of the size of the wind region.

\begin{figure}[b]
\begin{center}
 \includegraphics[width=3.4in]{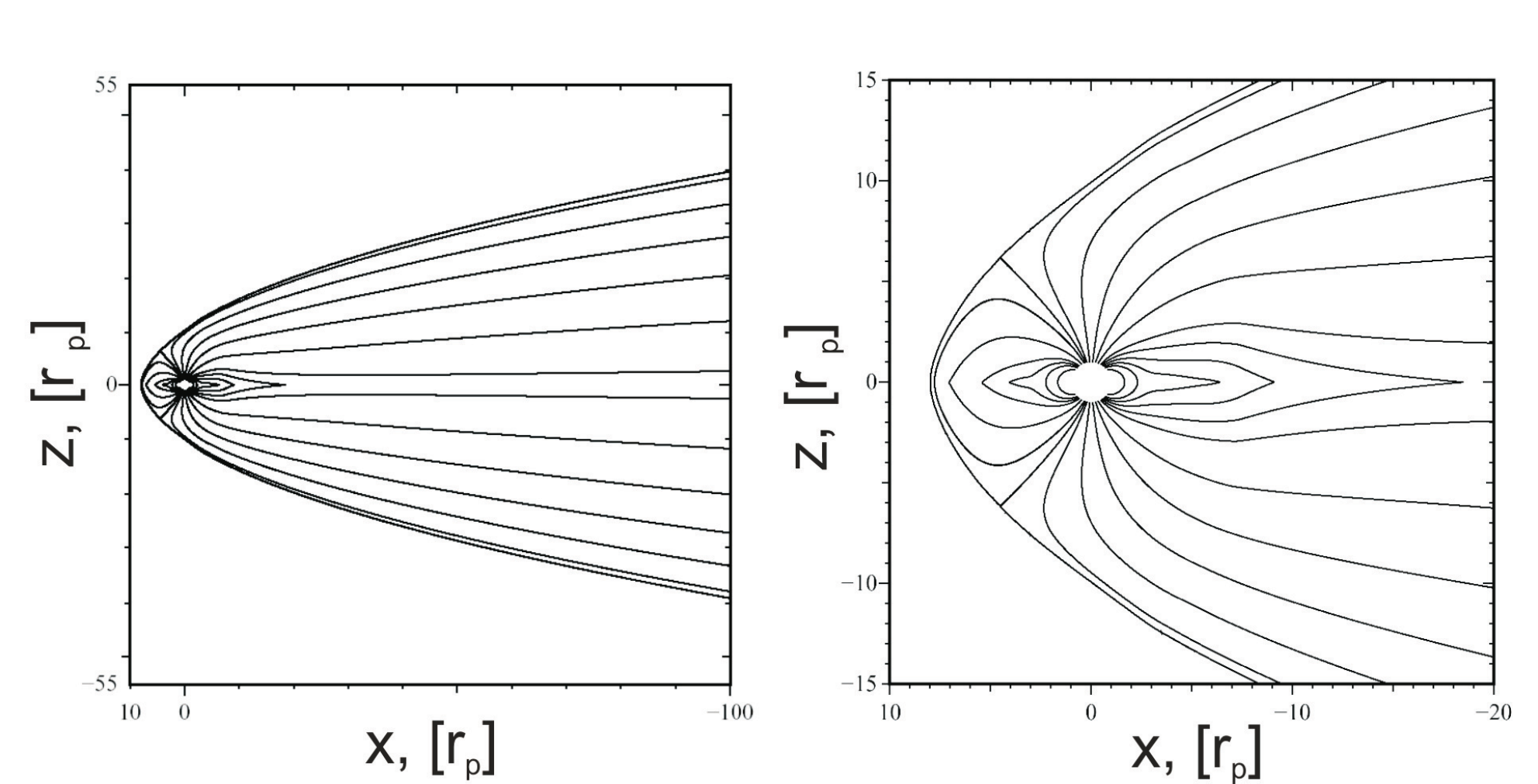}
 \caption{Typical view of a magnetodisk-dominated magnetosphere (adapted from \citet{Khodachenko-etal-2014}.)}
   \label{fig5}
\end{center}
\end{figure}

\subsection{Scaling of a magnetosphere with magnetodisk}

The global large-scale structure of a HJ's magnetosphere formed during the planet interaction with the stellar wind, may be investigated with the Paraboloid Magnetospheric Model (PMM), which enables judging on the shape and size of the planetary magnetosphere and its protective role \citep{Khodachenko-etal-2012}. PMM is a semi-analytical approach to the modeling of planetary magnetosphere structure \citep{Alexeev-etal-2003, Alexeev-Belenkaya-2005, Alexeev-et-al-2006, Khodachenko-etal-2012}. The name of the model is derived from its key simplifying assumption that the magnetopause of a planet may be represented by a paraboloid surface co-axial with the direction of the ambient stellar wind plasma. The PMM calculates the magnetic field generated by a variety of current systems located on the boundaries and within the boundaries of a planetary
magnetosphere. Besides the intrinsic planetary magnetic dipole and magnetopause currents, the PMM has, among the main sources of magnetic field,
also the electric current system of the magnetotail, and the induced ring currents of the magnetodisk. The model works without any restrictions
imposed on the values of interplanetary medium parameters, enabling therefore the description of the whole variety of possible magnetosphere
configurations caused by different intrinsic magnetic fields of exoplanets and various stellar wind conditions. The specifics of PMM consist of the substitution of the planetary magnetospheric plasma by the system of electric currents which reproduce the integral electrodynamic effects of plasma
As applied to HJs, PMM reveals that the electric currents induced in the plasma disk produce an essential effect on the overall magnetic field structure around the planet, resulting in the formation of a {\emph{magnetodisk-dominated}} magnetosphere of a Hot Jupiter. Due to the extension of the plasma disks around close-in exoplanets, the sizes of their magnetodisk-dominated magnetospheres are usually larger than those, followed from the traditional
estimates with the equation (\ref{eq-Rs}), based on taking into account only the screened planetary magnetic dipoles \citep{Griesseier-etal-2004, Khodachenko2007a}. In general, the role of the magnetodisk may be attributed to an expansion of a part of the dipole magnetic flux from the inner magnetosphere regions outwards and a resulting increase of the magnetosphere size. The magnetic field produced by the magnetodisk ring currents, dominates above the contribution of the intrinsic magnetic dipole of a HJ and finally determines the size and shape of the whole magnetosphere.

\citet{Khodachenko-etal-2012} provided an approximate formula for the estimation of the magnetopause stand-off distance taking into account the contribution of the \emph{rotation-driven} magnetodisk:
\begin{equation}
\frac{R_{\rm s}^{(dip+MD)}}{R_{\rm p}} \sim \frac{B_{d0J}^{1/2} (1 + \kappa^2)^{1/4}}{(2 \mu_{\rm 0} p_{\rm sw})^{1/4}} \left( \frac{R_{\rm AJ}}{R_{\rm p}} \right)^{-1/2} \times
\left( \frac{\omega_{\rm p}}{\omega_{\rm J}} \right)^{\frac{3 \alpha +1}{10}} \left( \frac{\mathrm{d} M_{\rm p}^{(th)}/\mathrm{d} t}{\mathrm{d} M_{\rm J} / \mathrm{d} t} \right)^{\frac{1}{10}} \:\:.
\label{eq-2}
\end{equation}
where $R_{\rm AJ}$, $\frac{\mathrm{d} M_{\rm J}}{\mathrm{d} t}$, and $B_{\rm d0J}$ are the known values corresponding to the Alfv\'{e}nic radius, mass load to the disk, and surface magnetic field for the solar system Jupiter. The parameter $\alpha$ appeared in the power index of the ratio of the planetary and Jovian rotation cyclic frequencies in (\ref{eq-2}) reflects the fact of proportionality of the planetary magnetic dipole moment $\mathcal{M}$ to certain power $\alpha$ of $\omega_{\rm p}$ (see equation (\ref{eq-1})).

The coefficient $\kappa \approx 2.44$ is an amplifying factor of the inner magnetospheric field at the magnetopause \citep{Alexeev-etal-2003}, which is required to take into account the contribution of the magnetopause electric currents (i.e. Chapman-Ferraro field) at the substellar point. It is connected with the form-factor $f_{\rm 0}$ from (\ref{eq-Rs}) as $\kappa = 2 f_{\rm 0}$. Therefore, according to (\ref{eq-2}), for a given ram pressure of stellar wind, $p_{\rm sw}$, the size of magnetosphere increases with the increasing planetary angular velocity $\omega_{\rm p}$ and$/$or thermal mass loss rate $\mathrm{d} M_{\rm p}^{(th)}/\mathrm{d} t$. In the case of a \emph{thermally-driven} magnetodisk a similar relatively simple analytic expression cannot be derived, because of the more complicated, as compared to "sling" mechanism \citep{Alexeev-etal-2003, Alexeev-Belenkaya-2005}, nature of the thermal planetary wind, which is described only by means of numerical simulations presented above. At the same time, the global effect, in the sense of the HJ's magnetosphere shaping, of the large-scale current system of the magnetodisk is basically the same for both, the {\emph{thermally-}} and {\emph{rotation-driven}} magnetodisks.

A slower, than the dipole-type decrease of magnetic field with distance comprises the essential specifics of the HJs' magnetodisk-dominated magnetospheres. According \citet{Khodachenko-etal-2012}, this results in their $40 - 70 \%$ larger scales, as compared to those traditionally estimated by taking into account only the planetary dipole. Such larger magnetospheres, extending well beyond the planetary exosphere height, provide better protection of close-in planets against the eroding action of extreme stellar winds \citep{Khodachenko2007a}.

Table~\ref{Table-magnetosph-scales} summarizes the values for the HJ's magnetopause stand-off distance $R_{\rm s}^{(dip+MD)}$ at different orbits around a full solar analogue star and gives for comparison the stand-off distance values $R_{\rm s}^{(dip)}$, obtained with equation (\ref{eq-Rs}), i.e. in the case when the contribution of magnetodisk is ignored (e.g., a pure dipole case). For the definiteness' sake, the planetary magnetic dynamo model by \citet{Stevenson-1983} with $\mathcal{M} \propto \omega_{\rm p}^{1/2}$ (i.e. $\alpha = 1/2$ by) has been taken for these calculations from the set (\ref{eq-1}).

\begin{table}
\caption{Hot Jupiter Alfv\'{e}nic radius, $R_{\rm A}$, and magnetopause stand-off distance for only a dipole controlled magnetosphere, $R_{\rm s}^{(dip)}$, and a magnetosphere with magnetodisk, $R_{\rm s}^{(dip+MD)}$, given by PMM. Full analog of the solar system Jupiter orbiting the Sun analog star at different orbits is considered. $^1$: Tidally locked. $^2$: Not tidally locked. $^3$: Jupiter}
\label{Table-magnetosph-scales}
\begin{center}
\begin{tabular}[t]{lccc}
\hline\noalign{\smallskip}
$d$ & $R_{s}^{(dip+MD)}$ & $R_{s}^{(dip)}$ & $R_{A}$ \\
               {[AU]} & {[$R_{\rm p}$]} & {[$R_{\rm p}$]} & {[$R_{\rm p}$]} \\
\noalign{\smallskip}\hline\noalign{\smallskip}
$0.045^1$    & $8.0$   & $5.76$  & $3.30$ \\
$0.1^1$      & $8.27$  & $6.16$  & $4.66$ \\
$0.3^2$      & $24.2$  & $15.0$  & $7.30$ \\
$5.2^3$      & $71.9$  & $41.8$  & $19.8$ \\
\noalign{\smallskip}\hline
\end{tabular}
\end{center}
\end{table}

A typical example of the magnetic field structure in the HJ's magnetosphere, obtained with PMM, is shown in Figure~\ref{fig5}.

\subsection{Conclusions}
\label{Conclusions}

To summarize this section we would like to repeat its major points and to outline the open-question areas which still require further investigation.
\\[0.3cm]
{\emph{\textbf{(1)}}} Stellar XUV radiation and stellar wind plasma flow strongly impact the environments of close-orbit exoplanets, such as terrestrial-type exoplanets in the HZs of low mass stars and HJs. In particular, the close location of a planet to its host star results in intense heating and ionization of the planetary upper atmosphere by the stellar X-ray and EUV radiation, which in their turn lead to the expansion of the ionized atmospheric material, contributing to the so-called planetary thermal mass loss. At higher altitudes, i.e. in the region of direct interaction of the expanding atmosphere with the stellar wind, the escaping planetary plasma is picked up by the stellar wind, resulting in a non-thermal mass loss. Both the thermal and non-thermal mass loss processes operating simultaneously contribute the total planetary mass loss and require their detailed description and quantifying. The planetary intrinsic magnetic field is known to be a crucial factor that influences the planetary mass loss and provides a protective effect for the upper atmospheric environment. It has two major aspects. First, the large-scale magnetic fields and electric currents, related to the planetary magnetism, form the planetary magnetosphere, which acts as a barrier for the upcoming stellar wind shielding the upper atmosphere against direct impact of the stellar plasmas. Second, the internal magnetic field of the magnetosphere also influences the streaming of the expanding planetary atmospheric plasma and therefore affects its escape.
\\[0.3cm]
{\emph{\textbf{(2)}}} The interaction of the outflowing partially ionized planetary plasma wind with the rotating planetary magnetic dipole field leads to
the development of a current-carrying magnetodisk. By this, two major regions with different topology of the magnetic field, so-called {\emph{dead-}} and {\emph{wind-zones}} are formed in the magnetosphere driven by the escaping plasma flow \citep{Mestel-1968, Adams-2011, Trammell-etal-2011, Trammell-etal-2014, Owen-Adams-2014, Khodachenko-etal-2015}. The field of the magnetodisk under typical conditions of a close-orbit HJ's,
by far exceeds the dipole field. Altogether, this leads to the development of a new type of magnetodisk-dominated magnetosphere \citep{Khodachenko-etal-2012}. Such expanded magnetospheres of HJs have been shown to be up to $40-70\%$ larger, as compared to the traditionally estimated dipole-type ones. That enabled the resolution of a problem of better magnetospheric protection of close-orbit HJs against the non-thermal erosive action of the stellar winds \citep{Khodachenko-etal-2012}. At the same time, the thermal mass loss due to the magnetically controlled escaping planetary wind has been shown to depend strongly on the size of the self-consistently formed {\emph{dead-zone}}, in which the significant portion of the planetary plasma stays locked and appears therefore excluded from the escaping material flow. Note, that the specifics and even existence of magnetodisks as well as the details of the structuring of the inner magnetospheres by lower mass exoplanets (e.g., Neptune- or Terrestrial type), under different stellar wind and radiation conditions, is still a subject for further study.
\\[0.3cm]
{\emph{\textbf{(3)}}} Investigation of key factors and physical mechanisms which determine the structure, topology and dynamics of exoplanetary magnetospheres at close orbits, form an important branch in the present day space physics. This topic is connected with the study of the whole complex of stellar-planetary interactions, as well as internal processes, including consideration of stellar radiation, plasma flows (e.g., stellar wind and coronal mass ejections), radiative energy deposition at the upper atmospheric layers, as well as planetary and stellar winds interaction. Special attention is paid to the development of a set of modelling tools and approaches for the description and simulation of exoplanetary magnetospheres and their basic elements, taking into account the specifics of exoplanetary conditions at close to host star orbits, such as the expanding upper atmospheric material (escaping planetary wind) heated and ionized by the stellar XUV, and formation in some cases under the conditions of planetary intrinsic magnetic dipole field, of an equatorial current-carrying plasma disk \citep{Khodachenko-etal-2012, Khodachenko-etal-2015}. While HD and MHD codes are fairly well developed and available, their use for simulation of expanding planetary plasma winds on exoplanets is a relatively novel area. A special challenge in that respect consists of the necessity to evaluate and then to include in the modeling the major underlying physics and most important factors that have been explored in recent years \citep{Yelle-2004, Erkaev-etal-2005, Erkaev-etal-2013, Tian-etal-2005, Garcia-Munoz-2007, Penz-etal-2008, Koskinen-etal-2010, Koskinen-etal-2013, Adams-2011, Guo-2011, Guo-2013, Trammell-etal-2011, Trammell-etal-2014, Owen-Adams-2014, Shaikhislamov-etal-2014, Khodachenko-etal-2015}. The development of a self-consistent model of an exoplanetary magnetosphere which properly includes all the basic physical effects and conditions still remains a challenging task.
\\[0.3cm]
{\emph{\textbf{(4)}}} An important aspect consists of the necessity to correctly take into account the effects of plasma partial ionization in the presence of a background magnetic field. So far, the dynamics of the expanding magnetized and partially ionized plasmosphere of an exoplanet has been considered in a simplified way, within an assumption of strong coupling, which ignores the relative motion of species in a restricted area of the inner planetary magnetosphere \citep{Trammell-etal-2011, Trammell-etal-2014, Owen-Adams-2014, Shaikhislamov-etal-2014, Khodachenko-etal-2015}. At the same time, it is well known that the different interactions of the moving electrons, ions and neutrals with the magnetic field and each other play a crucial role in the processes of electric current generation and energy dissipation. Because of the slippage between the ionized and neutral components, the magnetic field cannot be considered as being frozen into the material flow. The magnetic field causes essential anisotropy of the electric conductivity in the partially ionized plasmas. The increased Joule energy dissipation rate in the partially ionized magnetized plasmas is quantitatively described in terms of Cowling conductivity (\emph{Cowling} 1957). At the bottom of the radiatively heated upper atmosphere the material is weakly ionized. Then the ratio of neutral fraction to the ionized one decreases upwards, and at the heights of several $R_{\rm p}$ the planetary wind plasma becomes practically fully ionized \citep{Shaikhislamov-etal-2014, Khodachenko-etal-2015}. \citet{braginski65} gave the basic principles of transport processes in plasma including the effects of partial ionization. Neutral atoms change the plasma dynamics through collisions with charged particles, which may lead to different new phenomena in the plasma, for example increased Joule heating rates \citep{Khodachenko-Zaitsev-1992, Khodachenko-Zaitsev-2002, Khodachenko-1996, khodachenko04, forteza07} and the Farley-Buneman instability \citep{Khodachenko-Zaitsev-1992, fontenla2008}. The influence of a neutral flow on the electric current is significant when two conditions (so-called dynamo conditions) are satisfied. The first condition is that the electron cyclotron frequency, $\omega_{\rm e}$, should be much larger than electron collision frequency, $\nu_{\rm e}$. The second condition implies that $ 1<< (\omega_{\rm e} \omega_{\rm i})/(\nu_{\rm e} \nu_{\rm i})$, where $\omega_{\rm i}$, is the ion cyclotron frequency and $\nu_{\rm i}$ is the ion collision frequency \citep{Sen-White-1972, Khodachenko-Zaitsev-2002}. If these two conditions are satisfied then the flow of neutral particles across the ambient magnetic field may create electric currents, due to mechanisms similar to those of the Farley-Buneman instability, which will dissipate due to the ion-neutral collisions and convert part of the kinetic energy of the expanding material into the heat. That would finally influence the dynamics of the planetary wind.
\\[0.3cm]
{\emph{\textbf{(5)}}} The performed study also confirmed the importance of an appropriate account of the hydrogen chemistry at the upper atmospheric layer adjacent to the inner boundary of the simulation domain, where the most energetic photons are absorbed, as that affects the mass-loss rate (reduction by approximately a factor of two) and the whole numerical solution. That may be even more crucial for exoplanets at larger orbital distances than those typical for HJs, as the $H3^+$ cooling area tends to increase at lower atmospheric temperatures \citep{Chadney-etal-2015}. At the same time, the rates of the most important reactions included in our model were often extrapolated from the temperature range of 300 K at which they were measured to the higher-temperature conditions realized at HJs. Therefore, these, as well as the IR emission function of the $H3^+$ ion, are a subject for further investigation and revaluation with the development of more complex chemical models.

\begin{acknowledgements}
JLB and RS want to acknowledge financial support from MINECO AYA2014-54485-P and FEDER Funds, and from The Leverhulme Trust under grant IN-2014-016. RS acknowledges the ``Ministerio de Econom\'ia, Industria y Competitividad'' and the ``Conselleria d'Innovaci\'o, Recerca i Turisme del Govern Balear (Pla de ci\`encia, tecnologia, innovaci\'o i emprenedoria 2013-2017)'' for the ``Ram\'on y Cajal'' grant RYC-2014-14970. The results presented in this review by MK were obtained within the NFN project S116 ``Pathways to Habitability'' of the Austrian Science Foundation (FWF) and its related subprojects S11606-N16, S11607-N16. IA, IFS, and MK also acknowledge the support of the FWF projects I2939-N27, P25587-N27 and P25640-N27, Leverhulme Trust grant IN-2014-016, and the grants No.16-52-14006, No.14-29-06036 of the Russian Fund for Basic Research, as well as RAS presidium program N9 and SB RAS research program (project II.10.1.4, 01201374303). IIA was partially supported by Ministry of Education and Science of the Russian Federation grant 14.616.21.0084. TZ acknowledges support from FWF project  P26181-N27 and from the Georgian Shota Rustaveli National Science Foundation projects DI-2016-17 and 217146. Parallel computations crucial for the present study have been performed at the Supercomputing Center of the Lomonosov Moscow State University and at the SB RAS Siberian Super-Computing Center (SSCC) and Computation Center of Novosibirsk State University. EK and MCV are grateful for support by the Spanish Ministry of Science through the project AYA2014-55078-P and by the European Research Council in the frame of the FP7 Specific Program IDEAS through the Starting Grant ERC-2011-StG 277829-SPIA. All the authors want to thank ISSI for the support to the ISSI team on "Partially Ionized Plasmas in Astrophysics (PIPA)" and providing a collaborative environment for research and communication.
\end{acknowledgements}
%\bibliographystyle{aps-nameyear}      % American Physical Society (APS) style, author-year citations
%\bibliography{review}

\end{document}